\documentclass[twocolumn, amsmath,amssymb,
apsfloats]{revtex4}
\usepackage{dcolumn}
\usepackage{bm}
\usepackage[pdftex]{graphicx}
\usepackage{graphicx}
\usepackage{epstopdf}
\usepackage{float}
\usepackage{url}
\usepackage{hyperref}
\usepackage{xstring}
\usepackage{tabularx}
\usepackage{xcolor}

\def\KdS{Kerr--de~Sitter}

\def\be{\begin{equation}} 
\def\ee{\end{equation}}
\def\beq{\begin{equation}} 
\def\eeq{\end{equation}}
\def\bea{\begin{eqnarray}} 
\def\eea{\end{eqnarray}}
\def\din{\,\mathrm{d}} 
\def\dbe{\mathrm{d}}
\def\bet{\begin{tabular}} \def\ent{\end{tabular}}
 
\def\calk{{\cal K}}
\def\calr{{\cal R}}
\def\oder#1#2{\frac{\dbe#1}{\dbe#2}}
\def\dgr{^\circ}

\def\spo{\!S\!P\!O}

\begin{document}

\title{Light escape cones in the locally non-rotating reference frames of the \KdS\, superspinars and related superspinar shadows}

\author{Daniel Charbul\'{a}k}
\email	{daniel.charbulak@physics.slu.cz}
\author{Zden\v{e}k Stuchl\'{i}k}
\email	{zdenek.stuchlik@physics.slu.cz}

\address{Research Centre for Theoretical Physics and Astrophysics, 
	Institute of Physics, 
	Silesian University in Opava, 
	Bezru\v{c}ovo n\'{a}m. 13, CZ-746 01 Opava, Czech Republic}

\date{\today}

\begin{abstract}
The construction of local escape cones and their complementary cones, related to the locally non-rotating frames (LNRFs) orbiting superspinars with the external field described by the Kerr-de Sitter (KdS) naked singularity (NS) spacetimes, is described for all possible classes of the KdS NS spacetimes and all possible positions of the photon emitter. The notion of the local escape cones is then applied to construct the shadow of the KdS superspinars related to the distant static observers represented by the LNRFs located near the so-called static radius where the spacetime is close to the asymptotically flat region of the Kerr spacetimes. The shadow construction is focused on the KdS superpsinars limited by the cosmological and astrophysical restrictions on the spacetime parameters, namely the superspinar dimensionless spin and the dimensionless cosmological parameter relating the cosmological constant and the superspinar's mass. 

\end{abstract}

\maketitle

\section{Introduction}\label{intro}

Event Horizon Telescope (EHT) observations of the central regions of the galaxy M 87 \cite{Aki-etal:2019:ApJLa,Aki-etal:2019:ApJLb,Akiyama_2019:ApJLc,Akiyama_2019:ApJLd,Akiyama_2019:ApJLe,Akiyama:2019eap,Akiyama_2021:ApJLa,Akiyama_2021:ApJLb} and our Galaxy (source SgrA*) \cite{Aki-etal:2022:ApJLa,Aki-etal:2022:ApJLb,Aki-etal:2022:ApJLc,Aki-etal:2022:ApJLd,Aki-etal:2022:ApJLe,Aki-etal:2022:ApJLf} have opened an intensive study of the black hole (BH) shadow and the physics of processes close to the BH horizon. The observations are consistent with the expectation that the central object in both observed sources is a supermassive Kerr black hole. However, the Kerr geometry can also describe Kerr naked singularities (NSs), which contain a region of causality violation.  

The Penrose cosmic censorship hypothesis forbids existence of NSs, however, the hypothesis was not proved, and the validity of the weak cosmic censorship conjecture, which forbids the appearance of NSs clothing each one in an event horizon \cite{Pen:1969:NUOC2:}, is still an open question \cite{Shay-Dad:2023:JCAP:,Ong:2020xwv}. A variety of realistic naked singularity (NS) spacetimes have been found -- it has been shown that gravitational collapse can lead to NSs \cite{Dwi-Jos:1992:CQG:,Jos-Dwi:1993:PHYSR4:,Jos-Dad-Mar:2002:PHYSR4:,Jos-Gos-Dad:2004:PHYSR4:}. \footnote{The Big Bang itself represents a NS.} 
 
Therefore, the properties of Kerr NS spacetimes have been studied intensively in \cite{deFel:1974:ASTRA:,Stu:1980:BULAI:,Nob-Cal-deF:1980:NuoCim:}, while the optical phenomena related to the Kerr NS spacetimes were studied in \cite{Stu-Sche:2010:CQG:,Char-Stu:2017:EPJC:,Char-Stu:2018:EPJC:}, or for their generalizations in \cite{Kon-Stu-Zhi:2019:PRD:}.

The optical phenomena around Kerr-Newman BHs and NSs, along with the braneworld Kerr BHs and NSs, have been studied in \cite{Bal-Bic-Stu:1989:BAC:,Alam:2005pb,Sche-Stu:2009:IJMPD:,Sche-Stu:2009:GRG:,Stu-Kot-Tor:2008:CLAQG:,Stu-Cal:1991:GENRG2:,Bla-Stu:2016:PHYSR4:,Stu-Bla-Sche:2017:PHYSR4:,Sla-Stu:2020:EPJC:}, and for the regular BHs \cite{Bar:1968:ProcGR5:,Abd:2016:PHYSR4:,Fan-Wang:2016:PHYSR4:,Tos-etal:2018:PHYSR4:} constructed in the framework of GR in combination with various variants of nonlinear electrodynamics, these phenomena have been treated, e.g. in \cite{Abd-Tos-Sche-Stu-Ahm:2017:IJMPD:,Stu-Sche:2019:EPJC:,Stu-Sche-Ovch:2019:ApJ:}.

A modification of the Kerr NS spacetimes that does not exhibit the region of causality violations (time machine, see \cite{Calvani:1978xj}), located in the vicinity of the ring singularity, has been proposed, being based on the stringy solution corresponding to rapidly rotating compact objects, so-called Kerr superspinars, where the interior is assumed to be governed by the String theory, and the exterior is described by the standard Kerr NS geometry \cite{Gim-Hor:2009:PhysLetB:}. The pathological time machine region is then covered by correctly behaving "stringy region", which is expected to have a boundary at radius $\calr>0$. The Kerr superspinars can be considered as primordial, being remnants of the initial stringy phase of the evolution of the Universe \cite{Gim-Hor:2009:PhysLetB:,Boy-etal:2003:PHYSR4:}. Their properties have been extensively studied, yielding some extraordinary results \cite{Stu-Hle-Tru:2011:CLAQG:,Stu-Sche:2012:CQG:,Stu-Sche:2013:CLAQG:,Nak-etal:2018:PhysLetB:}. 

The optical phenomena associated with the Kerr superspinars have been studied, e.g., in \cite{Stu-Sche:2010:CQG:,Sche-Stu:2013:JCAP:} - it was shown that the superspinar shadow depends on the edge at $r=\calr$, where the external Kerr spacetime and the internal spacetime join each other, and is significantly different from the BH shadow. 

It is worth studying in detail the construction of the shadow of both the complete Kerr NS spacetimes and the Kerr superspinars external geometry, and in their extensions, to test their observational relevance, since we can expect that in the near future, due to the rapid development of the observational and data analysis techniques, we will be able to distinguish very subtle details of the compact objects, thus obtaining important information on various physical phenomena. 
 
Because of the cosmological observations related to both the CMB measurements and supernova tests that indicate the existence of dark energy with properties very close to the vacuum energy represented by the cosmological constant \cite{Dolgov-Zeldovich-Sazhin:1988:,Kam-etal:2010:}, it is important to clarify the role of the cosmological constant $\Lambda$ in astrophysical phenomena related to Kerr superspinars. The role of $\Lambda$ in accretion processes in active galactic nuclei has been discussed in \cite{Stu-etal:2020:Univ:,Stu-Sla-Kov:2009:CLAQG:PseNewSdS,Stu:2005:MODPLA:,Sla-Stu:2005:CLAQG:}, the motion of the gravitationally bound galaxies in \cite{Stu-Sche:2011:JCAP:}. The optical phenomena and the BH shadow have been studied for \KdS\ BHs in \cite{Stu-Char-Sche:2018:EPJC:}, namely for locally non-rotating reference frames (LNRFs), frames related to circular geodesic orbits, and for radially freely falling/escaping (geodesic) frames. For an alternative KdS BH solution and its shadow see \cite{Ovalle_2021,Omw-etal:2022:EPJC:}. Furthermore, the influence of the observer's state of motion on the BH shadow, related to that of the Sgr A* BH, is described in \cite{chang2021influence}. The shadows of BHs embedded in the expanding de~Sitter universe have been studied in \cite{Per-Bis-Tsu:2018:PHYSR4:,Cot_escu_2021,Roy:2020dyy}, and for charged and rotating BHs in \cite{Belhaj:2020kwv}, while the connection with the Friedman-Lemaitre-Robertson-Walker cosmological metric has been considered in \cite{Li_2020}. For \KdS\ NSs the shadow was treated, with inclusion of the region of negative radii, in \cite{Stu-Char:2024:PHYSR4:}, considering LNRFs, and radially freely falling/escaping (geodesic) frames.

In the present paper we focus attention on the shadow of the KdS superspinars, whose surface at $r=const=\calr > 0$ functions similarly to the BH horizon, thus excluding the possibility of reaching the causality violation region. We discuss in detail the construction of the light escape cones (LEC) related to the LNRFs, which are important from an astrophysical point of view \cite{Stu-etal:2020:Univ:}. The LECs are then applied in the construction of the KdS superspinar's shadow. 

Due to the variability and complexity of the KdS NS spacetimes, which represent the exterior of the KdS superspinars, 
we restrict our study of the shadows to the class of KdS spacetimes whose parameters are allowed by the observational constraints. We use the relic cosmological constant $\Lambda \sim 10^{-52}m^{-2}$ implied by cosmic observations \cite{Kra-Tur:1995:GENRG2:,Kra:1998:ASTRJ2:}, and the constraints on the mass and spin of the compact objects imposed by observations of active galactic nuclei in large galaxies or even in the galactic clusters \cite{Stu-Sla-Kov:2009:CLAQG:PseNewSdS,Stu-Hle-Nov:2016:}. This class of the KdS NS spacetimes allows the existence of static radii \cite{Stu:1983:BULAI:,Stu-Hle:1999:PHYSR4:,Stu-Char:2024:PHYSR4:} -- the spacetime at (and around) this radius represents in the KdS spacetimes the best local approximation to the asymptotically flat region of the Kerr spacetimes \cite{Stu-Hle:1999:PHYSR4:,Stu-Sla:2004:PHYSR4:}. Furthermore, the static radius also represents an upper limit on the extension of gravitationally bound systems (e.g., the dark matter halos) in spacetimes with the relic cosmological constant \cite{Stu-Hle-Nov:2016:,Stu-etal:2017:JCAP:}. At the static radius, the LNRF observers approximate the geodetic stationary observers at an unstable equilibrium point where the gravitational attraction of the BH or superspinar is just balanced by the cosmic repulsion.

The concept of LECs (and complementary light capture cones (LCC)) enables construction of the KdS superspinar shadow (silhouette) in the same way as in the case of the BH or NS shadows \cite{Stu-Sche:2010:CQG:,Stu-Bla-Sche:2017:PHYSR4:,Stu-Char-Sche:2018:EPJC:,Stu-Char:2024:PHYSR4:}, with respect to any observer, including those orbiting very close to the central object. Here, we restrict our attention to the standard case of a shadow (silhouette) cast by a KdS superspinar located between a very distant observer and an illuminating source much larger than the compact object \cite{Bar:1973:BlaHol:}. To construct the LEC and the shadow, we assume the light source at the position of the observer and determine the motion of the photons emitted in different directions. The results obtained for the construction of the LEC are then inverted to obtain the shadow (for details see \cite{Stu-Char:2024:PHYSR4:}). A special class of null geodesics is considered in the analysis, namely null geodesics corresponding to spherical photon orbits, which are now considered only at the radii corresponding to the external spacetime of the superspinars. The construction of the LEC and the complementary LCC for the superspinars is compared with the construction in the NS case.

\section{\KdS\, superspinar (NS) geometry} \label{sec2}

We assume that, similarly to pure Kerr case, the exterior of the \KdS\, superspinar spacetimes are determined by the standard \KdS\ naked singularity parameters, namely, the superspinar mass $M$, its spin $J$, which is replaced in a standard way by the dimensionless rotational parameter $a=J/M^2>1$, and the cosmological constant $\Lambda$, which is replaced by the dimensionless cosmological parameter $y = \frac{1}{3}\Lambda  M^2$.  Another parameter $\calr$ is important for the description of the \KdS\ superspinar, as it determines its boundary, separating its interior, which is assumed to be described by the string theory, from the exterior, where the stringy effects are assumed to be irrelevant and the spacetime is described by the standard \KdS\ geometry. The value of the boundary parameter $\calr$ can be set arbitrarily, at $\calr=r=const > 0$; usually it is assumed at $\calr=0$ or slightly above.

\par 
As usual, we use the Boyer-Lindquist spheroidal coordinates $t,r,\theta,\phi$ with geometric system of units ($c = G = 1$). We also introduce the dimensionless line element and the coordinates $s/M \rightarrow s$, $t/M \rightarrow t, r/M \rightarrow r$. This is equivalent to putting $M=1$. The line element then reads  
\bea
\dbe s^2 &=&
\frac{a^2\sin^2\theta \Delta_\theta-\Delta_r}{I^2 \rho^2}\dbe t^2 +\frac{\rho^2}{\Delta_r}\din r^2 + \frac{\rho^2}{\Delta_\theta}\din \theta^2  \\
&+&\frac{\sin^2\theta}{I^2 \rho^2}A\din \phi^2+\frac{2a\sin^2\theta}{I^2 \rho^2}\left[\Delta_r-\Delta_\theta\left(r^2 +   a^2\right) \right]\din t \din \phi ,\nonumber \label{ds2}
\eea
where
\bea
\Delta_r&=&\left(1 - y r^2\right)\left(r^2 + a^2\right) - 2r,\\
\Delta_\theta&=&1 + a^2 y \cos^2\theta,\\
I&=&1 + a^2 y,\\
\rho^2&=&r^2 + a^2\cos^2\theta. 
\eea
and
\be A=(r^2+a^2)^2\Delta_\theta-a^2\Delta_r\sin^2\theta. \label{A} \ee 

In some special cases, such as the situation in vicinity of the spacetime ring singularity, it is convenient to construct the figures in the so-called Kerr-Schild coordinates, which are related to the Boyer-Lindquist coordinates by the relations 
\bea
x^2 + y^2 &=& (r^2 + a^2)\sin^2\theta,\nonumber \\
z^2 &=& r^2\cos^2\theta. \label{KerrSchild}
\eea
Note that in the relations (\ref{KerrSchild}) $y$ denotes a Cartesian coordinate, not a cosmological parameter.

\subsection{Event horizons}

The roots of the quartic equation
\be
\Delta_r=0 \label{event_hor},
\ee
which can be rewritten in the form
\be
y=y_{h}(r; a^2)\equiv \frac{r^2-2r+a^2}{r^2(r^2+a^2)}, \label{yh(r,a2)} 
\ee
determine the coordinate pseudosingularities of the KdS geometry. Following the notation used in our previous paper \cite{Stu-Char:2024:PHYSR4:}, we denote them as $r^{-}_{c}, r_{i}, r_{o}, r_{c}$, where $r^{-}_{c}<0<r_{i}<r_{o}<r_{c}$, corresponding to the cosmological horizon in the negative radius region, the inner BH (Cauchy) horizon, the outer BH horizon and the cosmological horizon. The number and position of the horizons for a given parameter $y_{0}$ can be illustrated by the intersection of the horizontal $y_{0}$-line with the graph of the function $y_{h}(r; a^2)$, whose behaviour is shown  in Fig. \ref{fig_yh(r,a2)}a for different values of spin parameter $a^2$. It can be observed that the cosmological horizon $r^{-}_{c}$ in region of negative radii exists for any positive values of $y>0$ and $a^2>0$. However, we have to stress that in our study in this paper we consider only the KdS NS spacetimes excluding the region of $r<0$, since it is assumed to be irrelevant because of the assumption $R \geq 0$. Thus, it is clear that only the cosmological horizon in the region of $r>0$ is relevant  in our study. The number and nature of horizons in the $r>0$ region, which we discuss separately below, depends on the value of $y_{0}$ compared to the extremes $y_{min(h)}(a_{0}^2)$, $y_{max(h)}(a_{0}^2)$, for a given spin parameter $a_{0}^2>0$. Here and in the following enumeration the functions $y_{min/max(h)}(a^2)$ describe the minima/maxima of the function $y_{h}(r;a^2)$.

\begin{enumerate}
	\item Case $0\leq a_{0}^2\leq 1$
	\begin{description}
		\item[--] The function $y_{h}(r;a_{0}^2)$ has a positive maximum $y_{max(h)}(a_{0}^2)>0$ and non-positive minimum $y_{min(h)}(a_{0}^2)\leq 0$ (see the grey, red and blue curves in Fig. \ref{fig_yh(r,a2)}a); for $a_{0}^2=0$ the function $y_{h}(r;a^2_{0})$ has no minimum, since $y_{h}(r;0) \to -\infty$ as $r \to 0^+$, while $y_{max(h)}(a_{0}^2)=y_{crit(SdS)}\equiv 1/27 \doteq 0.03704$ (see the grey curve in Fig. \ref{fig_yh(r,a2)}a); $y_{crit(SdS)}$ denotes the critical value of the cosmological parameter that allows the Schwarzschild de-Sitter BH spacetimes to exist. For $a_{0}^2=1$, $y_{min(h)}(a_{0}^2)=0$, (see the blue curve in Fig. \ref{fig_yh(r,a2)}a).
	\end{description}
	\begin{itemize}
		\item $0< y_{0} < y_{max(h)}(a^2_{0})$
		\begin{description}
			\item[--] The geometry describes a BH spacetime; the radii  $0<r_{i}<r_{o}<r_{c}$, correspond respectively to the inner BH horizon, the outer BH horizon, and the cosmological horizon; the spacetime is dynamic in regions $r_{i}<r<r_{o}$.
		\end{description}
		\item $y_{0} = y_{max(h)}(a^2_{0})$
		\begin{description}
			\item[--]  Marginal NS spacetimes, where the outer BH and the cosmological horizons merge at $r_{o}=r_{c}$.
		\end{description}
	\end{itemize}
	
	\begin{itemize}
		\item $y_{max(h)}(a^2_{0})<y_{0}$
		\begin{description}
			\item[--] The geometry describes the NS spacetimes with one cosmological horizon $0<r_{c}$ in region of positive radii. 
		\end{description}
	\end{itemize}
	
	\item Case $1<a^2_{0}\leq a^2_{crit}\equiv 3(3+2\sqrt{3})/16 \doteq 1.21202$
	\begin{description}
		\item[--] Both local extrema  $y_{min(h)}(a^2_{0})$ and $y_{max(h)}(a^2_{0})$ of the function $y_{h}(r;a^2_{0})$ are  positive, $0<y_{min(h)}(a^2_{0})<y_{max(h)}(a^2_{0})\leq y_{crit(KdS)}\equiv 16(26\sqrt{3}-45)/9\doteq 0.05924$ (see the green curve in Fig. \ref{fig_yh(r,a2)}a) -- these extrema merge for $a^2_{0}=a^2_{crit}$ at the inflection point in $y=y_{crit(KdS)}$ (see the orange curve in Fig. \ref{fig_yh(r,a2)}a);  $y_{crit(KdS)}$ and $a^2_{crit}$ have the meaning of the critical values of the cosmological and spin parameters, respectively, that still allow the existence of \KdS\ BH spacetimes. 
	\end{description}
	\begin{itemize}
		
		\item $y_{0}<y_{min(h)}(a^2_{0})$ or $y_{max(h)}(a^2_{0})<y_{0}$
		\begin{description}
			\item[--] The geometry describes a NS spacetime with properties analogous to the previous NS case.
		\end{description}
	
		\item $y_{min(h)}(a^2_{0})<y_{0}<y_{max(h)}(a^2_{0})$
		\begin{description}
			\item[--] Geometry describes the BH spacetimes with properties analogous to the previous BH case. 
		\end{description}
	
		\item $y_{0}=y_{min(h)}(a^2_{0})$
		\begin{description}
		\item[--] Limit case of the extreme BH spacetimes with a merging inner and outer BH horizon. 
		\end{description}

		\item $y_{0}=y_{max(h)}(a^2_{0})$
		\begin{description}
			\item[--] Limit case of the marginal NS spacetimes with properties analogous to the previous marginal NS case. 
		\end{description}
	
		\item $a^2_{0}=a^2_{crit}$, $y_{0}=y_{crit(KdS)}$
		\begin{description}
			\item[--] Extreme case of geometry that describes the ultra-extreme NS spacetime with $r_{i}=r_{0}=r_{c}$. 
		\end{description}
	
	\end{itemize}
	
	\item Case $a^2_{0}>a^2_{crit(KdS)}$
	\begin{description}
		\item[--] There are no local extrema of the function $y_{h}(r;a^2_{0})$ (see the black curve in Fig. \ref{fig_yh(r,a2)}a), and for any value of $y_{0}$ the geometry corresponds to the NS spacetimes with properties analogous to the previous NS cases. 
	\end{description}
\end{enumerate}

\begin{figure}[htbp]
	\centering
	\begin{tabular}{c}
		\includegraphics[width=\linewidth]{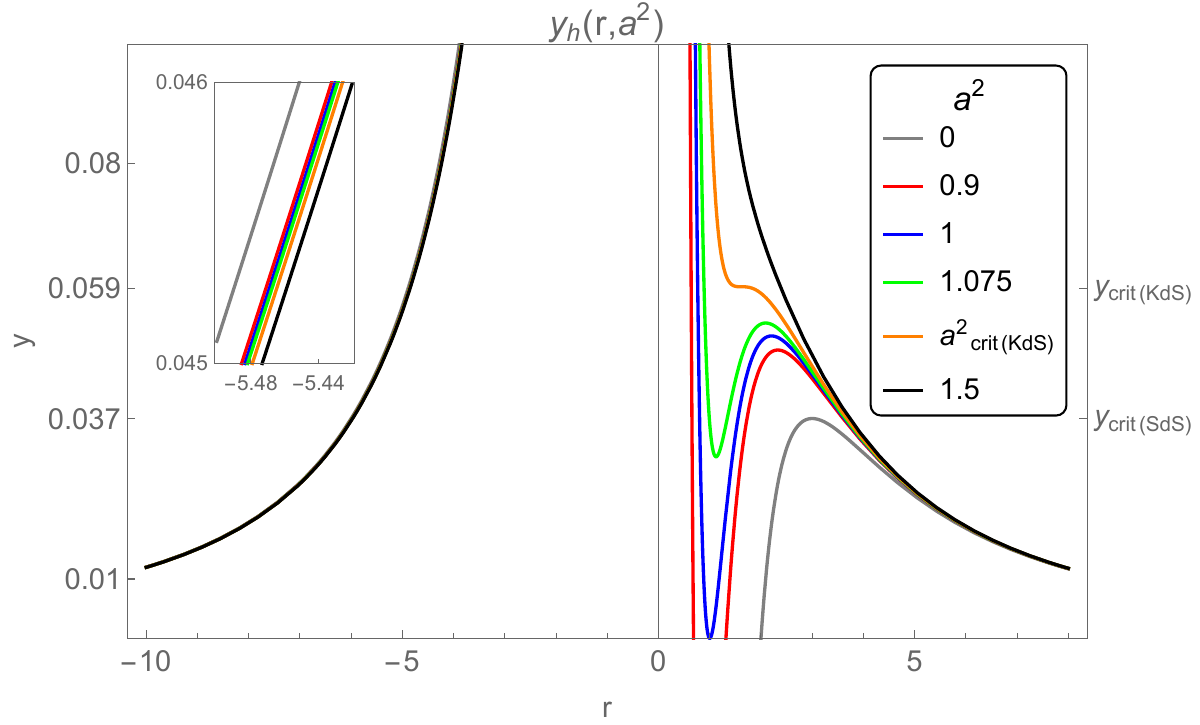}\\
		(a)\\
		\includegraphics[width=\linewidth]{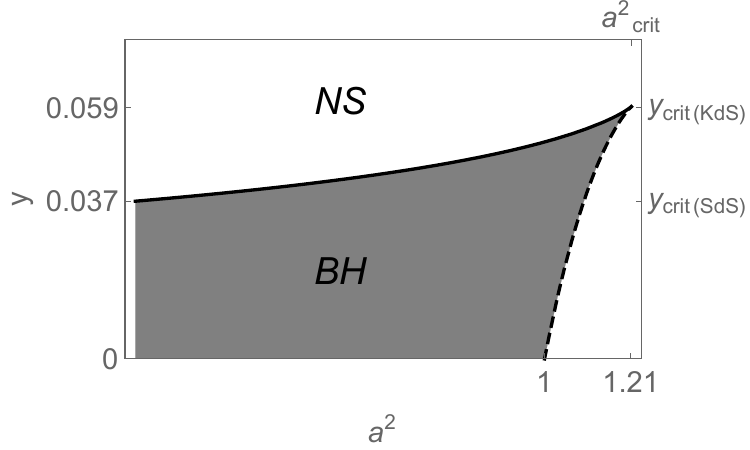}\\
		(b)		
	\end{tabular}
	\caption{(a) Behaviour of the function $y_{h}(r;a^2)$ for typical values of the spin parameter $a^2$. For a given cosmological parameter $y_{0}$, the intersections of the horizontal line $y=y_{0}$ with the curves $y=y_{h}(r;a^2)$ give the number and locations of the event horizons of the corresponding spacetime. (b) Behaviour of the functions $y_{max(h)}(a^2)$ (full black curve) and $y_{min(h)}(a^2)$ (dashed black curve) and the induced partitioning of the parameter plane $(a^2-y)$ into regions corresponding to the BH spacetimes (grey region) and NS spacetimes (white region).}\label{fig_yh(r,a2)}
\end{figure}

The behaviour of the functions $y_{min/max(h)}(a^2)$ is shown in Fig. \ref{fig_yh(r,a2)}b. They  can be expressed as follows: 
\be
y_{min(h)}(a^2)\equiv \frac{1}{12}[X-\frac{\sqrt{6}}{a^6}\sqrt{Y+Z}+\sqrt{3}z], \label{yminh(a2)}
\ee
and
\be
y_{max(h)}(a^2) \equiv \left\{
\begin{array}{l}
	y^{-}_{max(h)}(a^2),\quad \mbox{for}\quad 0 \leq a^2<a^2_{\mp (h)};\\
	\\ 
	y^{+}_{max(h)}(a^2),\quad \mbox{for}\quad a^2_{\mp (h)} \leq a^2<a^2_{crit}\\
\end{array}\right.
\label{ymaxh(a2)}
\ee
where $a^2_{\mp (h)}\equiv \frac{1}{48}(3+2\sqrt{3})=0.135$, $a^2_{crit}\equiv \frac{3}{16}(3+2\sqrt{3})=1.212$ is the maximum spin parameter allowing the existence of the BH spacetimes,
\be
y^{\pm}_{max(h)}(a^2)\equiv \frac{1}{12}[X+\frac{\sqrt{6}}{a^6}\sqrt{Y\pm Z}\pm \sqrt{3}z], \label{y^pm_maxh(a2)}
\ee
\bea
X&=&-\frac{3(1+4a^2)}{a^4},\nonumber \\
Y&=&3a^4(1-80a^2)-2\cdot 3^{1/3}u^{1/3}a^2-\frac{2\cdot 3^{2/3}p_{+}a^8}{u^{1/3}}, \nonumber \\
Z&=&\frac{3\sqrt{3}(768a^4-96a^2-1)}{z}, \label{XYZ}
\eea
\be
z=\sqrt{\frac{u^{1/3}(4\cdot 3^{1/3}u^{1/3}+3a^2-240a^4)+4\cdot 3^{2/3}p_{+}a^6}{u^{1/3}a^{10}}}, \label{z}
\ee
\be
u=1944a^{10}+18432 a^{14}+\sqrt{3}\sqrt{-p_{-}^3a^{18}}, \label{u}
\ee
and
\be
p_{\pm}=256a^4\pm288a^2-27. \label{ppm}
\ee

The exterior of KdS superspinar spacetimes is thus described by the KdS NS geometry. In the $(y-a^2)$ plane, the parameters $y, a^2$ of such spacetimes correspond to the region shown in white in Fig. \ref{fig_yh(r,a2)}b.

Since our intention here is to deal with astrophysically relevant objects, we admit that the mass parameter $M$ of the central object can be as large as $M\sim 10^{11} M _{\odot}$, as suggested by the case of the most massive BH observed so far, with mass of $M=7.10^{10}  M _{\odot}$ in the source TON 618. Here $M _{\odot}=1.5\cdot10^{3}m$ is the mass of the Sun in geometric units. Considering the current cosmological observations, which imply a very small value for the cosmological constant, $\Lambda \sim 1.1\cdot10^{-52} m^{-2}$, we find that the realistic estimates of the cosmological parameter parameter $y$ are at most $y \sim 10^{-34}$. However, for the sake of illustration, we consider here values of the cosmological parameter $y$, which are many orders of magnitude higher than the realistic ones, but which still correspond to Class IV KdS spacetimes, according to the classification introduced in \cite{Char-Stu:2017:EPJC:}, which are the main object of our investigation and  which give results qualitatively corresponding to the realistic values of $y$.

 The parameters of the KdS spacetimes of Class IV correspond to the case $y<y_{min(h)}(a^2)$ and we select parameters giving clear demonstration of the relevant optical phenomena.

\subsection{Superspinar boundary surface} \label{ssec_superspinar_surface}

We assume, in accordance with other related papers (\cite{Sche-Stu:2013:JCAP:, Stu-Hle-Tru:2011:CLAQG:, Stu-Sche:2010:CQG:, Stu-Sche:2012:CQG:}), that the boundary surface is formed by a sphere of constant radius $r(\theta)=\calr$ with properties similar to the BH event horizon. Although considerations that do not rule out possibility of non-zero emissivity or reflectivity can be found in the literature \cite{Car-Pan-Cad-Cav:2008:CLAQG:}, in the present paper we regard the superspinar surface acting as a one-way membrane, i.e., it absorbs all incoming radiation and matter, and does not radiate itself. Therefore, we demand that photons with radial turning point at $r=r_{t}$ can be detected by a distant observer if $r_{t}>\calr$. We assume that the stringy effects are appreciable in the region where the curvature of the spacetime is comparable to the Planck scale, hence we fix the superspinar surface $\calr \sim 0$ for our study. 

It is well known \cite{Car:1973:BlaHol:} that the complete Kerr NS spacetimes contain the causality violation region, where the axial Killing vector $\xi_{(\phi)}= \partial/\partial \phi$ becomes time-like and hence in principle allows the existence of closed time-like worldlines. As proved in \cite{Zan:2018:GRG:}, such pathological behaviour persists in presence of the cosmological constant in one of the special parts of the \KdS\ spacetimes, referred to as Carter's blocks \footnote{Carter's block is defined as a part of the \KdS\ spacetime with the Boyer-Lindquist radii in any open interval $(r_{i},r_{i+1})$, where $r_{i}, r_{i+1}$ are consecutive roots of $\Delta_{r}=0$ \cite{Zan:2018:GRG:}.}, which contain the ring singularity. It is shown that in this block there exists a non-empty region determined by 
\be
\xi^{\mu}_{(\phi)} \xi_{(\phi)\mu}=g_{\phi \phi}=\frac{\sin^2\theta }{I^2\rho^2}A<0, \label{gphiphi<0}
\ee
which acts as a so-called Carter's time machine (CTM), that generates closed timelike curves that can be extended through any point of the appropriate Carter's block. The analysis of the function $A$ defined in Eq. (\ref{A}) shows that the condition (\ref{gphiphi<0}) can be met at small negative radii $r<0$ in the vicinity of the ring singularity.

\subsection{Ergosphere} \label{ssec_ergos}

The ergosphere region is given by the inequality
\be
g_{tt}\equiv \frac{a^2\sin^2\theta \Delta_\theta-\Delta_r}{I^2 \rho^2}\geq 0, \label{gtt>=0}
\ee
where equality corresponds to the ergosphere boundary, i.e., the ergosurface. In the equatorial plane the radii of the ergosurface are given by the condition
\be
a^2-\Delta_{r}=0, \label{a2-Delta}
\ee
which is a quartic equation that always has a solution at $r=0$, and in the case of the KdS spacetimes with $a^2y<1$ that we are dealing with in this paper, it has three other real roots $r^{(-)}_{erg}$, $r^{\pm}_{erg}$, where $r^{(-)}_{erg}<$ $0<r^{+}_{erg}<r^{-}_{erg}$. The negative solution $r^{(-)}_{erg}$ delimits the ergoregion in the region of negative radii, which spreads between the radii $r^{-}_{c}\leq r\leq r^{(-)}_{erg}<0$ in the equatorial plane, where $r^{-}_{c}$ denotes the secondary cosmological horizon in the region $r<0$. For more details on the properties of the ergosphere and its extension beyond the equatorial plane at $r<0$, see \cite{Stu-Char:2024:PHYSR4:}. In the region of positive radii $r>0$, the inner radius $r^{+}_{erg}$ delimits the boundary of the inner ergosphere, connected to the ring singularity which extends between radii $0<r^{+}_{erg}$, while the outer radius $r^{-}_{erg}$ defines the boundary of the cosmological ergosphere, extending between the radii $r^{-}_{erg}<r_{c}$. These radii are located at

\be
r^{(-)}_{erg}=-2\sqrt{\frac{1-a^2y}{3y}}\cos[\frac{1}{3} \arccos \sqrt{\frac{27y}{(1-a^2y)^3}}] \label{r_erg<0}
\ee
and
\be
r^{\pm}_{erg}(a^2,y)\equiv 2\sqrt{\frac{1-a^2y}{3y}}\cos[\frac{\pi}{3}\pm \frac{1}{3} \arccos \sqrt{\frac{27y}{(1-a^2y)^3}}].\label{r_erg_pm} 
\ee

The parameters $a^2, y$ of the spacetimes in which the two distinct radii $r^{\pm}_{erg}$ exist are, according to Eq. (\ref{r_erg_pm}), related by the condition
	\be
	\frac{27y}{(1-a^2y)^3}\leq1. \label{exist_rerg_pm}
	\ee
The equality in the relation (\ref{exist_rerg_pm}) defines the function
\be
y_{erg-s}(a^2) \equiv \frac{1}{a^2}+\frac{3\cdot 2^{1/3}}{2a^{3}} \frac{2^{1/3}(\sqrt{4+a^2}-a)^{2/3}-2}{(\sqrt{4+a^2}-a)^{1/3}}, \label{yerg}
\ee
which in the $(a^2-y)$ parameter plane separates spacetimes with two separate ergospheres, the inner and the cosmological one, from spacetimes with a single one ergosphere extending from the ring singularity to the cosmological horizon. The spacetimes with a single/two separate ergoregions correspond to the region above/below the $y=y_{erg-s}(a^2)$ curve in the $(a^2-y)$ parameter plane. 
The function $y_{erg-s}(a^2)$ is also related to the existence of the static radius (see subsection below), which explains the label used.

Latitudinal coordinates $\theta_{erg}$ of the ergosurface outside the equatorial plane depend on the radial coordinate according to the equation
\be
\theta_{erg}(r)=\arccos \sqrt{\frac{a^2y-1+\sqrt{I^2-4y\Delta_{r}}}{2a^2y}}. \label{theta_erg}
\ee

\begin{figure}[htbp]
	\includegraphics[width=\linewidth]{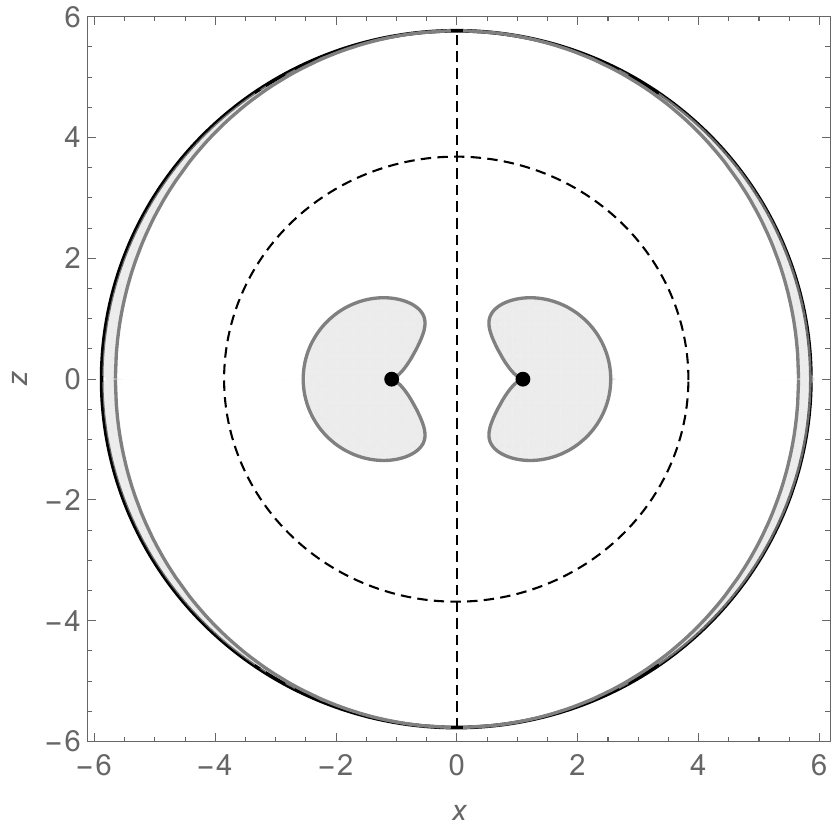}	
	\caption{Distribution of the ergosphere in KdS spacetime of Class IVa with parameters $a^2=1.2, y=0.02$, represented in the Kerr-Schild coordinates x,y,z. The figure shows a section through the plane y=0 (here y not to be confused with the cosmological parameter $y$). The ergosphere is depicted as a grey area. The dashed vertical line depict the spin axis, the black dots represent the incision through the ring singularity. The dashed ellipse depicts the static radius. It can be seen the inner  torus-like shaped ergosphere  connected to the ring singularity, and the cosmological ergosphere, which spreads among the outer ergosurface and the cosmological horizon, depicted as the outer black ellipse.       
	}\label{Fig_ergos}
\end{figure}

A detailed discussion of the properties of the ergosphere for relevant possible combinations of the KdS spacetime parameters $a^2, y$ has been presented in \cite{Stu-Char:2024:PHYSR4:} and will not be repeated here. We just recall that Eqs (\ref{r_erg<0}), (\ref{r_erg_pm}), (\ref{theta_erg}) hold assuming $a^2y<1$, which is the case of the Class IV KdS spacetimes considered in this paper. The distribution of the ergosphere in the Class IV KdS spacetimes is shown in Fig. \ref{Fig_ergos}, for other classes of the KdS spacetimes see \cite{Stu-Char:2024:PHYSR4:}.

\subsection{Static radius} \label{ssec_stac_rad}

In this paper we intend to construct the shadows of superspinars observed by observers located at LNRFs at the so-called static radius $r_{s}$, where the gravitational attraction due to the central mass $M$ is just compensated by the cosmological repulsion governed by the cosmological constant $\Lambda$. At and near the static radius the KdS spacetimes are closest to the asymptotically flat region of the Kerr spacetimes, and the LNRFs are close to the frames of static observers \cite{Stu-Hle:1999:PHYSR4:,Stu-Sla:2004:PHYSR4:}.

In dimensional units, the static radius depends on the mass and the cosmological constant as $r_{s}=(3GM^2/\Lambda c^2)$ \cite{Stu:1983:BULAI:}, which can be simplified to the form 
\be
       r_{s}=y^{-1/3} 
\ee
in the dimensionless quantities we use. From the Eq. (\ref{r_erg_pm}) it follows that the static radius coincides with the radius where the two radii $r^{\pm}_{erg}$ coalesce in the borderline case of spacetimes with $y=y_{erg-s}(a^2)$. Since at the static radius the only non-zero component of the freely falling observer's four-momentum $P^{\mu}$ is its time component $P^{t}$, while the other spatial components are zero, $P^{i}=0$ \cite{Sla-Stu:2020:EPJC:}, it follows that in the spacetimes with $y>y_{erg-s}(a^2)$ the static radius is limited outside the equatorial plane, since there is an ergosphere where necessarily $P^{\phi}>0$.

\section{Equations of photon motion} \label{sec_phot_mot}

Since the components of the metric tensor $g_{\mu \nu}$ determined by the line element (\ref{ds2}) are independent of time and azimuthal coordinates, $(t,\phi)$, the conjugate momenta
\bea
k_{t}&=&g_{t \alpha}k^{\alpha} \equiv -E, \label{E}\\
k_{\phi}&=&g_{\phi \alpha} k^{\alpha} \equiv =\Phi \label{Phi}
\eea
of the photon 4-momentum $k^{\mu}\equiv \din x^{\mu}/\dbe \lambda$, $x^{\mu} \in (t, r, \theta, \phi)$, are constants of motion. Here $\lambda$ is the affine parameter and $E$, $\Phi$ are covariant 'energy' and covariant 'axial angular momentum' -- in the pure Kerr case $y=0$, where the spacetime is asymptotically flat, these are the energy and axial angular momentum as measured by distant static observers \cite{Car:1973:BlaHol:}, while for $y>0$ the spacetime is asymptotically de~Sitter and these quantities are close to those measured by static observers at the static radius and its vicinity \cite{Stu:1983:BULAI:,Stu-Hle:1999:PHYSR4:}. In addition to another motion constant, the rest mass $m$ of the particle ($m=0$ for photons), there is one more, the so-called fourth Carter constant $\calk$, related to the hidden symmetry of the geometry, which must be non-negative,  $\calk \geq 0$. The only case where $\calk=0$ occurs is for photons whose trajectories coincide with the spin axis $\theta=0,\pi$, or for the so-called Principal Null Congruence (PNC) photons whose trajectories lie on the cones $\theta=const$ \cite{Bic-Stu:1976:BULAI:}.

As in previous works \cite{Stu:1983:BULAI:,Stu-Hle:2000:CLAQG:, Char-Stu:2017:EPJC:, Stu-Char-Sche:2018:EPJC:,Stu-Char:2024:PHYSR4:}, for $E\neq 0$ it is convenient to introduce the modified motion constants $X, q$ by the relations 
\be 
X\equiv \ell-a, \label{def_X}
\ee
where 
\be
\ell \equiv \Phi/E  \label{def_ell}
\ee
is the impact parameter, and the modified fourth Carter constant 
\be
q\equiv \calk/I^2E^2-X^2, \label{def_q}
\ee 
which must vanish for the motion confined to the equatorial plane. Using the above motion constants, the Carter equations for photon motion become

\bea
\frac{\rho^2}{I^2E} k^{t} &=&\frac{(r^2+a^2)(r^2-aX)}{\Delta_r}\nonumber \\
&+&\frac{a(a\cos^2\theta+X)}{\Delta_\theta },\label{CarterT}\\
\frac{\rho^2}{I^2E^2} k^{r}&=&\pm \sqrt{R}, \label{CarterR}\\
\frac{\rho^2}{I^2E^2} k^{\theta}&=&\pm \sqrt{W}, \label{CarterW}\\
\frac{\rho^2}{I^2E} k^{\phi} &=&\frac{a(r^2-aX)}{\Delta_r}\label{CarterPhi}+\frac{(a\cos^2\theta+X)}{\Delta_\theta \sin^2\theta},
\eea
where
\bea
R(r)&\equiv& \left(r^2 - aX\right)^2 -\Delta_r\left(X^2 + q\right),\label{R(r)} \\
W(\theta)&\equiv& (X^2+q)\Delta_\theta-\frac{(a\cos^2\theta+X)^2}{\sin^2\theta}. \label{W(theta)}
\eea

A detailed discussion of these equations of motion related to the both the KdS BH and NS spacetimes can be found in \cite{Char-Stu:2017:EPJC:,Stu-Char:2024:PHYSR4:}.

\subsection{Allowed regions of motion constants} \label{ssec_allow_reg}

The photon motion is fully governed by the motion constants $X,q$, which in turn are related to the directional angles of a photon with respect to an emitter fixed to an observer (frame), and which govern the construction of the LEC related to the observer. Therefore, we have to determine for which values of these motion constants the motion of a photon released at given position of the emitter, described by the pair $(r_{e},\theta_{e})$ of its radial and latitudinal coordinates, is actually allowed. Starting point are the reality conditions

\be W(\theta_{e};X,q)\geq0 \label{W>0}
\ee
 
\be R(r_{e};X,q)\geq0,\label{R>0}
\ee

Using the linear dependence of the functions $R(r;X,q),W(\theta;X,q)$ on the motion constant $q$, similar to \cite{Stu-Char-Sche:2018:EPJC:,Stu-Char:2024:PHYSR4:}, we can rewrite the conditions (\ref{W>0},\ref{R>0}) in the form 
\be
q_{min}(X;\theta_{e})\leq q \leq q_{max}(X;r_{e}),\label{real_conditions}
\ee
where
\be
q_{min}(X;\theta_{e})\equiv \frac{\cot^2\theta_{e}}{\Delta_{\theta_{e}}}[(1-a^2y\sin^2\theta_{e})X^2+2aX+a^2\cos^2\theta_{e}] \label{qmin(X,th)}
\ee 
and
\be
q_{max}(X;r_{e})\equiv \frac{(r_{e}^2-a X)^2}{\Delta_{r_{e}}}-X^2. \label{qmax(X,r)}
\ee
It follows from Eq.\ref{real_conditions} that the motion constants of a photon reaching the surface of the superspinar at $r=R$ must satisfy the inequality 
\be
q\leq q_{max}(X;R). \label{q<qmax(X,R)}
\ee
We now briefly outline some properties of the functions $q_{min}(X;\theta)$, $q_{max}(X;r)$ for later reference.

\subsubsection{Latitudinal function  $q_{min}(X;\theta_{e})$ }\label{ssec_lat_func}

\begin{figure}[htbp]
	\includegraphics[width=\linewidth]{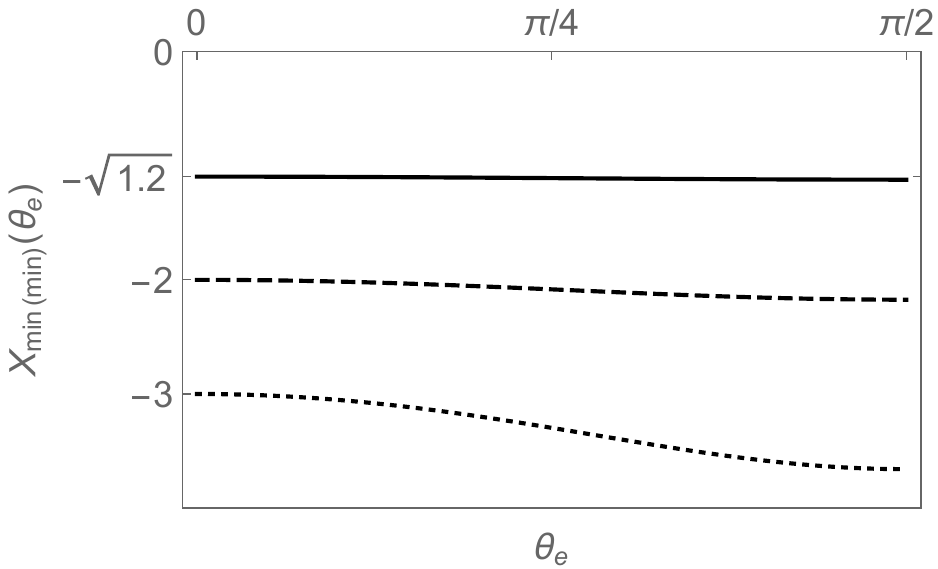}
	\caption{Behaviour of the function $X_{min(min)}(\theta_{e})$ for fixed parameter $y=0.02$ and varying parameter $a^2=1.2$ (full curve), $a^2=4$ (dashed curve) and $a^2=9$ (dotted curve).}\label{Fig_Xminmin}
\end{figure}

\begin{figure}[htbp]
	\includegraphics[width=\linewidth]{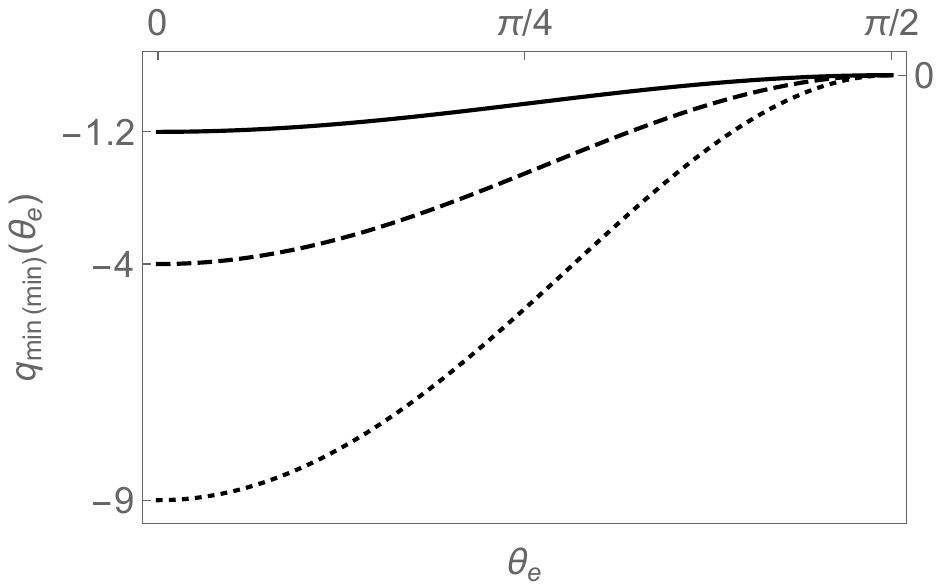}
	\caption{Behaviour of the function $q_{min(min)}(\theta_{e})$ with the same styling and corresponding parameters as used in Fig. \ref{Fig_Xminmin}. }\label{Fig_qminmin}
\end{figure}

For the spacetime parameters satisfying $a^2y<1$, the function $q_{min}(X;\theta_{e})$ is convex for $0<\theta_{e}<\pi/2$, while for $\theta_{e}=\pi/2$ it holds $q_{min}(X;\theta_{e}=\pi/2)\equiv 0$. \footnote{Since the spacetime is symmetric with respect to the reflection $\theta \to \pi-\theta$, we can restrict our considerations to the range $0\leq\theta_{e}\leq\pi/2$.} The global minimum of the function $q_{min}(X;\theta_{e})$ is located at 
\be
X=X_{min(min)}(\theta_{e})\equiv \frac{-a}{1-a^2y\sin^2\theta_{e}}, \label{Xex(min)}
\ee 
and its value is
\be
q=q_{min(min)}(\theta_{e})\equiv \frac{-a^2 \cos^2\theta_{e}}{1-a^2y\sin^2\theta_{e}}, \label{qex(min)}
\ee
where $0<\theta_{e}<\pi/2$. For $\theta_{e}\to 0$ the following relations hold: $X_{min(min)}(\theta_{e})\to -a$, $q_{min(min)}(\theta_{e})\to -a^2$, while for $\theta_{e}=0$ (the motion along the spin axis) the reality condition $W(\theta;X,q)\geq 0$ reduces to 
\bea
X&=&-a\quad(\ell=0), \label{X=-a} \\
q&\geq&-a^2.\label{q>-a2}
\eea 
The behaviour of the both functions $X_{min(min)}(\theta_{e}), q_{min(min)}(\theta_{e})$ is shown in Figs. \ref{Fig_Xminmin}, \ref{Fig_qminmin}. A detailed discussion of the properties of the function $q_{min}(X;\theta)$ can be found e.g. in \cite{Char-Stu:2017:EPJC:,Stu-Char-Sche:2018:EPJC:,Stu-Char:2024:PHYSR4:}.

\subsubsection{Radial function $q_{max}(X;r_{e})$}

In general, the function $q_{max}(X;r_{e})$ can be convex or concave depending on the sign of the coefficient $q_{max(2)}$ of the quadratic term, where
\be
q_{max(2)}=\frac{a^2-\Delta_{r_{e}}}{\Delta_{r_{e}}}. \label{qmax(2)}
\ee
As follows from Eq.(\ref{qmax(2)}), at radii $r>0$ that do (do not) interfere with the ergosphere, the function $q_{max}(X;r_{e})$ is convex (concave). In the first case, as shown in \cite{Char-Stu:2017:EPJC:}, the part of the plane of the $(X-q)$ plane of motion constants described by the inequalities (\ref{real_conditions}) and lying below the rising part of the function $q_{max}(X;r_{e})$ corresponds to locally counter-rotating photons with positive impact parameter $X$ and with negative energy, which are of course trapped by the gravitational potential. The parts of the $(X-q)$ plane of motion constants described by the inequalities (\ref{real_conditions}) and lying below the descending part of the $q_{max}(X;r_{e})$ function correspond to photons with $E>0$ which can be trapped or escaping. In the latter case, the parts of the $(X-q)$ plane of motion constants satisfying the inequalities  (\ref{real_conditions}) correspond only to photons with $E>0$, which again can be escaping or trapped. 

According to the relation (\ref{qmax(2)}), the function $q_{max}(X;r_{e})$ is concave at $r_{(erg)-}<r<0$, and at $r=0$ it is $q_{max}(X;0)=0$.

\subsection{Important cases of photon trajectories}

The specific types of photon trajectories that have a major impact on the design of the resulting LECs have been discussed in detail in our previous work \cite{Stu-Char:2024:PHYSR4:}. Here we briefly recall their properties. Some of the concepts introduced here are needed for the basic classification of the null geodesics in section \ref{class_null_geo}.

\subsubsection{Spherical photon orbits}
The spherical photon orbits (SPOs) are determined by the conditions 
\be
R(r)=0,\quad \oder{R}{r}=0,\label{cond_spo}
\ee
which, when solved simultaneously, give the motion constants for the photons on the SPOs
\be
X=X_{\spo}(r) \equiv \frac{r[(1-ya^2)r^2-3r+2a^2]}{a[2yr^3+(ya^2-1)r+1]}, \label{Xsph}
\ee
or, alternatively
\be
\ell=\ell_{\spo}(r) \equiv \frac{Ir^3-3r^2+a^2Ir+a^2}{a[2yr^3+(ya^2-1)r+1]} \label{lsph}
\ee
and
\bea
q&=&q_{\spo}(r) \\
 &\equiv& -\frac{r^3}{a^2}\; \frac{y^2a^4r^3+2ya^2r^2(r+3)+r(r-3)^2-4a^2}{[2yr^3+(ya^2-1)r+1]^2}.\nonumber \label{qsph}
\eea
The zeros of the function $q_{\spo}(r)$ give the radii of the equatorial circular photon orbits (ECPOs). It can be shown that in the case of the NS spacetimes there is only one ECPO at $r=r^{-}_{ph}$ in the region of positive radii, which is locally counterrotating in the LNRFs and unstable with respect to the radial perturbations. It gives the boundary of the region of the SPOs at $r>0$. However, its expression is rather complex and therefore will not be presented here.

If for $a^2y<1$, the functions $X_{\spo}(r)$, $\ell_{\spo}(r)$, $q_{\spo}(r)$ have common points of divergence at radii $r^{(-)}_{d(\spo)}$, $r^{\pm}_{d(\spo)}$, where
\be
r^{(-)}_{d(\spo)}=-2\sqrt{\frac{1-a^2y}{6y}}\cos[\frac{1}{3}\arccos \sqrt{\frac{27y}{2(1-a^2y)^3}}], \label{rdexm}
\ee
and
\be
r^{\pm}_{d(\spo)}=2\sqrt{\frac{1-a^2y}{6y}}\cos[\frac{\pi}{3}\pm \frac{1}{3}\arccos \sqrt{\frac{27y}{2(1-a^2y)^3}}], \label{rdexpm}
\ee
for which the inequality $r^{(-)}_{d(\spo)}<0<r^{+}_{d(\spo)}<r^{-}_{d(\spo)}$ holds.

The radii $r^{\pm}_{d(\spo)}$ coalesce for $y=y_{d(\spo)}(y)$, where
\be
y_{d(\spo)}(a^2) \equiv \frac{1}{a^2}-\frac{6}{\sqrt{2}a^3}\sinh [1/3 \sinh^{-1} \frac{\sqrt{2}a}{2}]. \label{yd(spo)}
\ee
Note that the function $y_{d(\spo)}(y)$ in the $(y - a^2)$ plane determines the boundary between the KdS spacetime Classes VI and VII, 
as introduced in \cite{Char-Stu:2017:EPJC:}. For $y\lessgtr y_{d(\spo)}(y)$, the functions $X_{\spo}(r)$, $q_{\spo}(r)$ have two/no points of divergence.
\footnote{Obviously, there is always zero of the function $q_{\spo}(r)$ at $r=0$, but such an orbit cannot be regarded as an orbit of any physical particles, since it coincides with the physical singularity of the KdS spacetimes.}

\begin{figure}[htbp]
	\centering
	\begin{tabular}{c}
			\includegraphics[width=\linewidth]{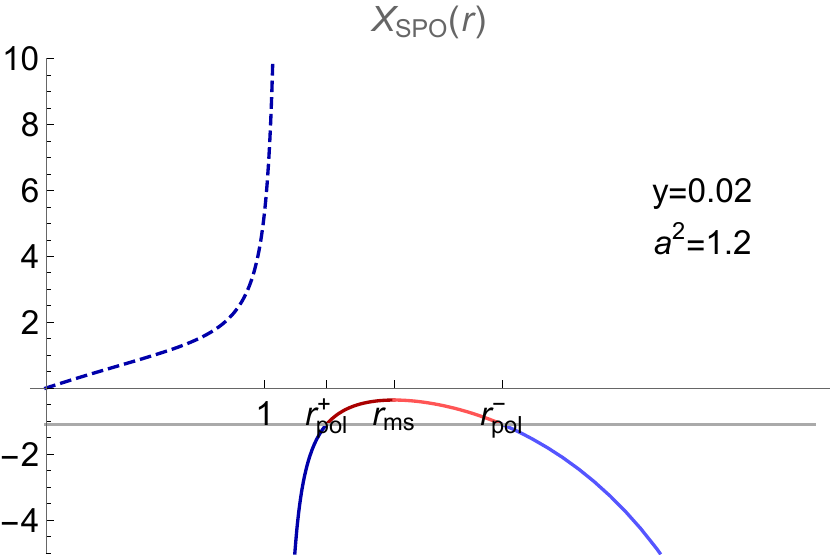}\\
			(a)\\
			\includegraphics[width=\linewidth]{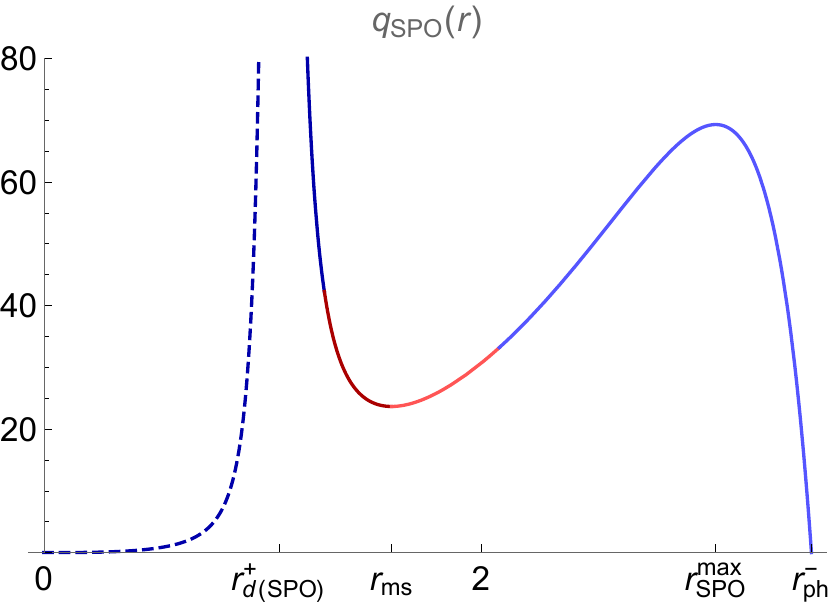}\\
			(b)
	\end{tabular}
	\caption{Typical behaviour of the functions $X_{\spo}(r)$ and $q_{\spo}(r)$ for parameters corresponding to Class IV spacetime. The solid horizontal grey line in the figure (a) indicates the value of $X=-a$ ($\ell=0$). The dark blue dashed parts of these curves determine the motion constants of photons with  negative covariant energy $E<0$, delimited by the point of divergence $r^{+}_{d(SPO)}$ of both functions. The red/blue parts correspond to locally corotating/counterrotating SPOs, delimited by the radii $r^{\pm}_{pol}$ of the polar SPOs, which are the zeros of the function $\ell_{\spo}$. The dark/bright colours correspond to the stable/unstable SPOs, that are separated by the radii $r_{ms}$ of the marginally stable SPOs, which are the loci of the local maximum/minimum of the function $\ell_{\spo}(r)$/$q_{\spo}(r)$. The region of the SPOs is delimited by the radii of the equatorial circular photon orbit  $r^{-}_{ph}$, which is a zero of the function $q_{\spo}(r)$ (for more details see \cite{Stu-Char:2024:PHYSR4:}).}\label{Fig_ell_qspo}
\end{figure}

The condition for unstable spherical orbits 
\be
\mathrm{d}^2R/\mathrm{d}r^2<0 \label{d2R}
\ee
implies the inequality
\be
8r(3yr^4-I^2r^3+3(1-a^2y)r^2-3r+a^2)>0, \label{stab_cond}
\ee 
to be evaluated at the radii of the appropriate spherical orbit $r=r_{\spo}$.  Typical behaviour of the functions $\ell_{\spo}(r)$, $q_{\spo}(r)$ for spacetime parameters corresponding to the Class IV, and the related significant radii of the SPOs, are shown in Fig. \ref{Fig_ell_qspo}.

The curve $(X_{\spo}(r_{\spo}),q_{\spo}(r_{\spo}))$ (hereafter $q_{\spo}(X_{\spo})$) parametrized by the radii $r_{\spo}$ of the unstable spherical photon orbits plays a crucial role in the construction of LECs, since it gives the critical locus (cf. e.g. \cite{Vie:1993:ASTRA:}) in the $X-q$ plane of the motion constants that determinine the boundary between escaping and trapped photons.

The critical locus has essential importance in section \ref{shadows}, where we construct the KdS superspinar shadow for an observer at radius $r_{o}>r_{ph}^{-}$. This curve then corresponds to an arc or circle on the observer's celestial sky.
 
In the vicinity of stable SPOs there are bounded orbits where the photons, which we will call 'trapped' (see section \ref{class_null_geo}), oscillate radially between two radial turning points. These orbits can be occupied if the source is located between these radii. Such a situation can occur if there is an overlap of the SPOs with the equatorial circular orbits of the test particles, allowing the possible occurrence of a thin Keplerian disk. In such a case the construction of LECs is important for the study of special optical effects such as self-illumination and self-eclipse, or light echo \cite{Char-Stu:2018:EPJC:}. 

\subsubsection{Polar spherical photon orbits}

\begin{figure}
	\includegraphics[width=\linewidth]{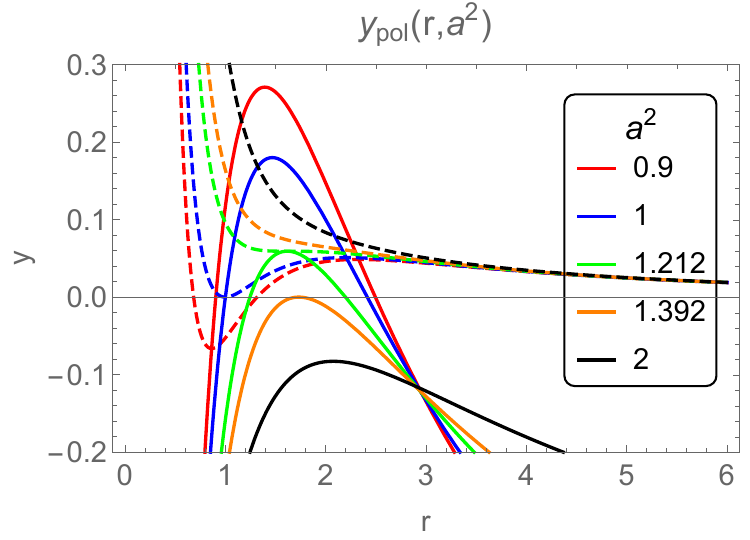}
	\caption{Behaviour of the function $y_{pol}(r;a^2)$ (full curves) compared to the function $y_{h}(r;a^2)$ (dashed curves) for some representative values of the cosmological parameter $a^2$. For a given parameter $a^2$, both curves have the same colour. It can be seen that the graph of the function $y_{pol}(r;a^2)$ intersects the graph of function $y_{h}(r;a^2)$ at its local extrema, if any. } \label{fig_ypol}
\end{figure}

\begin{figure}
	\includegraphics[width=\linewidth]{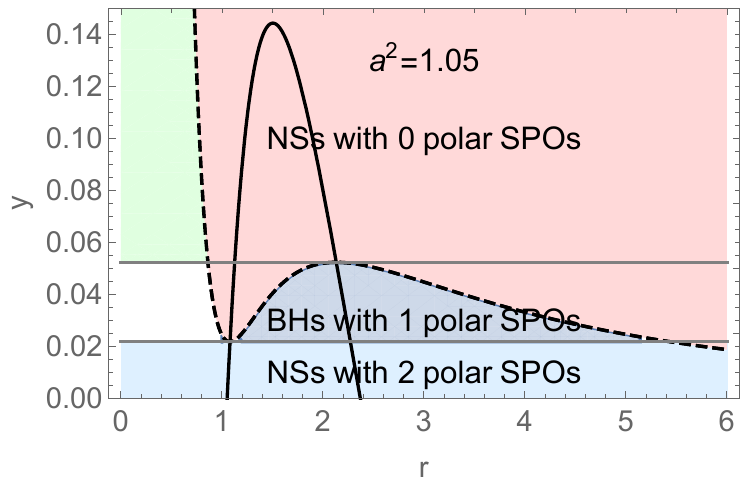}
	\caption{Behaviour of the function $y_{pol}(r;a^2)$ (black full curve) for the spin parameter corresponding to the case $a_{SdS}^2 < a^2 < a^2_{crit}$. The horizontal grey lines represent the values $y_{min(h)}(a^2)$, $y_{max(h)}(a^2)$ of the local minimum and maximum, respectively, of the function $y_{h}(r;a^2)$ (black dashed curve). For $0<y<y_{min(h)}(a^2)$ (blue region), the function $y_{pol}(r;a^2)$ determines two polar SPOs (the inner one stable, the outer one unstable) in the NS spacetimes belonging to Class IVa or VIa. For $y_{min(h)}<y<y_{max(h)}(a^2)$ it determines one polar SPO (unstable) in the stationary region (grey area) of the BH spacetimes of Class II or III. For $y_{max(h)}(a^2)<y$ the curve $y=y_{pol}(r;a^2)$ is irrelevant, since this part lies completely in the dynamic region (red area) of the NS spacetimes of Class VII or VIII. The white/green area corresponds to the stationary region below the inner (Cauchy)/ cosmological horizon of the appropriate BH/N spacetime.} \label{fig_ypol_a2=1.05}
\end{figure}

\begin{figure}
	\includegraphics[width=\linewidth]{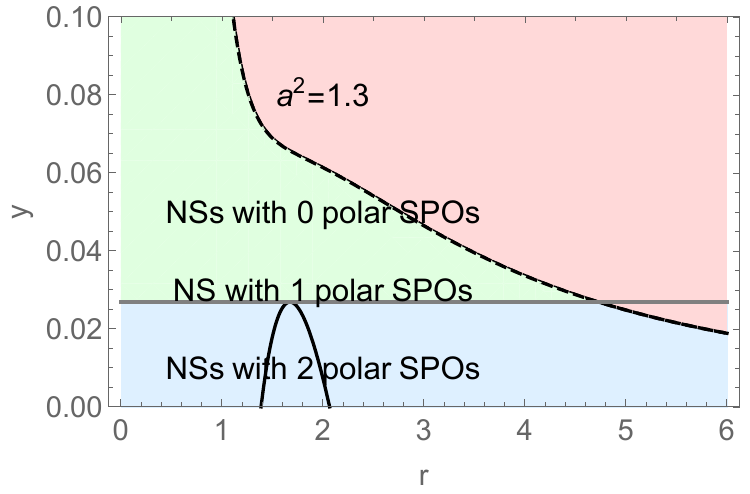}
	\caption{Behaviour of the function $y_{pol}(r;a^2)$ (full curve) for the spin parameter corresponding to the case $a^2_{crit}<a^2<a^2_{max(pol)K}$. The horizontal grey line represents the value $y_{max(pol)}(a^2)$ of the local maximum of the function $y_{pol}(r;a^2)$. The loci of this maximum represent the radius $r_{max(pol)}(y)$ of the marginally stable polar SPO. This line separates the spacetimes with 2 polar SPOs, which belong to the Class IVa or VIa, from the spacetimes with no polar SPOs, which correspond to the Class IVb or VIb. For $a^2_{max(pol)K}<a^2$, $y_{max(pol)}(a^2)<0$, so there are no polar SPOs in the spacetimes with $a^2_{max(pol)K}<a^2$. The colouring has the same meaning as in the previous figure. The division of the parameter plane $(y-a^2)$ into regions corresponding to the different types of KdS spacetimes discussed above is shown in Fig. \ref{fig_logya2plane}. } \label{fig_ypol_a2=1.3}
\end{figure}

The polar spherical photon orbits are the special case of the SPOs orbiting over the poles, crossing the symmetry axis. They are determined by the condition 
\be
\ell_{\spo}=0,\label{zero_ell}
\ee
which can be expressed by the relation
\be
y=y_{pol}(r;a^2)\equiv \frac{r^2(3-r)-a^2(1+r)}{ra^2(r^2+a^2)}. \label{ypol}
\ee
The behaviour of the function $y_{pol}(r;a^2)$ is shown in Fig. \ref{fig_ypol} in comparison with the behaviour of the function $y_{h}(r;a^2)$ for some appropriately selected values of the spin $a^2$. For given parameters $a_{0}^2$, $y_{0}$, the intersection of the line $y=y_{0}$ with the curve $y=y_{pol}(r;a_{0}^2)$ gives the radii $r^{\pm}_{pol}$ of the polar SPOs, which, as we have found in \cite{Stu-Char:2024:PHYSR4:}, reads
\be
r^{\pm}_{pol}=\frac{1}{I}[1+2\sqrt{1-\frac{a^2I^2}{3}}\cos(\frac{\pi\pm\psi}{3})], \label{rpolpm}
\ee
where
\be
\cos\psi=\frac{a^2I^2-1}{(1-\frac{a^2I^2}{3})^{3/2}}.\label{cosvarphi}
\ee
Fig. \ref{fig_ypol_a2=1.05} shows that there are KdS superspinar (NS) spacetimes with two polar SPOs, an inner stable one, and an outer unstable one, which is the case of the superspinar (NS) spacetimes of Classes IVa, VIa, introduced in \cite{Char-Stu:2017:EPJC:}, BH spacetimes with one unstable SPO, which corresponds to Classes I-III, and superspinar (NS) spacetimes without polar SPOs, which correspond to Classes VII, VIII. Fig. \ref{fig_ypol_a2=1.3} shows the limiting case of the superspinar (NS) spacetime, separating the spacetimes of Classes IVa, VIa, from the superspinar (NS) spacetimes with no polar SPO that correspond to Classes IVb, V, VIb. In this limiting case, the two polar SPOs merge. The radius $r_{max(pol)}$ of this single polar SPO is given by the local maximum of the function $y_{pol}(r;a^2)$, which, as follows from the analysis of Eqs.(\ref{rpolpm}), (\ref{cosvarphi}), increases linearly with the spin $a$ as
\be
r_{max(pol)}(a)= a \sqrt{1+2/\sqrt{3}}.\label{rpol_max} 
\ee
The minimum value for $a=a_{crit(KdS)}$, $y=y_{crit(KdS)}$ is the same as the merging event horizons (see the green curves in Fig. \ref{fig_ypol}), i.e., 
\be
r_{max(pol)}(a_{crit(KdS)})=r_{i}= r_{o}=r_{c}=\frac{1}{4}(3+2\sqrt{3})=1.616, \label{rmaxpol(acrit)}
\ee
while the maximum value occurs for $a=a_{max(pol)K}\equiv \sqrt{6\sqrt{3}-9}=1.180$, $y=0$, and reads 
\be
r_{max(pol)}(a_{max(pol)K})=1+\sqrt{4-2\sqrt{3}}=1.732, \label{rmaxpol(aK)}
\ee
where $a_{max(pol)K}$ denotes the maximum spin that allows the existence of the polar SPOs, corresponding to the Kerr spacetime $(y=0)$ (cf. \cite{Char-Stu:2018:EPJC:}). Its square is $a^2_{max(pol)K}=6\sqrt{3}-9=1.3923$. The parameters of the superspinar (NS) spacetimes with the single polar SPO at radii $r_{max(pol)}$ are tied by the condition
\be
\cos \psi =1 \label{cosvarphi_1}
\ee
in (\ref{cosvarphi}), which can be written in the form
\be
y_{max(pol)}(a^2)\equiv a^{-2}(\sqrt{\frac{a^2_{max(pol)K}}{a^2}}-1). \label{ypolmax}
\ee
The curve $y_{max(pol)}(a^2)$ defines a boundary in the $(y-a^2)$-plane between the spacetimes of Classes IVa, IVb, that correspond to the superspinar (NS) spacetimes with two polar SPOs, and the spacetimes of Classes VIa, VIb, corresponding to the superspinar (NS) spacetimes with no polar SPOs, as introduced in \cite{Char-Stu:2017:EPJC:}. 

\begin{figure}
	\includegraphics[width=\linewidth]{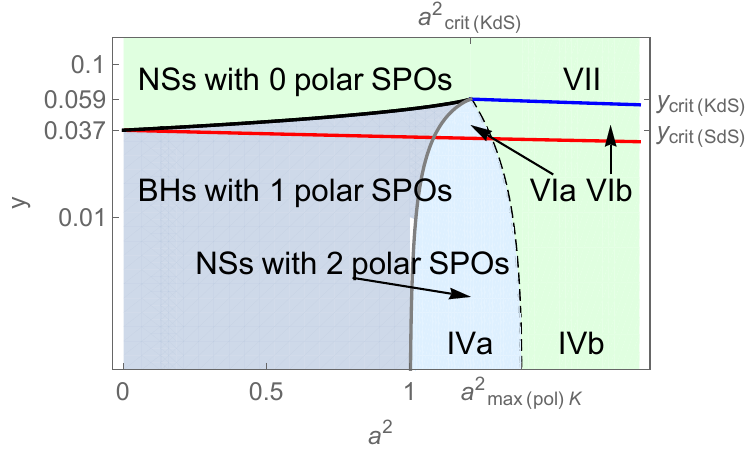}
	\caption{Functions $y_{max(h)}(a^2)$ (black curve), $y_{min(h)}(a^2)$ (grey curve), $y_{max(pol)}(a^2)$ (black dashed curve), $y_{erg-s}(a^2)$ (red curve), $y_{d(\spo)}(a^2)$ (blue curve) and the corresponding division of the parameter plane $(y-a^2)$ into sections corresponding to BH and NS spacetimes with appropriate number of polar SPOs. The colouring of each region is chosen to be consistent with Figs. \ref{fig_ypol_a2=1.05}, \ref{fig_ypol_a2=1.3}. The red and blue curves represent the subdivision of the parameter plane according to the classification introduced in \cite{Char-Stu:2017:EPJC:}, which is recalled below and in the commentary to Fig. \ref{Fig_eff_pots1}. Note that the values on the $y$-axis are plotted on a logarithmic scale. } \label{fig_logya2plane}
\end{figure}

In this paper, we do not address all the properties that trigger this classification, except for the presence/absence of polar SPOs. In Fig. \ref{fig_logya2plane} we therefore present the classification of spacetimes only in terms of the number of horizons and polar SPOs.

\subsection{Classification of the KdS superspinar (NS) spacetimes} \label{class_null_geo}

The classification of KdS superspinar (NS) spacetimes into Classes IV-VIII, some of which are indicated in Fig. \ref{fig_logya2plane}, is determined by the nature of the radial motion of the photons. This motion is determined by the behaviour of the effective potentials  $X_{\pm}(r;q,y,a^2)$, where the equality
\be
X=X_{\pm}(r;q,y,a^2)\equiv \frac{ar^2\pm\sqrt{\Delta_r\left[r^4+q(a^2-\Delta_r)\right]}}{a^2-\Delta_r},\label{Xpm} 
\ee
gives the turning points in the radial direction for photons with the impact parameter $X$. Eq. (\ref{Xpm}) follows from the inequality (\ref{R>0}) by inserting an equal sign and expressing the parameter $X$.

 The curves $X=X_{\pm}(r;q,y,a^2)$ delimit the region of forbidden values of the impact parameter $X$ in the $(r-X)$ plane, where the inequality (\ref{R>0}) is not met, acting as a repulsive potential barrier of the radial photon motion. It follows from the definition in Eq. (\ref{Xpm}) and from the discussion made in subsection \ref{ssec_ergos} that the function $X_{+}(r;q,y,a^2)$ can have up to two divergent points in the region $r>0$ \footnote{the function $X_{-}(r;q,y,a^2)$ cannot diverge}, that coincide with the solution of the Eq.  (\ref{a2-Delta}), i. e., with radii given by Eq.(\ref{r_erg_pm}). Note that these radii depend only on the spacetime parameters $a^2, y$. If these two real solutions exist, we have the KdS spacetimes with the divergent repulsive barrier (DRB) of the radial photon motion. Thus, it further follows from subsection \ref{ssec_ergos} that spacetimes with DRB correspond to spacetimes with two separate ergospheres - the inner one and the cosmological one. On the other hand, if there is no real solution of Eq.(\ref{a2-Delta}), the function $X_{+}(r;q,y,a^2)$ has no points of divergence and we have the KdS spacetimes with the restricted repulsive barrier (RRB) of the radial photon motion. It therefore corresponds to the spacetimes with a single ergosphere. The limiting cases between the KdS spacetimes with DRB and RRB, where the radii given by Eq.(\ref{r_erg_pm}) merge, have spacetime parameters restricted by the relation $y=y_{erg-s}(a^2)$, which is represented by the red curve in Fig. \ref{fig_logya2plane}. The spacetimes with DRB/RRB correspond to the region below/above the curve $y=y_{erg-s}(a^2)$ in the $(a^2-y)$ parameter plane. 

\begin{figure*}[h]
	\centering
	\begin{tabular}{cc}
		\includegraphics[width=0.5\textwidth]{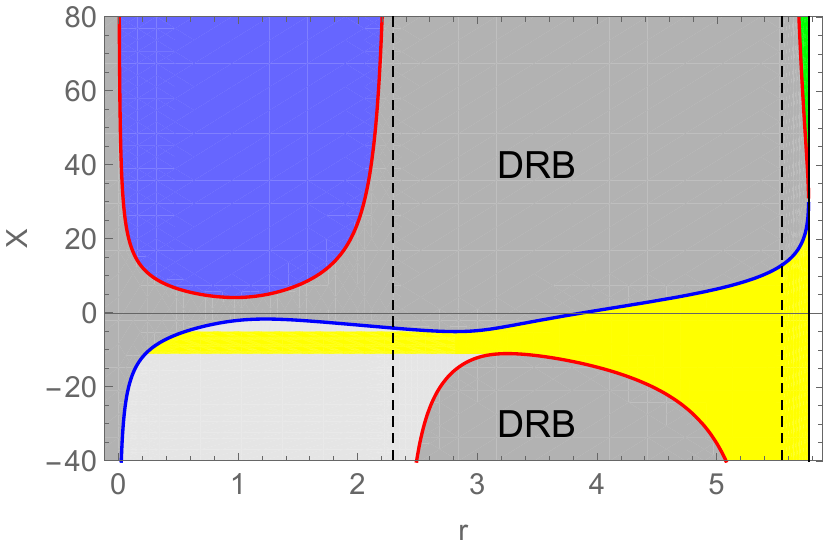} & \includegraphics[width=0.5\textwidth]{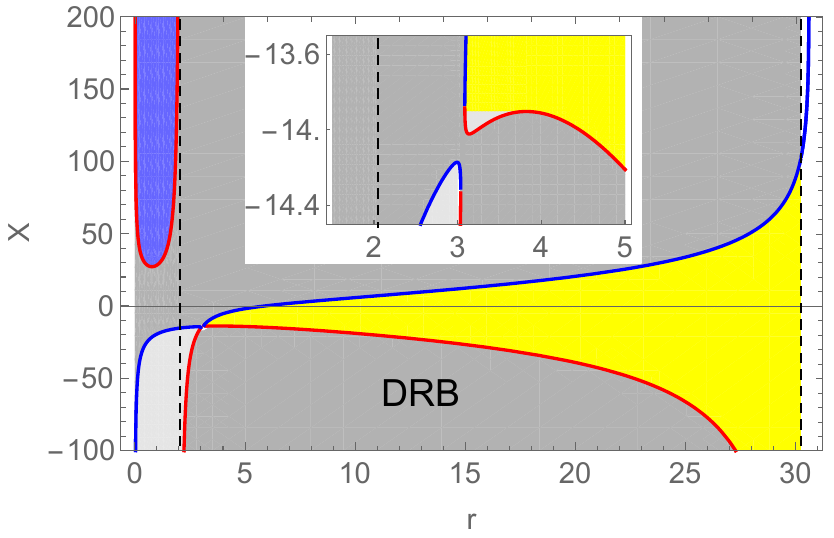}\\
		(a) IVa: $y=0.02$, $a^2=1.2$; $q=60$ & (b) V: $y=0.001$, $a^2=20$\\
		\includegraphics[width=0.5\textwidth]{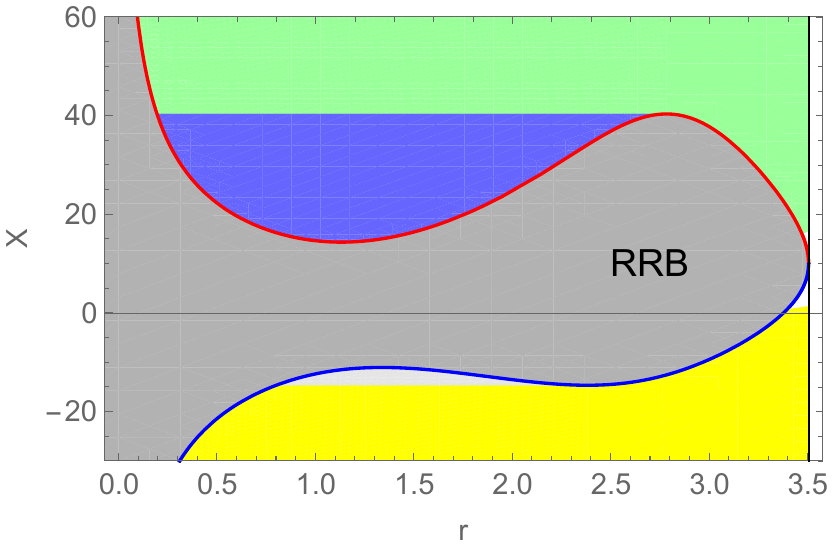} & \includegraphics[width=0.5\textwidth]{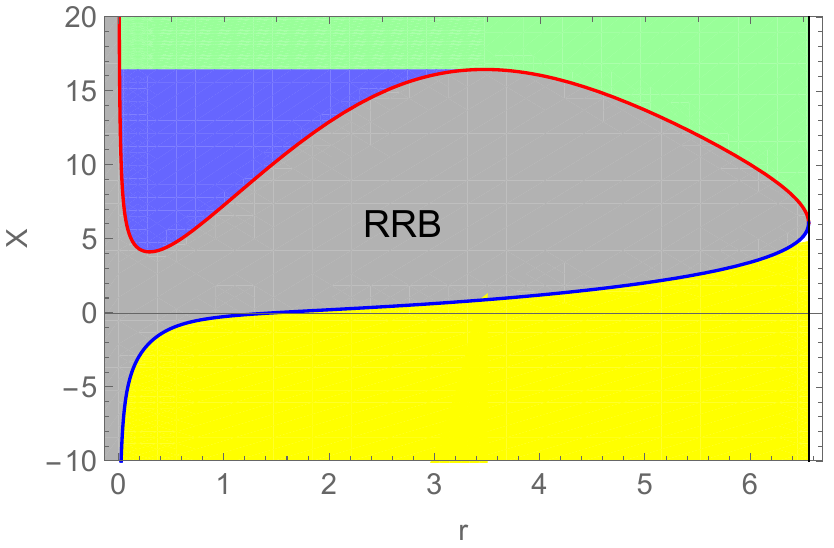}\\
		(c) VIb: $y=0.04$, $a^2=1.5$; $q=500$ & (d) VII: $y=0.02$, $a^2=50$; $q=0.1$ 
	\end{tabular}
	\caption{Basic types of photon orbits in individual classes of the KdS superspinar (NS) spacetimes with appropriately chosen parameters $y, a^2$ in relation to the effective potentials $X_{+}(r;q,y,a^2)$ (red curve) and $X_{-}(r;q,y,a^2)$ (blue curve). The motion constant $q$ is chosen to highlight the different types of orbits. The curves $X_{\pm}$ delimit the region of forbidden values of the impact parameter $X$ (dark grey region), acting as a barrier of radial motion, which can be divergent (DRB) or restricted (RRB). The blue region corresponds to bound orbits occupied by trapped photons with negative covariant energy $E<0$ (see \cite{Char-Stu:2017:EPJC:} for details), the green region corresponds to unbound orbits occupied by photons with $E<0$ but escaping towards the cosmological horizon. Light grey regions mark the trapped photons with $E>0$, yellow regions mark the escaping photons with at most one turning point, if initially directed inwards. Black full vertical lines at the common points of the functions $X_{\pm}(r)$ mark the loci of the cosmological horizon. Black dashed vertical lines are the asymptotes of the function $X_{+}(r)$ at $r=r^{\pm}_{erg}$, delimiting the radial extent of the ergosurface in the equatorial plane. Only Class IV is relevant from an astrophysics perspective, since the other classes correspond to enormously high values of a cosmological parameter that is inconsistent with observations and will therefore not be discussed in detail. The range and structure of the photon orbits in the case of Class IVa are as follows. The local minima of the function $X_{+}(r)$ correspond to the stable spherical photon orbits with $E<0$, which appear to be retrograde in the LNRFs. Varying the motion constant $q$, these minima correspond to the blue dashed part of the curve $X_{\spo}(r)$ in Fig. \ref{Fig_ell_qspo}a. Their radii $r^{E<0}_{\spo s}$ are in the range $0<r^{E<0}_{\spo s}<r^{+}_{d(\spo)}$. The local maxima of the function $X_{-}(r)$ correspond to the stable SPOs with $E>0$, which can be either retrograde or prograde in the LNRFs, with the polar spherical orbit at $r=r^{+}_{pol}$ as being the boundary between these directions. Varying the motion constant $q$, these local maxima correspond to the dark blue and red parts of the curve  $X_{\spo}(r)$ in Fig. \ref{Fig_ell_qspo}a). The radii $r^{E>0}_{\spo s}$ are in the range $r^{+}_{d(\spo)}<r^{E>0}_{\spo s}<r_{ms}$. The local maxima of the function $X_{+}(r)$ and the local minima of the function $X_{-}(r)$ represent the unstable SPOs. With varying motion constant $q$, these extremes correspond to the light blue and red parts of the $X_{\spo}(r)$ curve in Fig. \ref{Fig_ell_qspo}a. These extrema meet each other at $r=r^{max}_{\spo}$, where $r^{max}_{\spo}$ is the local maximum of the function $q_{\spo}(r)$. Varying the motion constant $q$, the projection of these maxima onto the $(q-X)$ motion constant plane forms the critical locus. The radii $r_{\spo}$ of the unstable SPOs are in the range $r_{ms}<r_{\spo}<r^{-}_{ph}$, where $r^{-}_{ph}$ is the radii of the equatorial circular photon orbit, which is counter-rotating. The structure of the photon orbits in other cases can be found e.g. in \cite{Char-Stu:2017:EPJC:,Stu-Char:2024:PHYSR4:}.
	}\label{Fig_eff_pots1}
\end{figure*}

 A detailed analysis of the $X_{\pm}(r;q,y,a^2)$ functions has been carried out in \cite{Char-Stu:2017:EPJC:} for $r>0$ and will not be repeated here. Using the typical behaviour of the potentials shown in Fig. \ref{Fig_eff_pots1}, we briefly outline here the basic characteristics of the individual KdS superspinar (NS) spacetimes for the following suitably chosen representatives:
\begin{itemize}
	\item  Class IVa, represented by the spacetime with parameters $y=0.02$, $a^2=1.2$ (see Fig. \ref{Fig_eff_pots1}a), corresponding to the spacetimes with DRB and the regions of trapped photon orbits with both positive and negative covariant energy (see the comment on Fig. \ref{Fig_eff_pots1}).  The label "a" marks the presence of two polar SPOs, which is manifested by the fact that local extrema $X_{ex(-)}$ of the function $X_{-}(r;q,y,a^2)$ can take the value $X_{ex(-)}= X_{\spo}=-a$. The behaviour of the effective potentials corresponding to Class IVb, not shown in Fig. \ref{Fig_eff_pots1}, where the label "b" corresponds to the absence of the polar SPOs, is qualitatively the same as for Class IVa. The absence of the polar SPOs  is manifested by the extrema $X_{ex(-)}<-a$.   
	\item Class V, represented by the spacetime with parameters $y=0.001$, $a^2=20$ (see Fig. \ref{Fig_eff_pots1}b), corresponding to the spacetimes with the same properties of the radial photon motion as in the previous case, but besides that it contains a small region of unusual bound orbits with $X_{+}<X<0$ and $E>0$ (see detail in the cut-out) -- for $r_{e}<r^{max}_{\spo}$ the critical locus $X_{\spo}(q_{\spo})$ is partly hidden in the forbidden region.
	 
	\item  Class VIb, represented by the spacetime with parameters $y=0.04$, $a^2=1.5$ (see Fig. \ref{Fig_eff_pots1}c), corresponding to the spacetimes with RRB of the radial photon motion and the regions of trapped photon orbits with both positive and negative covariant energy. The label "b" again corresponds to the absence of the polar SPOs. The behaviour of the effective potentials for the both Classes VIa, VIb differs in the same manner as in the cases of IVa, IVb. 
	\item Class VII, represented by the spacetime with parameters $y=0.02$, $a^2=50$ (see Fig. \ref{Fig_eff_pots1}d), corresponding to the spacetimes with no trapped photon orbits with positive covariant energy. The other properties of the photon motion are the same as for Class VI.
	
\end{itemize}

Class VIII corresponds to spacetimes with extremely high spacetime parameters, for which  $a^2y>1$. It is not shown in Fig. \ref{Fig_eff_pots1}, since the behaviour of the effective potentials is qualitatively the same as for Class VII. However, it differs from the Class VII by the properties of the latitudinal photon motion (for more details see \cite{Char-Stu:2017:EPJC:}).

In the present paper we do not construct LECs for the KdS spacetime Classes VI-VIII because of the extremely outlying values of the spacetime parameters relative to the parameters corresponding to cosmological observations of the relict cosmological constant \cite{Rie-etal:2004:ASTRJ2:}, for which any connection with real observations is excluded.

\clearpage

\section{Light escape cones and KdS superspinar shadows in the LNRFs} \label{sec_LECs_shadows}

In this section, we present the resulting LECs/shadows for some selected KdS superspinar (NS) spacetimes that belong to Classes IV, V, with astrophysically acceptable values of the cosmological parameter. However, in order to highlight the optical effects, we use a cosmological parameter many orders of magnitude above the realistic values, which qualitatively does not change the result. Emitters' positions are chosen to give qualitatively different resulting LECs. 

The shape of the LEC for a given spacetime depends on the latitudinal and radial position of the emitter.  We present the construction of the LEC for the special latitudinal positions, namely on the spin axis ($\theta_{e}=0\dgr$), in the equatorial plane ($\theta_{e}=90\dgr$), and we choose characteristic latitudinal coordinates of $\theta_{e}=60\dgr$ or $\theta_{e}=70\dgr$ representing a general position of the observer, depending on the better representation of the shadow structure. In order to clearly display a continuous progression of the shape of the LEC as the latitude increases, some specific latitudes are used in addition to those mentioned above. 

The LECs are constructed in greatest detail for accordingly chosen radii in the case of the spacetimes of Class IVa, which has the most complex structure of null geodesics according to the radial photon motion. The outstanding radii, which give some qualitative changes of the LECs, are presented in the attached tables \ref{tab_IVa_orbs}-\ref{tab_V_orbs}.

\subsection{Directional angles of photons in LNRFs}

Using the locally measured photon four-momentum components $k^{(a)}$, the standard ($\alpha, \beta, \gamma$) directional angles can be introduced using the relations (\cite{Stu-Sche:2010:CQG:}) 
\bea
\cos \alpha&=&k^{(r)}/k^{(t)},\label{cosalpha}\\
\sin \alpha \cos \beta&=&k^{(\theta)}/k^{(t)},\label{sinalphacosbeta}\\
\sin \alpha \sin \beta&=&\cos \gamma=k^{(\phi)}/k^{(t)}. \label{sinalphasinbeta}
\eea

The frame and the coordinate components are related by the general formula
\be
k^{(a)}=\omega^{(a)}_{\mu}k^{\mu},\label{k(a)}
\ee
where the differential 1-forms of the LNRFs are given by relations
\bea
\omega^{(t)}&=&\sqrt{\frac{\Delta_{r} \Delta_{\theta} \rho^2}{I^2 A}} \din t, \\
\omega^{(r)}&=&\sqrt{\frac{\rho^2}{\Delta_{r}}} \din r,\\
\omega^{(\theta)}&=&\sqrt{\frac{\rho^2}{\Delta_{\theta}}} \din \theta,\\
\omega^{(\phi)}&=&\sqrt{\frac{A \sin ^2 \theta}{I^2 \rho^2}}(\din \phi-\Omega_{LNRF} \din t) \label{1-forms} 
\eea
and the coordinate four-momentum components $k^{\mu}=\oder{x^{\mu}}{\lambda}$, $x^{\mu}\in
\{t,r,\theta,\phi\}$ are given by the equations (\ref{CarterT})-(\ref{W(theta)}).

For completeness, we present the orthonormal tetrad of basis vectors:
\bea
e_{(t)}&=&\sqrt{\frac{I^2A}{\Delta_r \Delta_{\theta} \rho^2}}\left( \frac{\partial}{\partial t}+\Omega_{LNRF} \frac{\partial}{\partial \phi}\right) ,\\
e_{(r)}&=&\sqrt{\frac{\Delta_{r}}{\rho^2}}\frac{\partial}{\partial r},\\
e_{(\theta)}&=&\sqrt{\frac{\Delta_{\theta}}{\rho^2}}\frac{\partial}{\partial \theta},\\
e_{(\phi)}&=&\sqrt{\frac{I^2 \rho^2}{A \sin^2\theta}}\frac{\partial}{\partial \phi}. \label{e_i}
\eea

Here $\Omega_{LNRF}$ is the angular velocity of the LNRF relative to distant static observers, which is given by the relation 
\be
\Omega_{LNRF}=-\frac{g_{t\phi}}{g_{\phi \phi}}=\frac{a[\Delta_{\theta}(r^2+a^2)-\Delta_r]}{A}.\label{OLNRF}
\ee    

Through the Eqns. (\ref{cosalpha})-(\ref{1-forms}) and the associated relations, there is a correspondence between the triad of the directional angles  ($\alpha, \beta, \gamma$) and the double of the motion constants ($X, q$). In fact, any two directional angles from this triad are sufficient; the standard choice is the pair ($\alpha, \beta$). The construction of the LECs can thus be regarded as a transformation of the ($X-q$) plane of the motion constants onto the directional angles \cite{Stu-Char-Sche:2018:EPJC:}. To give an insight into the notion of the motion constants, we present the corresponding ($X-q$) - plane together with each constructed LEC.

\subsection{Representation of the LECs}

In the present paper, we represent the LECs as two-spheres represented in 3D, following the work \cite{Stu-Char-Sche:2018:EPJC:}, where we were more concerned by their performance in 2D-mode. These spheres consist of the endpoints of photon's three-momentum vector of unit magnitude $k^{(t)}=1$ located at its centre, where the emitter resides; the directional angles are given by the relations (\ref{cosalpha})-(\ref{sinalphasinbeta}) corresponding to all the motion constants $(X,q)$ allowing the motion iat the considered emitter's location. This embodiment allows to show more details of a given situation without distortion, but the disadvantage is that we need two views for each situation so that no hidden parts of the spheres remain. Therefore, in the figures below we present for each situation the plane of motion constant $(X-q)$ and the corresponding two-sphere shown from two different views.

\subsection{Complementary cones to the LECs}

When constructing the LECs, we can distinguish the photons that escape towards the cosmological horizon from those that do not, thus forming a complementary cone to the LEC. While the concept of the LEC is clear for both black holes and superspinars (NSs), in the case of complementary cones there are differences between the BH, superspinar and NS spacetimes.

In the case of BH spacetimes the situation is much simpler due to the presence of the BH event horizon, which acts as a one-way membrane, and due to the absence of stable SPOs and related bound photon orbits -- the photon trajectories in BH spacetimes can have at most one radial turning point. The total light flux from a light emitter consists of photons escaping to the cosmological horizon, either directly or after passing through some pericentre, forming thus the LEC, and photons absorbed by the black hole, which again reach the event horizon either directly or after passing through some apocentre; these photons form the complementary cone to the LEC, i.e., the captured cone. The boundary between these cones is given by the photons emitted towards the unstable SPOs.

\subsubsection{Classification of the complementary cones in superspinar (NS) spacetimes} \label{ssec_class_compl_cones}

In the case of the superspinar (NS) spacetimes, the notion of the LECs is the same as in the case of BH spacetimes. However, the cone complementary to the LEC has a more complex character even in the case of KdS superspinars, where the surface is assumed to serve as a one-way membrane, similar to the BH spacetimes, because of the existence of stable SPOs. The case of the KdS NSs is even more complex, as there is no surface serving as a one way membrane. We propose the following classification of the complementary cones to the LECs in the case of superspinars and NSs. 

KdS superspinars:
\begin{itemize}
	\item 'trapped' cones, formed by bound trajectories of photons that oscillate between two radial turning points $r_{1}$, $r_{2}$, $0<r_{1}<r_{2}<+\infty$, i.e., the pericentre and apocentre, around some stable SPO;
	in this case, the light source must be in a region of stable SPOs;
	\item 'engrossed' cones, formed by trajectories terminating on the superspinar surface; 
\end{itemize}
	
KdS NS spacetimes:
\begin{itemize}		
	\item 'repelled escape cones', consisting of photons passing through the interior of the ring singularity at $r=0, 0\leq \theta<\pi/2$ into the region of negative radii and, after passing through a certain radial turning point at $r_{t}<0$, finally repelled back to $r>0$;	
	\item  'oppositely escape cones' composed of photons passing through the interior of the ring singularity at $r=0, 0\leq \theta<\pi/2$ into the region of negative radii and continuing to the other infinity $r \to -\infty$. 	
\end{itemize}

In the present article we will not discuss the last two classes (NS spacetime) of light cones in detail, as we have covered them in \cite{Stu-Char:2024:PHYSR4:}. Here we assume that light does not reach the negative radius region, since the surface of the superspinar prevents it. 

\subsubsection{Negative energy light cones}

It should be added that there is another criterion for classifying photon orbits according to the sign of their covariant energy $E$. As we have found in \cite{Char-Stu:2017:EPJC:}, in each KdS spacetime there exist orbits, where the photons have positive impact parameter $X$, but appear to move in retrograde direction as seen in the LNRFs. These orbits must be located in the ergosphere, and the photons must have negative covariant energy $E<0$. \footnote{Note that such photons play a crucial role in the new version of the Penrose process of extraction of energy from rotating BH or superspinars, introduced in \cite{Kol-Tur-Stu:2021:PHYSR4:,Stu-Kol-Tur-Gal:2024:Universe:}.} In \cite{Char-Stu:2017:EPJC:} we found that their impact parameter $X$ must satisfy $X>X_{+}(r_{e};q\geq0)$. In spacetimes with DRB of the radial photon motion, i.e., in the case of Classes IV,V, there are two spatially separated ergoregions (see Fig. \ref{Fig_ergos}). The inner ergoregion is occupied by the trapped negative energy photons, which we have shown in Fig. \ref{Fig_eff_pots1}(a), (b) as the blue part of the ($X-r$) plane, while the cosmological ergoregion (see subsection \ref{ssec_ergos}) is occupied by the negative energy photons escaping towards the cosmological horizon, which we have depicted as the green part of the ($X-r$) plane in Fig. \ref{Fig_eff_pots1}(c), (d).

In spacetimes with RRB of the radial photon motion, i.e., in the case of Classes VI - VIII, the ergoregion spreads along whole equatorial plane. For $r_{e}<r^{-}_{d(\spo)}$ there are photons with $E<0$ which can be trapped or escape, for $r^{-}_{d(\spo)}<r_{e}<r_{c}$ there exist only escaping photons with $E<0$ (see Fig. \ref{Fig_eff_pots1}(c), (d)).

\subsection{Description of the light cones}

We give the structure of the light cones for the KdS superspinars, and for comparison we also include the parts corresponding to the KdS NSs, where the photon can reach the region of negative radii. 
  
\begin{figure*}[h]
	\centering
	\begin{tabular}{|c|c|}
		
		\hline
		\multicolumn{2}{|c|}{$r_{e}=2.65$}\\
		\hline
		\multicolumn{2}{|c|}{}\\
		\multicolumn{2}{|c|}{	\includegraphics[width=0.4\textwidth]{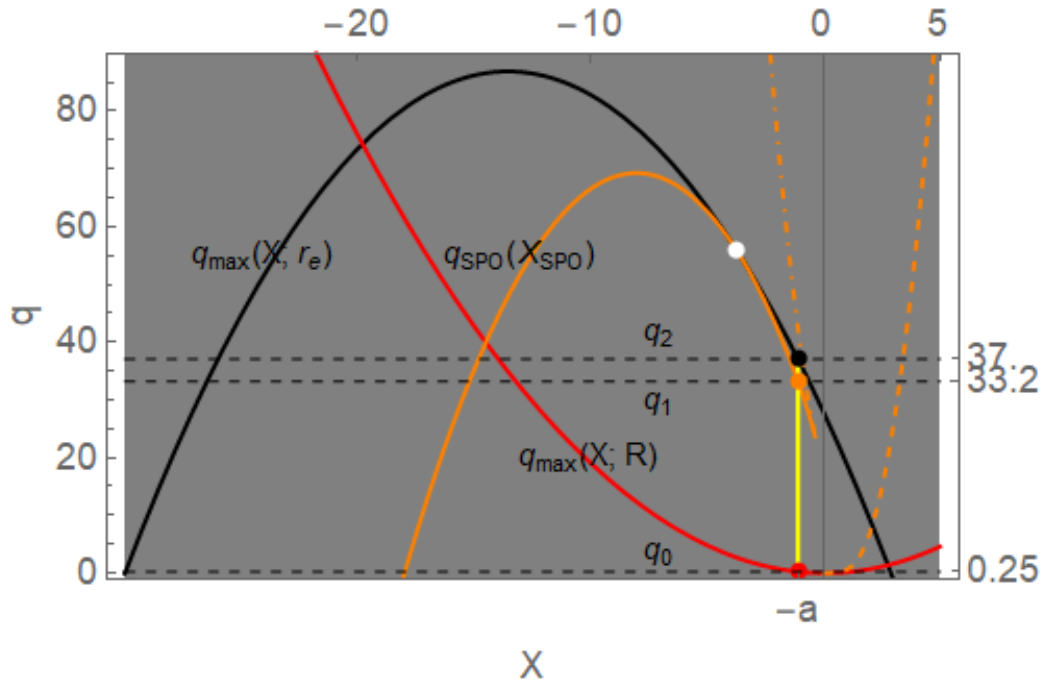}\includegraphics[width=0.4\textwidth]{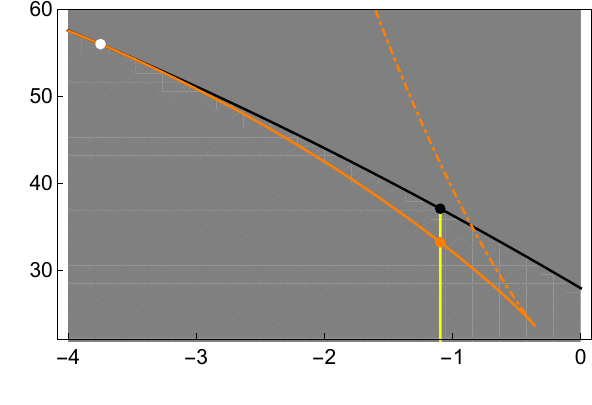}}\\
		\multicolumn{2}{|c|}{(a): $\theta_{e}=0\dgr$}\\
		\hline
		\multicolumn{2}{|c|}{\includegraphics[width=0.4\textwidth]{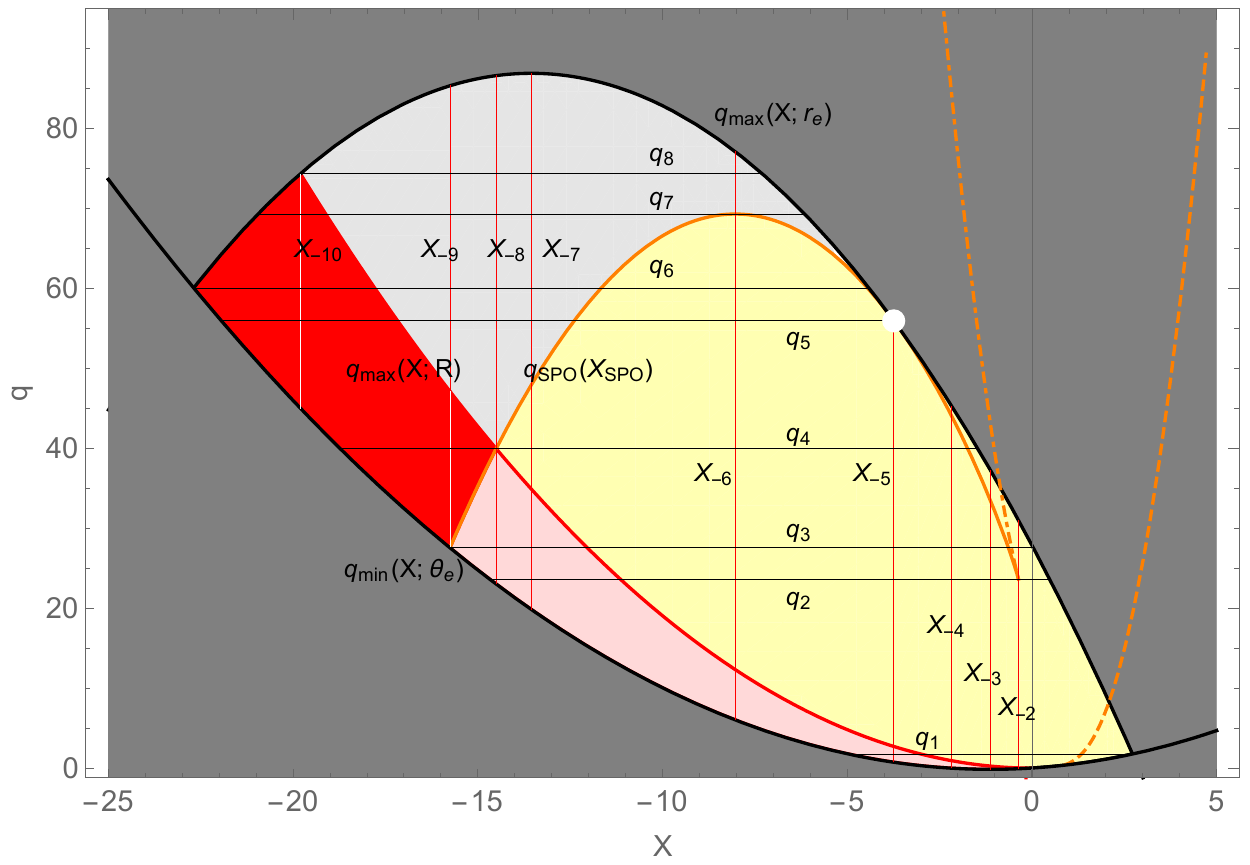}	\includegraphics[width=0.4\textwidth]{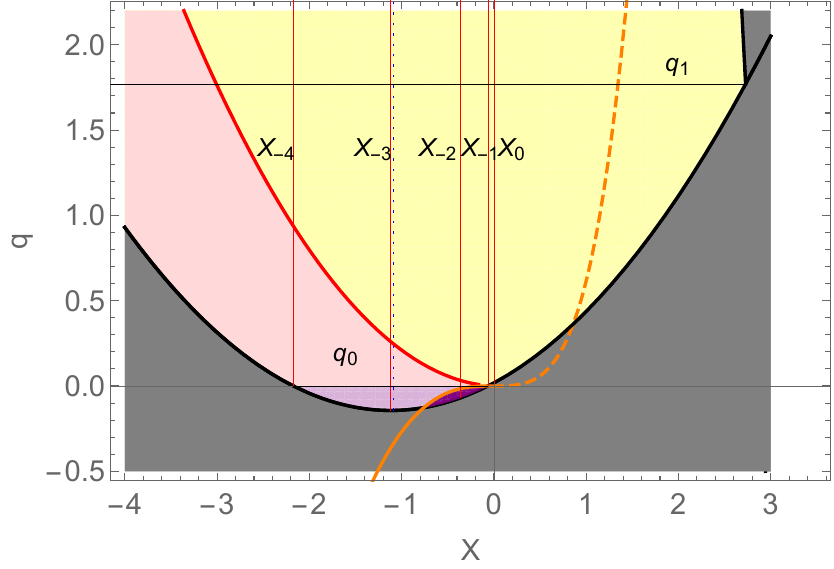}}\\
		\multicolumn{2}{|c|}{(b): $\theta_{e}=70\dgr$}\\
		\hline
		\multicolumn{2}{|c|}{\includegraphics[width=0.4\textwidth]{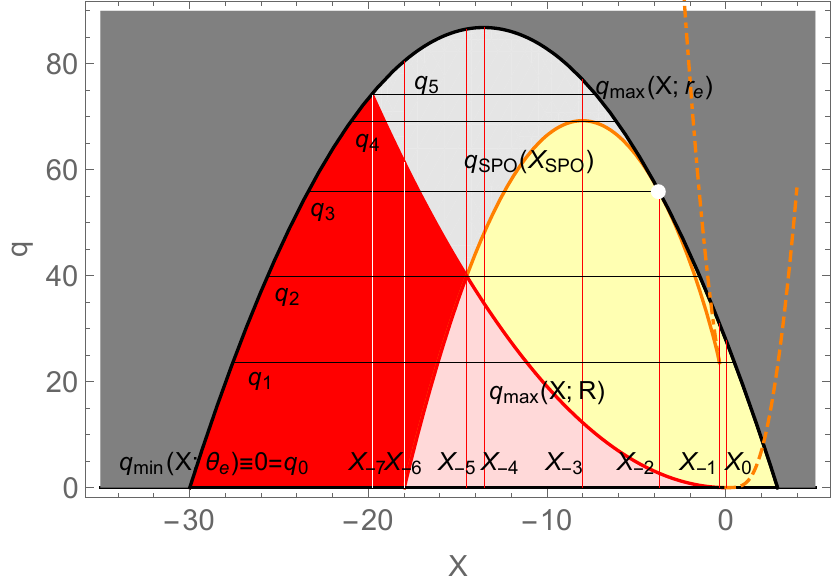}}\\
		\multicolumn{2}{|c|}{(c): $\theta_{e}=90\dgr$}\\
		\hline
	\end{tabular}	
	\caption{Planes of the motion constants $(X-q)$ for the case $a^2=1.2$, $y=0.02$, $r_{e}=2.65$. The latitudinal coordinates correspond to special cases on the spin axis, in the equatorial plane and in the general position. The radial coordinate corresponds to case $r^{+}_{erg}<r_{e}<r^{-}_{ph}$, i. e. the region of unstable counterrotating SPOs with positive covariant energy $E>0$ outside the ergosphere, hence, the function $q_{max}(X;r_{e})$ is now concave. The label $X_{-3}$ in the detail on the right denotes two close lines, namely, $X=-1.116$ (red), which is local minimum of $q_{min}(X;\theta)$ and $X=a=-1.095$, i. e., $\ell=0$ (blue, dotted).   
	}\label{Fig_Xq_IVa_re_2.65}
\end{figure*}

The escape cones in Figs. \ref{Fig_cones_IVa_1.25_0_2p4deg}-\ref{Fig_cones_V_2.5_60deg} are shown as the yellow part of the two-sphere, which we can regard as the emitter's local celestial sphere (LCS), the trapped cones are shown in grey. The blue parts correspond to the trapped photons with negative covariant energy $E<0$, which occurs when the source is in the inner ergosphere. The green parts depict the photons with negative covariant energy escaping towards the cosmological horizon when the source is in the outer ergosphere. The boundary between the trapped and escape light cones, given by a beam directed towards the unstable spherical photon orbits, i. e. with motion constants $X_{\spo}$, $q_{\spo}$, is illustrated by full or dashed orange curves. The dashed curve denotes that the photon has a radial turning point before it reaches the unstable SPO. The darker purple parts correspond to the escape cones to the other infinity, the lighter purple parts correspond to the repelled photons with radial turning points at negative radii $r<0$. The engrossed cones are depicted in red; in Figs. \ref{Fig_cones_IVa_1.25_0_2p4deg}-\ref{Fig_cones_V_2.5_60deg} the surface radius of the superspinar is set to r=0.1.

Note that the colouring in Figs. \ref{Fig_cones_IVa_1.25_0_2p4deg}-\ref{Fig_cones_V_2.5_60deg} is chosen to match the colouring in the attached ($X-q$) planes and the corresponding parts in the ($X-r$) plots in Fig. \ref{Fig_eff_pots1}.

\subsection{Dependence of LEC's shape on the emitter's position}  

\subsubsection{Latitudinal position on the spin axis}
\begin{figure*}[h]
	\centering \textbf{Class IVa: $y=0.02,\quad a^2=1.2$}\\
	
	\begin{tabular}{|c|c|}
		\hline
		\multicolumn{2}{|c|}{$r_{e}=0.9$}\\
		\hline \\
		\raisebox{2cm}[0pt]{\bet{c}	\includegraphics[height=3.5cm]{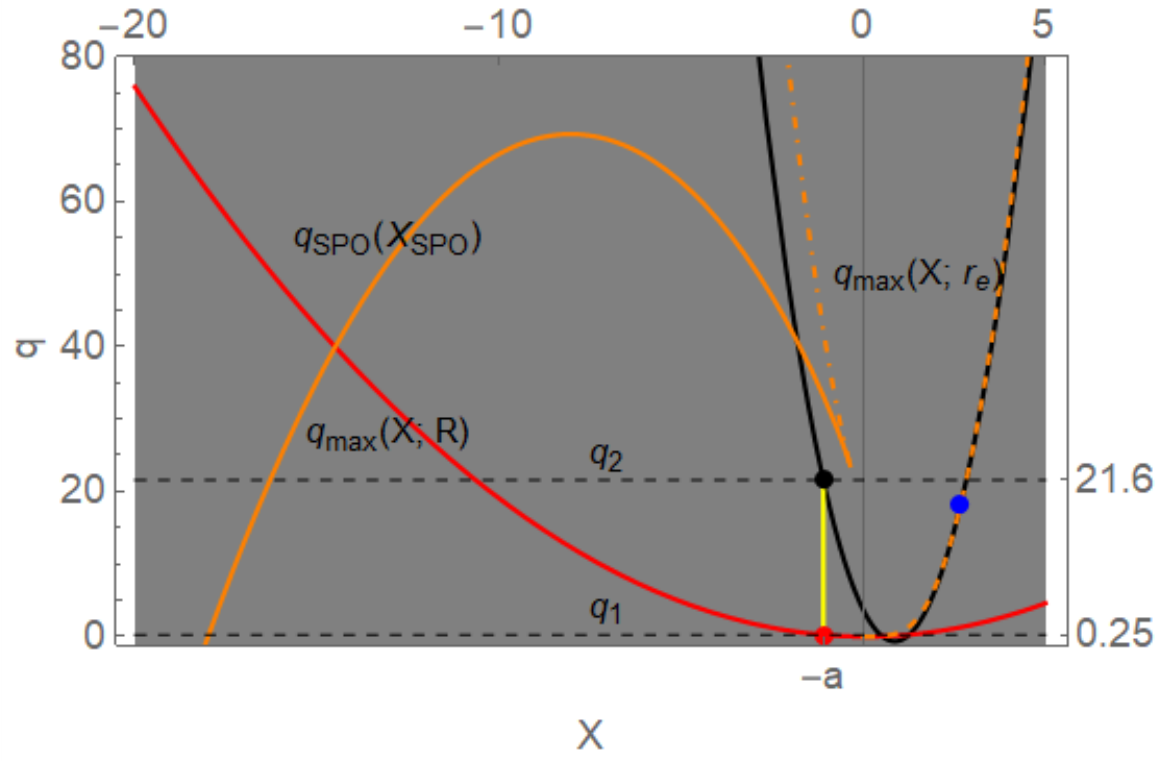}\\ \includegraphics[height=3.5cm]{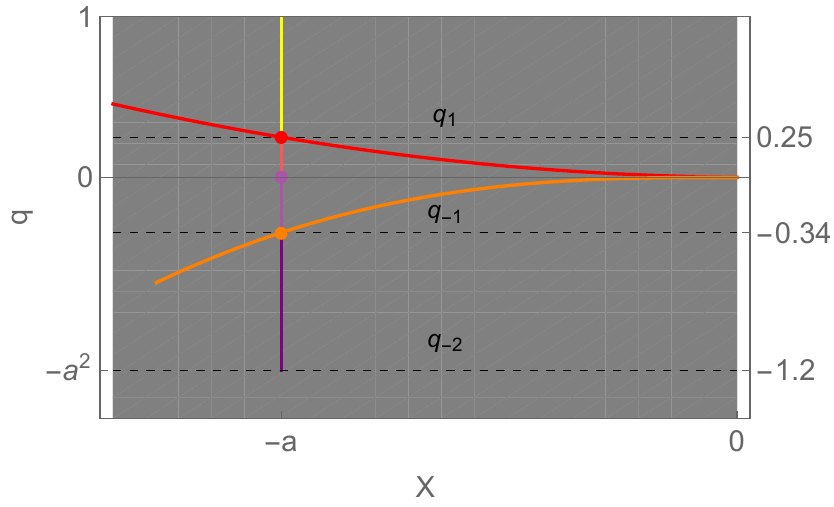} \ent }&\includegraphics[width=0.4\textwidth]{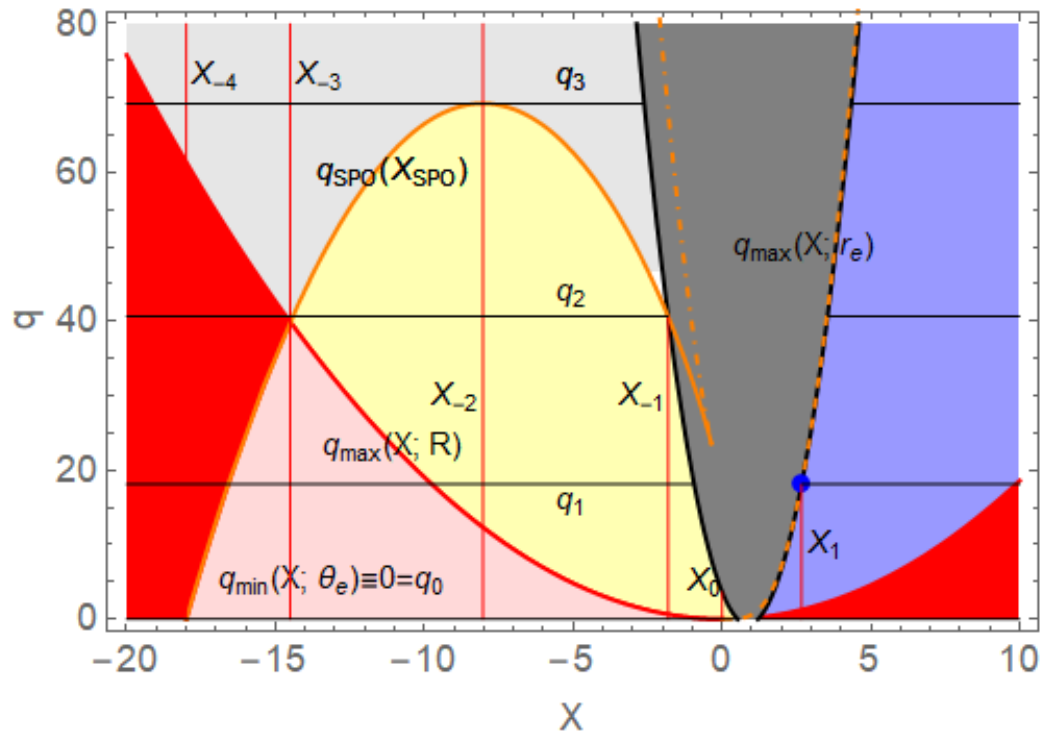}\\
		(a): $\theta_{e}=0\dgr$&(b): $\theta_{e}=90\dgr$\\
		\hline
		\multicolumn{2}{|c|}{}\\
		\multicolumn{2}{|c|}{\includegraphics[width=0.4\textwidth]{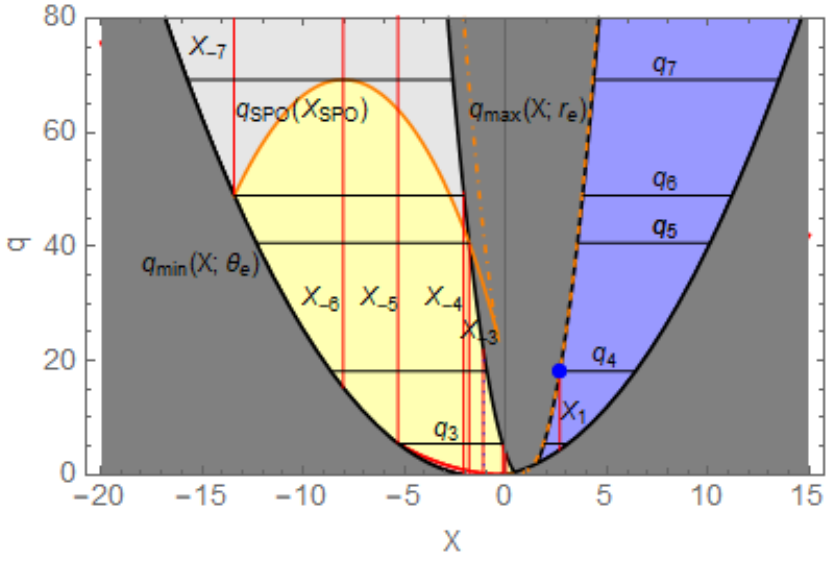} \includegraphics[width=0.4\textwidth]{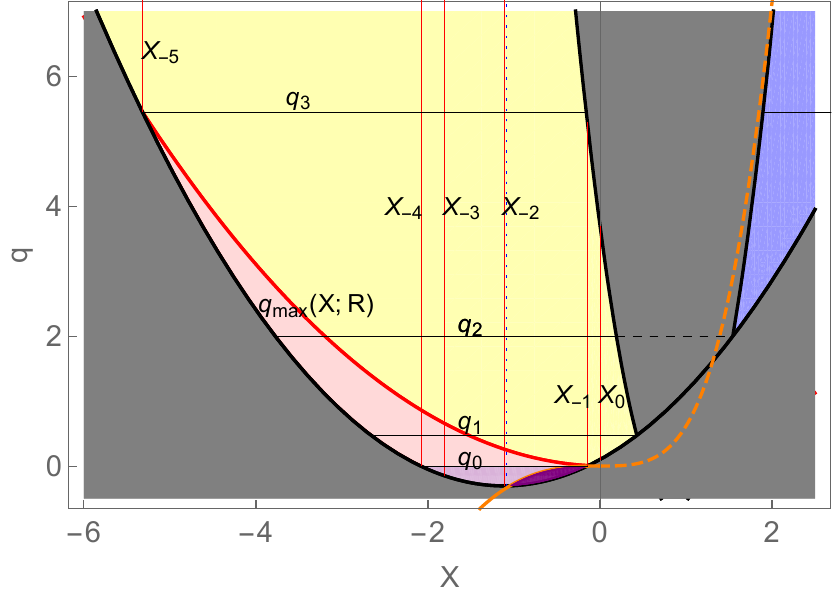}}\\
		\multicolumn{2}{|c|}{(c): $\theta_{e}=60\dgr$}\\
		\hline
	\end{tabular}	
	\caption{Planes of motion constans ($X-q$) for the displayed latitudinal coordinates $\theta_{e}$ corresponding to special location on the spin axis, in the equatorial plane and one general location in between. The chosen radial coordinate fulfils $r_{e}<r^{+}_{d(sph)}$, which corresponds to the region of the inner ergosphere with stable locally counterrotating SPOs  with negative covariant energy $E<0$. The colouring reflects different properties of the photon motion in the same manner as in Fig. \ref{Fig_eff_pots1}: yellow parts distinguish the motion constants of photons escaping  to $r \to +\infty$, either directly, or with one turning point if inwards directed; light grey parts correspond to the captured photons with positive covariant energy $E$, blue parts correspond to the captured photons with negative covariant energy $E<0$, dark grey parts mark the forbidden regions due to the reality conditions (\ref{W>0}), (\ref{R>0}). For $q<0$, the light purple part corresponds to photons entering the region of negative radii $r<0$ and repelled back to positive radii, the dark purple part corresponds to photons escaping to $r \to -\infty$. The light red part corresponds to photons engrossed by the surface of the superspinar if they are initially directed inwards, or to escaping photons if they are initially directed outwards. The rich red part corresponds to photons absorbed by the surface of the superspinar independent of their initial radial motion. Full orange curves designate the critical locus corresponding to the unstable SPOs, dashed orange curve corresponds to the stable SPOs with $E<0$, dot-dashed curve corresponds to the stable SPOs with $E>0$. The stable SPOs are always hidden for photons by the potential barrier, the only exception being the photon emitted with the motion constants highlighted by the blue point, which denotes the touch point of the displayed curves. It stays on the stable SPO with the current radius $r_{e}$. For $\theta_{e}=0\dgr$, the condition (\ref{W>0}) implies $X=-a$, hence the allowed region has shrunk to an abscissa $X=-a$, $-a^2\leq q \leq q_{max}(-a;r_{e})$. For a better understanding of the meaning of the motion constants, Figs. \ref{Fig_cones_IVa_0.9}, \ref{Fig_cones_IVa_2.65_90deg} explicitly show the photon directions corresponding to the prominence curves $X=const.$ and $q=const.$, marked here by the red vertical lines and the black horizontal lines. Note that with the same labels they can have different meanings for different latitudes $\theta_{e}$.  It should be emphasised that here, and in all other figures below, the surface of the superspinar at radius $\calr=0.1$ is considered.      
	}\label{Fig_Xq_IVa_re_0.9}
\end{figure*}

 If the emitter is located on the spin axis, then all emitted photons have the impact parameter $X=-a$ $(\ell=0)$ and the motion constant $q$ can take the values $-a^2\leq q \leq q_{max}(X=-a;r_{e})$ (see Figs. \ref{Fig_Xq_IVa_re_0.9}(a), \ref{Fig_cones_IVa_1.25_0_2p4deg}(a), \ref{Fig_cones_IVa_1.8_6.4deg}(a), \ref{Fig_cones_IVa_2.2_3.3deg}(a), \ref{Fig_Xq_IVa_re_2.65}(a), \ref{Fig_cones_IVa_5.75_90deg}(a), \ref{Fig_cones_VIb_rs_22}(a). The inward directed photons with $-a^2\leq q < 0$ enter the region of negative radii $r<0$, from which the photons with $-a^2\leq q\leq q_{\spo}(r_{\spo}<0)$ escape to the other infinity ($r\to -\infty$), photons with $q_{\spo}(r_{\spo}<0)<q<0$ have the radial turning point at $r_{t}<0$ and are repelled back to the $r>0$-region, photons with $q=0$ hit the ring singularity. For $0<q\leq q_{max}(X=-a;R)$, the radiation is engrossed by the superspinar surface at $r=R$. Other photons may escape or be trapped depending on the radial position with respect to the radii of the outer polar unstable SPO at $r^{-}_{pol}$, if it exists.
 
In Figs \ref{Fig_Xq_IVa_re_0.9}(a), \ref{Fig_cones_IVa_0.9}(a) the outstanding values of the motion constant $q$, i.e. $-a^2$,  $q_{\spo}(r_{\spo}<0)$, $0$, $q_{max}(X=-a;R)$, $q_{max}(X=-a;r_{e})$ are denoted as $q_{-2}$, $q_{-1}$, $q_{0}$, $q_{1}$, $q_{2}$ respectively.
 
If $r_{e}<r^{+}_{pol}$ is small enough such that $q_{max}(X=-a;r_{e})<q_{\spo}(X_{\spo}=-a)$ (Fig. \ref{Fig_Xq_IVa_re_0.9}(a) left) or there are no polar SPOs (Fig. \ref{Fig_cones_VIb_rs_22}(a) left), then all photons with $q>0$ escape (Figs. \ref{Fig_cones_IVa_0.9}(a), \ref{Fig_cones_VIb_rs_22}(a) right).
 
For larger values of $r_{e}$, but still satisfying the condition $r_{e}<r^{-}_{pol}$, such that $q_{max}(X=-a;r_{e})>q_{\spo}(X_{\spo}=-a)$ (cf. black and orange points of the $(X-q)$ motion constant plane in Figs. \ref{Fig_cones_IVa_1.25_0_2p4deg}(a), \ref{Fig_cones_IVa_1.8_6.4deg}(a)), there are trapped cones, which are depicted as a spherical belt. They are delimited by photons emitted directly outwards to $r^{-}_{pol}$, represented by full orange circles, and indirectly inwards with radial turning point $r_{t}$, $r_{t}<r_{e}$, which are represented by the dashed orange circles (Figs. \ref{Fig_cones_IVa_1.25_0_2p4deg}(a), \ref{Fig_cones_IVa_1.8_6.4deg}(a)). The trapped cones then exist for any latitude.
  
If $r_{e}>r^{-}_{pol}$, there are no trapped photons, but there are inward directed photons terminating on the outer polar SPO  (see the orange circle in Figs. \ref{Fig_cones_IVa_2.2_3.3deg}(a), \ref{Fig_Xq_IVa_re_2.65}(a), \ref{Fig_cones_IVa_2.65_90deg}(a)).
 
In the spacetimes without polar SPOs, all photons either escape or are engrossed by the superspinar surface, independently of the radial position of the emitter (see Fig. \ref{Fig_cones_VIb_rs_22}a). 
 
 \subsubsection{General latitudinal position}
  
Depending on the latitude of the emitter, the trapped cones can take the shape of a spherical belt (see Fig. \ref{Fig_cones_IVa_1.25_0_2p4deg}, \ref{Fig_cones_IVa_1.8_6.4deg}a), bi-angle (\ref{Fig_cones_IVa_1.25_5_66deg}a, \ref{Fig_cones_IVa_1.8_6.4deg}b, c), spherical canopy (Fig. \ref{Fig_cones_V_2.5_60deg}), or cover substantial part of the emitter's sky on a hemisphere oriented in the direction opposite to that of the LNRF's motion (see Figs. \ref{Fig_cones_IVa_0.9}(b), \ref{Fig_cones_IVa_1.25_5_66deg}b, \ref{Fig_cones_IVa_1.25_90deg}, \ref{Fig_cones_IVa_2.2_90deg}). The boundary of the cone is made up of two lines, one corresponding to photons with no radial turning point before reaching the unstable SPOs, which we present as a full orange curve, and the other corresponding to photons with one radial turning point, which we present as a dashed orange curve.
 
If the emitter is in the region of stable SPOs, i.e., under the radii $r_{ms}$ in spacetimes of Classes IV-VI, or under the radii $r^{max}_{\spo}$ in the case of Classes VII, VIII (see the comment to Fig. \ref{Fig_eff_pots1}), then all the unstable SPOs are located above the emitter, and for every outward photon that directly reaches an unstable SPO, there exists an inward photon with the same motion constants $X_{\spo}$, $q_{\spo}$ that reaches the same orbit after passage through some radial turning point. The two lines forming the boundary of the trapped cone are then placed symmetrically on the emitter's sky with respect to the $(\mathbf{e_{\theta}}-\mathbf{e_{\phi}})$-plane of his local vector basis (see Figs. \ref{Fig_cones_IVa_1.25_0_2p4deg}-\ref{Fig_cones_IVa_1.25_90deg}). The photon sent on the actual stable SPO, thus maintaining the actual radius $r_{e}$, is represented by a blue dot. 

If the emitter is located in the region of unstable SPO's, i.e. at $r_{ms}<r_{e}<r^{-}_{ph}$ for Class IVa, some unstable SPO's are located below the emitter's location and can only be reached directly by photons emitted in the radial inward direction, while the unstable SPO's located above can be reached directly by outwardly directed photons, or indirectly by inwardly directed photons, as described above. The full orange curve then exceeds the dashed one and forms an arc (see Figs. \ref{Fig_cones_IVa_1.8_6.4deg}c, \ref{Fig_cones_IVa_2.2_3.3deg}c, \ref{Fig_cones_IVa_2.2_90deg}, \ref{Fig_cones_IVa_2.65_90deg}). The trapped cone appears for latitudes exceeding some critical value $\theta_{max(circ)}$ and has a shape of spherical bi-angle. The tips of the bi-angles, depicted as white dots, represent the directions of photons sent into the currently occupied unstable SPO.
 
For the location of the emitter above the region of unstable SPOs, i.e. at $r_{e}>r^{-}_{ph}$, only inwardly directed photons can reach the unstable SPOs, hence the dashed orange line and the trapped cone disappear. The full orange curve then forms a circle for $0\leq \theta_{e}<\theta_{max(circ)}$ or an arc for $\theta_{max(circ)}\leq \theta_{e}\leq \pi/2$ about the engrossed cone (see Figs. \ref{Fig_cones_IVa_4_90deg}, \ref{Fig_cones_IVa_5.75_90deg}).
  
We can emphasize that in the spacetimes with the polar SPOs there exist some maximum angle $\theta_{max(circ)}$ for which the circular edges of the trapped cones making the bi-angles disconnect (Figs. \ref{Fig_cones_IVa_1.25_5_66deg}a, \ref{Fig_cones_IVa_1.8_6.4deg}a, \ref{Fig_cones_IVa_2.2_90deg}c), \ref{Fig_cones_IVa_2.2_3.3deg}c, or the circle around the engrossed cone splits into an arc (c.f. Figs. \ref{Fig_cones_IVa_4_90deg}a,b). 
 
In the spacetime with no polar SPOs, there exists some minimum angle $\theta_{min(arc)}$ which must be exceeded for occurrence of the arc (c.f. Figs.\ref{Fig_cones_VIb_rs_22}b,c). 
   
Since the LECs are closely related to the superspinar shadows, it is more interesting from an astrophysical point of view that the angle $\theta_{max(circ)}$ can be regarded as a maximum angle for which an observer in a spacetime with the polar SPOs at $r_{o}>r^{-}_{ph}$ can see a distortion of the celestial background along a circular contour on his celestial sphere, which changes to an arc at larger latitudes. Observing the circle can thus give upper bounds either on the latitude of the observer or on the spin of the observed object for a known value of the second parameter.
 
Accordingly, the angle $\theta_{min(arc)}$ can be regarded as the minimum angle at which an observer at $r_{o}>r^{-}_{ph}$ can see the arc in spacetimes without polar SPOs. Observing an arc with a small central angle $\xi$ can give a lower limit on the spin parameter $a^2$ (see section \ref{shadows}).
 
Note that in all types of the KdS superspinar (NS) spacetimes there is a limiting angle $\theta_{max(ell)}$ for which the shadow of a superspinar observed by a distant observer, i.e., at $r_{e}>r^{-}_{ph}$, has the shape of an ellipse (Fig. \ref{Fig_cones_IVa_4_90deg}). At larger observer latitudes, the shadow gradually transforms into a wedge shape. 
  
There is an outstanding region of the stable SPOs separated by the forbidden region in the spacetimes of Class V (see Fig. \ref{Fig_cones_V_2.5_60deg}), located under the radii $r^{max}_{\spo}$, while the unstable orbits are in the range as in the Classes VII, VIII. For emitters located at radii $r_{e}\lesssim r^{max}_{\spo}$, the cone of trapped photons is delimited by a circle on the two-sphere, which occurs for latitude greater than a certain minimum, $\theta_{e}>\theta_{min(circ)}$; for $\theta_{e}=\theta_{min(circ)}$ this cone shrinks to an infinitesimal extent in the negative $\phi$-direction. For details about Class V spacetime, see \cite{Char-Stu:2017:EPJC:}.
 
 \subsubsection{Position in the equatorial plane} 
 
In the most interesting case $r_{e}>r^{-}_{ph}$ the photons escaping towards the unstable SPOs always make an arc, which central angle $\xi$ we investigate below. We also investigate the central angle $\eta$ of the wedge-shaped shadow of the superspinar. 
 
In the NS spacetimes, the ring singularity is reached by photons with motion constants $q=0$ and $X_{\spo}(r^{-}_{ph})\leq X \leq 0$. They are depicted in magenta as an arc in the equatorial plane.     

\subsection{Relation between the LECs and the superspinar shadows} \label{shadows}

The LECs constructed in the 3D mode presented below can also be used as a representation of the observer's local sky including the superspinar shadow. The construction of the shadows in the 3D mode would proceed in an analogous way using the analysis of photon motion above. Each incoming photon is associated with a point on the celestial sphere where the associated ray of light would pass through the sphere. Since with the same motion constants $X, q$, the incoming photon would pass through the sphere symmetrically with respect to the apparent image of the spin axis as related to the leaving photon, the emitter's LEC and the observer's celestial sky are symmetric overall. However, what appears to be symmetric from the centre looks as if it were seen from the outside of the sphere. Hence, the below constructed 3D LECs can be viewed as the observer's star globes, depicting his celestial sky viewed from the point of his residence. This is the another advantage of this embodiment.

Whatever is shown on the LECs has a counterpart on the globe. For instance, the engrossed cones correspond to the superspinar shadow, since no photons with the appropriate motion constants can come to the observer from far distances from the corresponding directions. Furthermore, if there were no source of radiation in the directions of the trapped cones at $0<r_{e}<r^{-}_{ph}$, the observer at $r_{o}=r_{e}$ should see a dark sky in directions that are axially symmetric with respect to the apparent image of the spin axis. On the other hand, if there were a source, e.g. hot matter from the accretion disk, there would be a large amount of radiation coming from these directions.

If the observer is located at $r=r_{o}>r^{-}_{ph}$, then he sees an arc as a part of the superspinar shadow. The arc itself is infinitesimally thin, but it is conspicuous due to the distortion of star patterns, which images lie along the arc, as the photons with motion constants $X\approx X_{\spo}$, $q \approx q_{\spo}$ emitted from the far stars must go many times around some unstable SPO corresponding to the arc before they reach the observer.

When constructing the shadow of a superspinar, we consider for comparison the situation where there is a naked singularity in its position. Apart from the arc, there is a direct silhouette of the NS given by the null geodesics connecting the ring singularity and its interior with the observer. The dark spot is seen in the directions where the photons escape to the other infinity (at negative radii sheet), its boundary is determined by photons with the motion constants $X_{\spo}(r_{\spo}<0),q_{\spo}(r_{\spo}<0)$. The null geodesics from the remaining part have a turning point in region $r<0$, so if there is a source of illumination, the region on the observer's sky corresponding to photons coming from the directions between the ring singularity and the boundary of the dark spot could be visible.

\subsection{Relevance of the polar SPOs with construction of the LECs/shadows} \label{ssec_relevance polar SPOs_LECs}

\begin{figure*}[h]
	\centering \textbf{Class IVa: $y=0.02,\quad a^2=1.2$}\\
	\begin{tabularx}{\textwidth}{|XXX|}
		\hline
		\multicolumn{3}{|c|}{$r_{e}=1.8$}\\
		\hline
		\raisebox{2.0cm}[0pt]{\bet{c}	\includegraphics[width=0.25\textwidth]{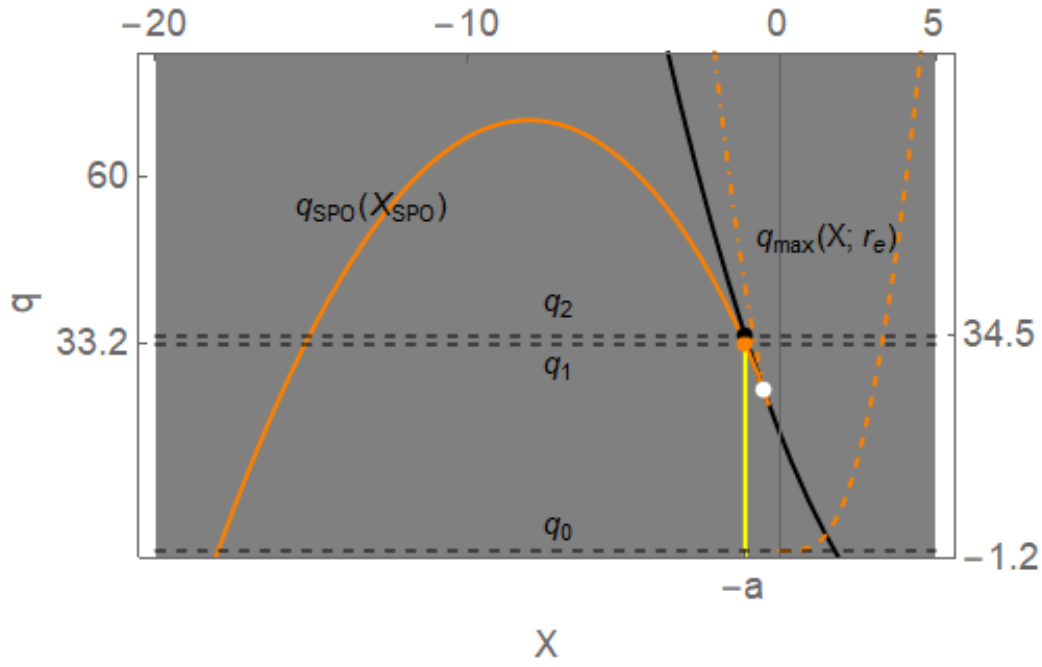}\\ \includegraphics[width=0.25\textwidth]{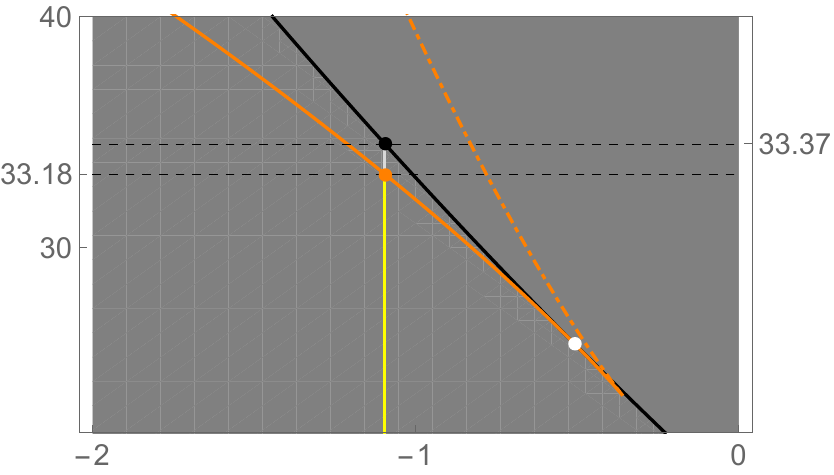} \ent }&\includegraphics[width=0.25\textwidth]{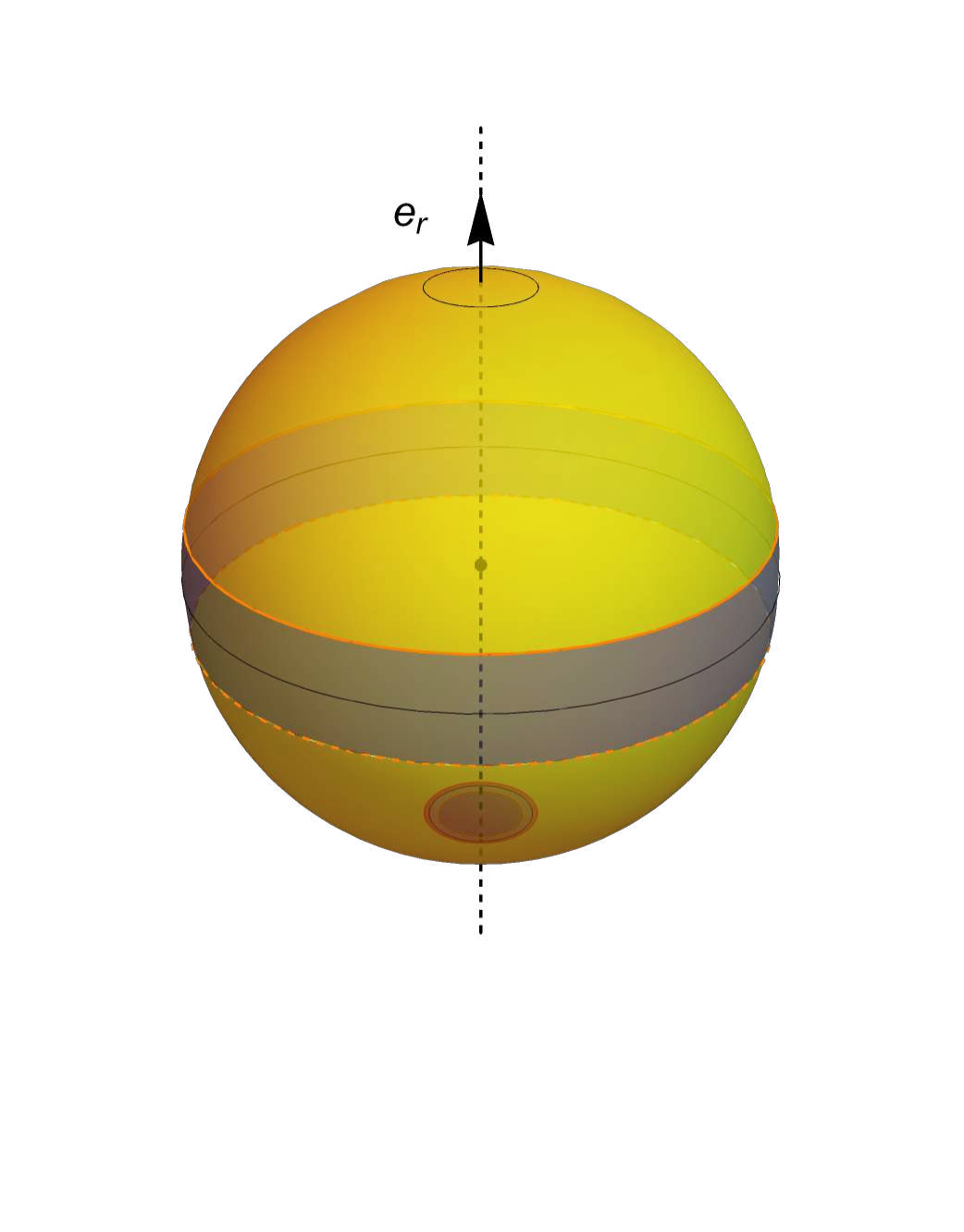}&\includegraphics[width=0.25\textwidth]{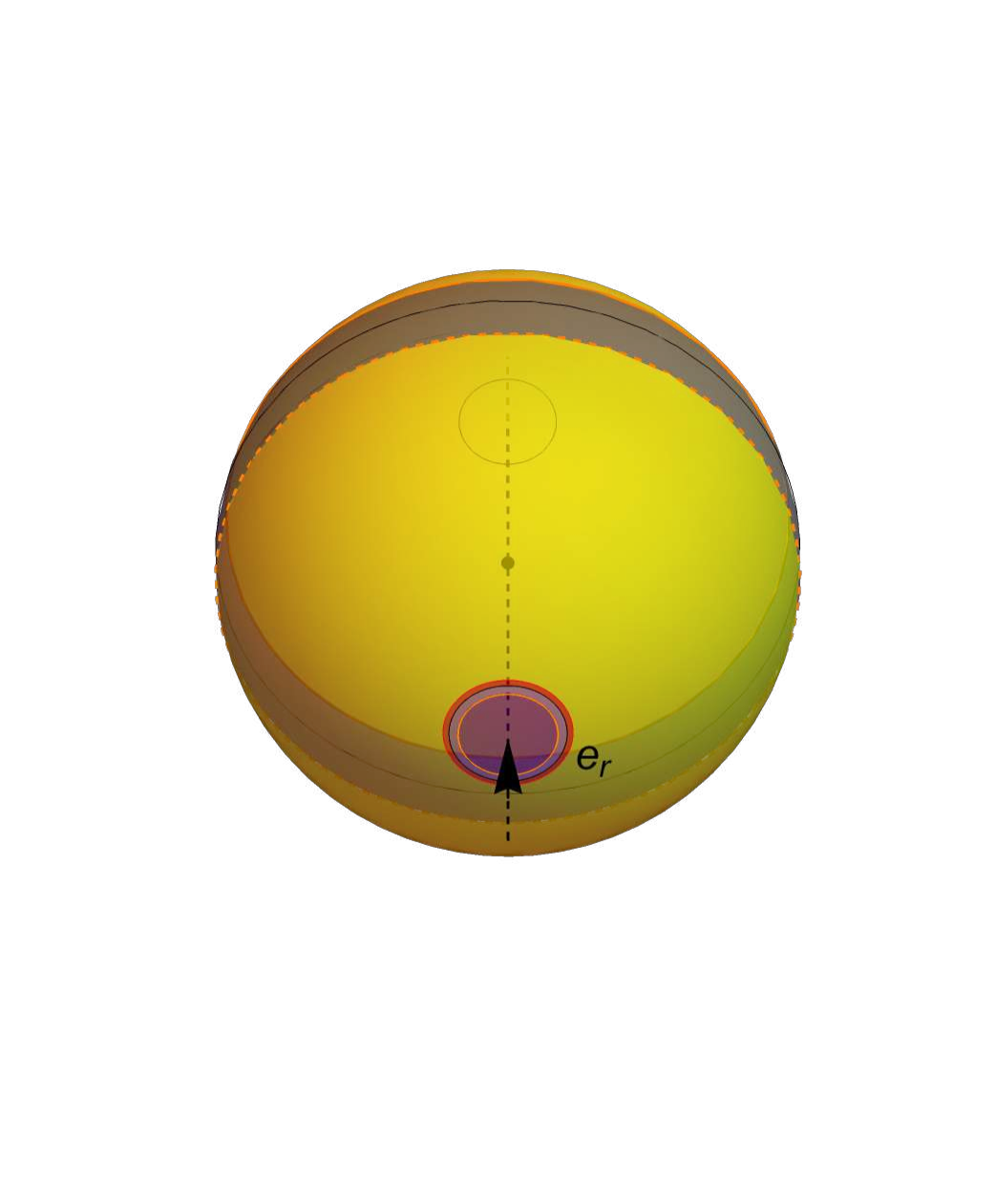}\\
		\multicolumn{3}{|c|}{(a): $\theta_{e}=0\dgr$}\\
		\hline
	
		\includegraphics[width=0.3\textwidth]{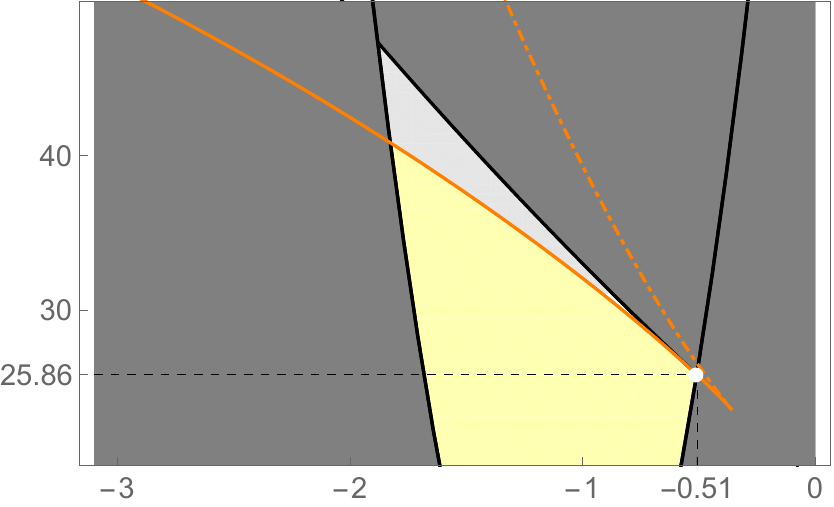}&
		\includegraphics[width=0.3\textwidth]{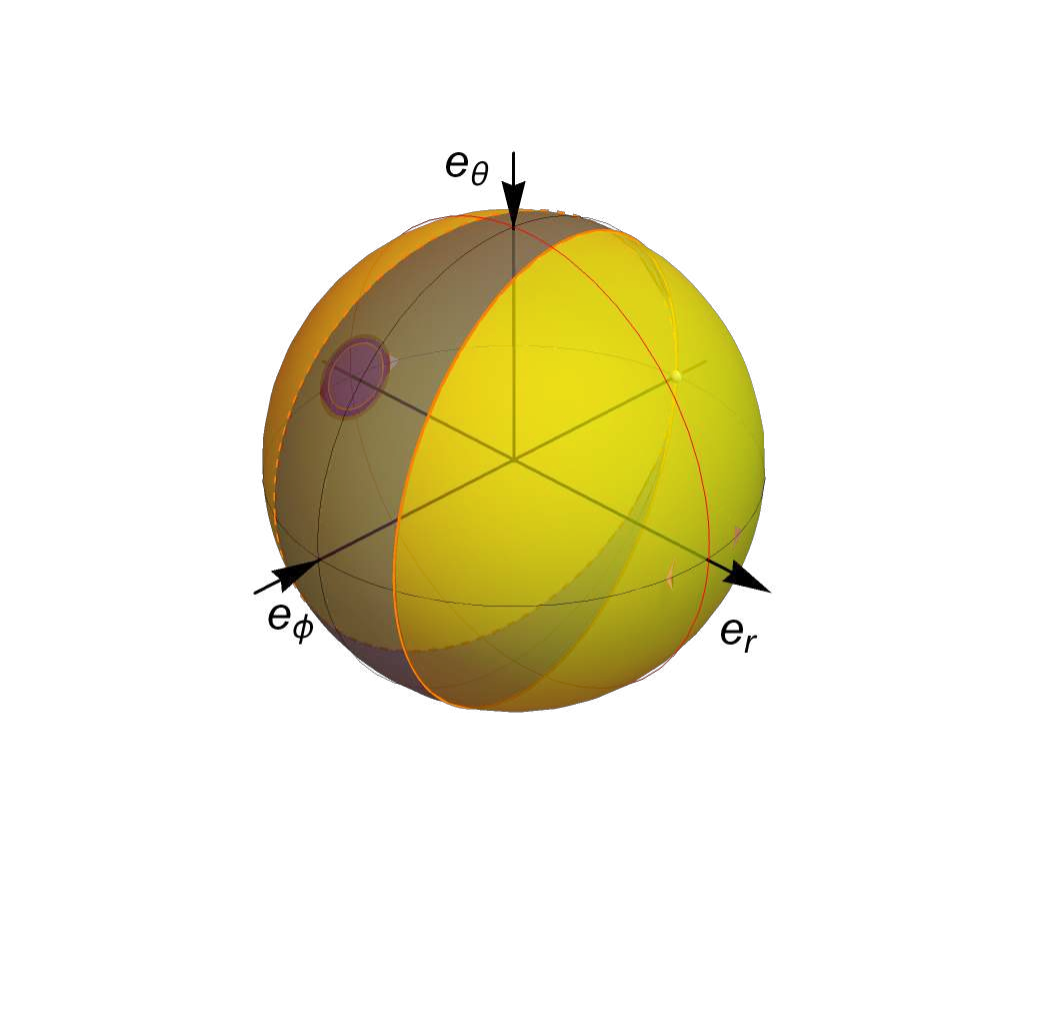}&\includegraphics[width=0.3\textwidth]{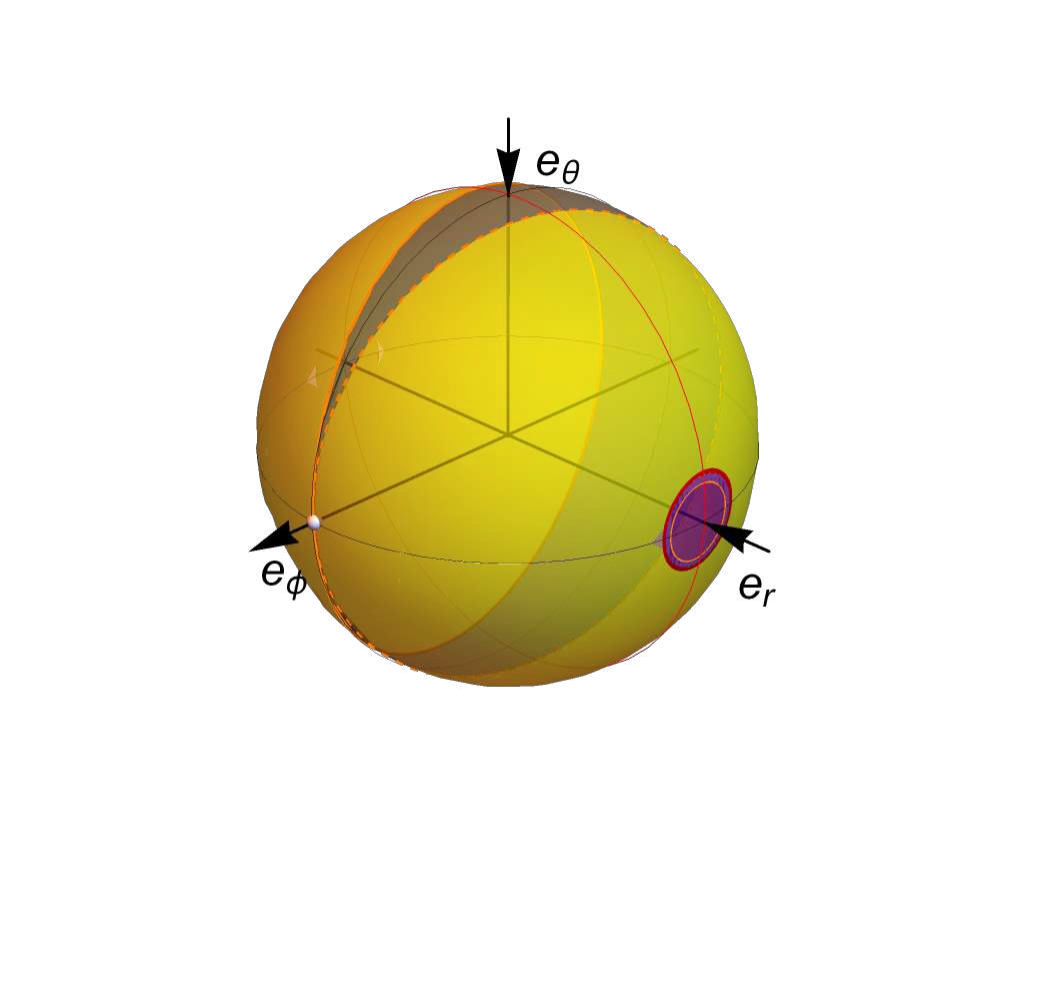}\\
		\multicolumn{3}{|c|}{(b): $\theta_{e}=6.36\dgr$}\\
		\hline
		\includegraphics[width=0.3\textwidth]{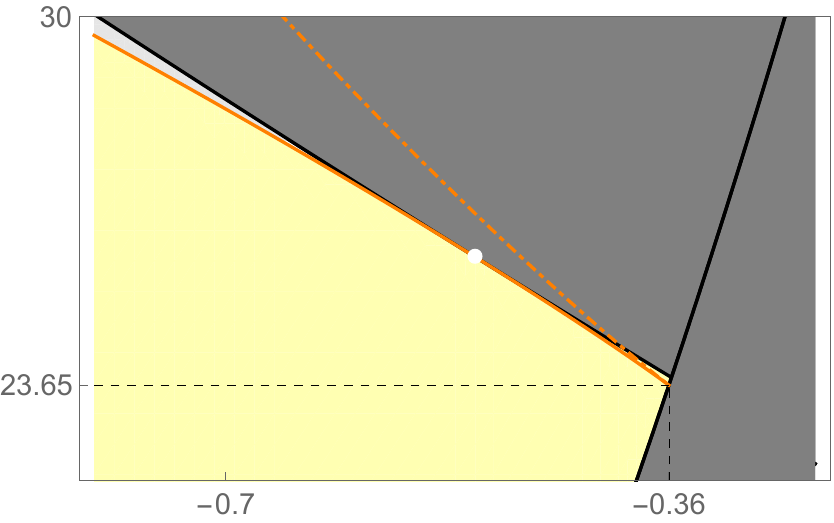}&
		\includegraphics[width=0.3\textwidth]{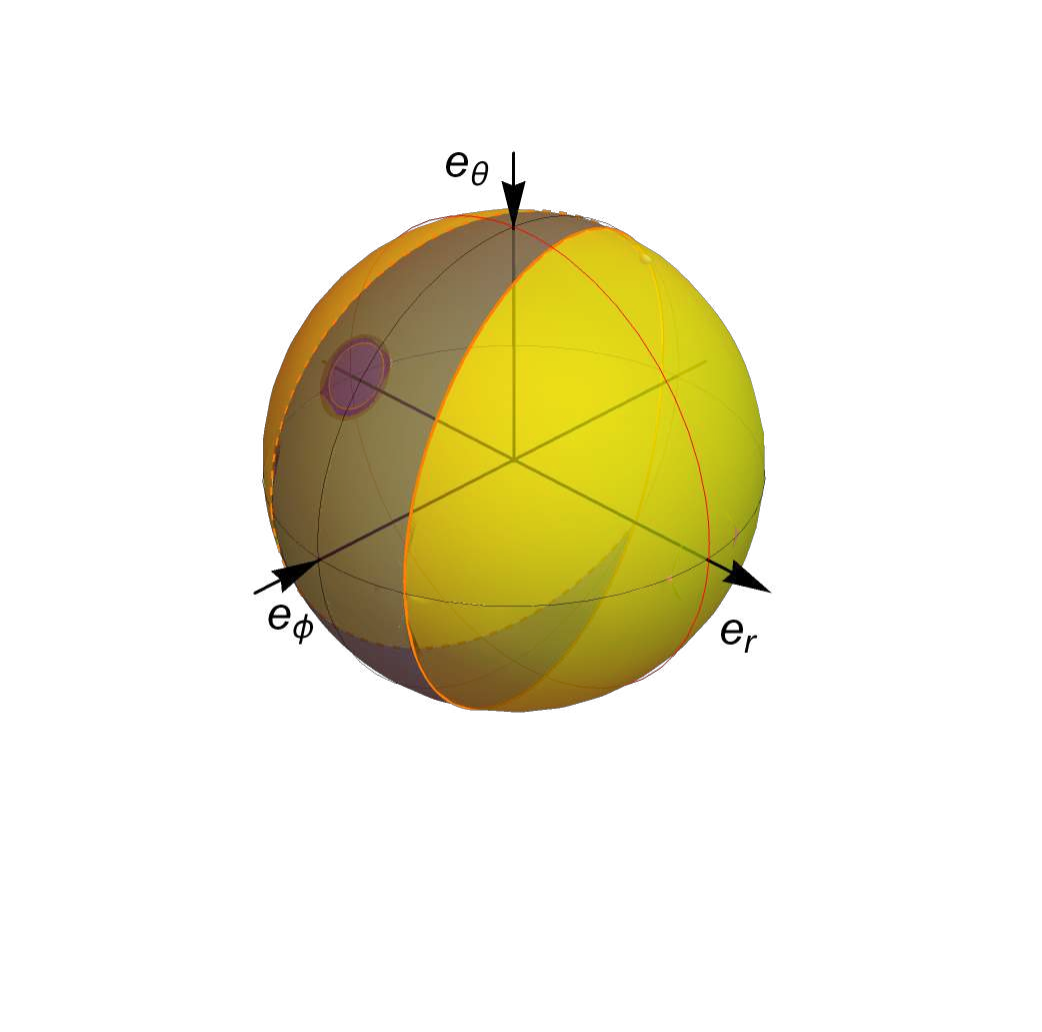}&\includegraphics[width=0.3\textwidth]{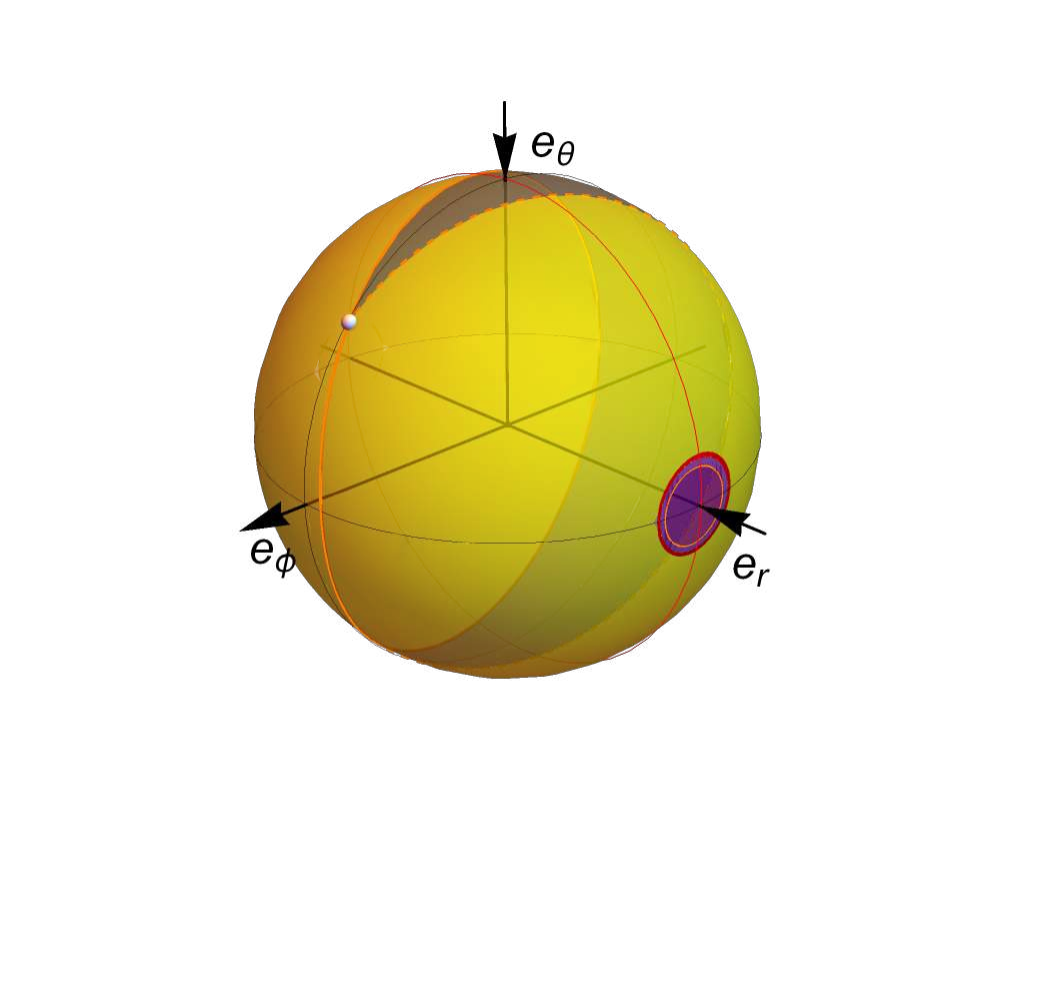}\\
		\multicolumn{3}{|c|}{(c): $\theta_{e}=\theta_{max(circ)}=8.29\dgr$}\\
		\hline
		\end{tabularx}	
\caption{LECs for $r_{e}=1.8$ representing the case $r_{ms}<r_{e}<r^{-}_{pol}$, corresponding to the region of the inner ergosphere with unstable corotating SPOs with positive covariant energy $E>0$. For $\theta_{e}=0\dgr$ the there exist a trapped cone as in previous case. Since $r_{e}<r^{-}_{pol}$, the trapped cone exists for any latitude.  A photon emitted in the direction represented by the white point follows the unstable SPO at the current coordinate $r_{e}$. As a result of emitter's radial position, the lower SPOs are reached only by inward photons, so the full orange curve exceeds the dashed one and for latitude $\theta_{max(circ)}$ it disconnects and an arc appears. 
}\label{Fig_cones_IVa_1.8_6.4deg}
\end{figure*}

\begin{figure*}[h]
	\centering
	\begin{tabularx}{\textwidth}{|XX|}
	\hline
	\multicolumn{2}{|c|}{$r_{e}=4$}\\
	\hline
	
	\includegraphics[width=0.3\textwidth]{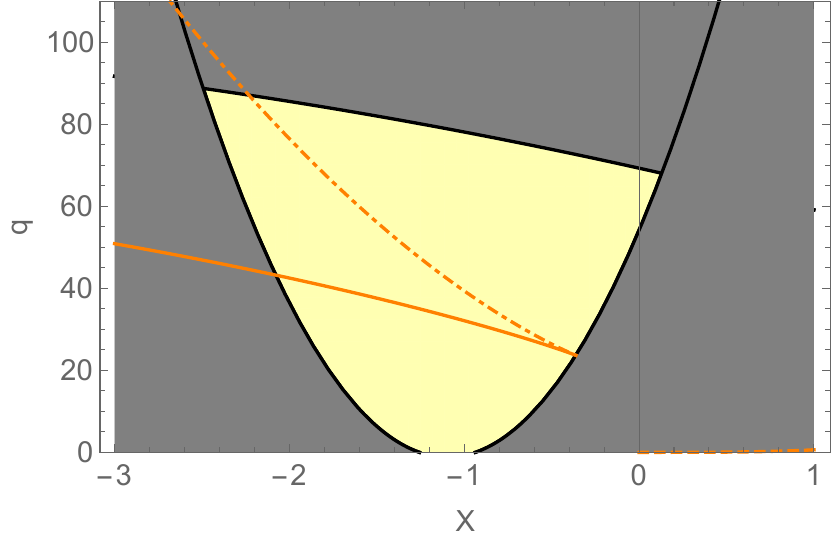}&
	\includegraphics[width=0.3\textwidth]{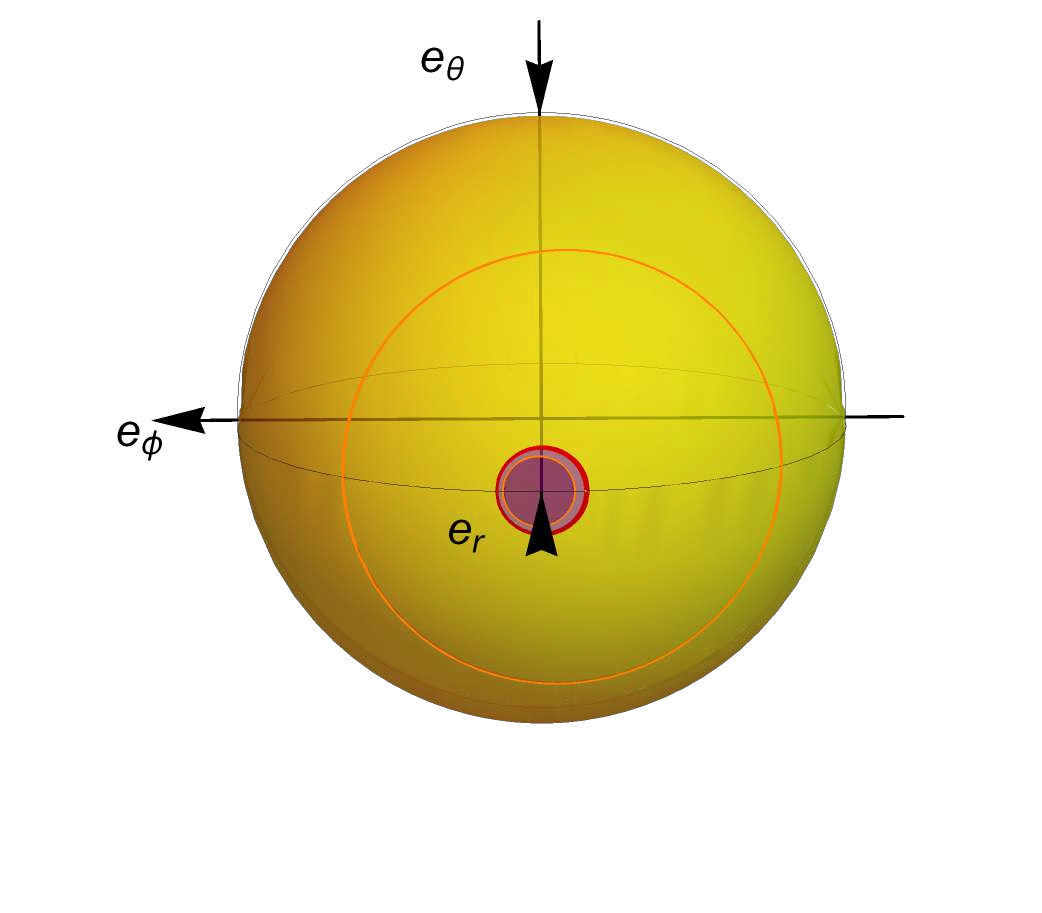}\\
	\multicolumn{2}{|c|}{(a): $\theta_{e}=\theta_{max(circ)}=8.29\dgr$}\\
	\hline
	\includegraphics[width=0.3\textwidth]{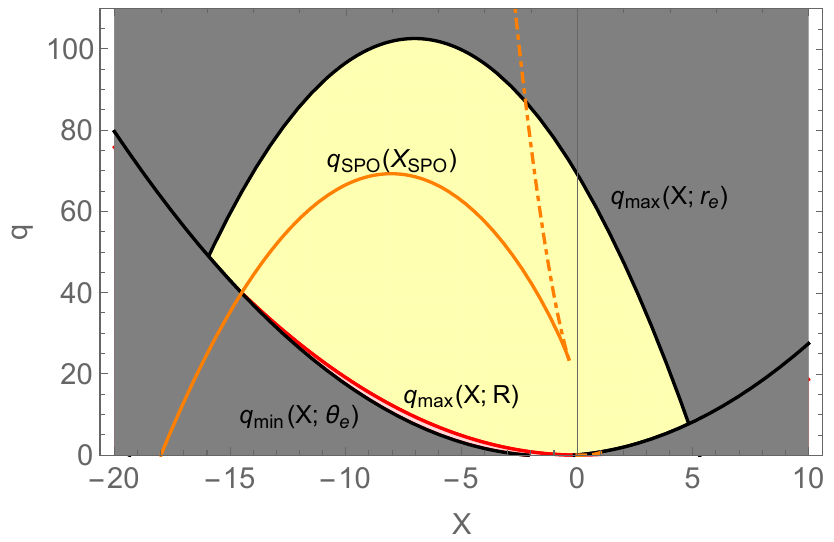}&
	\includegraphics[width=0.3\textwidth]{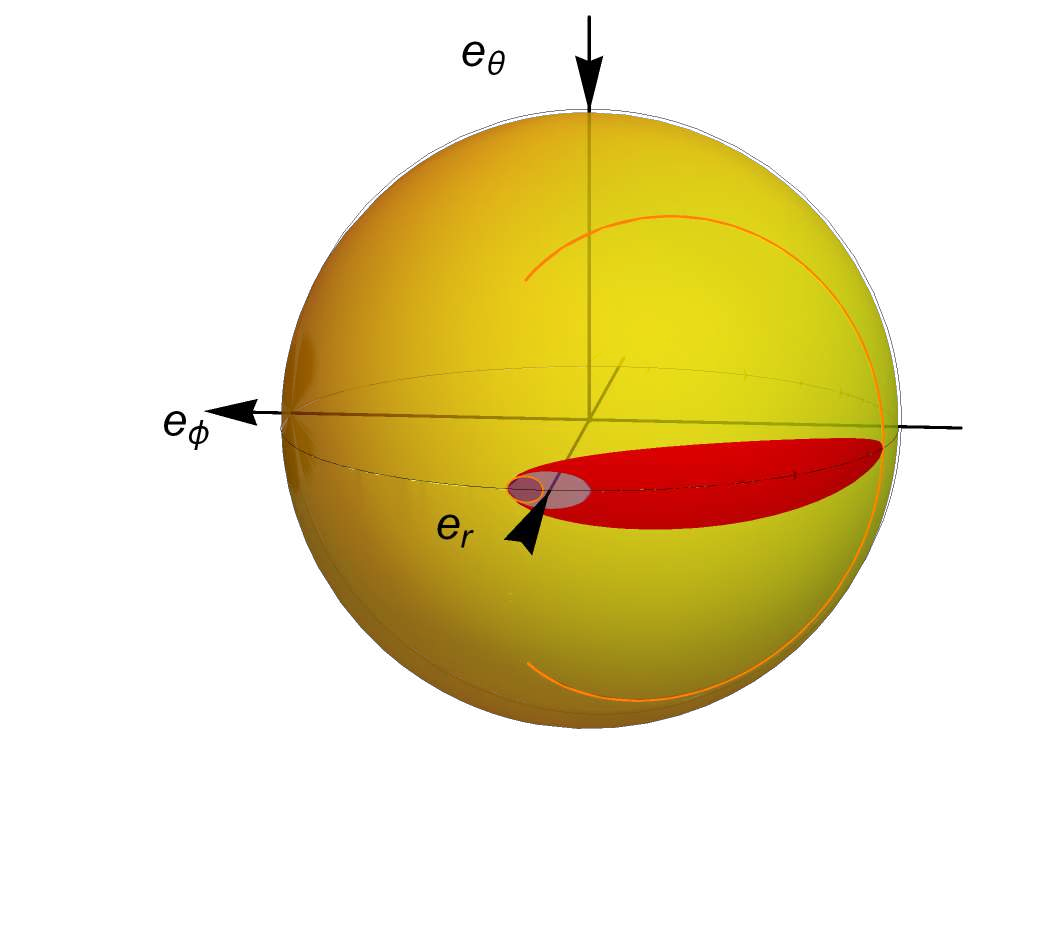}\\
	\multicolumn{2}{|c|}{(b): $\theta_{e}=\theta_{max(el)}=64.39\dgr$}\\
	\hline
	\includegraphics[width=0.3\textwidth]{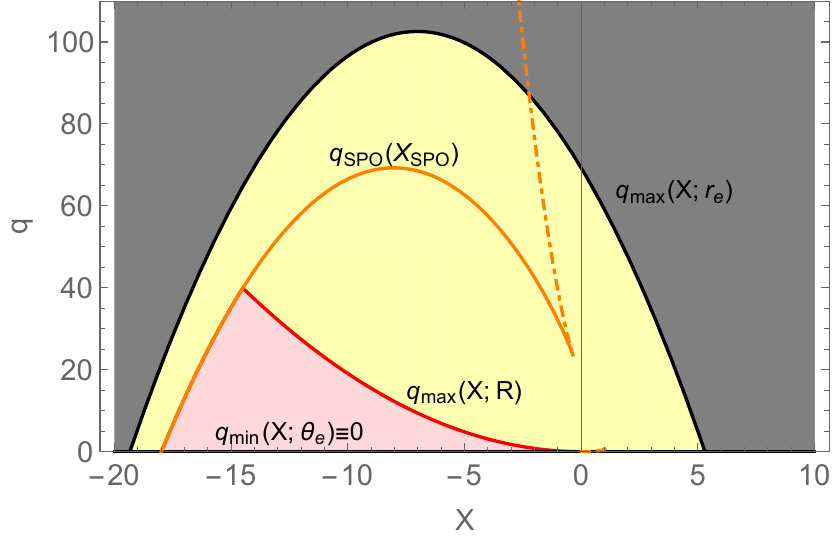}&
	\includegraphics[width=0.3\textwidth]{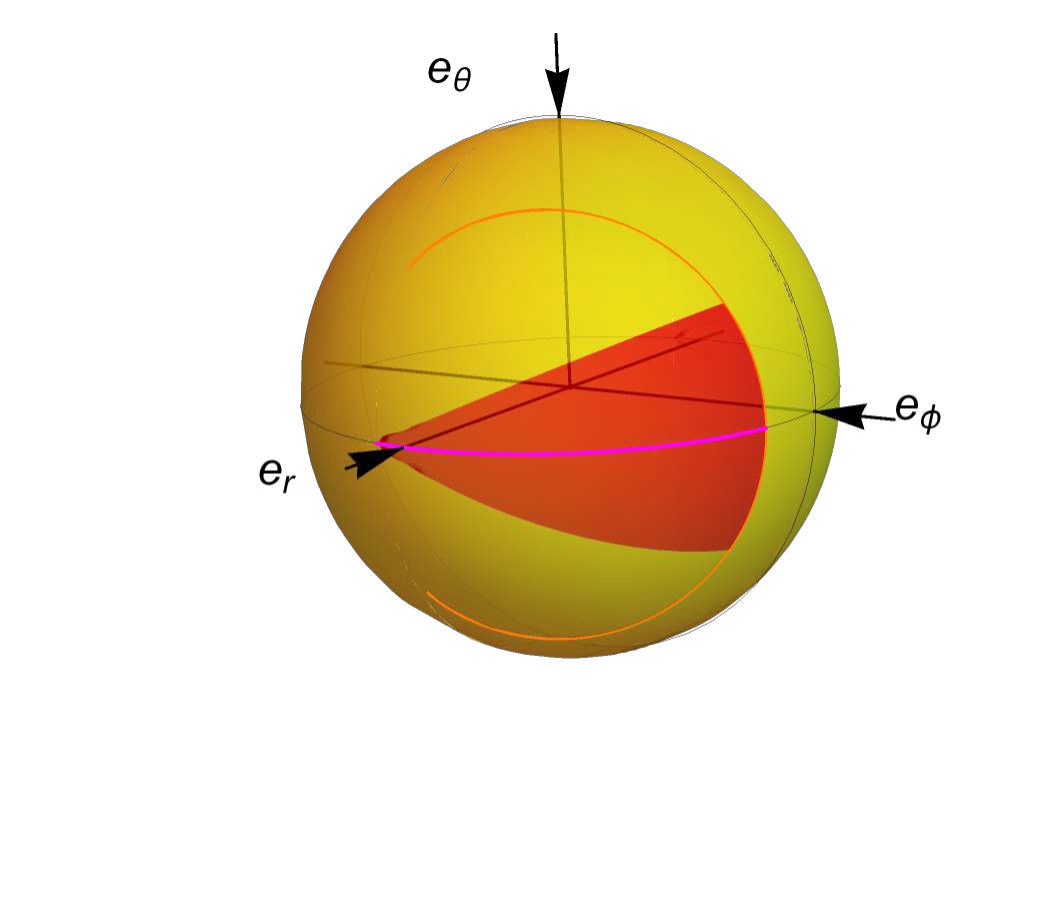}\\
	\multicolumn{2}{|c|}{(c): $\theta_{e}=90\dgr$}\\
	\hline
	\end{tabularx}	
	\caption{LECs for the case $r^{-}_{ph}<r_{e}<r^{-}_{erg}$, i.e the region outside the ergosphere and SPOs and close to the static radius at $r_{s}=3.68$. There are no trapped cones and an arc appears for latitudes large enough, $\theta_{e}>\theta_{max(circ)}=8.29\dgr$. The middle figure introduces the limiting angle $\theta_{max(ell)}$, for which the cone of absorbed photons is still elliptical. In the equatorial plane in the figure below, this cone transforms into a wedge shape. The cone of light escaping to $r\to - \infty$ and the repelled cone degenerate into the shape of a line in the equatorial plane, shown in magenta. 
	}\label{Fig_cones_IVa_4_90deg}
\end{figure*}

From the above we can summarize that the relevance of the polar SPOs with the construction of the LECs and superspinar/NS shadows is as follows. If a spacetime is endowed with the polar SPOs, then the critical locus $q_{\spo}(X_{\spo})$ in the corresponding $(X-q)$ plane of motion constants is defined for $X_{\spo}=-a$ $(\ell_{\spo}=0)$. It then intersects the degenerated graph of the function $q_{min}(X;\theta_{e}\to 0)$ defined by the relations (\ref{X=-a}), (\ref{q>-a2}). As the photon emitter moves away from the spacetime symmetry axis, i.e. as $\theta_{e}$ grows, the parabola $q_{min}(X;\theta_{e})$ cuts an arc at the critical locus. According to the radial position of the emitter relative to the outer unstable spherical photon orbit $r^{-}_{pol}$, and the equatorial circular photon orbit $r^{-}_{ph}$, this may cause a spherical belt or bi-angle to form on the emitter's celestial sphere, where the emitted photons are trapped, or just a circle separating the two cones of the escaping photons (see e.g. Figs. \ref{Fig_cones_IVa_1.25_5_66deg}, \ref{Fig_cones_IVa_1.8_6.4deg}, \ref{Fig_cones_IVa_2.2_3.3deg} and the comments). Then there exist a critical angle $\theta_{max(circ)}$ for which the circle disconnects and an arc appears (cf. e.g. Figs. \ref{Fig_cones_IVa_2.2_90deg}a, b, or Figs. \ref{Fig_cones_IVa_4_90deg}b, c).

\begin{figure*}[h]
	\centering \centering \textbf{Class IVb: $y=0.02,\quad a^2=2$}\\
	\begin{tabularx}{\textwidth}{|XX|}
		\hline
		\multicolumn{2}{|c|}{$r_{e}=r_{s}=3.68$}\\
		\hline
		\raisebox{2.4cm}[0pt]{\bet{c}	\includegraphics[width=0.3\textwidth]{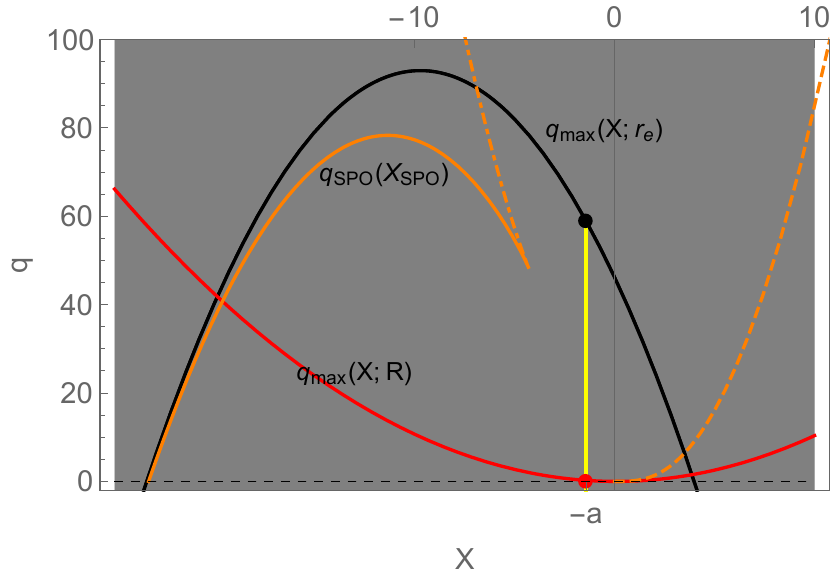}\\ \includegraphics[width=0.3\textwidth]{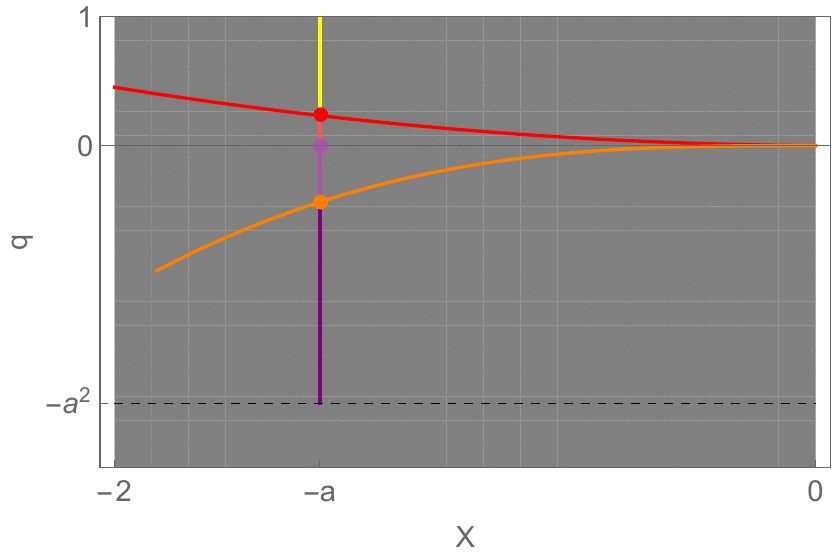} \ent }&
		\includegraphics[width=0.3\textwidth]{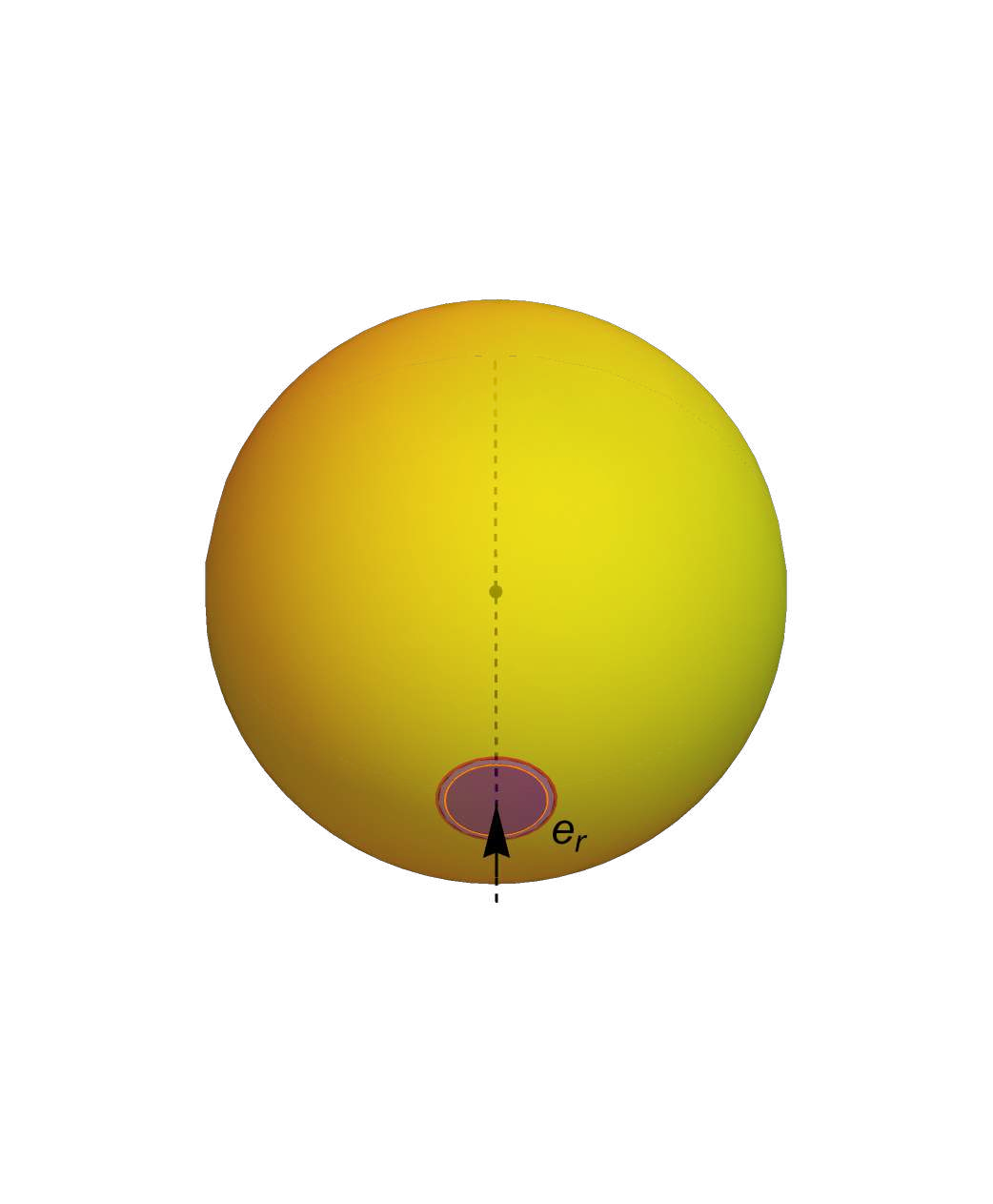}\\
		\multicolumn{2}{|c|}{(a): $\theta_{e}=0\dgr$}\\
		\hline
		\includegraphics[width=0.3\textwidth]{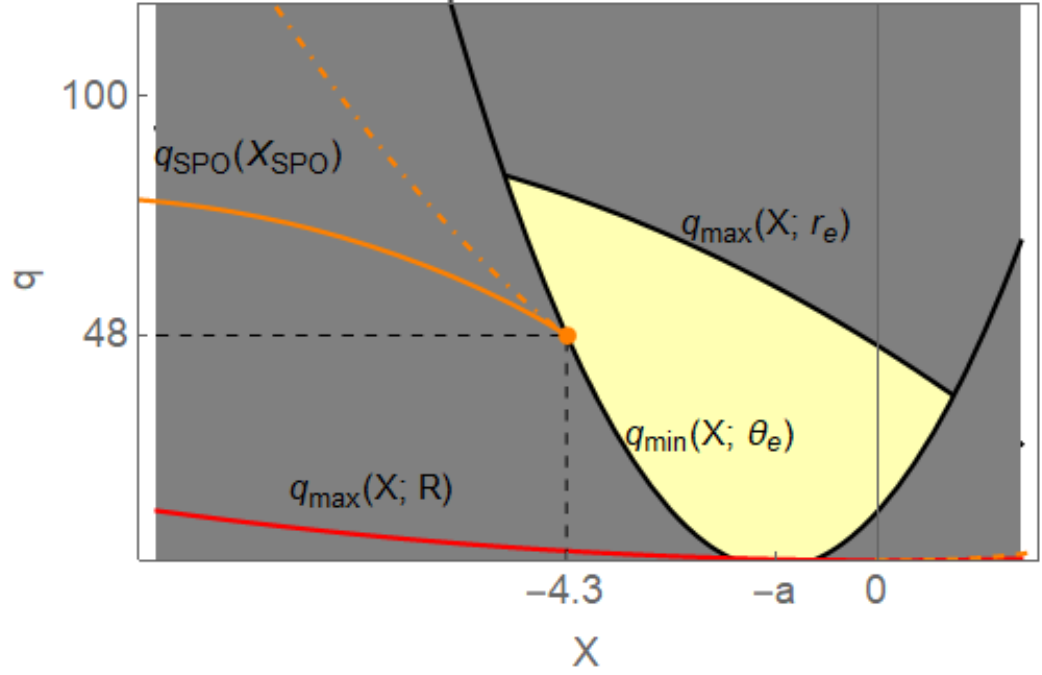}&
		\includegraphics[width=0.3\textwidth]{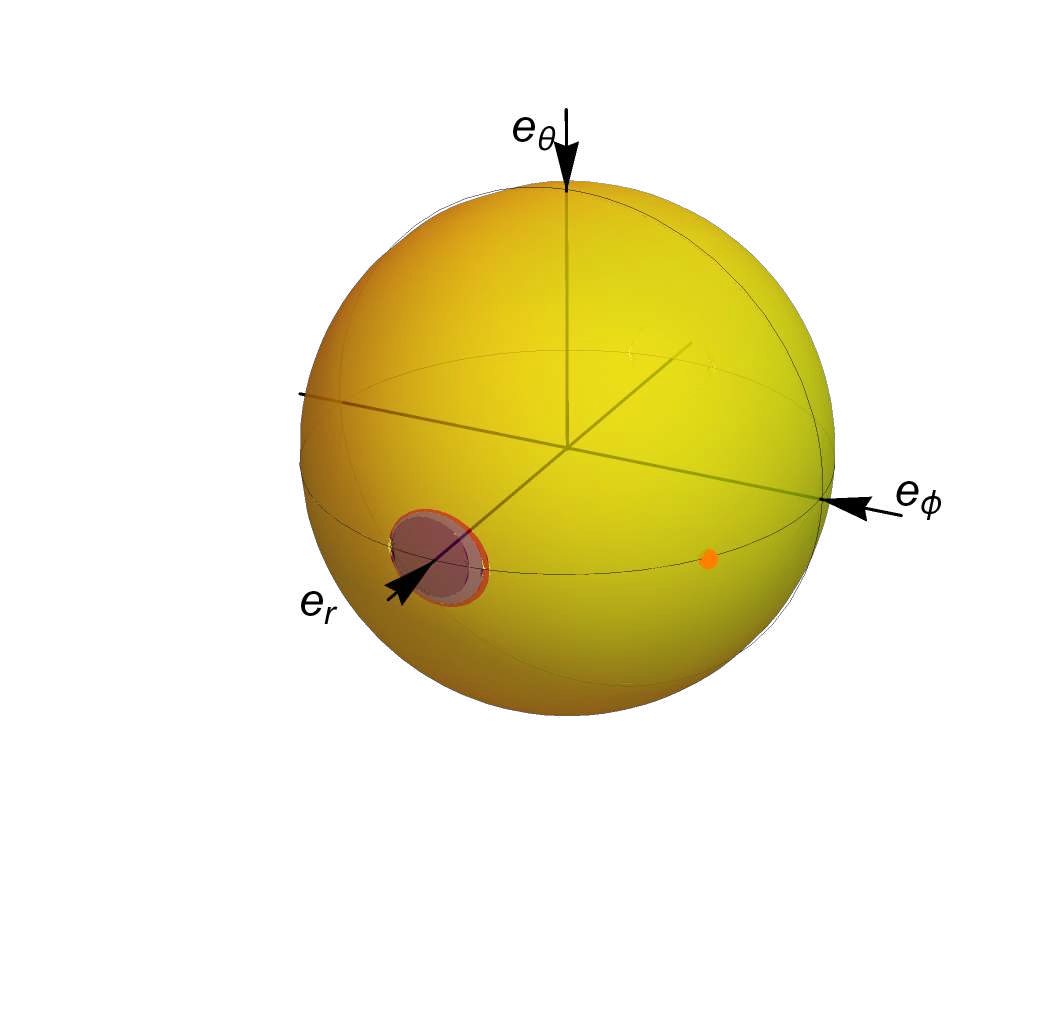}\\
		\multicolumn{2}{|c|}{(b): $\theta_{e}=\theta_{min(arc)}=21.79\dgr$}\\
		\hline
		\includegraphics[width=0.3\textwidth]{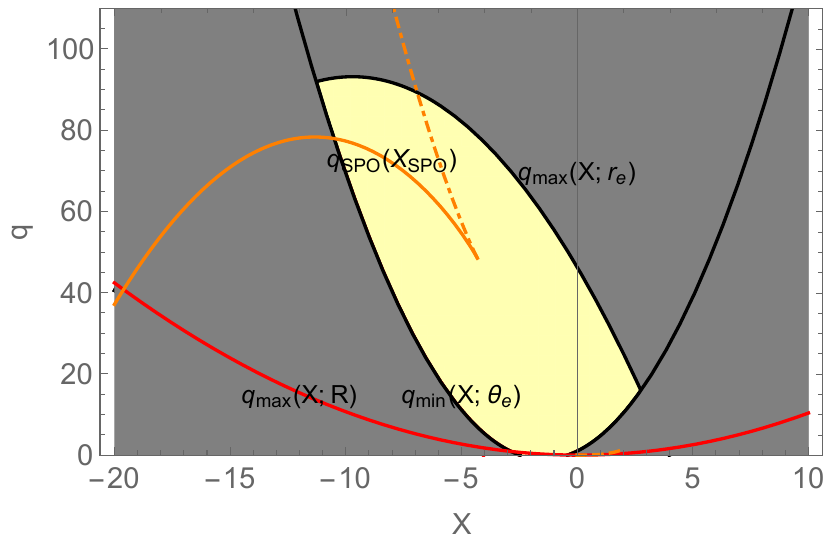}&
		\includegraphics[width=0.3\textwidth]{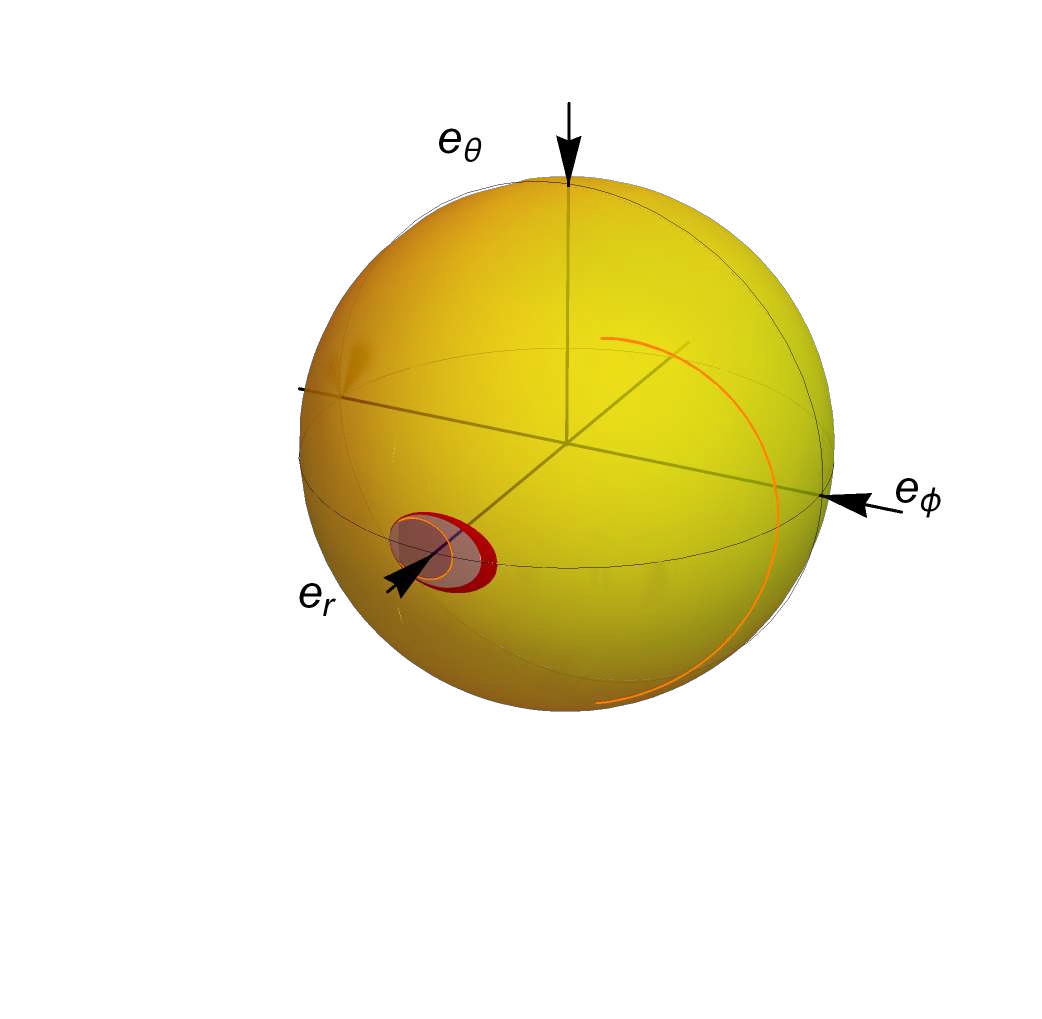}\\
		\multicolumn{2}{|c|}{(c): $\theta_{e}=45\dgr$}\\
		\hline
	\end{tabularx}	
	\caption{LECs in spacetime with DRB and no polar SPOs. The radii of the observers are chosen to be at the static radius $r_{s}=3.68$. Since there are no polar SPOs, there is neither a trapped cone nor critical locus for an observer located on the polar axis, but it appears as an arc for $\theta_{e}>\theta_{min(arc)}=21.79\dgr$. Otherwise, the appearance of the LECs for $\theta_{e}>\theta_{min(arc)}$ remains similar to the case of the spacetimes of the Class IVa  and angles $\theta_{e}>\theta_{max(circ)}$ as presented in the figures above. 
	}\label{Fig_cones_VIb_rs_22}
\end{figure*}

\begin{figure*}[h]
	\centering \centering \textbf{Class V: $y=0.001,\quad a^2=20$}\\
	\begin{tabularx}{\textwidth}{|XX|}
		\hline
		\includegraphics[width=0.3\textwidth]{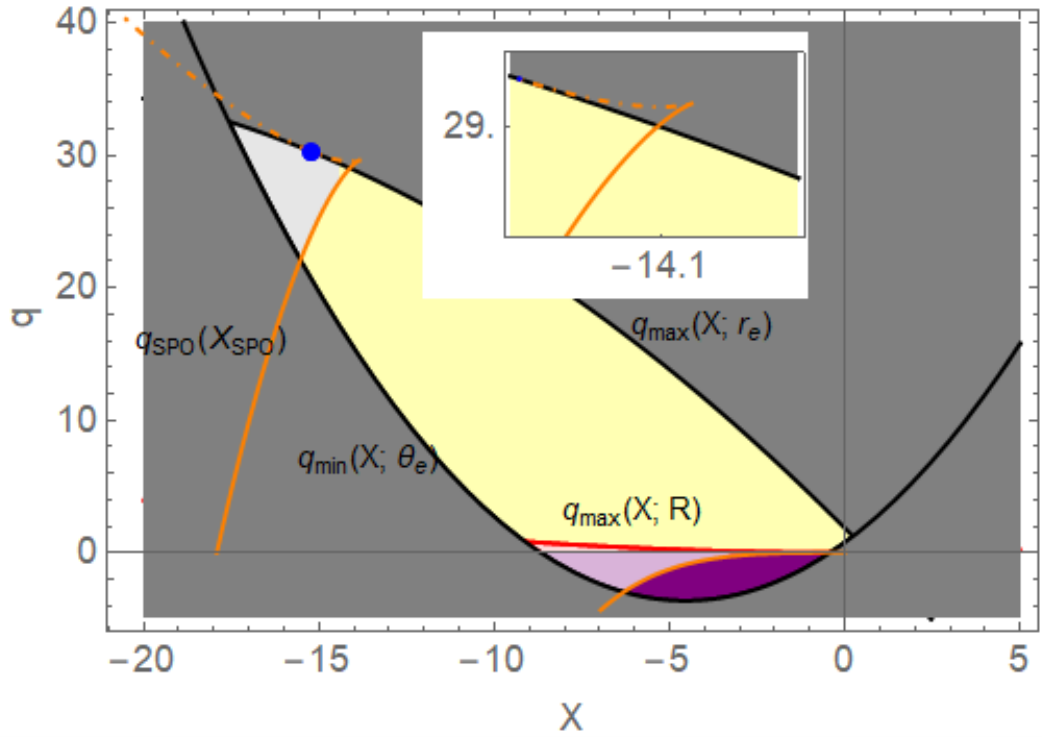}&\includegraphics[width=0.3\textwidth]{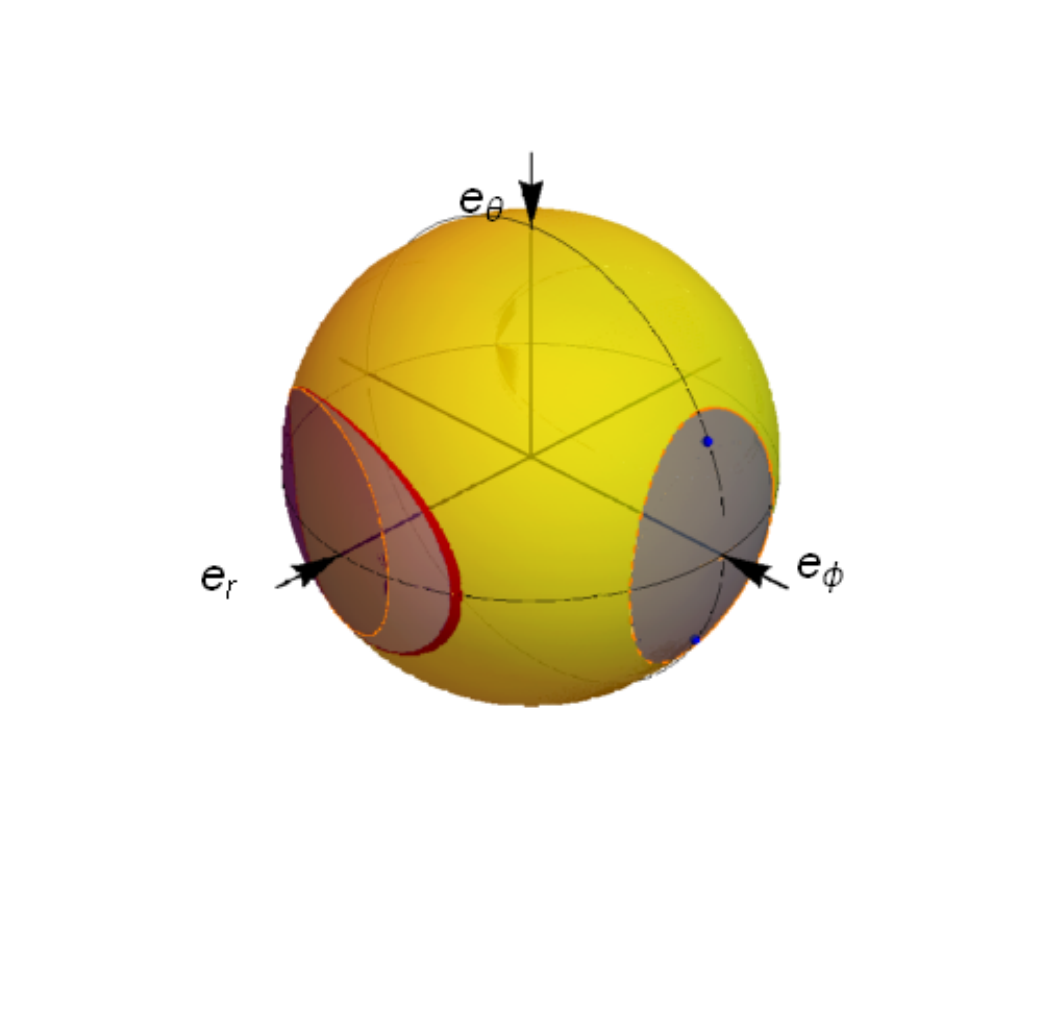}\\
		\multicolumn{2}{|c|}{$r_{e}=2.5$, $\theta_{e}=65\dgr>\theta_{min(circ)}=60.7\dgr$}\\
		\hline
		
	\end{tabularx}	
	\caption{LEC in the case of Class V spacetime with small region of unusual bound orbits with $X_{+}<X<0$ and $E>0$ (see the detail in Fig. \ref{Fig_eff_pots1}b). It causes that for $r_{e}\lesssim r^{max}_{\spo}$ the critical locus $X_{\spo}(q_{\spo})$ is partly hidden in the forbidden region (detail in the figure left). As a consequence, there exists a minimal angle $\theta_{min(circ)}$, such that for $\theta_{e}>\theta_{min(circ)}$, here $\theta_{min(circ)}=60.7\dgr$, a light cone of the trapped photons appears, which is bounded by a circle-like shaped rim on the sphere (figure right). For $\theta_{e}=\theta_{min(circ)}$ this cone shrinks to an infinitesimal extent directed in the negative $\phi$-direction.  
	}\label{Fig_cones_V_2.5_60deg}
\end{figure*}

On the other hand, if there are no polar SPOs in the spacetime, then the intersection of the  graph of the function $q_{min}(X;\theta_{e})$ with the critical locus appears after a certain minimum angle is attained. We denoted it $\theta_{min(arc)}$ in the spacetimes of Class IVb or VIb, since for $\theta_{e}> \theta_{min(arc)}$ an arc occurs (see Fig. \ref{Fig_cones_VIb_rs_22}), and $\theta_{min(circ)}$ for the spacetimes of Class V, since for $\theta_{e}> \theta_{min(circ)}$ a circle-like shaped boundary of the trapped cone occurs (see Fig.\ref{Fig_cones_V_2.5_60deg} ). Then there is neither a belt nor an arc for $\theta_{e}<\theta_{min(arc/circ)}$.

\renewcommand{\arraystretch}{2.0}
\begin{table*}[h]
	\caption{Significant radii related to the photon motion in the KdS spacetime of Class \textbf{IVa}.}\label{tab_IVa_orbs}
	\begin{tabularx}{\textwidth}{|ccc|XXXXXXXXX|}
		\hline		
	$y$&&$a^2$&$r^{+}_{d(\spo)}$&$r^{+}_{pol}$&$r_{ms}$&$r^{-}_{pol}$&$r^{+}_{erg}$&$r^{-}_{ph}$&$r_{s}$&$r^{-}_{erg}$&$r_{c}$\\
		\hline
	$0.02$&&$1.2$&$1.08$&$1.28$&$1.59$&$2.08$&$2.30$&$3.51$&$3.68$&$5.55$&$5.77$\\
		\hline

	\end{tabularx}	
\end{table*}

\begin{table*}[h]
	\caption{Significant radii related to the photon motion in the KdS spacetime of Class \textbf{IVb}.}\label{tab_IVb_orbs}
	\begin{tabularx}{\textwidth}{|ccc|XXXXXXX|}
		\hline		
		$y$&&$a^2$&$r^{+}_{d(\spo)}$&$r_{ms}$&$r^{+}_{erg}$&$r^{-}_{ph}$&$r_{s}$&$r^{-}_{erg}$&$r_{c}$\\
		\hline
		$0.02$&&$2$&$1.10$&$1.93$&$2.36$&$3.58$&$3.68$&$5.44$&$5.81$\\
		\hline

	\end{tabularx}	
\end{table*}

\begin{table*}[h]
	\caption{Significant radii related to the photon motion in the KdS spacetime of Class \textbf{V}.}\label{tab_V_orbs}
	\begin{tabularx}{\textwidth}{|ccc|XXXXXXXX|}
		\hline		
		$y$&&$a^2$&$r^{+}_{d(\spo)}$&$r^{+}_{erg}$&$r^{max}_{\spo}$&$r_{ms}$&$r^{-}_{ph}$&$r_{s}$&$r^{-}_{erg}$&$r_{c}$\\
		\hline
		$0.001$&&$20$&$1.02$&$2.05$&$3.06$&$3.57$&$6.24$&$10$&$30.23$&$30.59$\\
		\hline

	\end{tabularx}	
\end{table*}

\begin{figure*}[h]
	\centering
	\begin{tabular}{|ccc|}
		\hline	
		\multicolumn{3}{|c|}{$r_{e}=0.9$}\\
		\hline
		\includegraphics[width=0.3\textwidth]{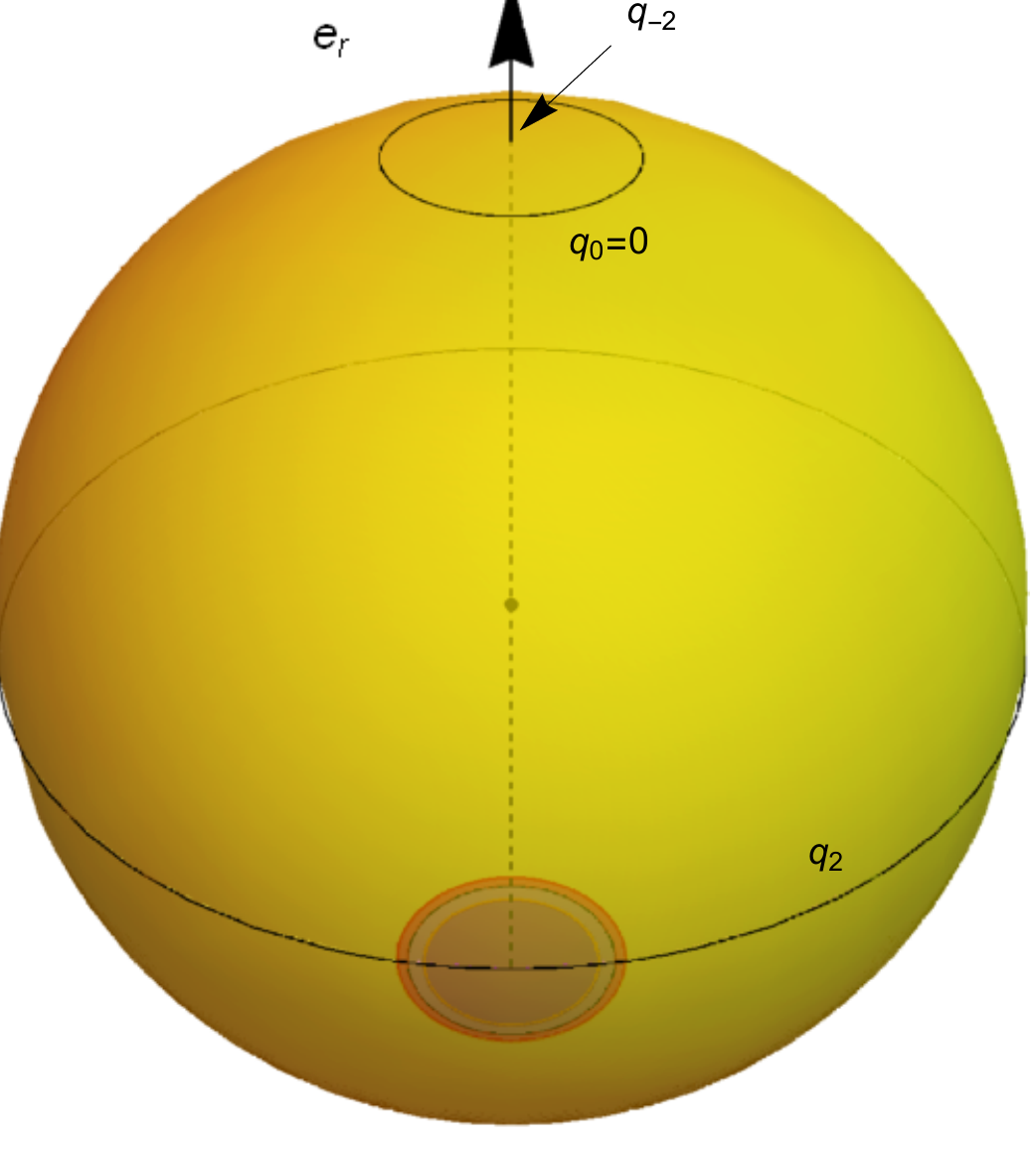}&(a): $\theta_{e}=0\dgr$&	\includegraphics[width=0.3\textwidth]{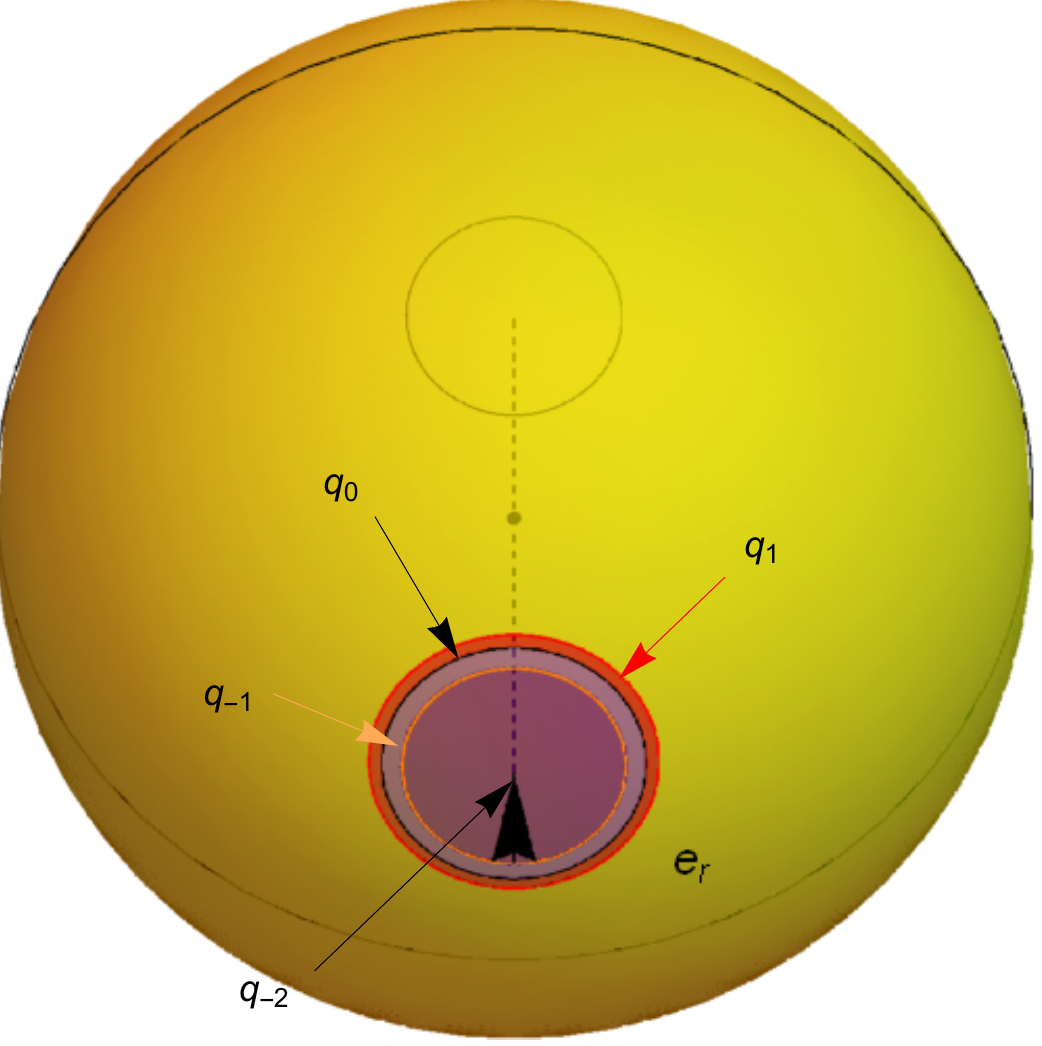}\\
		\hline
		\includegraphics[width=0.3\textwidth]{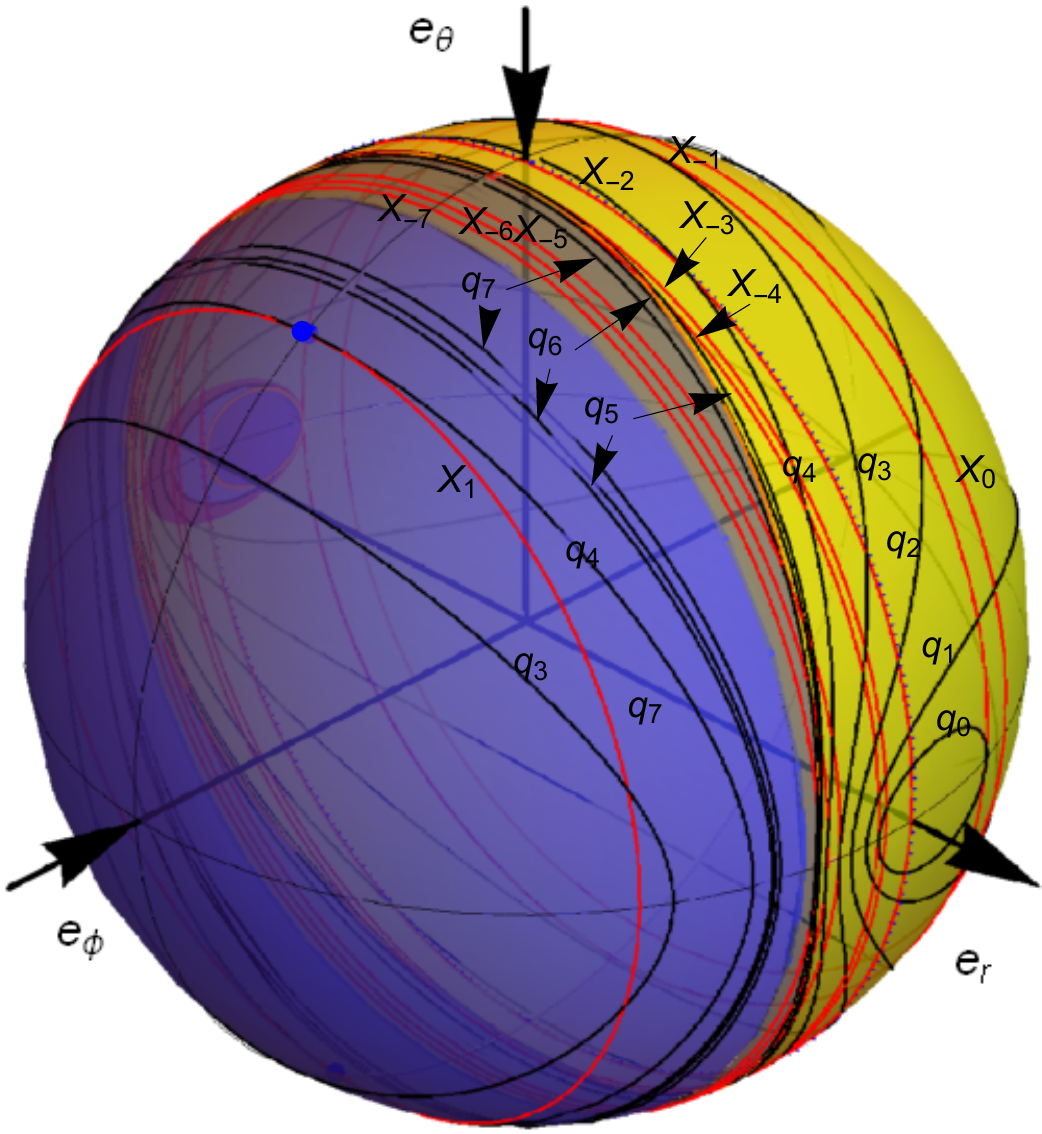}&(b): $\theta_{e}=60\dgr$&\includegraphics[width=0.3\textwidth]{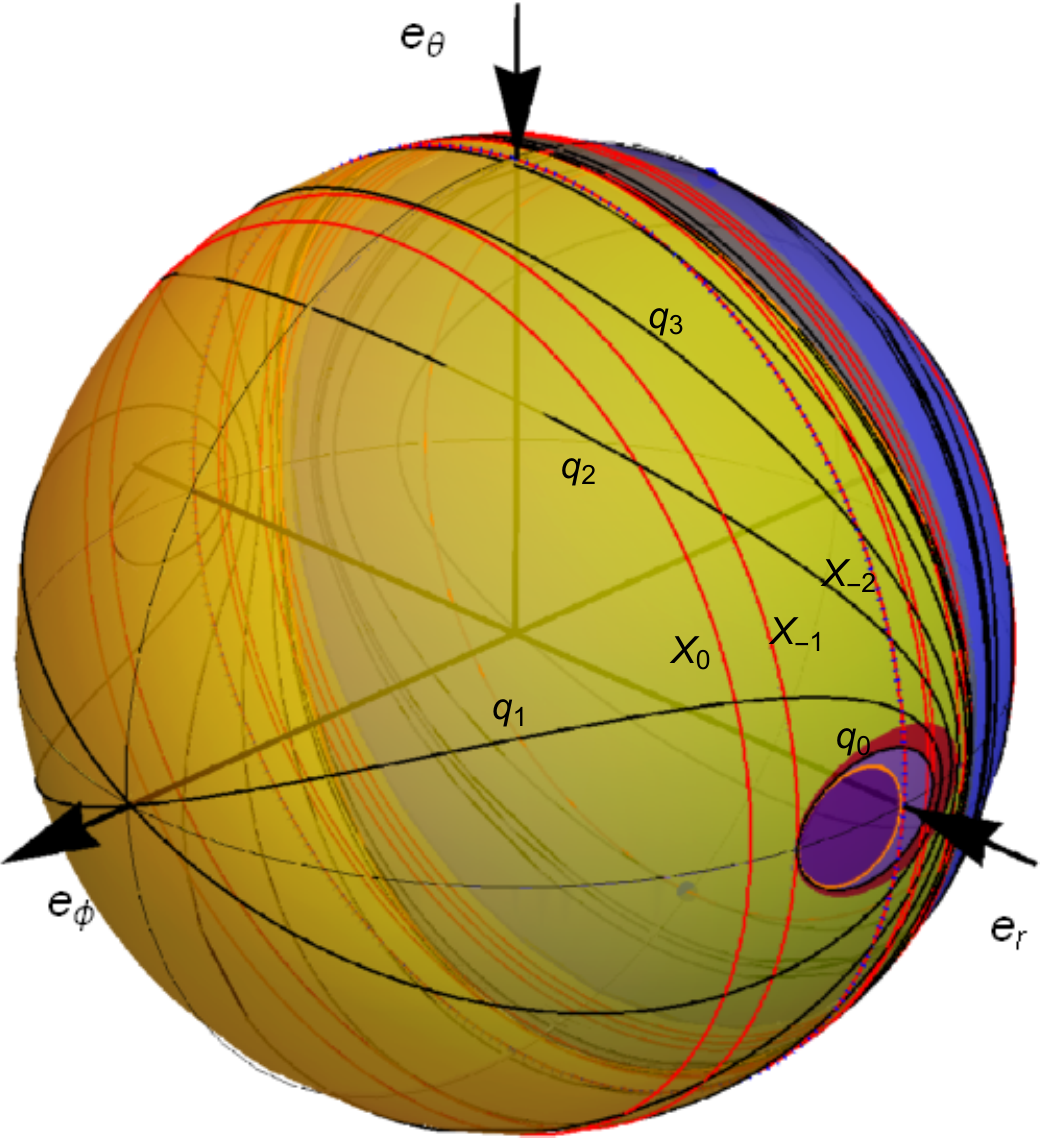}\\
		\hline
		\includegraphics[width=0.3\textwidth]{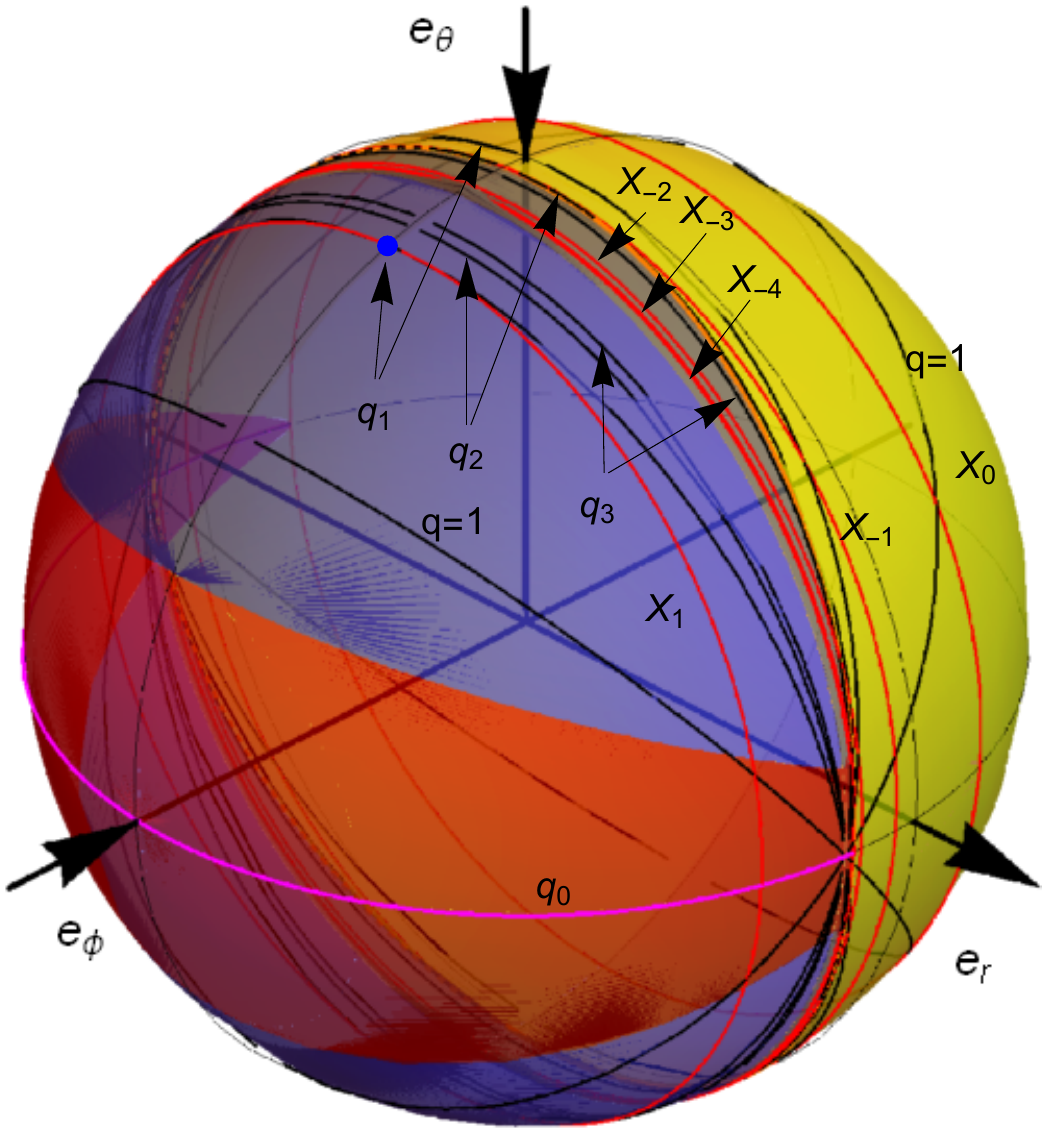}&(c): $\theta_{e}=90\dgr$&\includegraphics[width=0.3\textwidth]{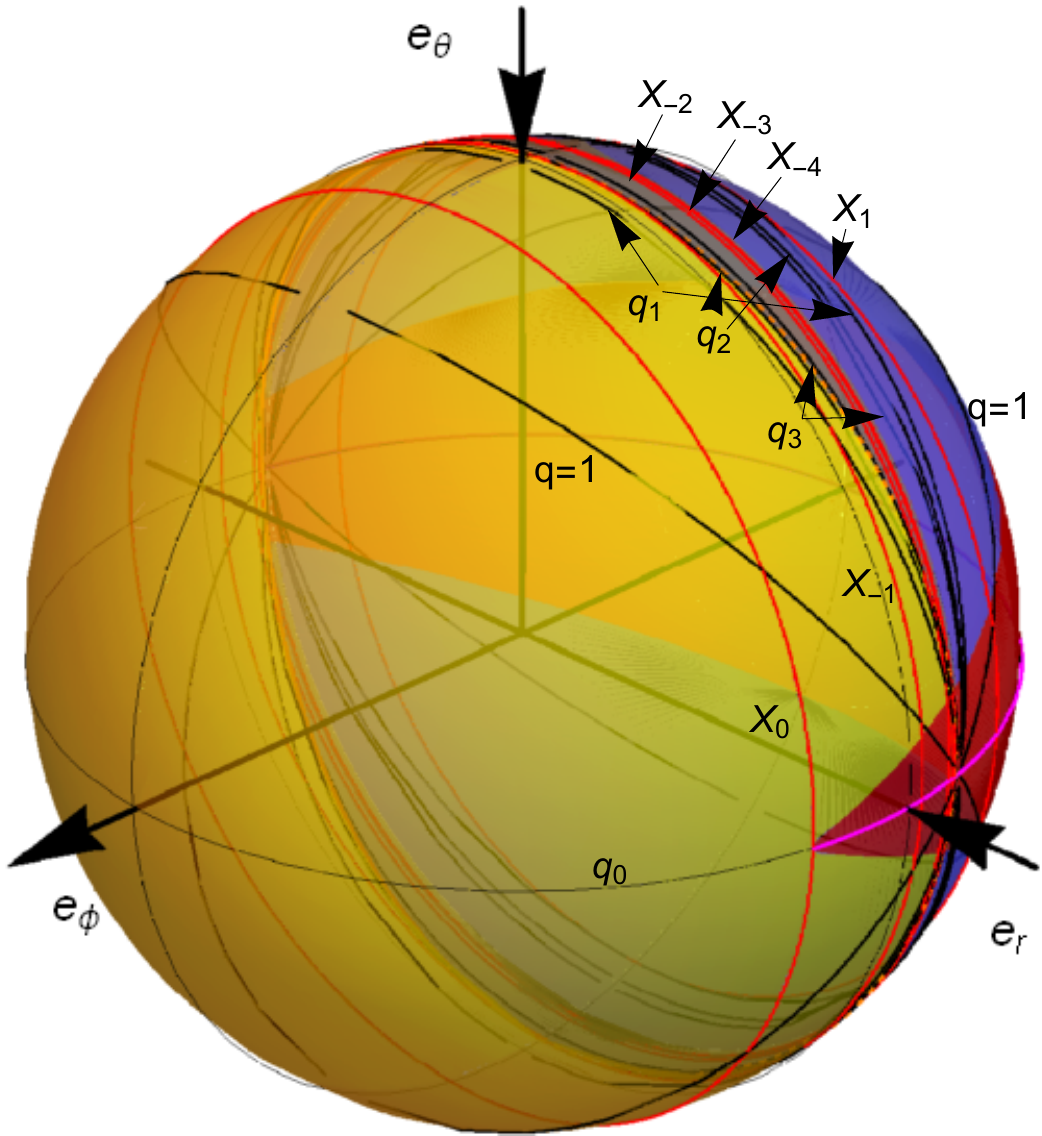}\\
		\hline

	\end{tabular}	
	\caption{Light escape cones represented by the spherical surface formed by the endpoints of the unit three-momentum vectors of the photons emitted by the source located at its centre, corresponding to Fig. \ref{Fig_Xq_IVa_re_0.9}, i.e., the chosen radial coordinate $r_{e}$ corresponds to the region of the inner ergosphere with negative covariant energy SPOs. The angle of view is apparent from the orientation of the local basis vectors. The meaning of the individual spherical sectors, curves, points and their colouring can be read in the comments to Fig. \ref{Fig_Xq_IVa_re_0.9}. The light purple part corresponds to photons entering the region of negative radii $r<0$ and repelled back to positive radii, the dark purple part corresponds to photons escaping to $r \to -\infty$. The boundary of this part represents a photons that hit the ring singularity. For $\theta_{e}=90\dgr$ it shrinks to an arc in the equatorial plane depicted in magenta.  A photon emitted in the direction represented by the blue point follows the stable SPO at the current coordinate $r_{e}=0.9$. The actual observed shadow of the superspinar corresponds to the red region, while the shadow of the NS corresponds to the purple region - the dark purple represents a dark spot flanked by a circle (Fig. (a)) or sickle-shaped (Fig. (b)) bright fringe, represented by the light purple region. The shadow of the NS observed from the equatorial plane corresponds to the arc in magenta, but it is practically unobservable due to its infinitesimal thickness. The same conclusions apply to the shadows in the following images.
	}\label{Fig_cones_IVa_0.9}
\end{figure*}

\begin{figure*}[h]
	\centering \textbf{Class IVa: $y=0.02,\quad a^2=1.2$}\\
	\begin{tabularx}{\textwidth}{|XXX|}
		\hline
		\multicolumn{3}{|c|}{$r_{e}=1.25$}\\
		\hline
		
		\includegraphics[width=0.28\textwidth]{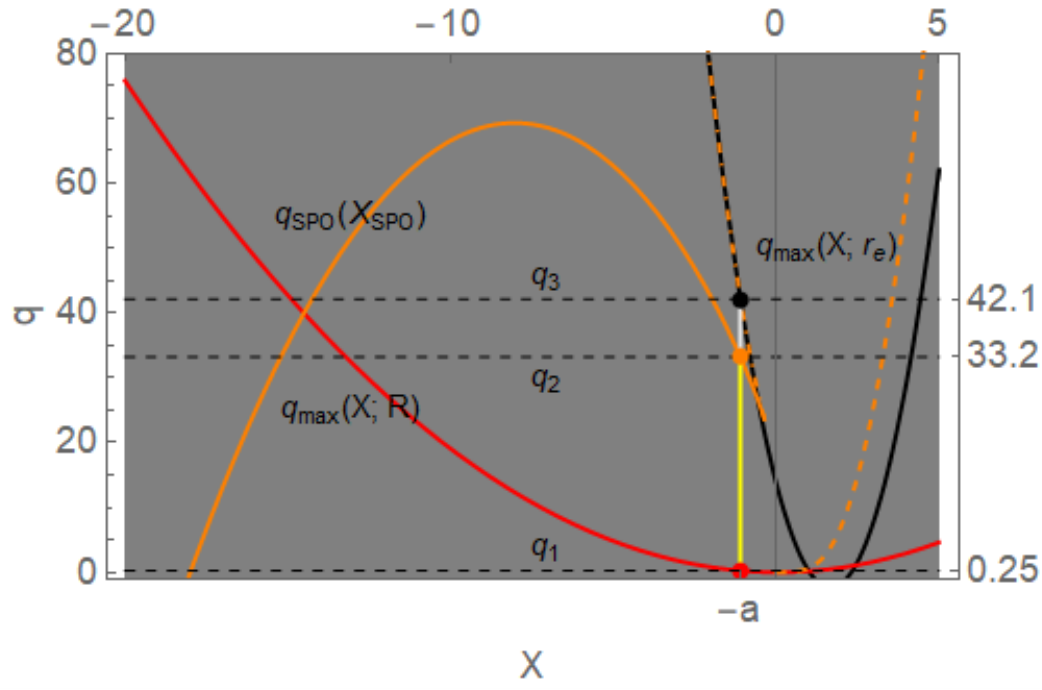}&
		\includegraphics[width=0.28\textwidth]{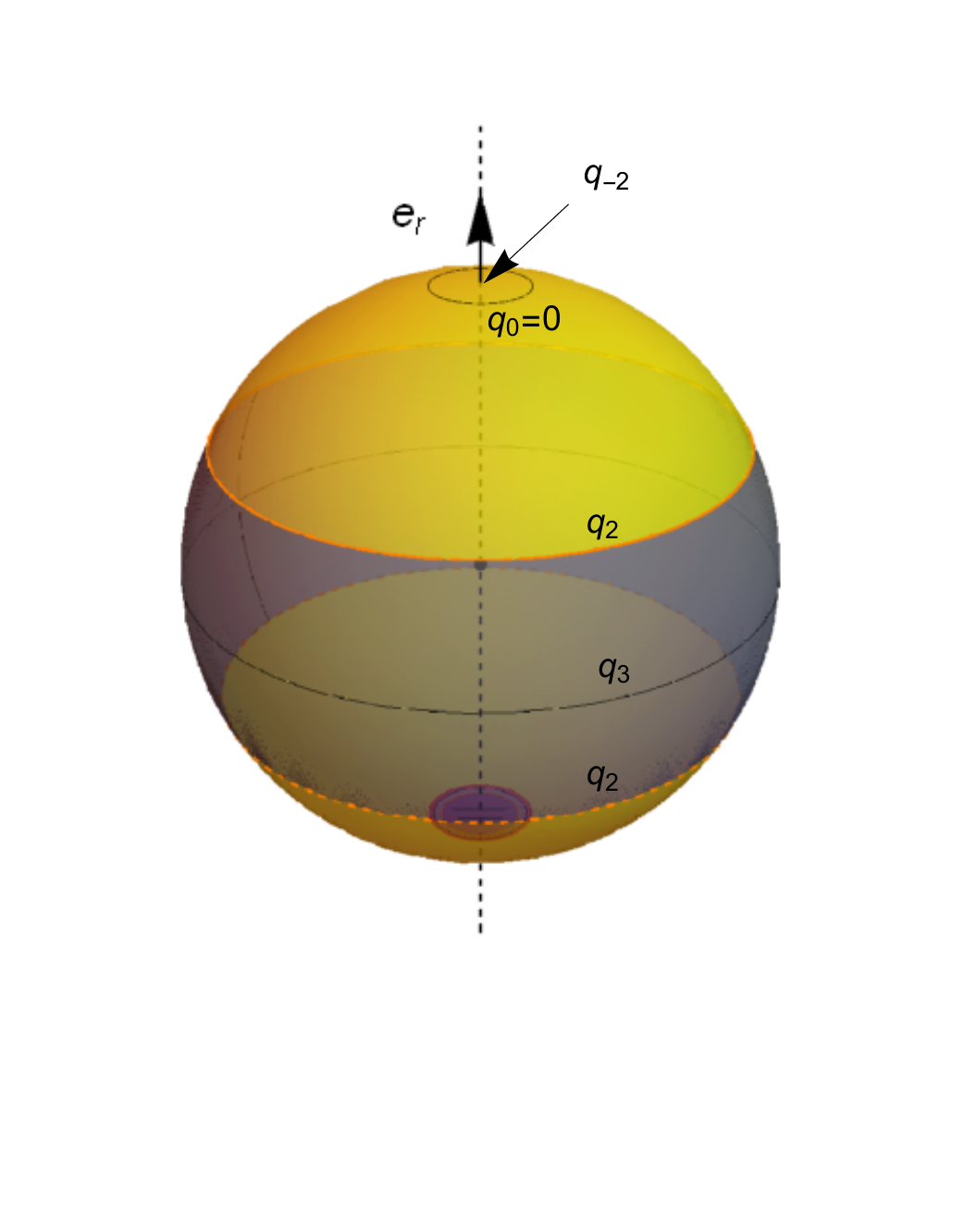}&\includegraphics[width=0.28\textwidth]{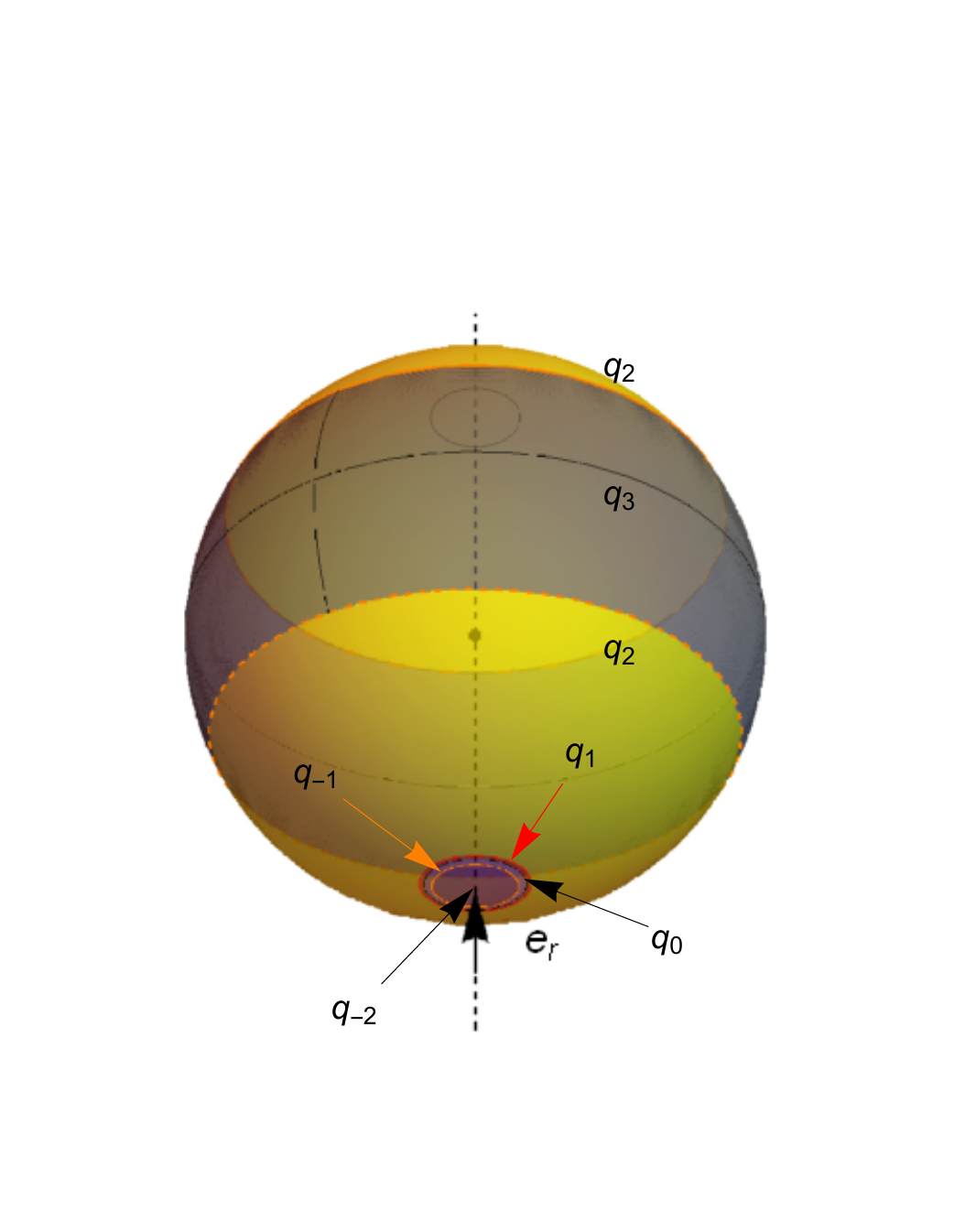}\\
		\multicolumn{3}{|c|}{\raisebox{1cm}[1cm]{(a): $\theta_{e}=0\dgr$}}\\
		\hline
		\raisebox{1.0cm}[0pt]{\bet{c} \includegraphics[width=0.25\textwidth]{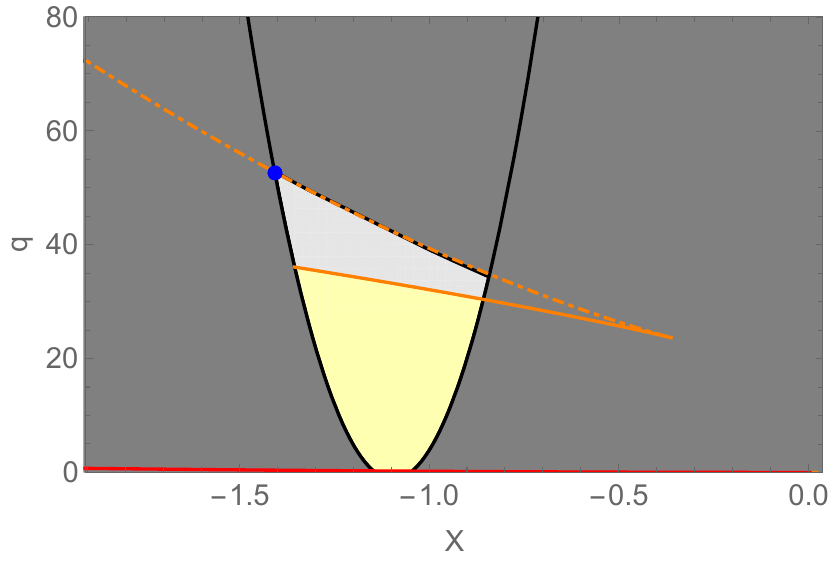}\\
		\includegraphics[width=0.25\textwidth]{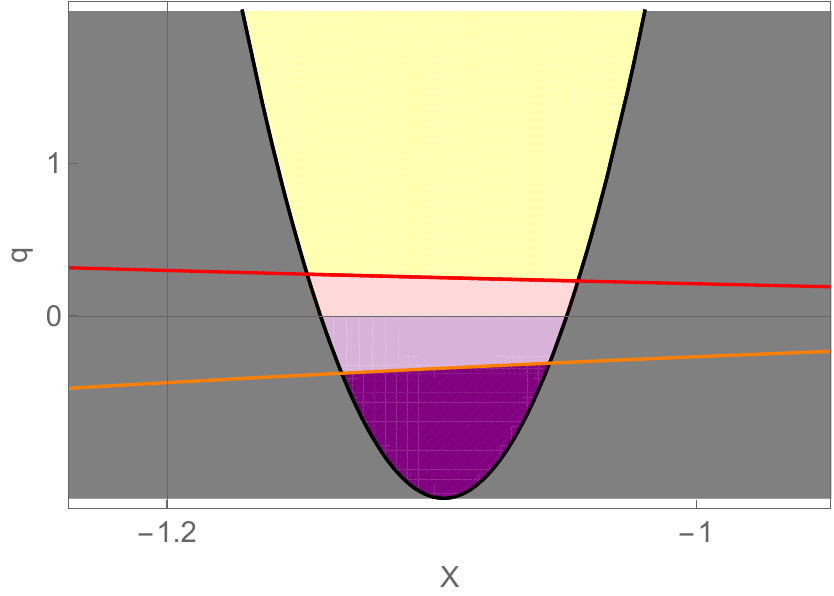} \ent }	&
		\includegraphics[width=0.28\textwidth]{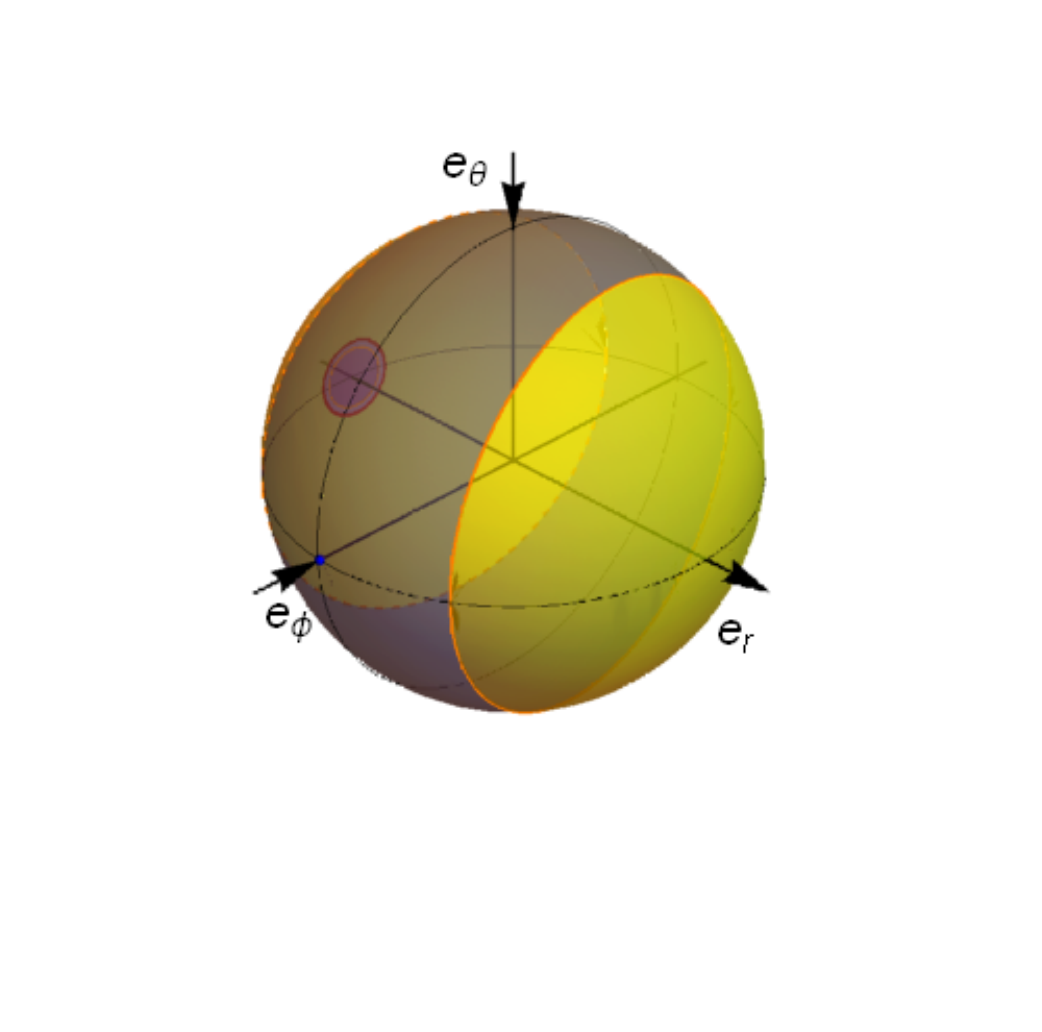}&\includegraphics[width=0.28\textwidth]{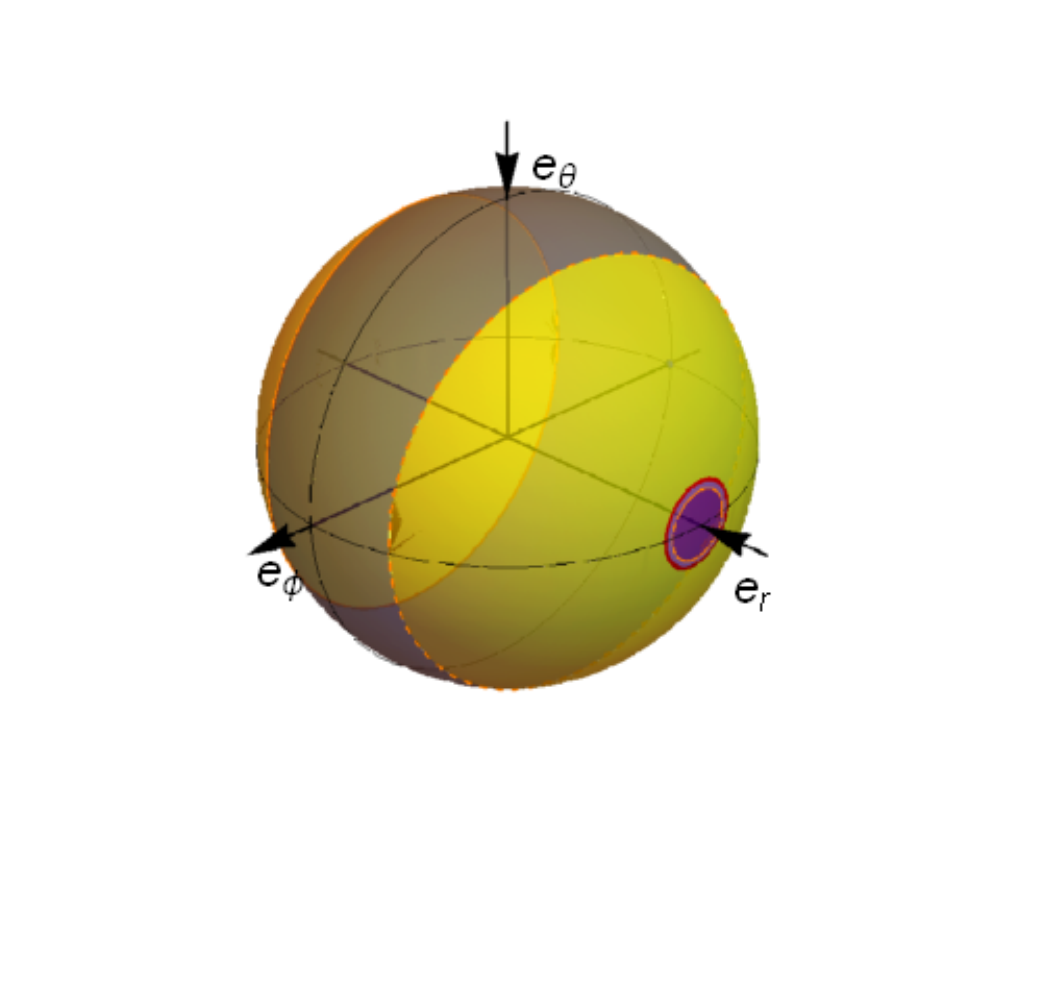}\\
		\multicolumn{3}{|c|}{(b): $\theta_{e}=2.40\dgr$}\\
		\hline
		
	\end{tabularx}	
\caption{Planes of motion constants ($X-q$) and affiliated LECs for $r^{+}_{d(\spo)}<r_{e}=1.25<r^{+}_{pol}$ corresponding to the region of the inner ergosphere with stable locally counterrotating SPOs with positive covariant energy $E>0$. In the left column, the full/dashed orange curves in the ($X-q$) plane for $q>0$ correspond to the unstable/stable SPOs in the region $r>0$, the full orange curves with $q<0$ correspond to the unstable SPOs at $r<0$. The meaning of the other curves follows from the captions. The yellow, light grey, light red, light purple and purple region depicts the motion constants of the escaping, trapped, engrossed, repelled and escaping to the other infinity photons, respectively. The colouring of the individual parts of the local celestial spheres of the emitter in the middle and right columns are chosen correspondingly. Dashed orange curves  correspond to photons which reach some radial turning point and finally wind to the unstable polar SPO, while the full orange curve matches the photons that reach it directly. In the case $\theta_{e}=0\dgr$, the unstable SPO is the polar SPO at $r^{-}_{pol}=2.08$. The blue dot in the figures below represents a photon maintaining the current stable spherical orbit.   
}\label{Fig_cones_IVa_1.25_0_2p4deg}
\end{figure*}

\begin{figure*}[h]
	\centering \textbf{Class IVa: $y=0.02,\quad a^2=1.2$}\\
	\begin{tabularx}{\textwidth}{|XXX|}
		\hline
		\multicolumn{3}{|c|}{$r_{e}=1.25$}\\
		\hline
		\includegraphics[width=0.25\textwidth]{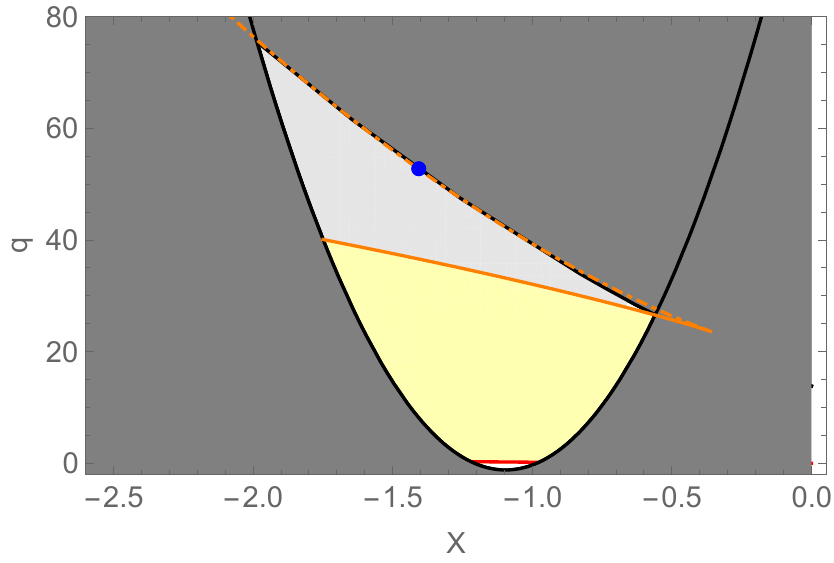}&
		\includegraphics[width=0.28\textwidth]{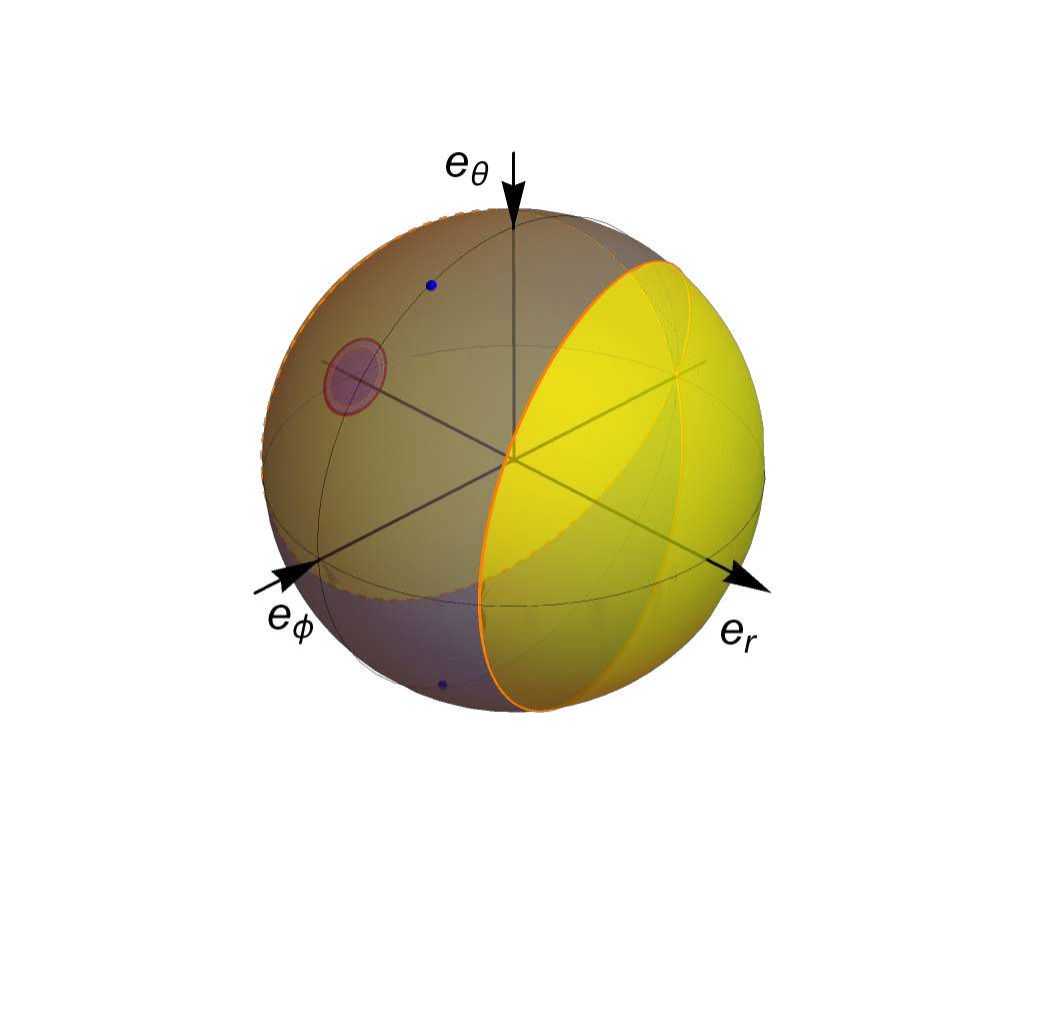}&\includegraphics[width=0.28\textwidth]{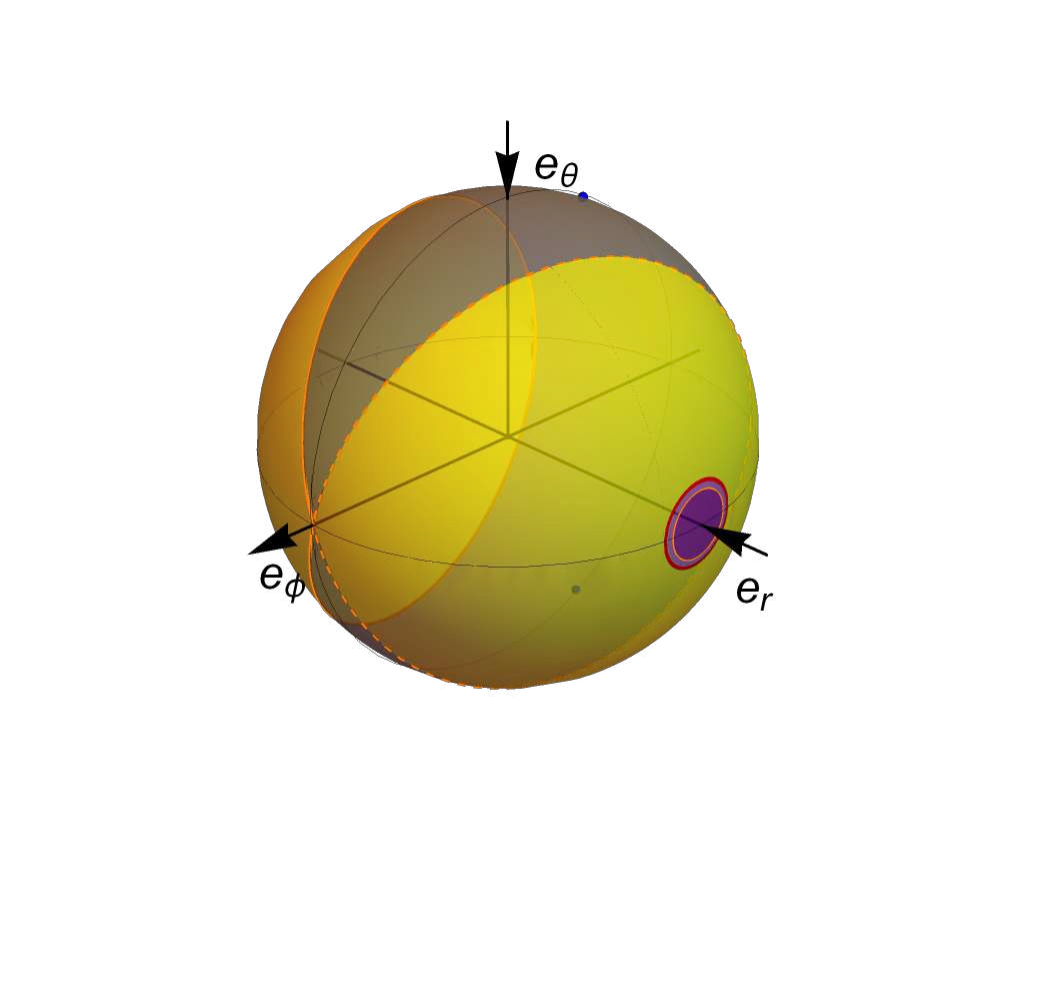}\\
		\multicolumn{3}{|c|}{(a): $\theta_{e}=\theta_{max(circ)}=5.74\dgr$}\\
		\hline
		\raisebox{1.0cm}[0pt]{\bet{c} \includegraphics[width=0.25\textwidth]{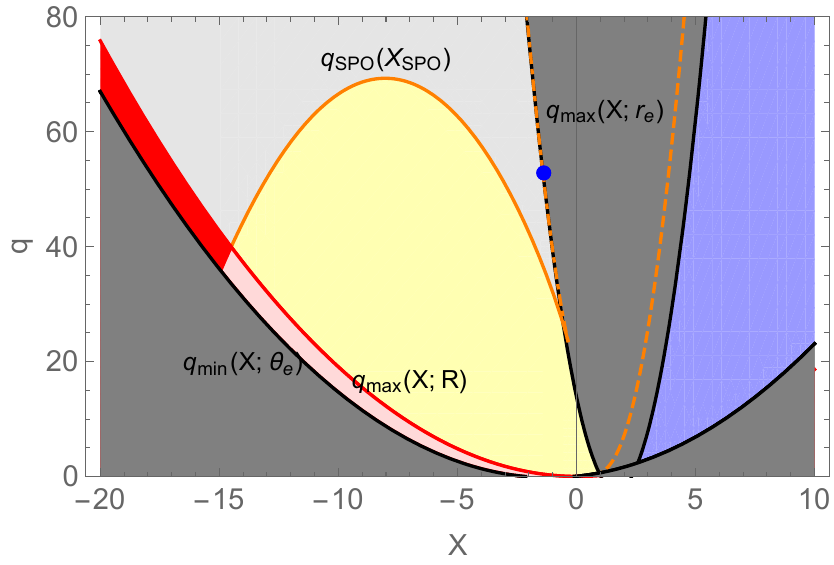}\\
			\includegraphics[width=0.25\textwidth]{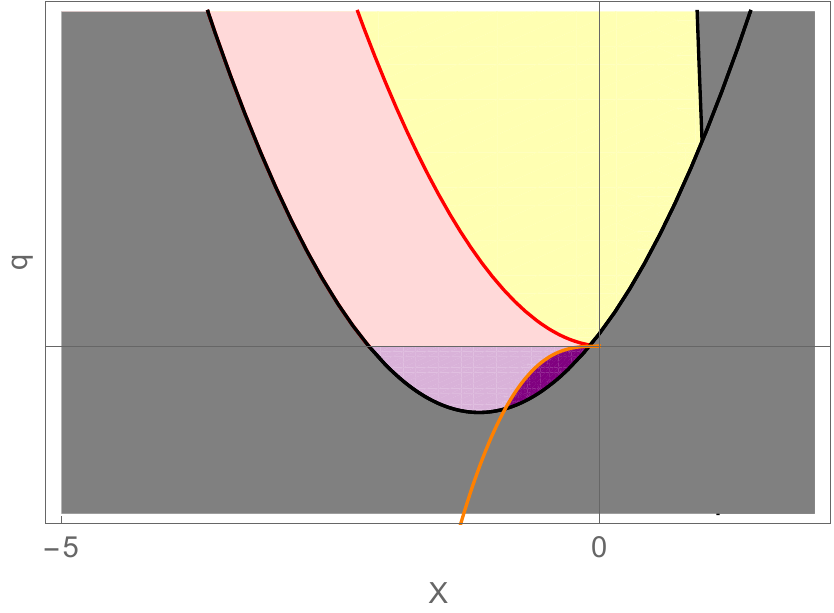} \ent }&
		\includegraphics[width=0.28\textwidth]{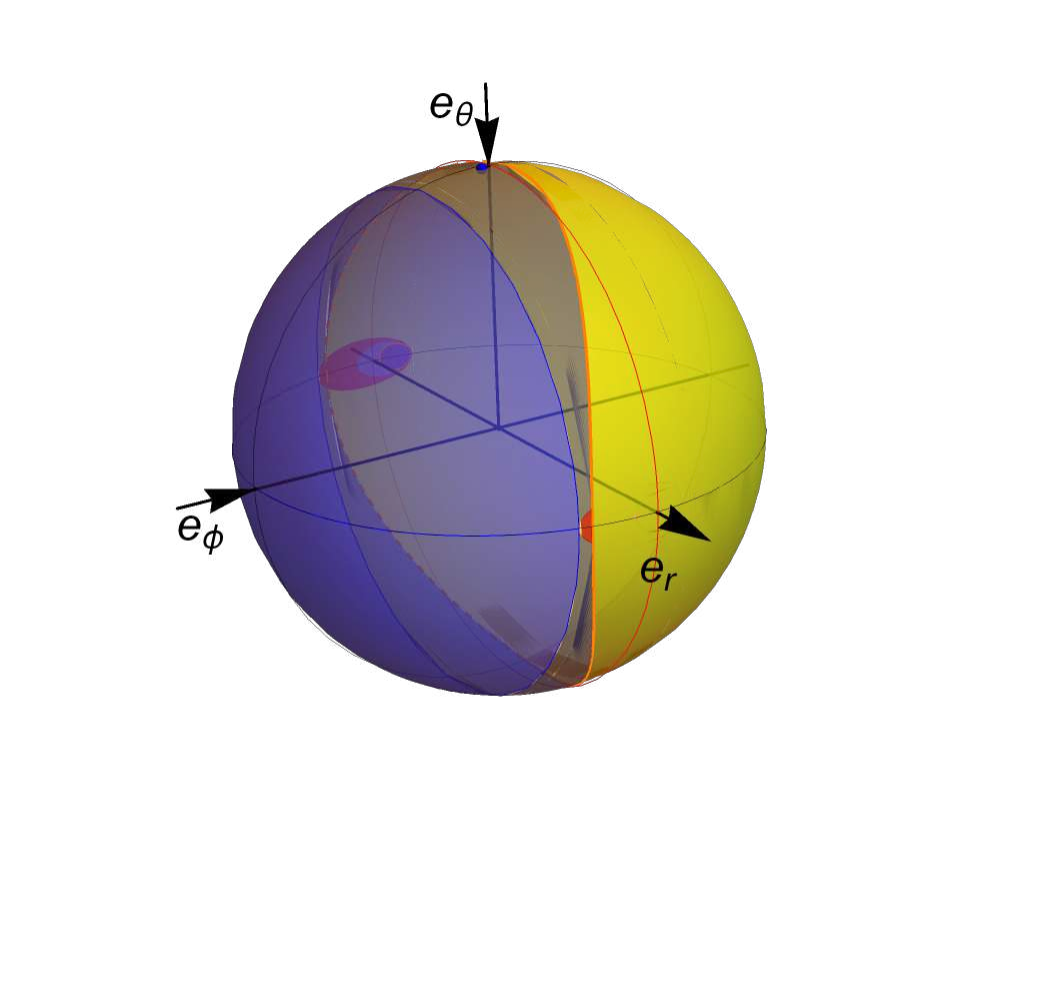}&\includegraphics[width=0.28\textwidth]{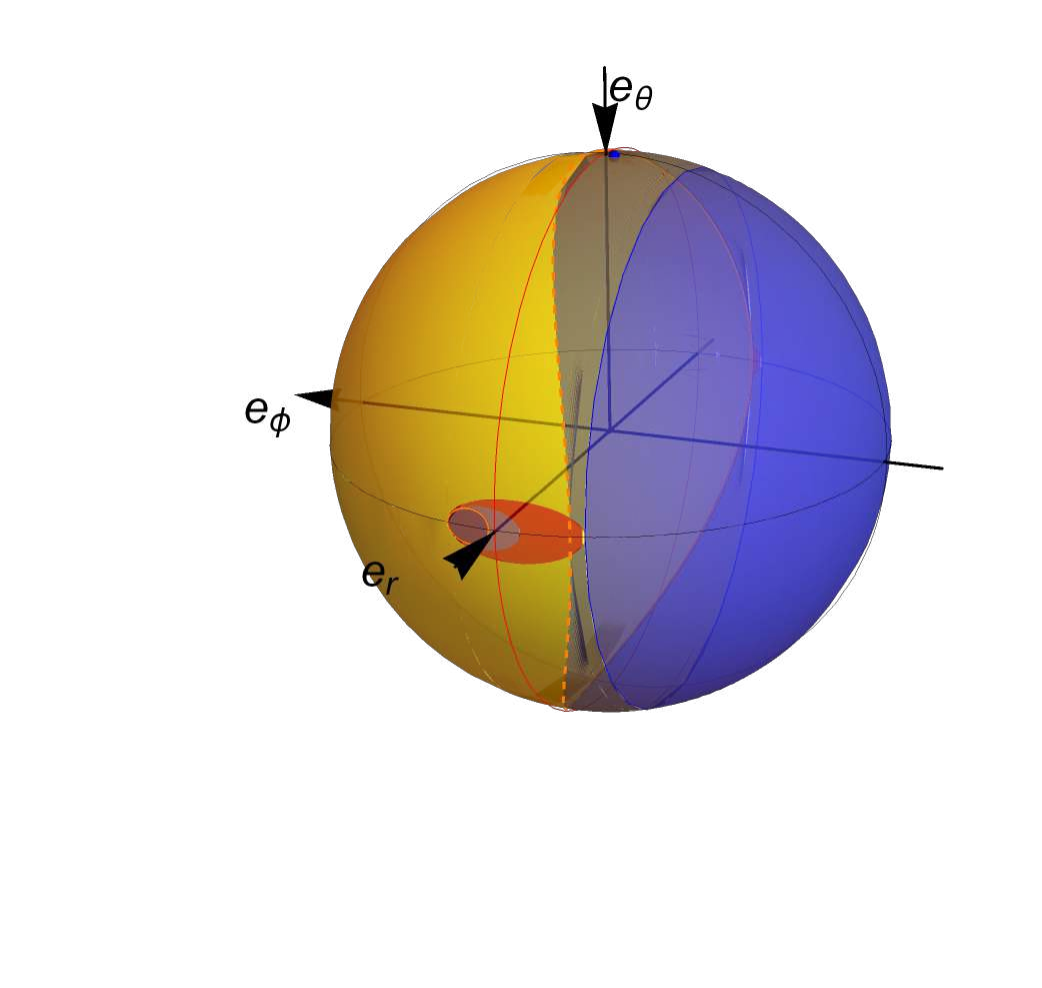}\\
		\multicolumn{3}{|c|}{(b): $\theta_{e}=66.28\dgr$}\\
		\hline
	\end{tabularx}	
	\caption{Continuation of Fig. \ref{Fig_cones_IVa_1.25_0_2p4deg}.  The latitude in the upper figure is chosen so that a continuous transition of the shape of the trapped cone from the spherical belt to the spherical bi-angle can be seen, whose circular edges finally disconnect for $\theta_{max(circ)}$.  In the figure below, the source is inside the ergoregion $\theta_{e}>\theta_{erg}=22.3\dgr$, so there are photons detected with $E<0$, corresponding to the blue regions on the ($X-q$) -plane and on the celestial sphere. The light/rich red parts of the ($X-q$) plane correspond to engrossed photons reaching the superspinar surface directly/after passing through a radial turning point. The exact value of the latitude corresponds to the marginal case, when negative energy photons start to reach the superspinar surface.   
	}\label{Fig_cones_IVa_1.25_5_66deg}
\end{figure*}

\begin{figure*}[h]
	\centering \textbf{Class IVa: $y=0.02,\quad a^2=1.2$}\\
	\begin{tabularx}{\textwidth}{|XXX|}
		\hline
		\multicolumn{3}{|c|}{$r_{e}=1.25$}\\
		\hline
		\includegraphics[width=0.28\textwidth]{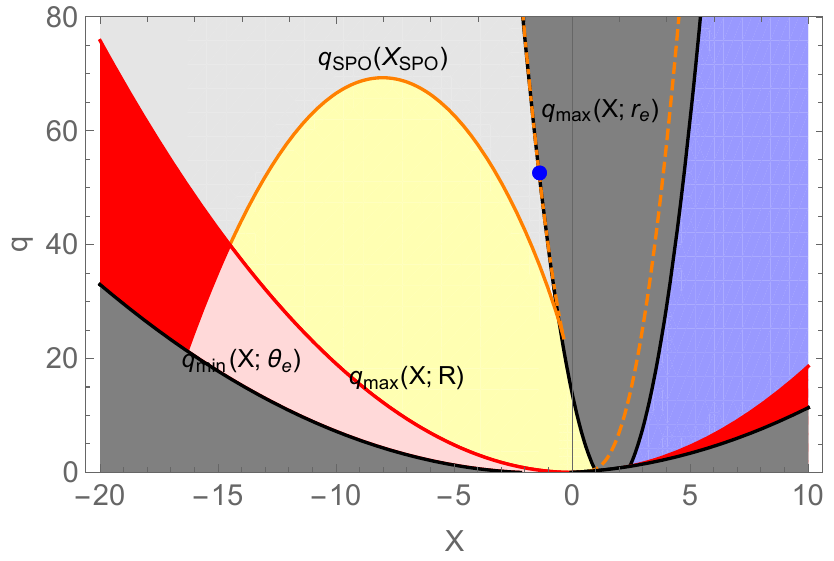}&
		\includegraphics[width=0.28\textwidth]{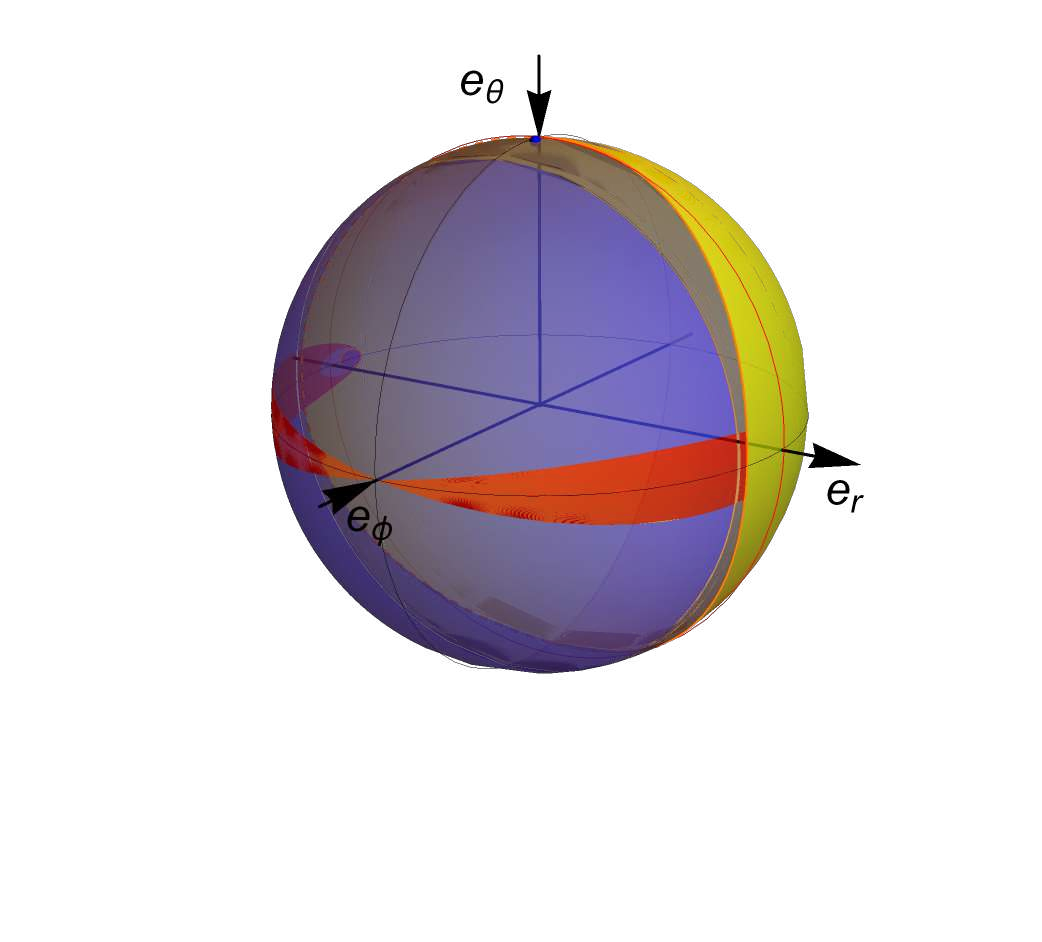}&\includegraphics[width=0.28\textwidth]{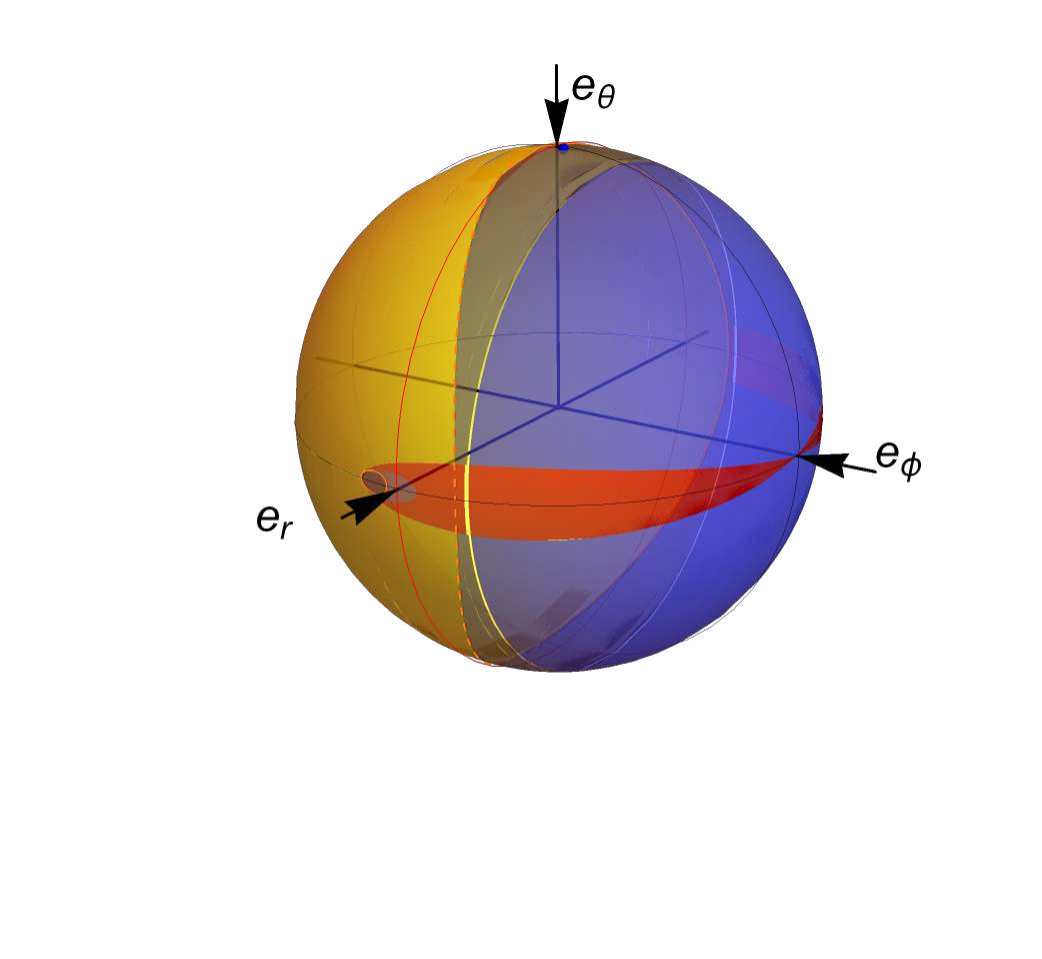}\\
		\multicolumn{3}{|c|}{(a): $\theta_{e}=72.86\dgr$}\\
		\hline
		\includegraphics[width=0.28\textwidth]{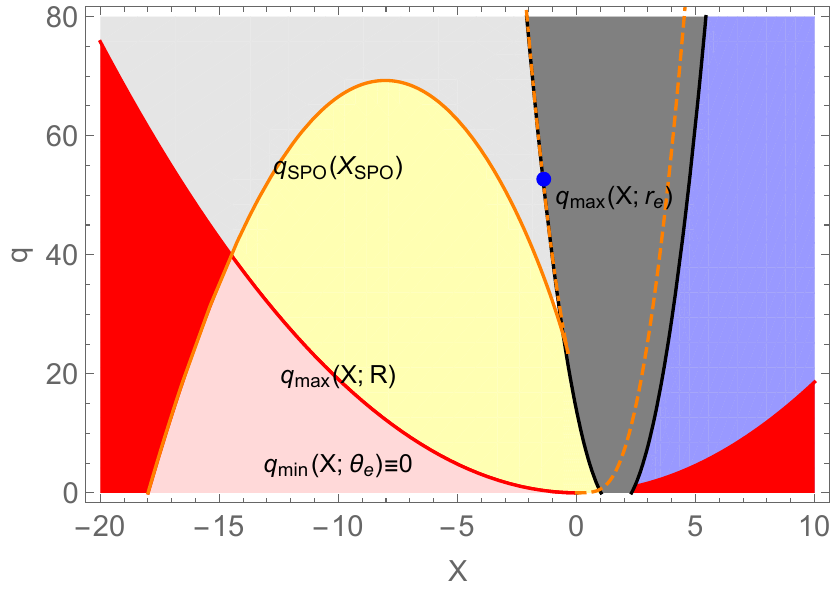}&
		\includegraphics[width=0.28\textwidth]{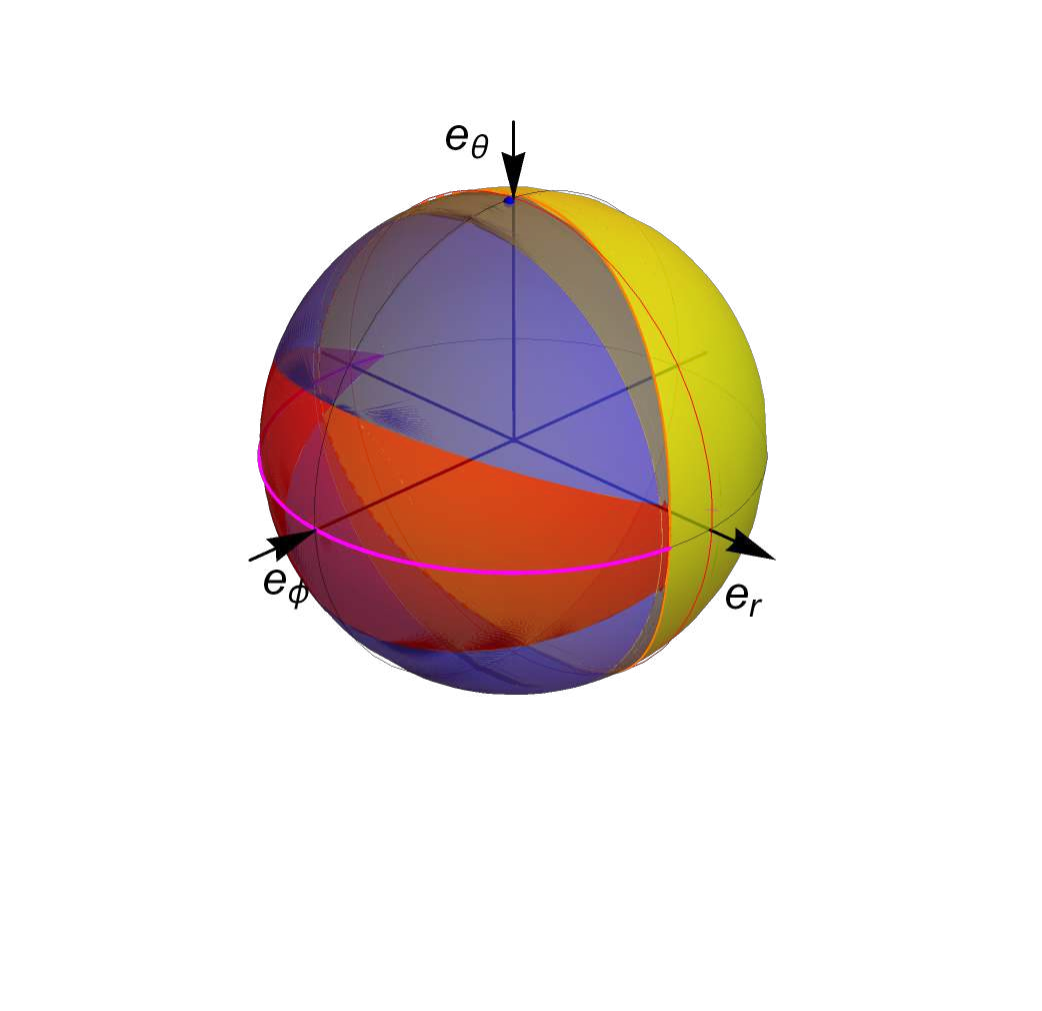}&\includegraphics[width=0.28\textwidth]{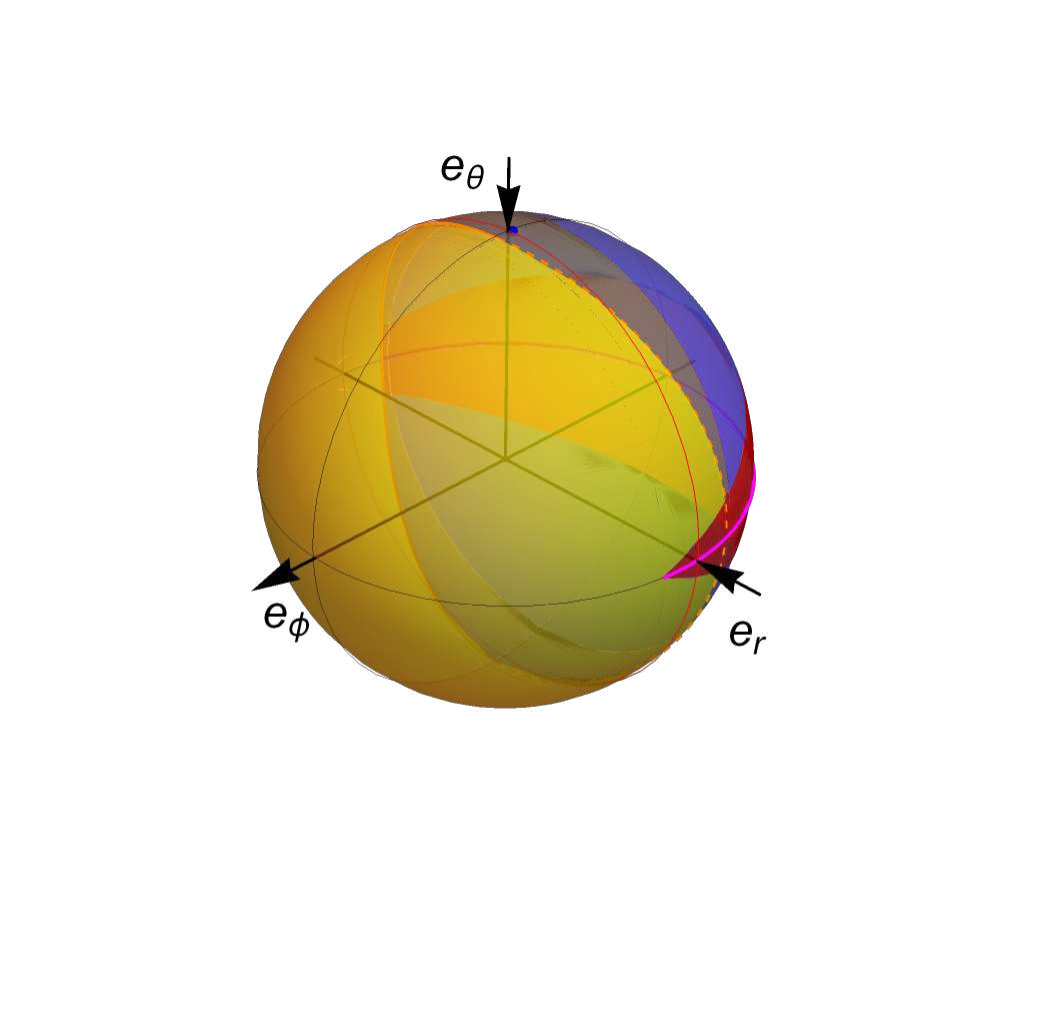}\\
		\multicolumn{3}{|c|}{(b): $\theta_{e}=90\dgr$}\\
		\hline
		
	\end{tabularx}	
	\caption{Continuation of Fig. \ref{Fig_cones_IVa_1.25_5_66deg}. The surface of the superspinar with increasing latitude towards the equatorial plane is successively impacted by the negative energy photons in both positive ($k^{(r)}>0$) and negative ($k^{(r)}<0$) locally measured radial directions. The meaning of the colouring is the same as in the previous figures. 
	}\label{Fig_cones_IVa_1.25_90deg}
\end{figure*}

\begin{figure*}[h]
	\centering
\begin{tabularx}{\textwidth}{|XXX|}
	\hline
	\multicolumn{3}{|c|}{$r_{e}=2.2$}\\
	\hline
	\raisebox{2.8cm}[0pt]{\bet{c}	\includegraphics[width=0.25\textwidth]{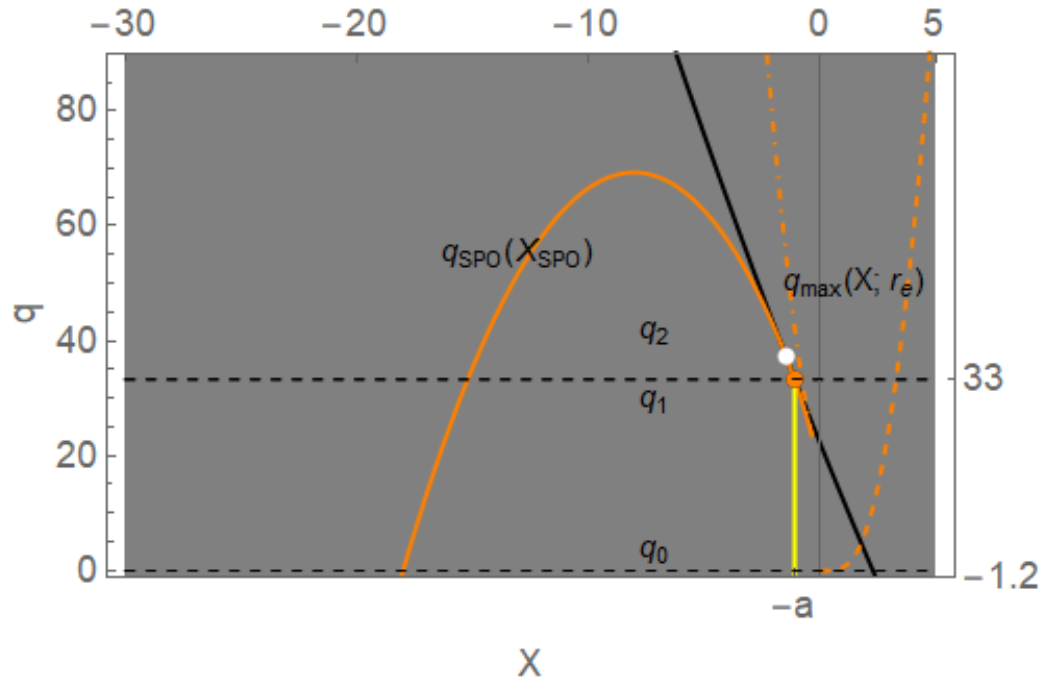}\\ \includegraphics[width=0.25\textwidth]{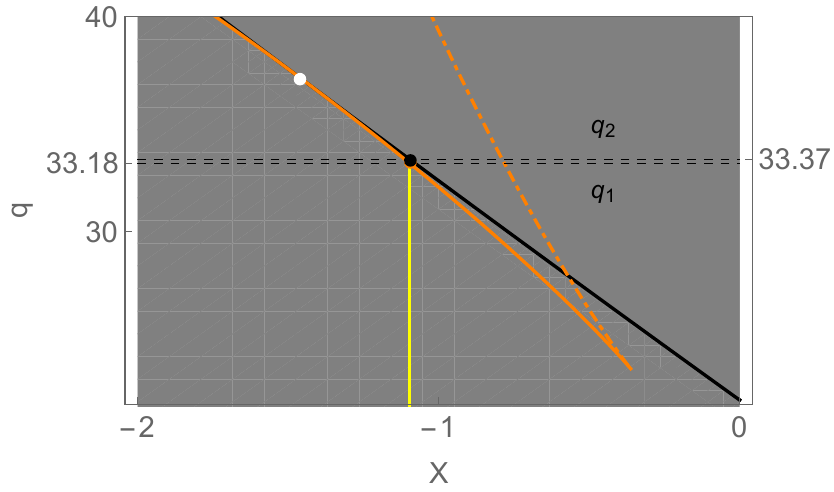} \ent } &
	\raisebox{-0.5cm}[0pt]{\includegraphics[width=0.3\textwidth]{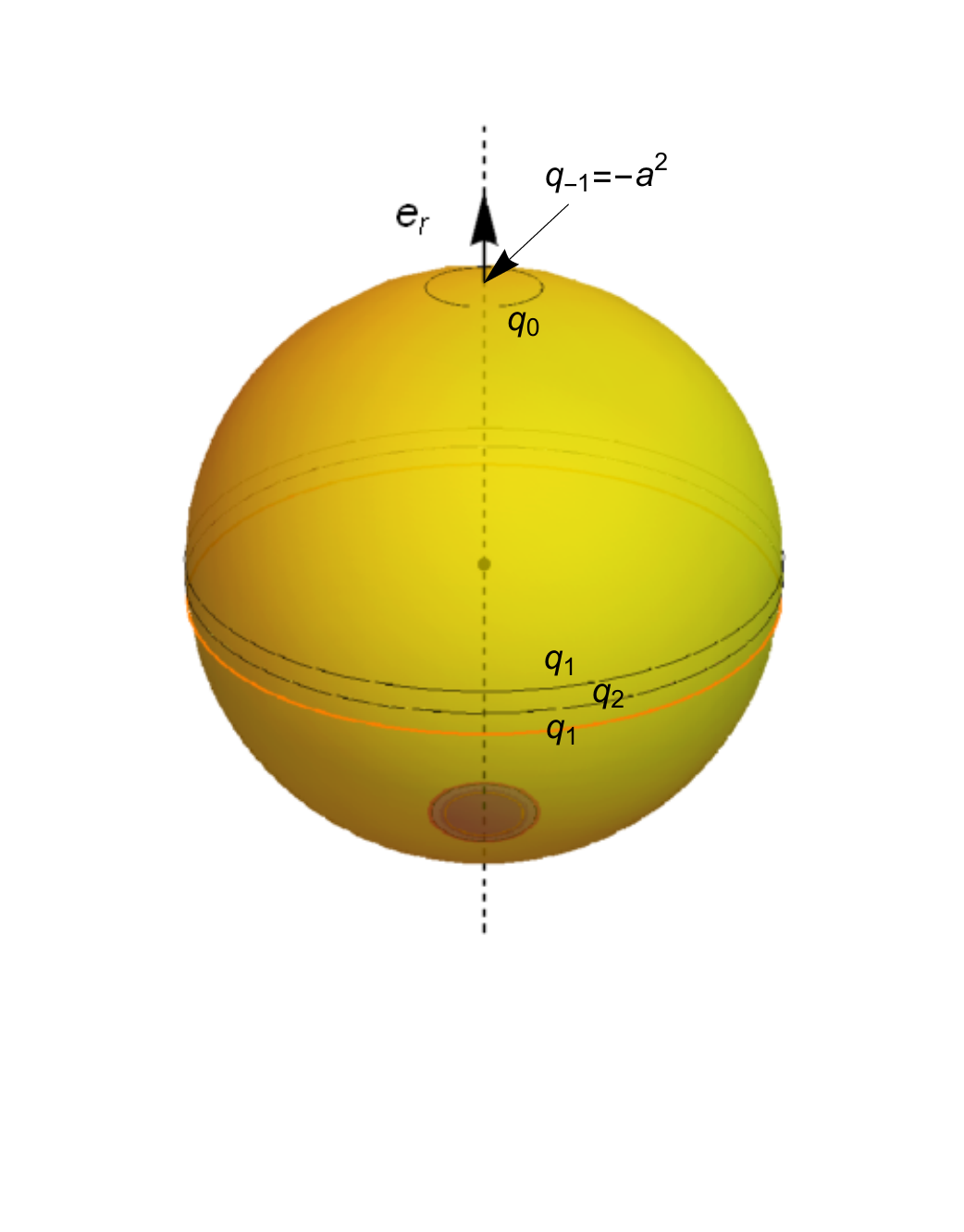}}	&\includegraphics[width=0.3\textwidth]{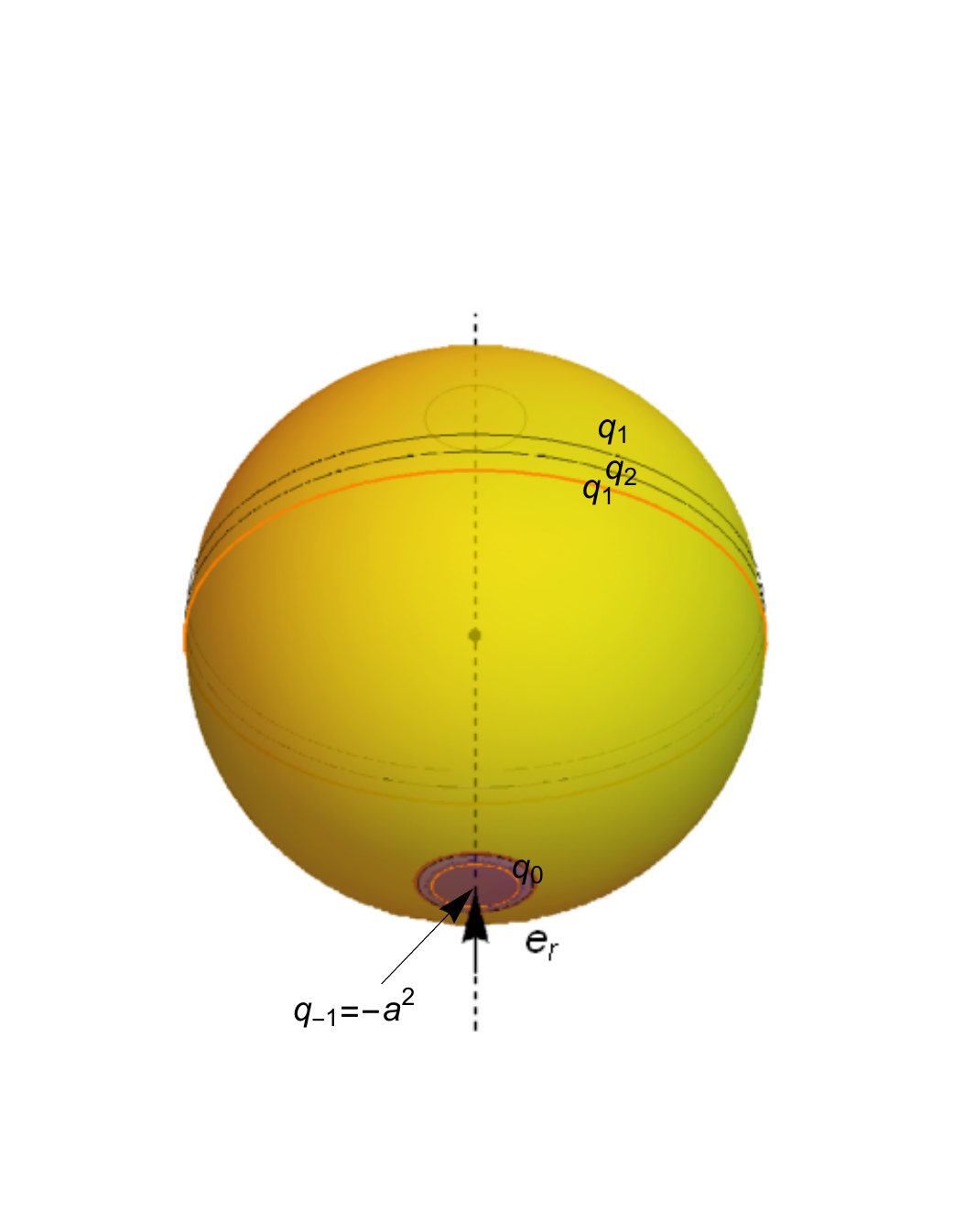}\\
	\multicolumn{3}{|c|}{(a): $\theta_{e}=0\dgr$}\\
	\hline
	
	\includegraphics[width=0.3\textwidth]{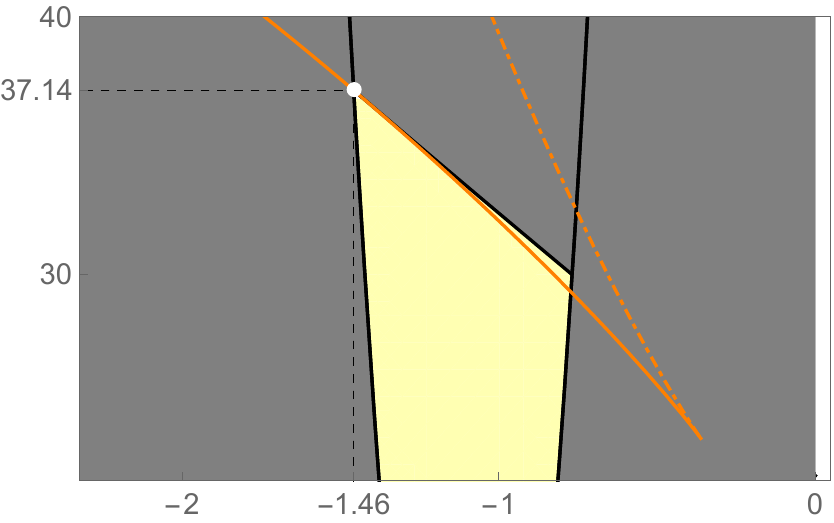}&
	\includegraphics[width=0.3\textwidth]{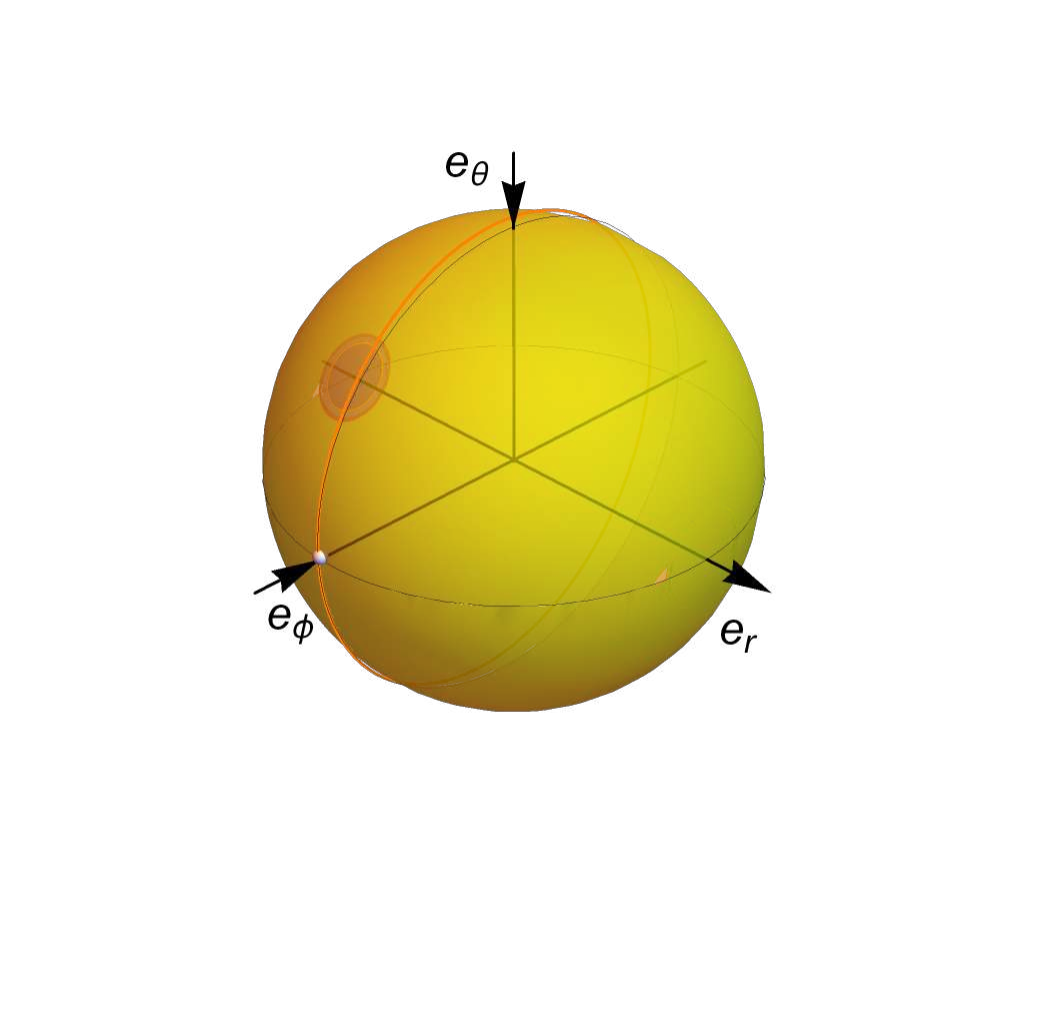}&\includegraphics[width=0.3\textwidth]{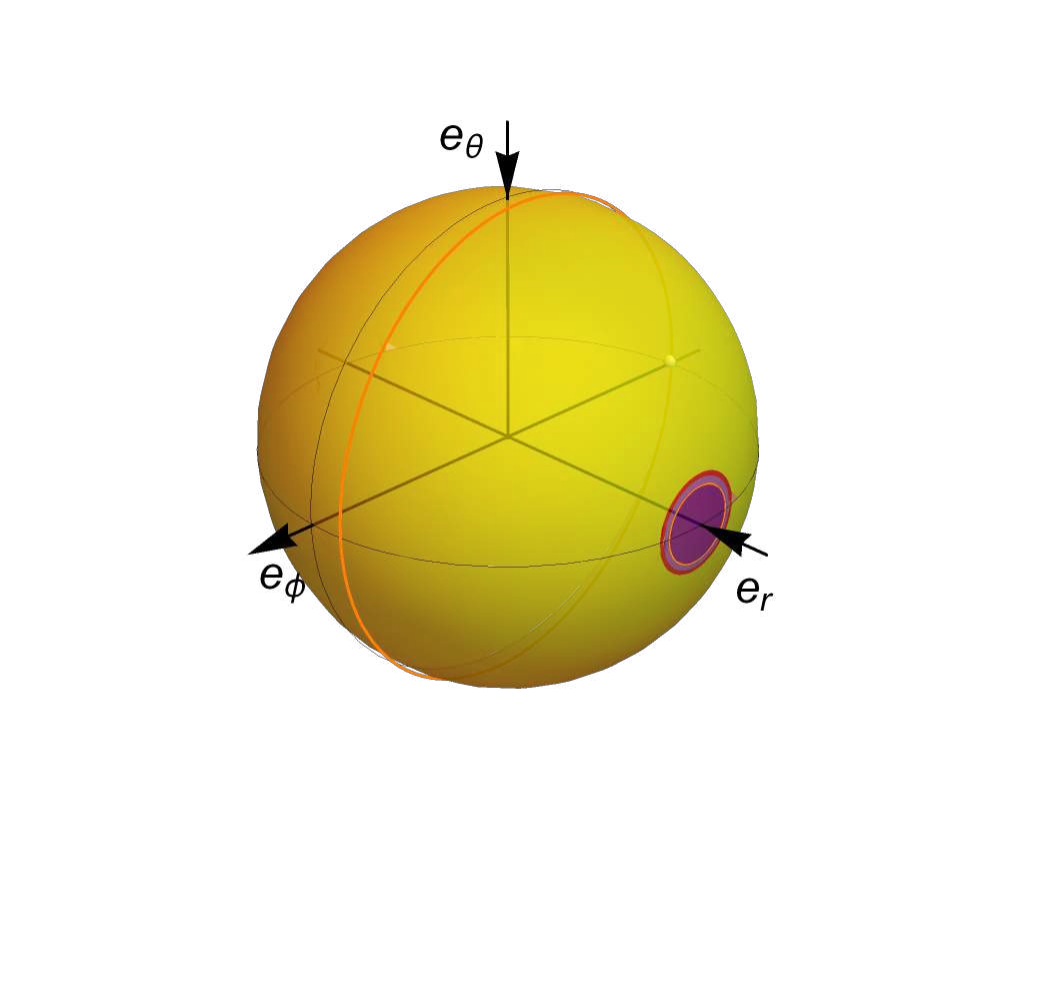}\\
	\multicolumn{3}{|c|}{(b): $\theta_{e}=3.32\dgr$}\\
	\hline
		\includegraphics[width=0.3\textwidth]{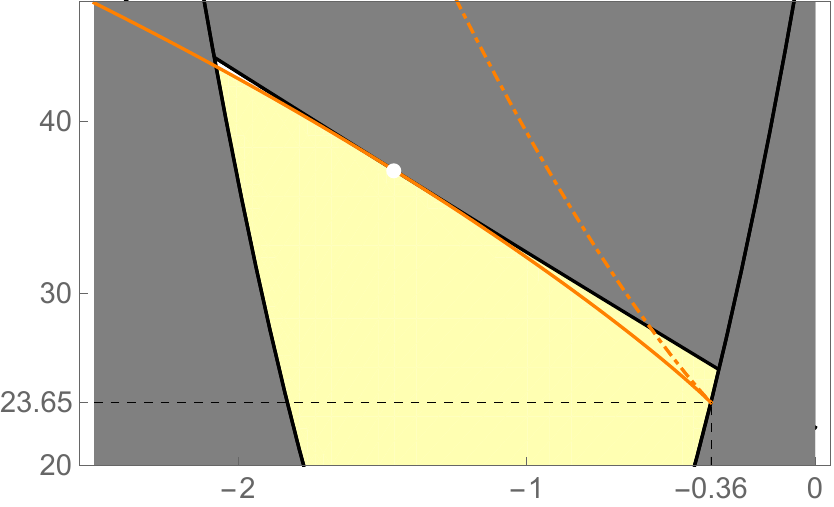}&
	\includegraphics[width=0.3\textwidth]{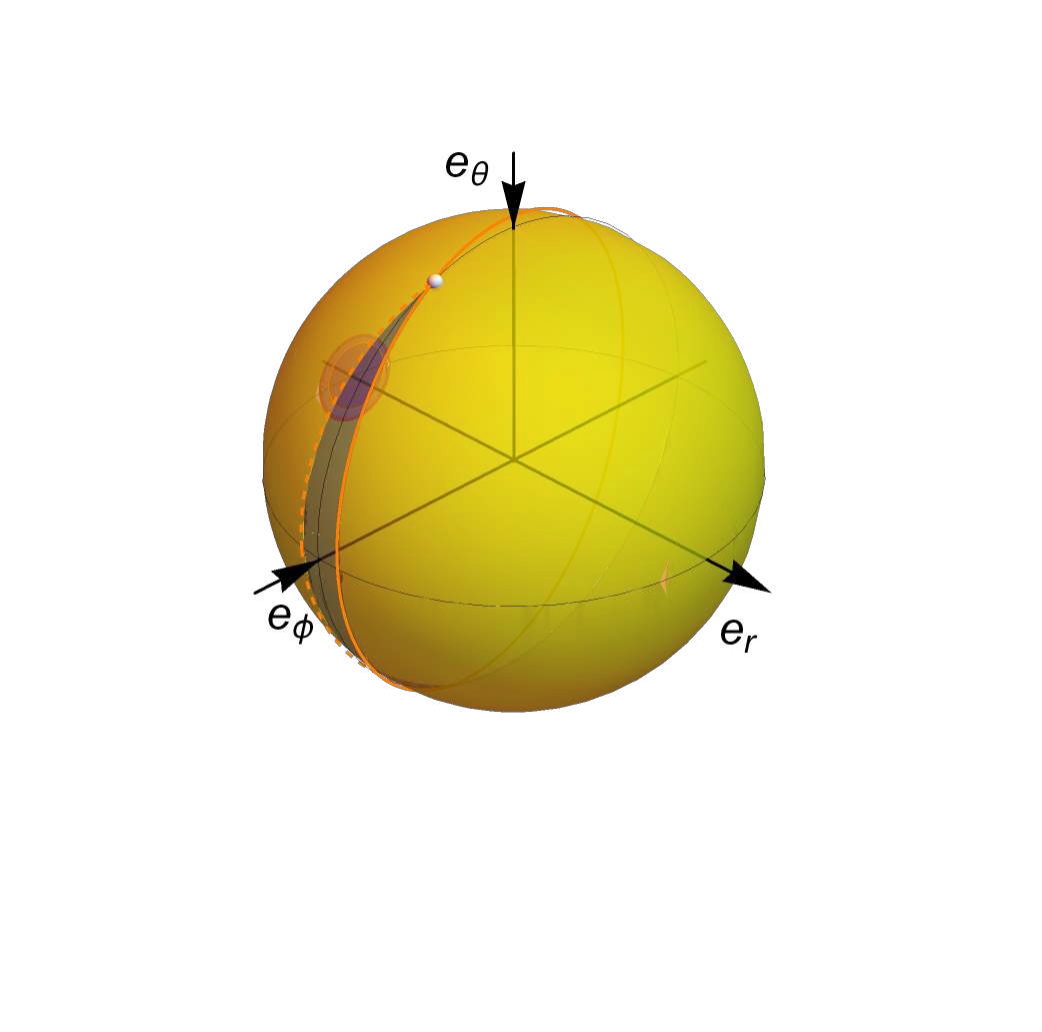}&\includegraphics[width=0.3\textwidth]{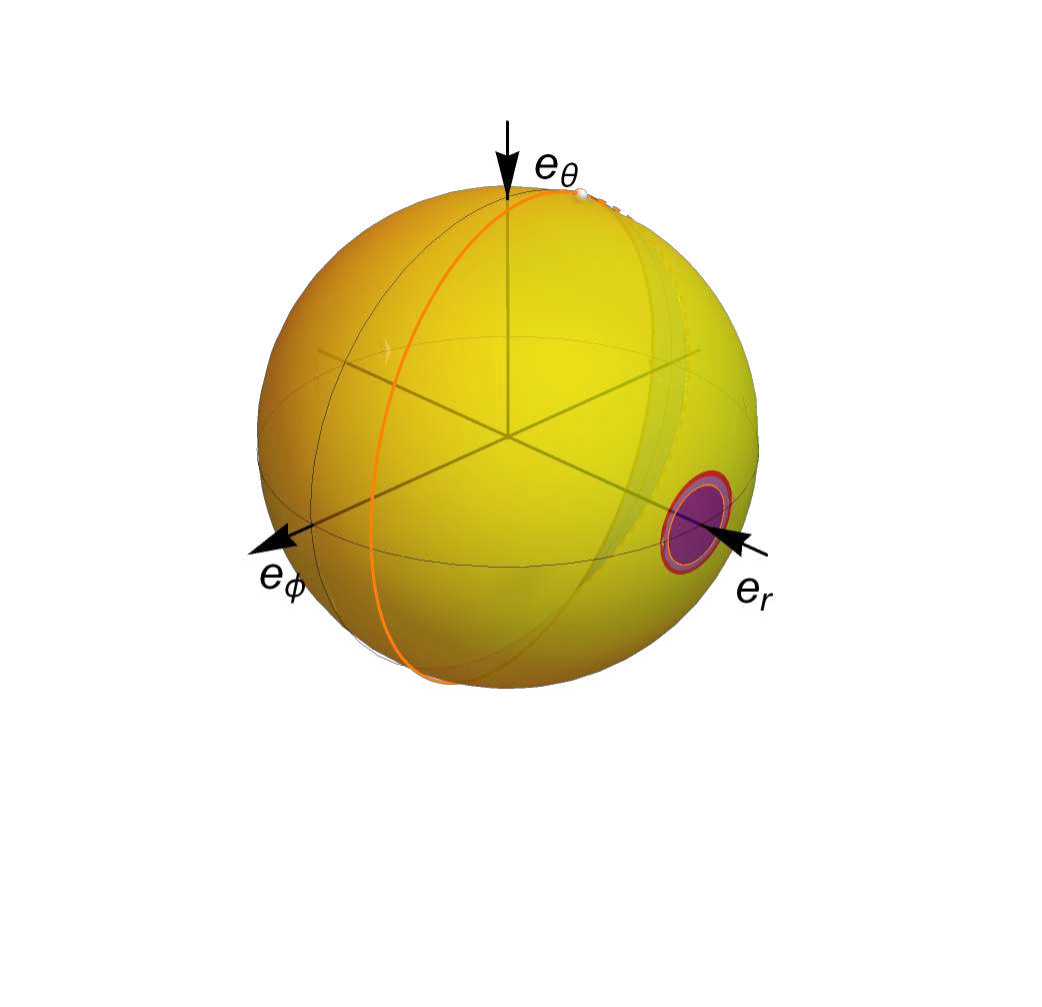}\\
	\multicolumn{3}{|c|}{(c): $\theta_{e}=\theta_{max(circ)}=8.29\dgr$}\\
	\hline
	\end{tabularx}	
\caption{LECs for $r_{e}=2.2$ representing the case $r^{-}_{pol}<r_{e}<r^{+}_{erg}$, corresponding to the region of the inner ergosphere with unstable counterrotating SPOs. There are no trapped cones on the spin axis, since all photons emitted to the polar SPO are directed inwards. For $\theta_{e}=3.32\dgr$ there is photon directed on the actual unstable SPO, represented by the white dot. For $\theta_{e}>\theta_{max(circ)}=8.29\dgr$ the orange curve disconnects and an arc is formed.}
\label{Fig_cones_IVa_2.2_3.3deg}
\end{figure*}	

\begin{figure*}[h]
	\centering
\begin{tabularx}{\textwidth}{|XXX|}
	\hline
	\multicolumn{3}{|c|}{$r_{e}=2.2$}\\
	\hline
	\includegraphics[width=0.3\textwidth]{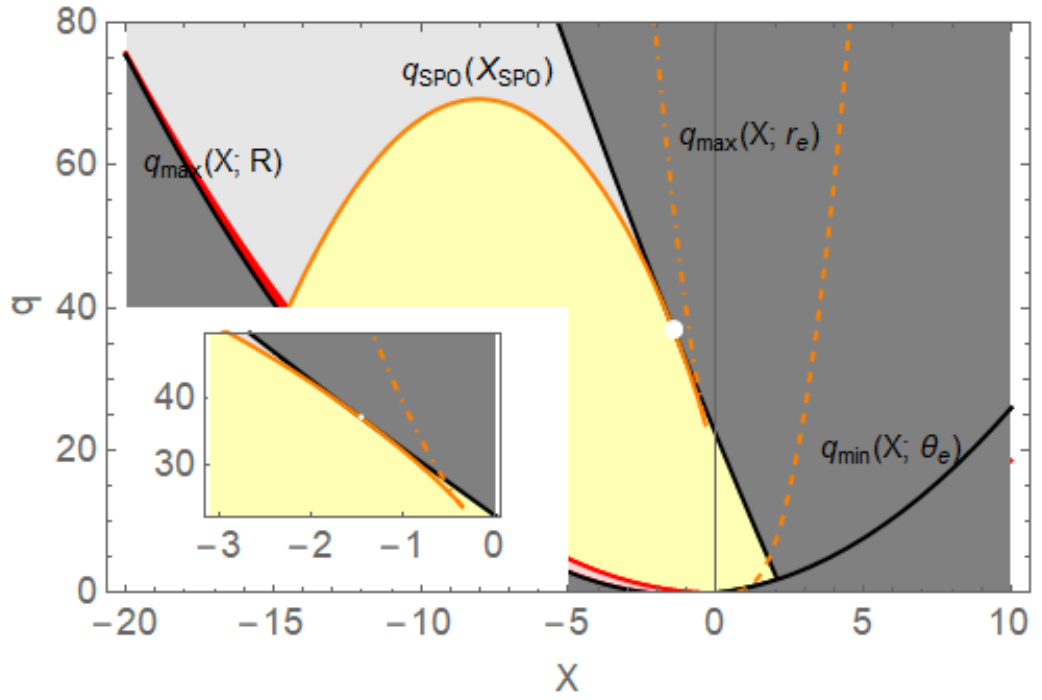}&
	\includegraphics[width=0.3\textwidth]{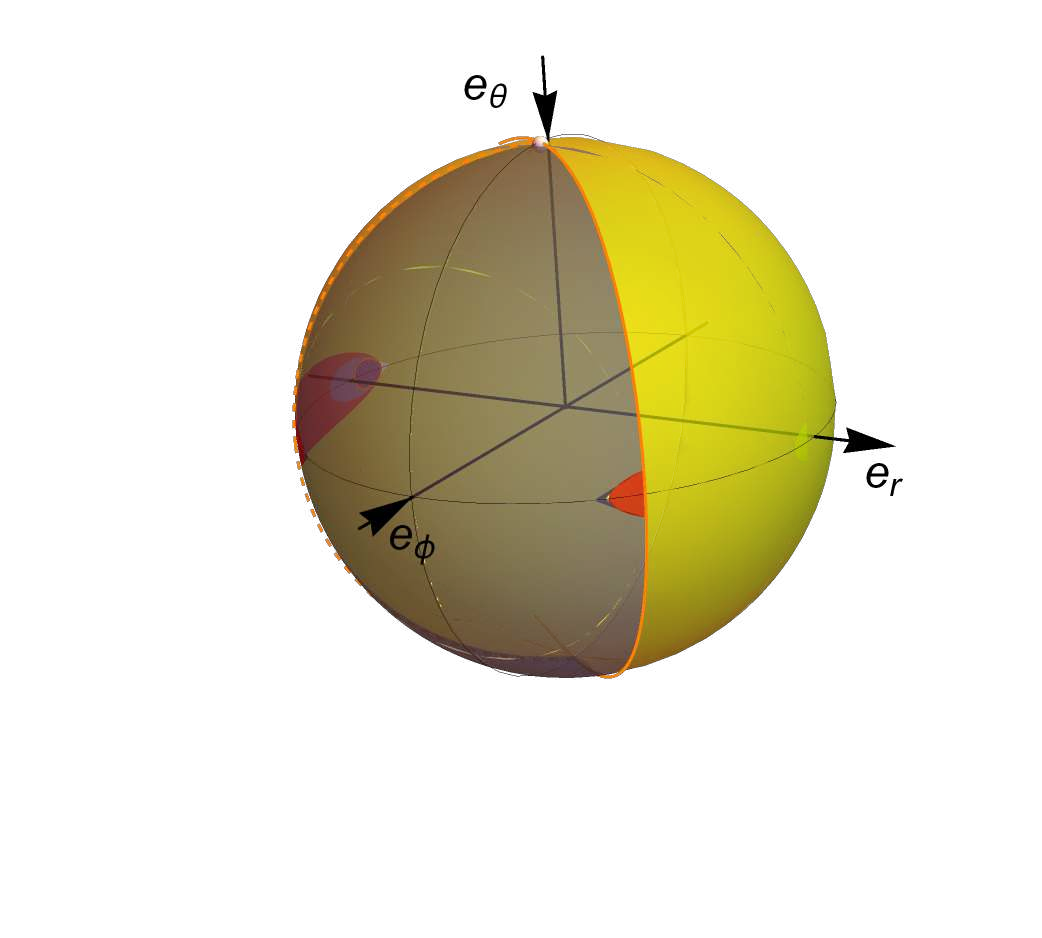}&\includegraphics[width=0.3\textwidth]{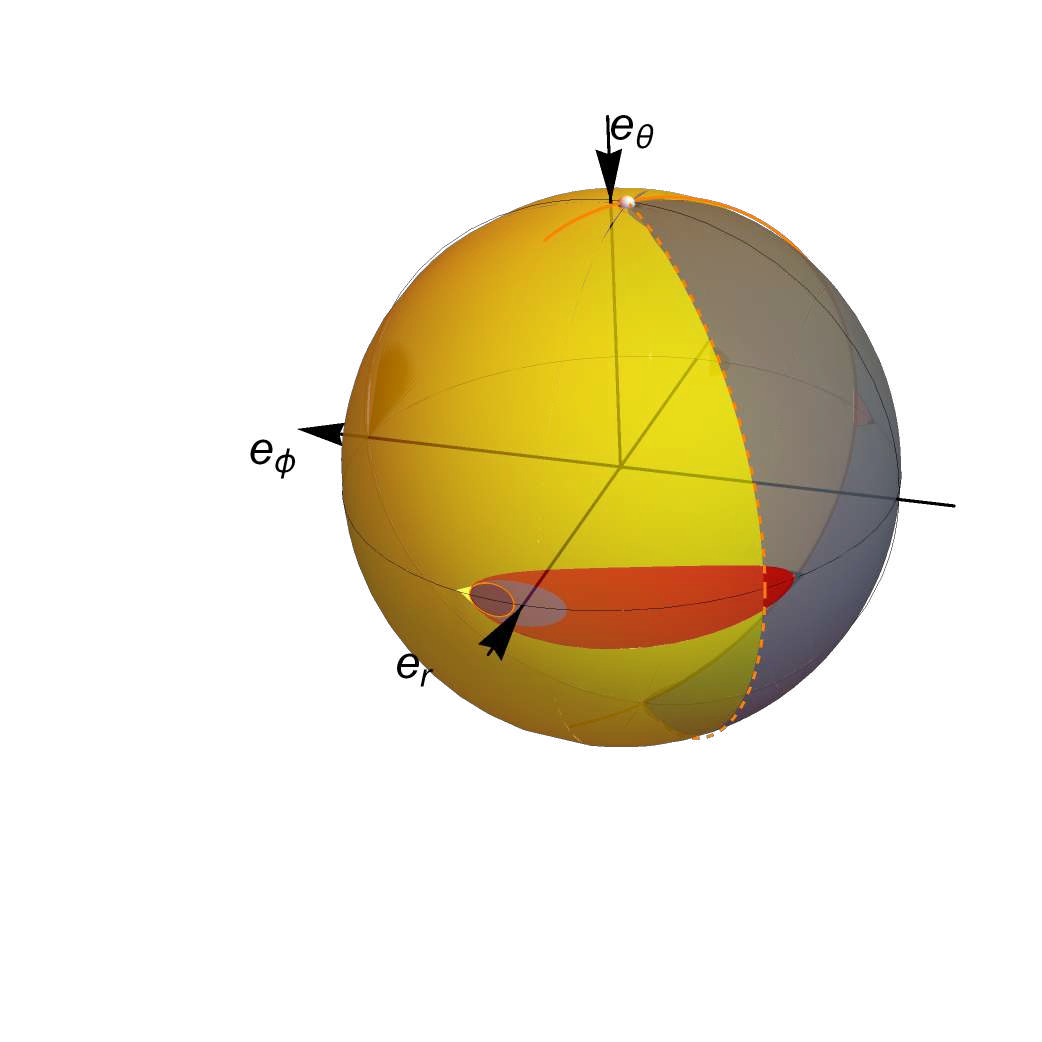}\\
	\multicolumn{3}{|c|}{(a): $\theta_{e}=65\dgr$}\\
	\hline
	\includegraphics[width=0.3\textwidth]{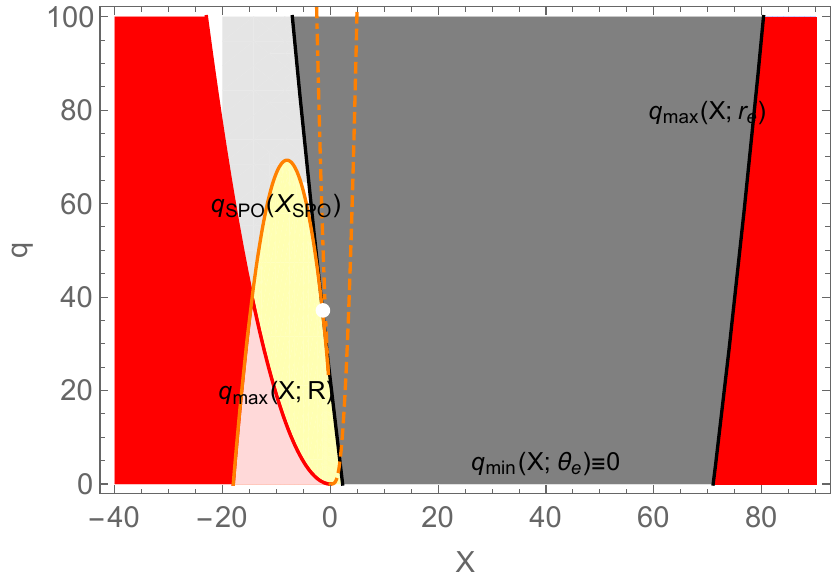}&
	\includegraphics[width=0.3\textwidth]{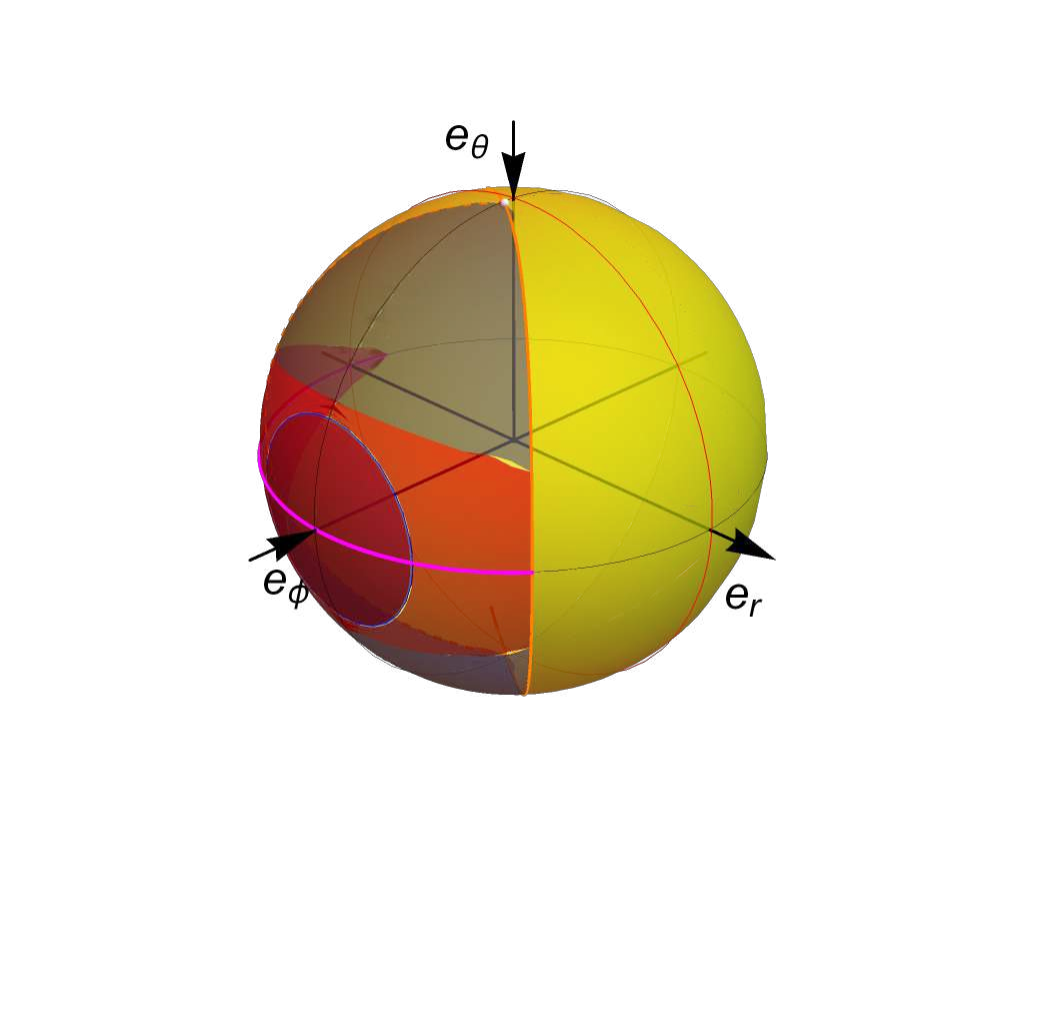}&\includegraphics[width=0.3\textwidth]{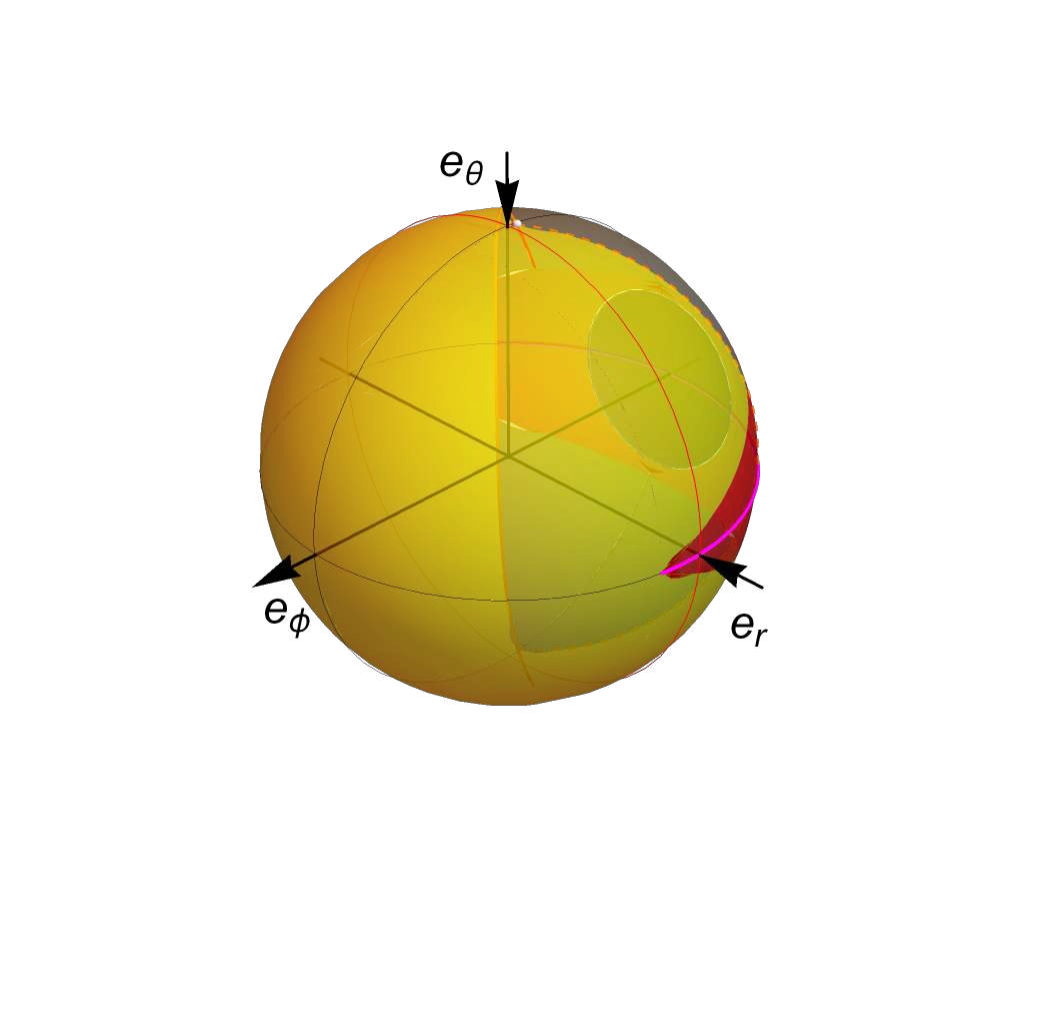}\\
	\multicolumn{3}{|c|}{(b): $\theta_{e}=90\dgr$}\\
	\hline
	\end{tabularx}	
	\caption{Continuation of Fig. \ref{Fig_cones_IVa_2.2_3.3deg}. Since $r_{e}<r^{-}_{ph}$ there is a trapped cone for sufficiently large latitudes. Furthermore, due to the position of the emitter in the ergosphere, there is a marginal latitude $\theta_{erg}(r)=69.5\dgr$, so that for $\theta_{e}>\theta_{erg}(r)$ there is a flux of trapped photons with $E<0$. In the figure below, all photons with negative energy are absorbed by the surface of the superspinar, which is shown by the blue region being covered by the red.    
	}\label{Fig_cones_IVa_2.2_90deg}
\end{figure*}

\begin{figure*}[h]
	\centering
	\begin{tabular}{|ccc|}
		\hline	
		\multicolumn{3}{|c|}{$r_{e}=2.65$}\\
		\hline
		\includegraphics[width=0.3\textwidth]{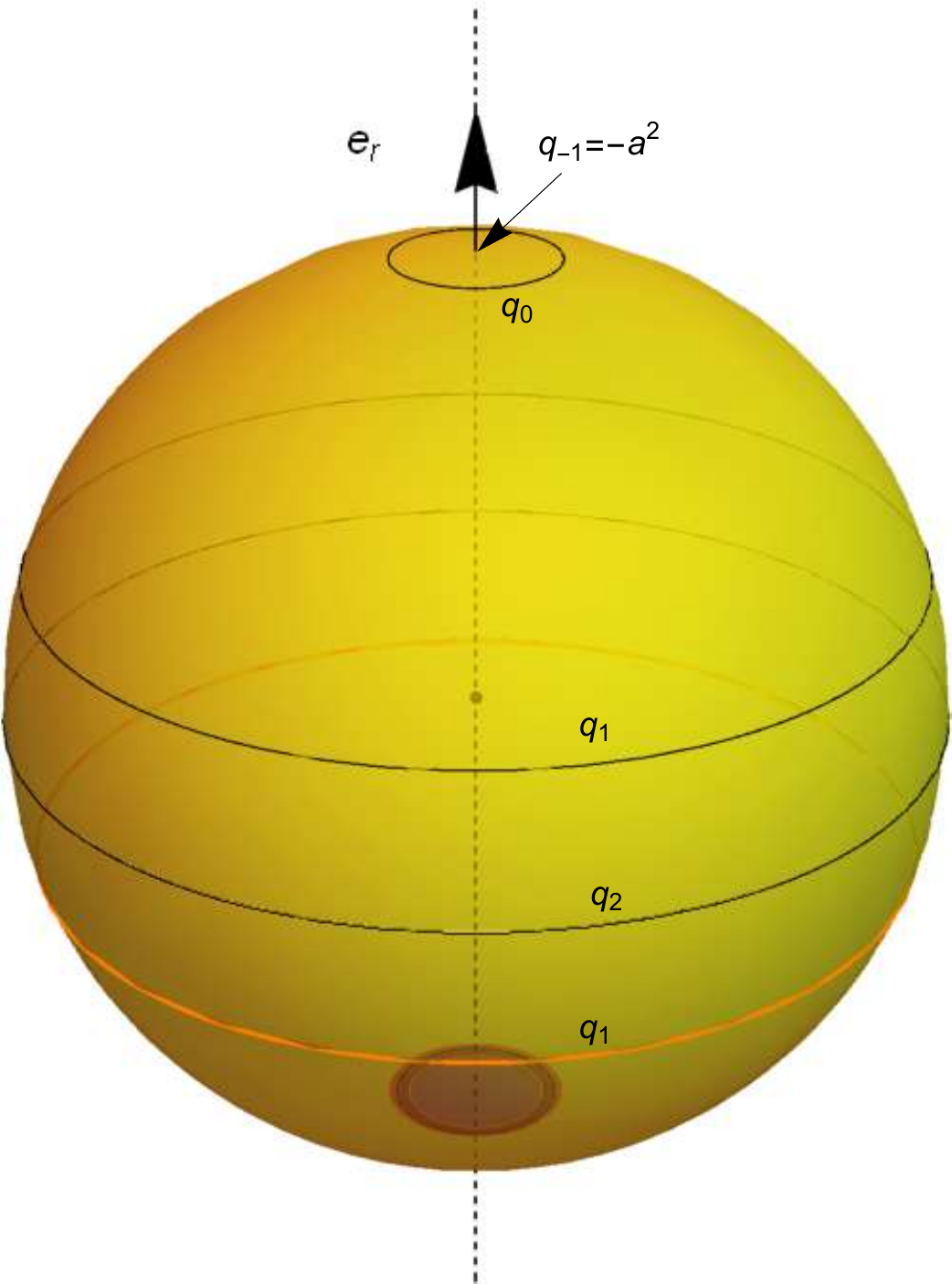}&(a): $\theta_{e}=0\dgr$&\includegraphics[width=0.3\textwidth]{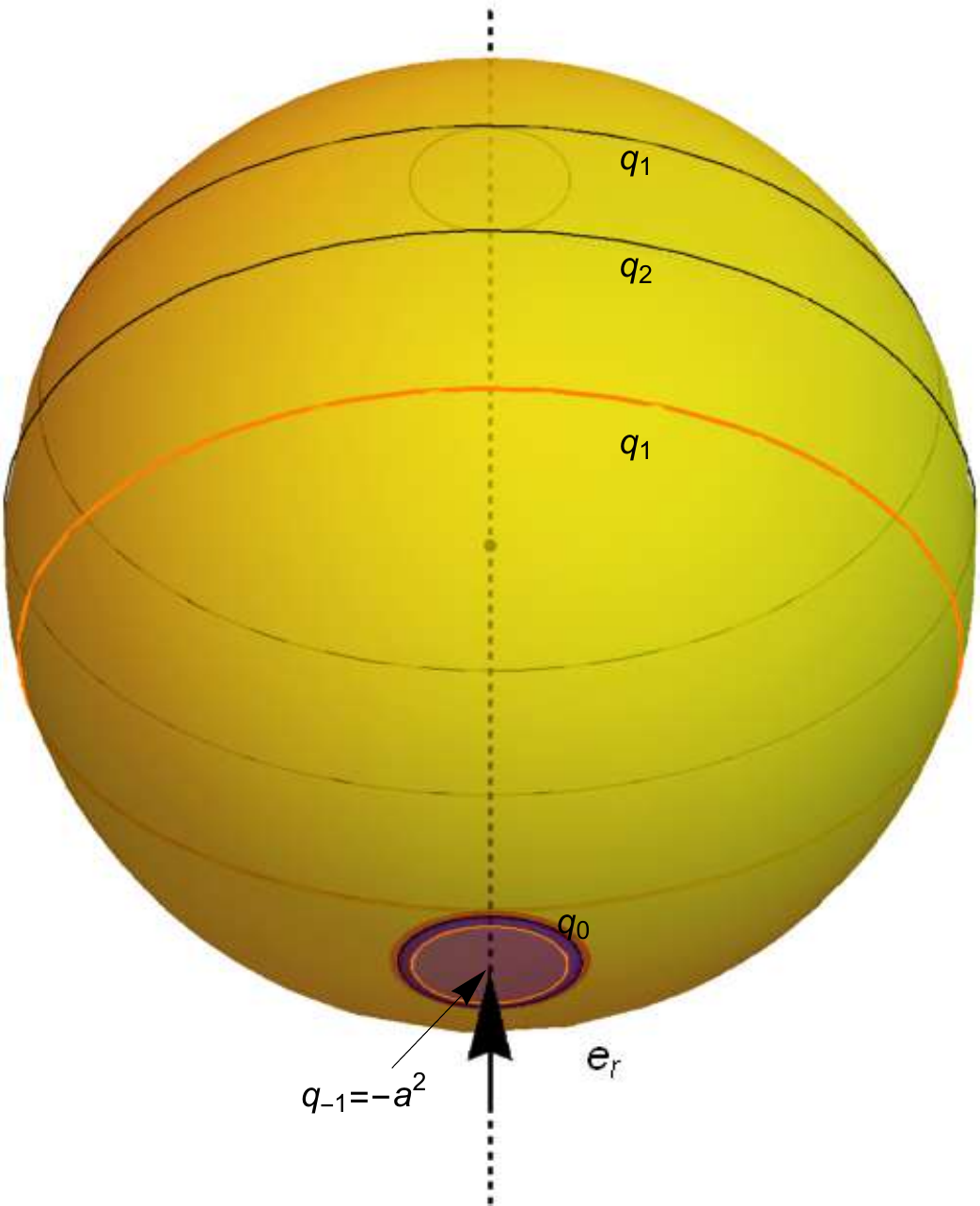}\\
		\hline
		\includegraphics[width=0.3\textwidth]{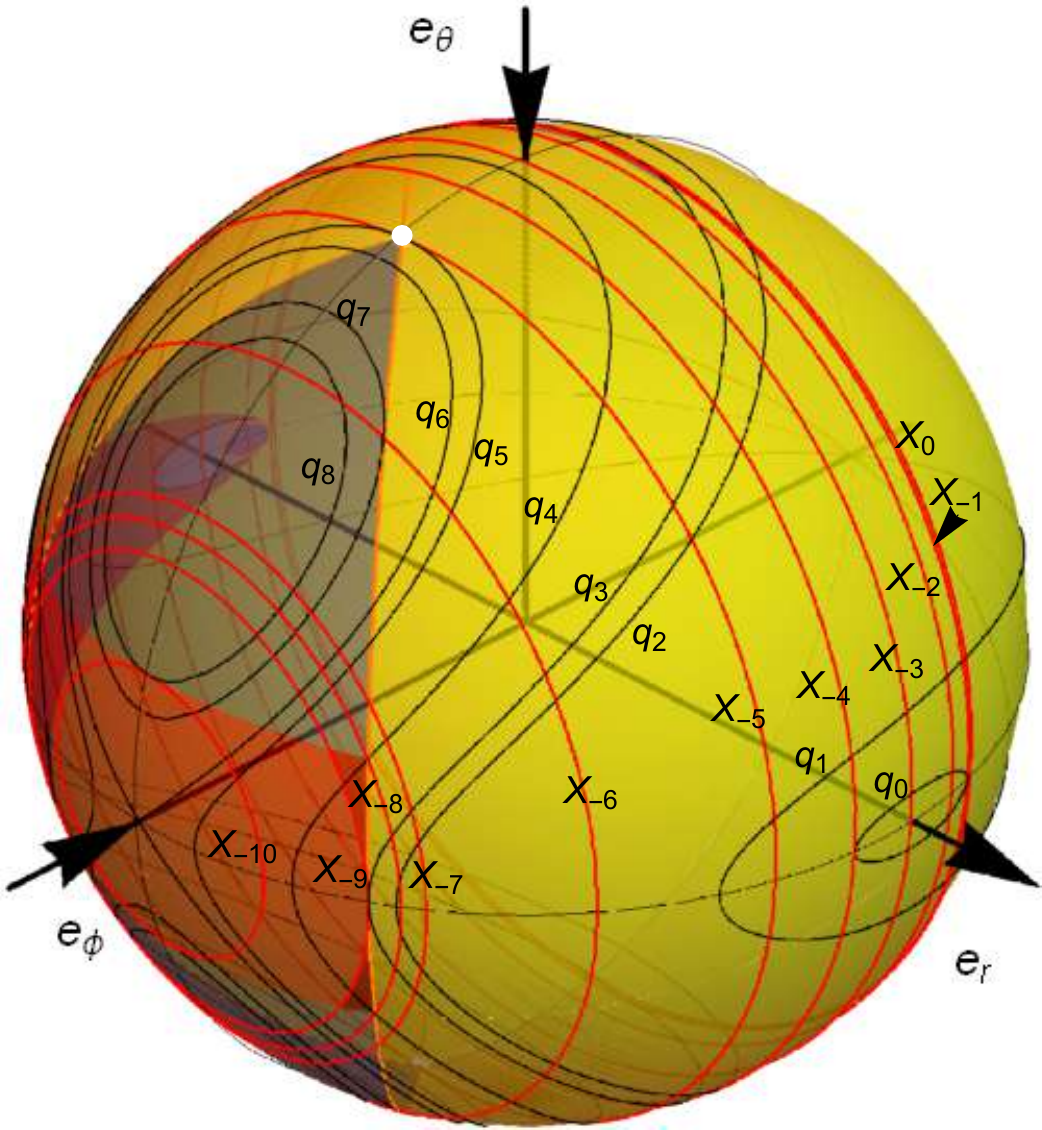}&(b): $\theta_{e}=70\dgr$&\includegraphics[width=0.3\textwidth]{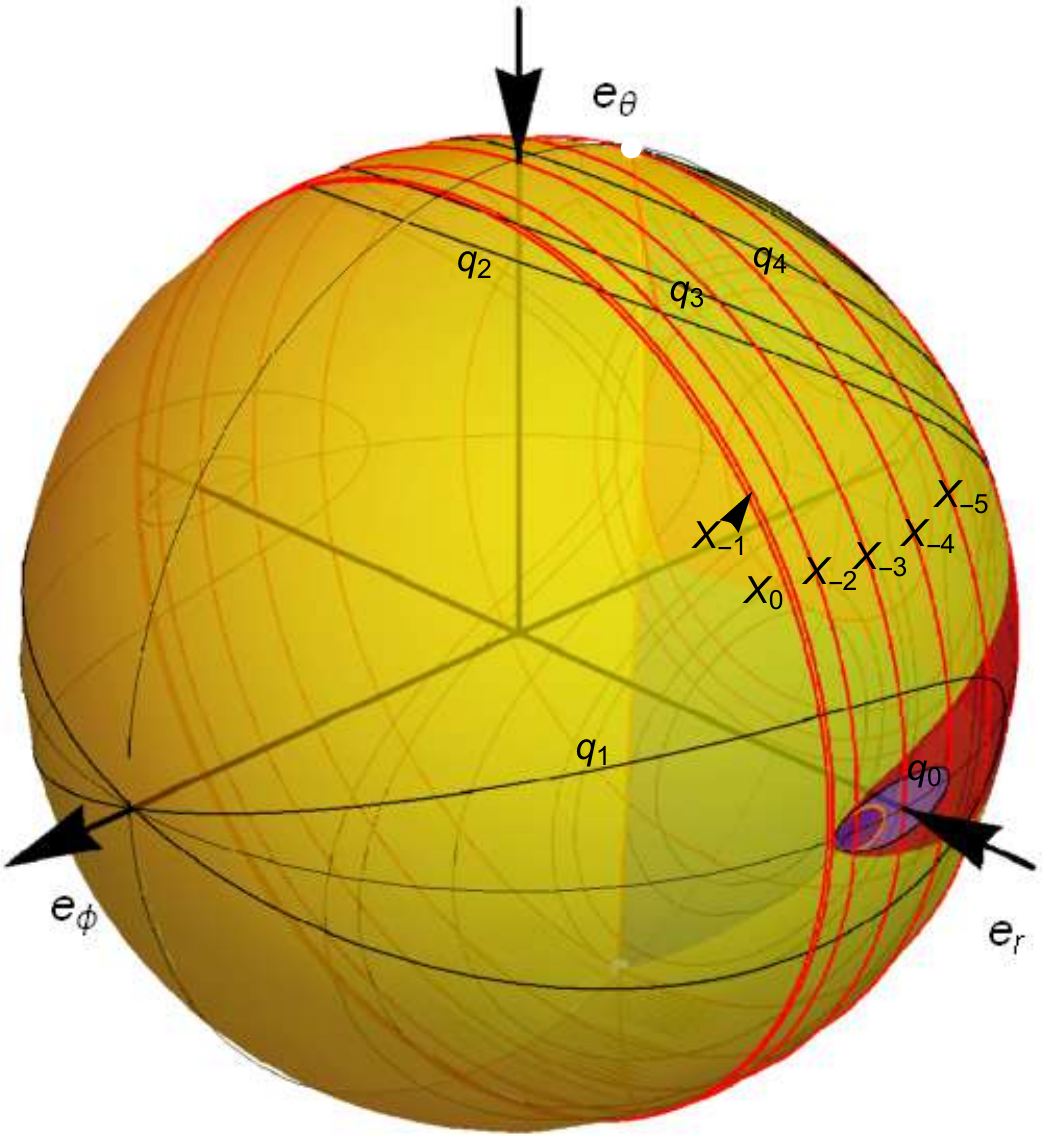}\\
		\hline
		\includegraphics[width=0.3\textwidth]{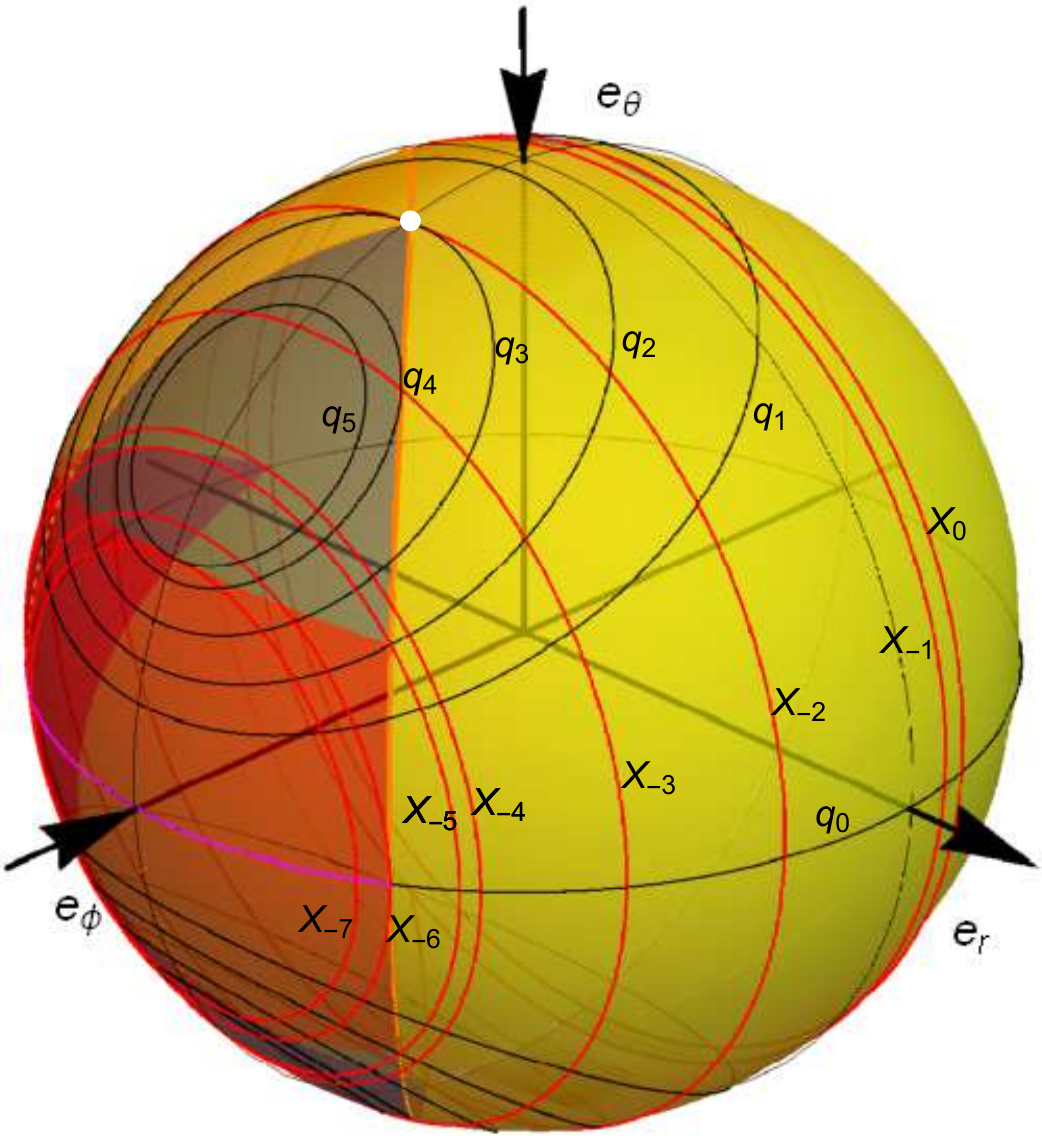}&(c): $\theta_{e}=90\dgr$&\includegraphics[width=0.3\textwidth]{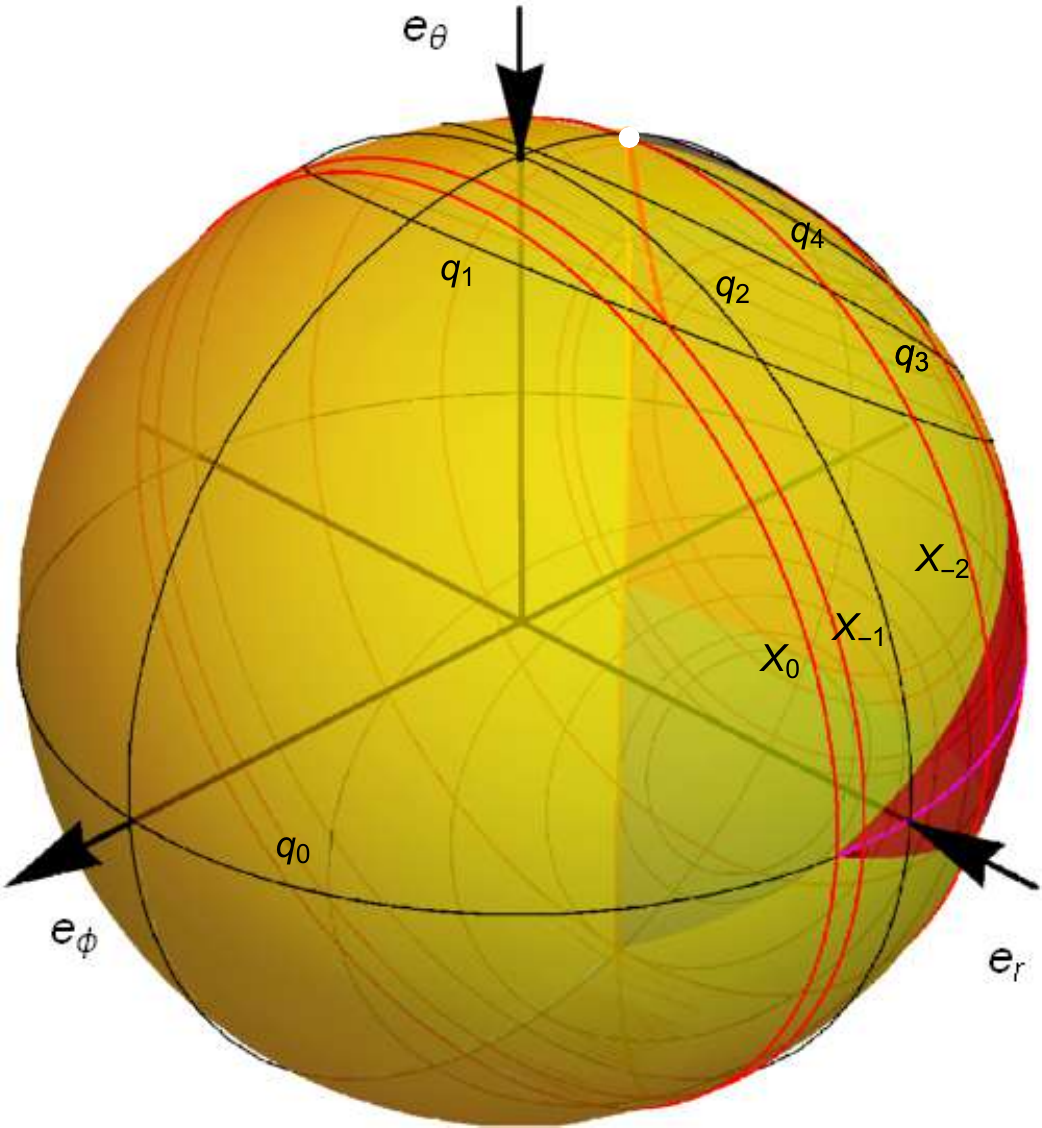}\\\hline

	\end{tabular}	
	\caption{Light escape cones associated with the planes of motion constants $(X-q)$ in Fig. \ref{Fig_Xq_IVa_re_2.65}.  
	}\label{Fig_cones_IVa_2.65_90deg}
\end{figure*}

\begin{figure*}[h]
	\centering
	\begin{tabularx}{\textwidth}{|XXX|}
		\hline
		\multicolumn{3}{|c|}{$r_{e}=5.75$}\\
		\hline
		\renewcommand{\arraystretch}{0.8}
		\raisebox{2.5cm}[0pt]{\bet{c}	\includegraphics[width=0.25\textwidth]{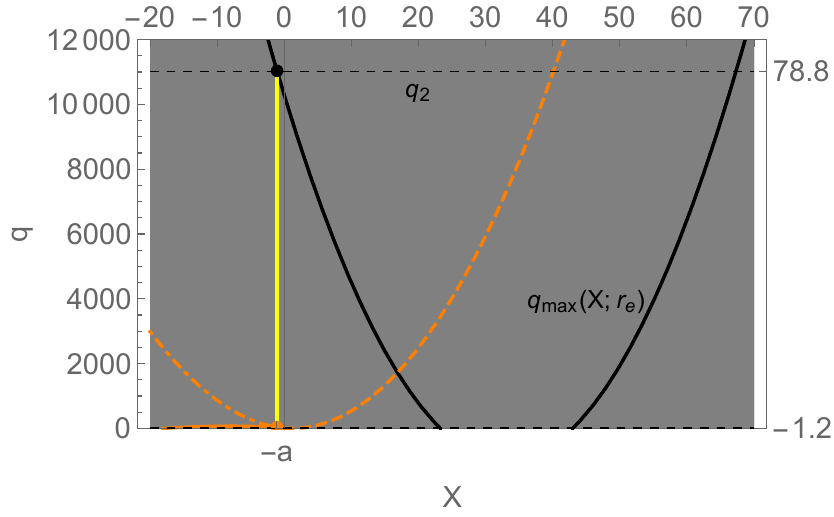}\\ \includegraphics[width=0.28\textwidth]{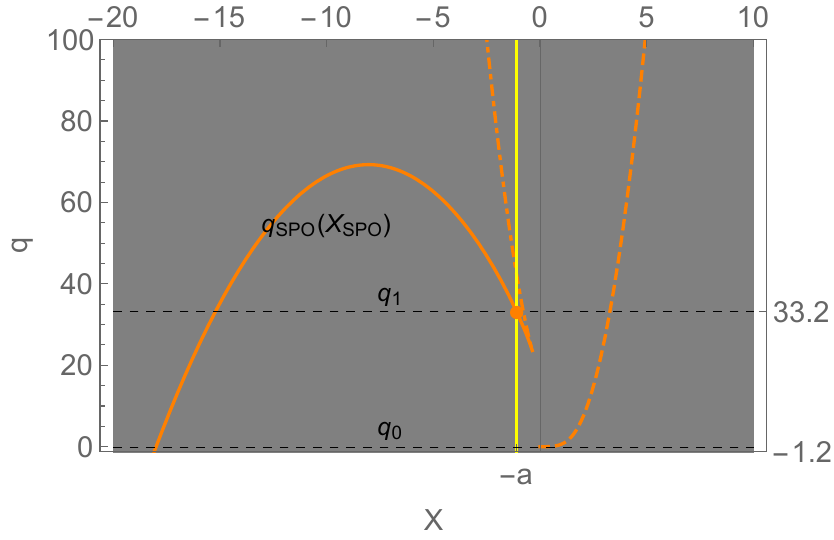} \ent }&
		\raisebox{3cm}[2cm]{\bet{c}\includegraphics[width=0.3\textwidth]{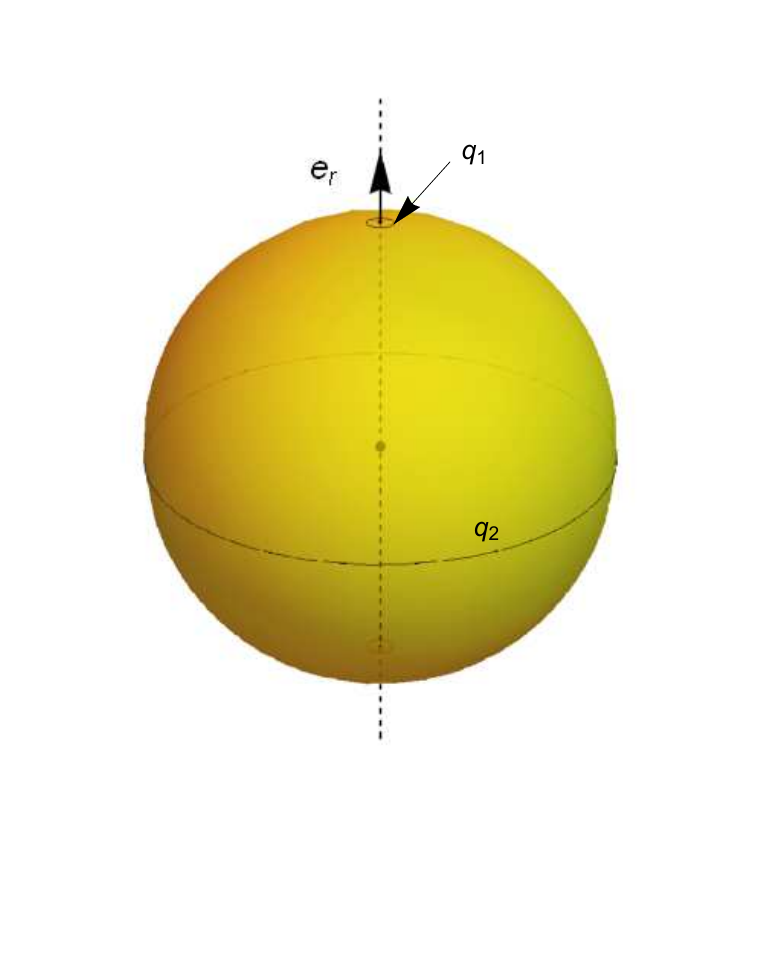}\\[-1.5cm] (a): $\theta_{e}=0\dgr$ \ent } &\includegraphics[width=0.3\textwidth]{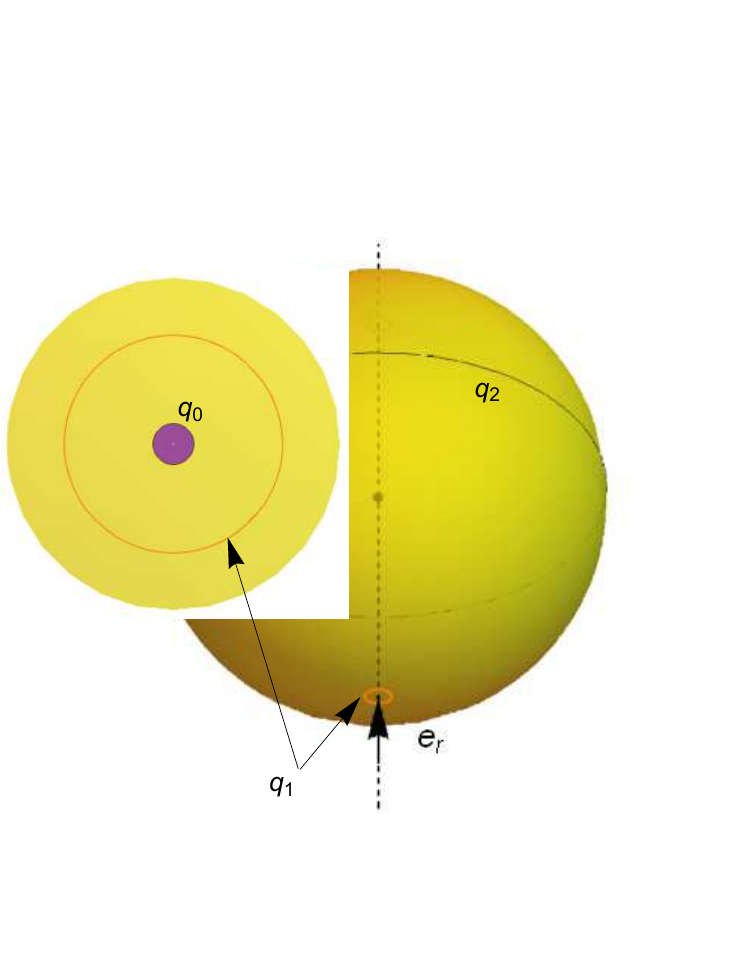}\\
		\hline		
		\raisebox{0.8cm}[5cm]{\bet{c}	\includegraphics[width=0.25\textwidth]{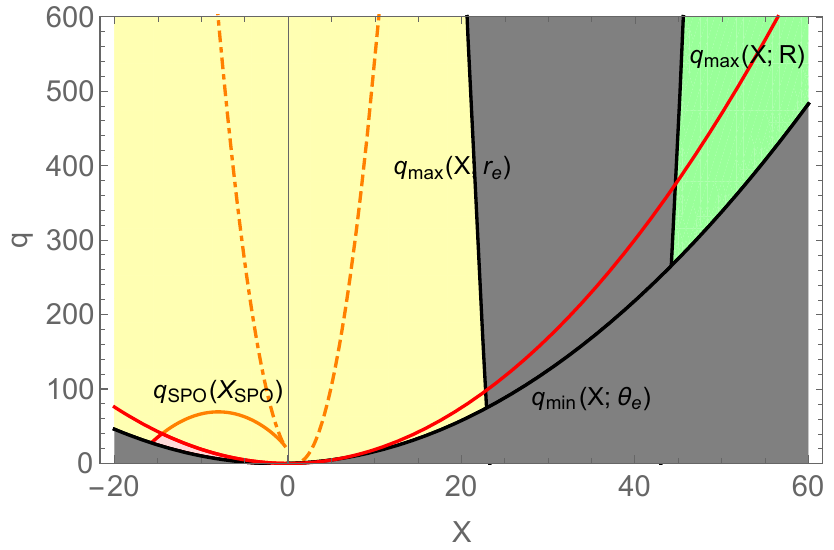}\\ \includegraphics[width=0.28\textwidth]{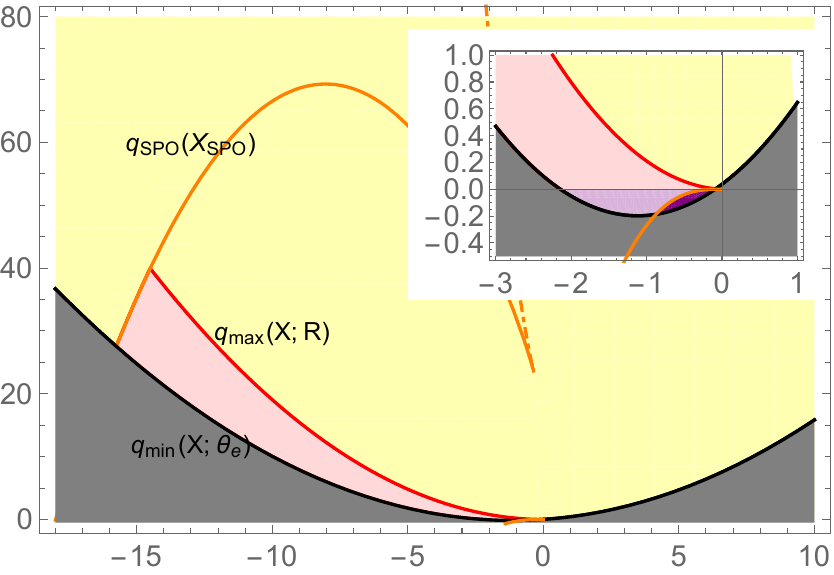} \ent } &\raisebox{1.5cm}[0pt]{\bet{c}\includegraphics[width=0.28\textwidth]{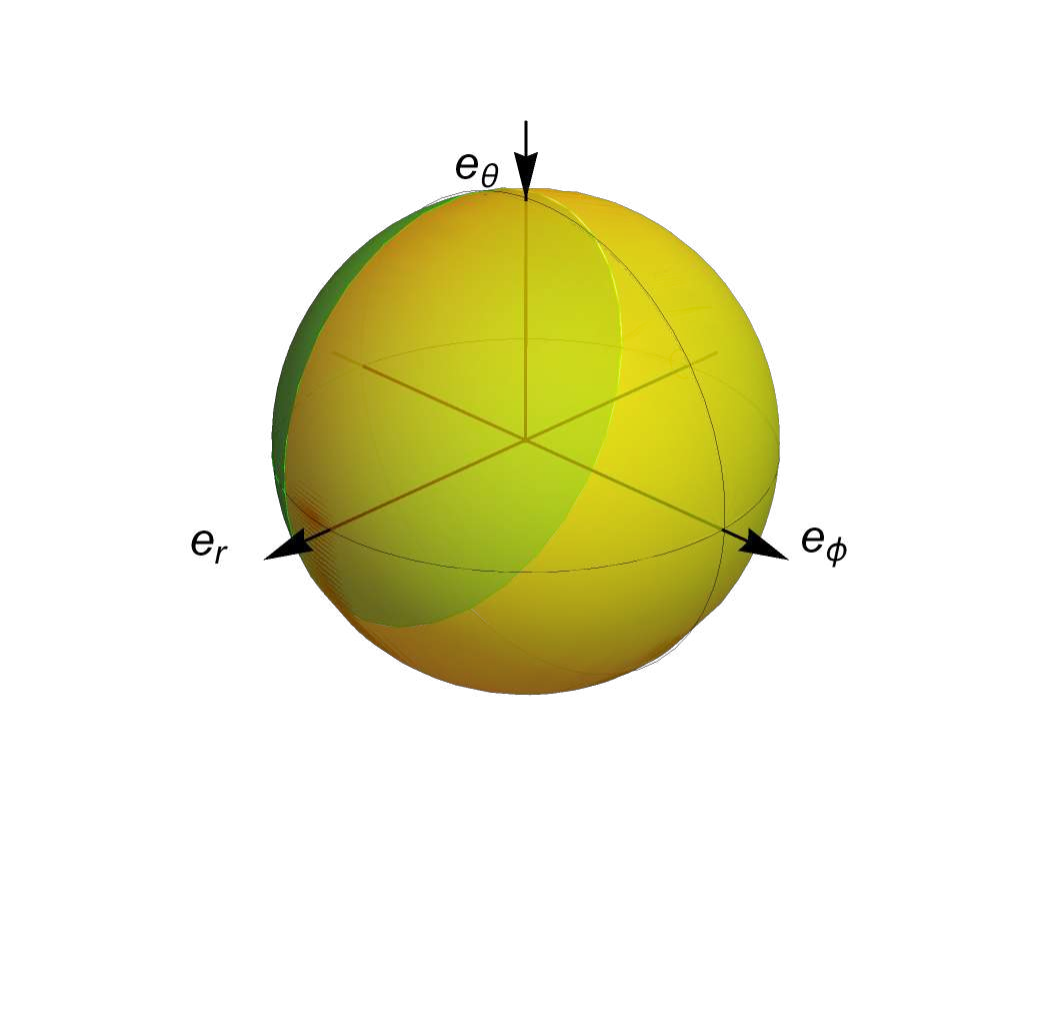}\\[-1cm] (b): $\theta_{e}=70\dgr$ \ent}&
		\includegraphics[width=0.35\textwidth]{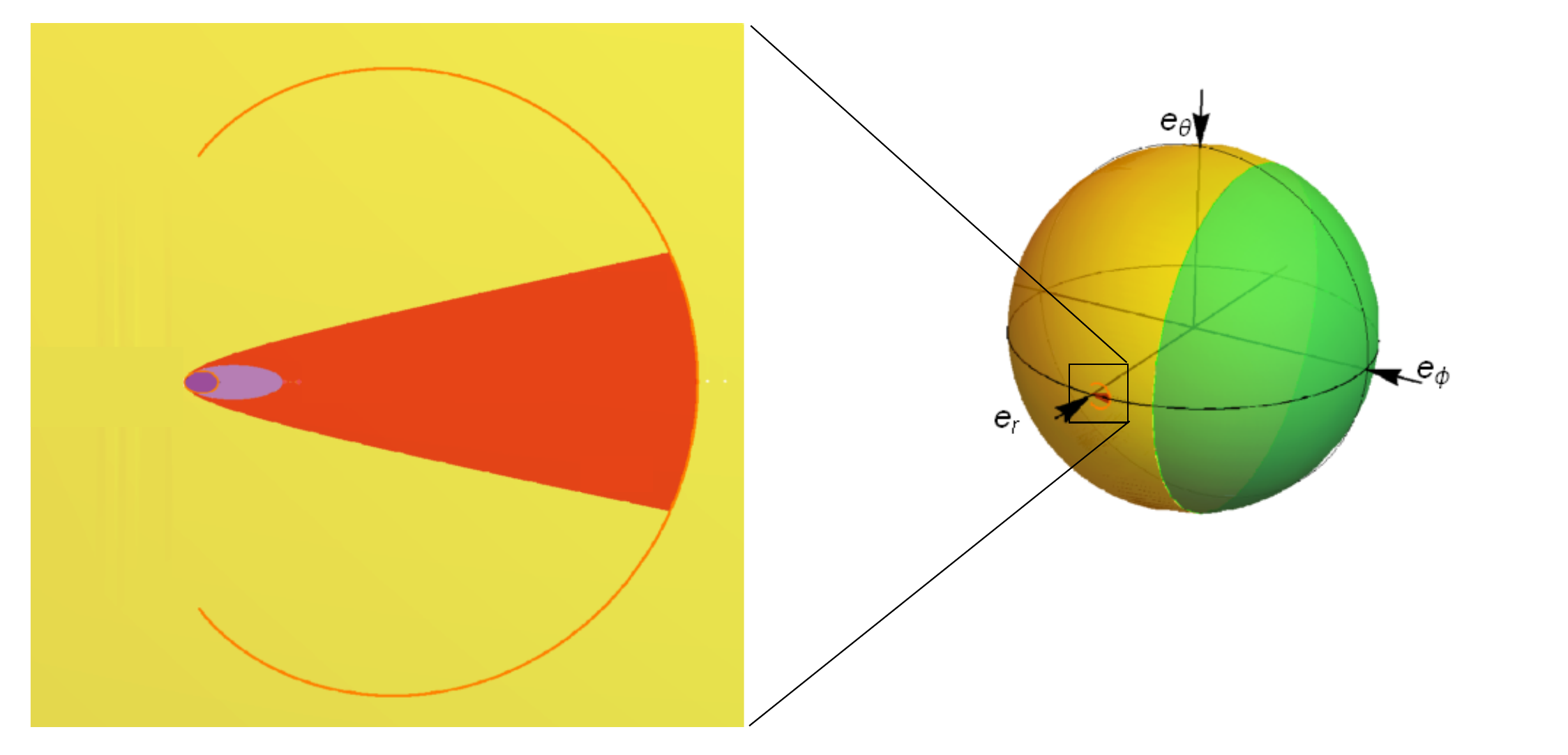}\\		
		\hline
		\raisebox{0.8cm}[5cm]{\bet{c}	\includegraphics[width=0.25\textwidth]{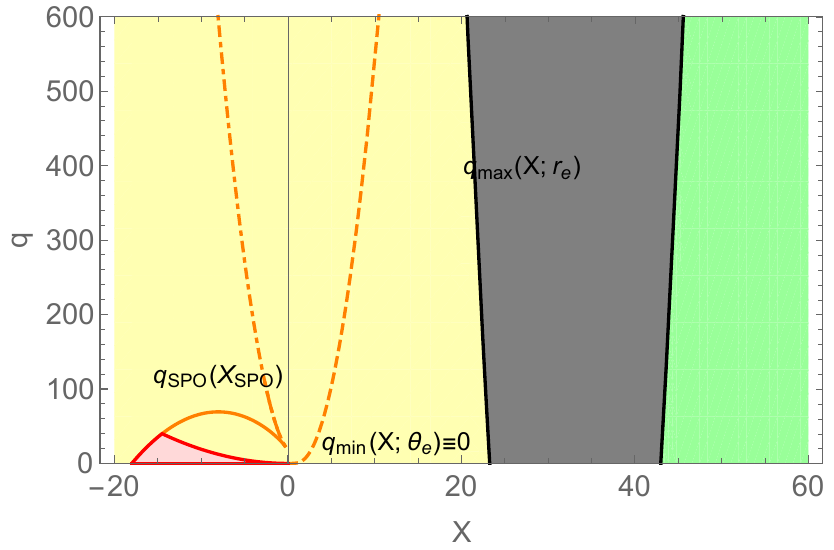}\\ \includegraphics[width=0.28\textwidth]{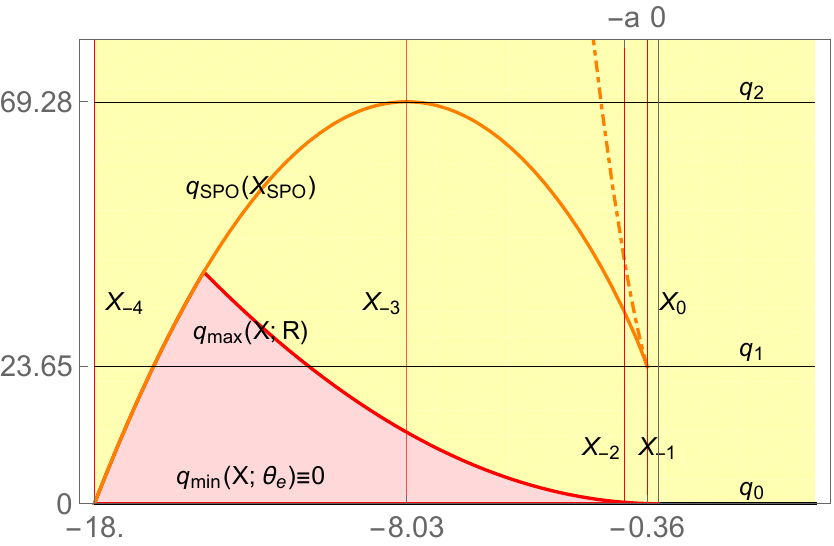} \ent } &\raisebox{1.5cm}[0pt]{\bet{c}\includegraphics[width=0.28\textwidth]{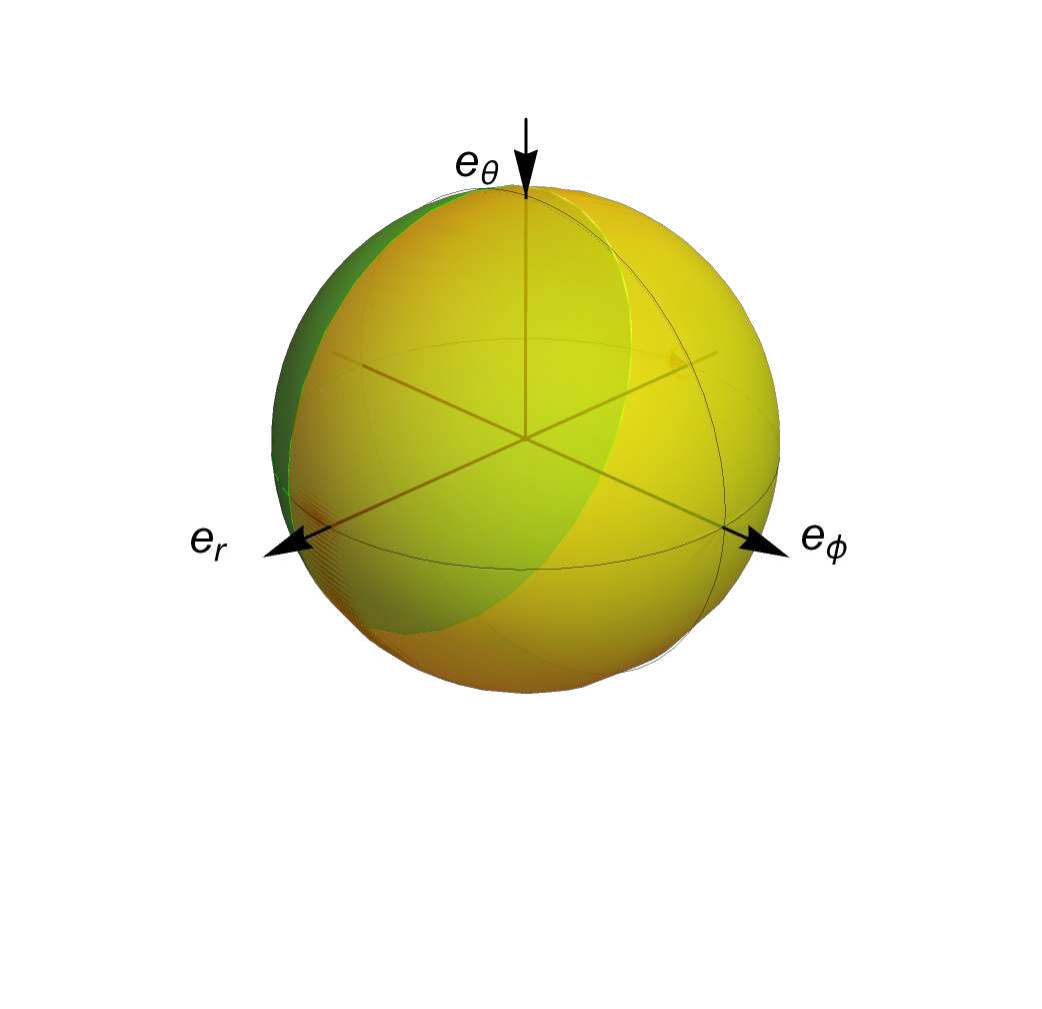}\\[-1cm] (c): $\theta_{e}=90\dgr$ \ent}&
		\includegraphics[width=0.35\textwidth]{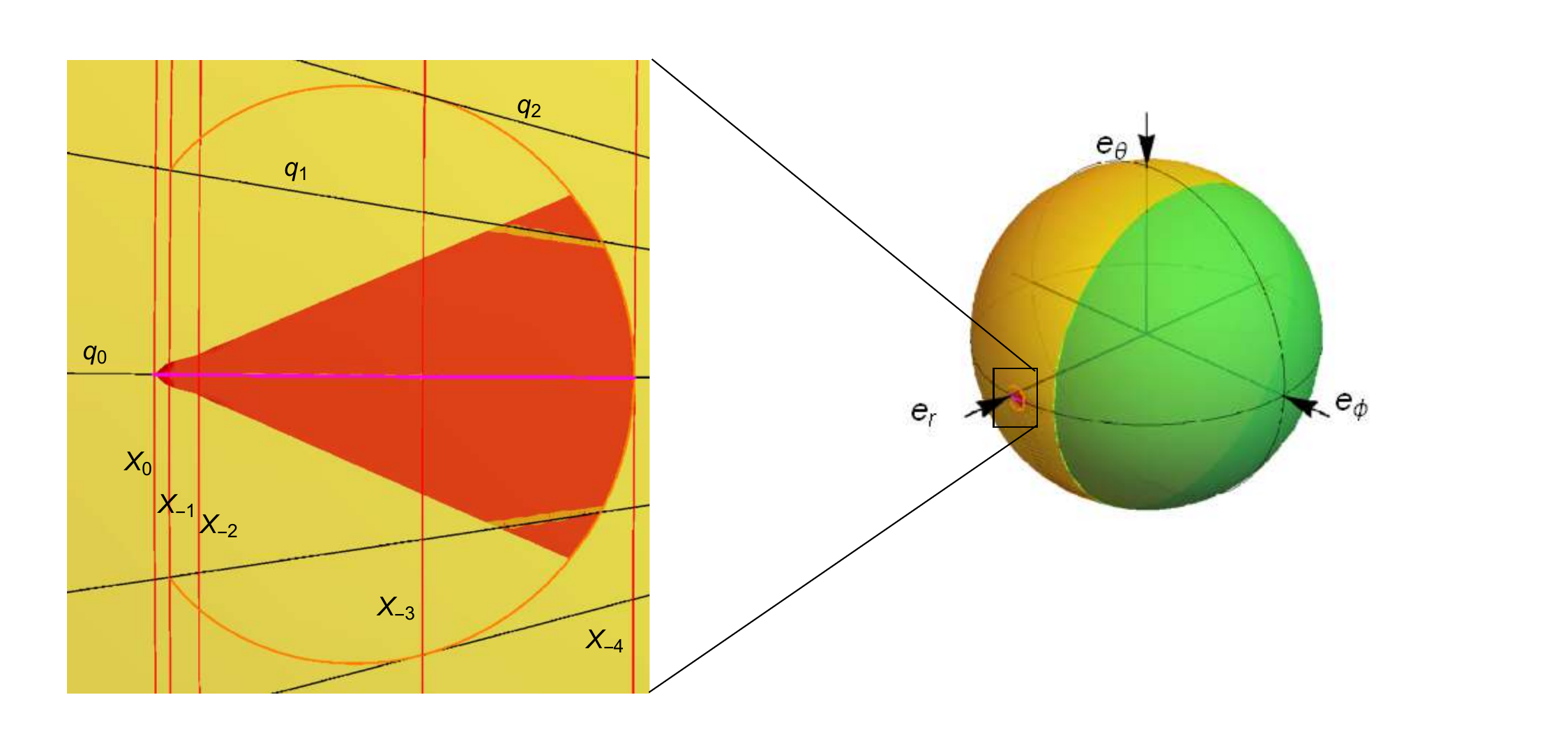}\\		
		\hline
	\end{tabularx}	
	\caption{LECs for the case $r^{-}_{erg}<r_{e}<r_{c}$, corresponding to the cosmological ergosphere, where the photons with negative locally measured azimuthal component $k^{(\phi)}<0$ of their 4-momentum and with negative covariant energy $E<0$ occur. These photons correspond to the regions depicted in green. The meaning of the remaining colouring is the same as in the previous case. 
	}\label{Fig_cones_IVa_5.75_90deg}
\end{figure*}

\clearpage

\section{Astrometric observables}

We can now relate our constructions to the notion of the KdS superspinar shadow to the astrophysical context, namely to coordinates used by nearly static observers at a large distance from the superspinar.  

\subsection{Celestial coordinates}

In the following we shall focus on the detail of the observer's celestial sphere containing the superspinar shadows for an observer at a large distance. For our purposes it is convenient to introduce the celestial coordinates $\tilde{\alpha}, \tilde{\beta}$, which are related to the directional angles $\alpha, \beta$ by the relations
\bea
\tilde{\alpha} & = &-\sin \alpha \sin \beta =- k^{(\phi)}/k^{(t)} \nonumber \\
\tilde{\beta} & = & \sin \alpha \cos \beta = k^{(\theta)}/k^{(t)}.\label{stereograph}
\eea

These angles are, as in the Kerr case, $y=0,$ proportional to $1/r$, however, here we cannot use the usual formula (see e.g.
\cite{Bar:1973:BlaHol:,Sche-Stu:2009:GRG:,Abd-Ami-Ahm-Gho:2016:PHYSR4:})

\bea
\tilde{\alpha}&=& \lim_{r_{o} \to \infty}( -r_{o} k^{(\phi)}/k^{(t)}) \label{alphaK} \\
\tilde{\beta}&=& \lim_{r_{o} \to \infty}( r_{o} k^{(\theta)}/k^{(t)}), \label{betaK}
\eea
as we cannot place the observer at arbitrary distances due to the presence of the cosmological horizon at $r=r_{c}.$ We shall deal with a distant observer, by which we mean an observer located outside the region of the SPOs, i.e. at $r^{-}_{ph}<r_{o}<r_{c}$.  Following our previous work \cite{Stu-Char:2024:PHYSR4:}, we choose the family of LNRFs near the static radius in Kerr-de Sitter spacetimes, since they exhibit characteristics similar to those of static observers at large distances in asymptotically flat spacetimes \cite{Stu:1983:BULAI:,Stu-Hle:1999:PHYSR4:}.

\subsection{Observable quantities}

\begin{figure*}[h]
	\centering
	\begin{tabular}{ccc}
		\includegraphics[width=0.32\textwidth]{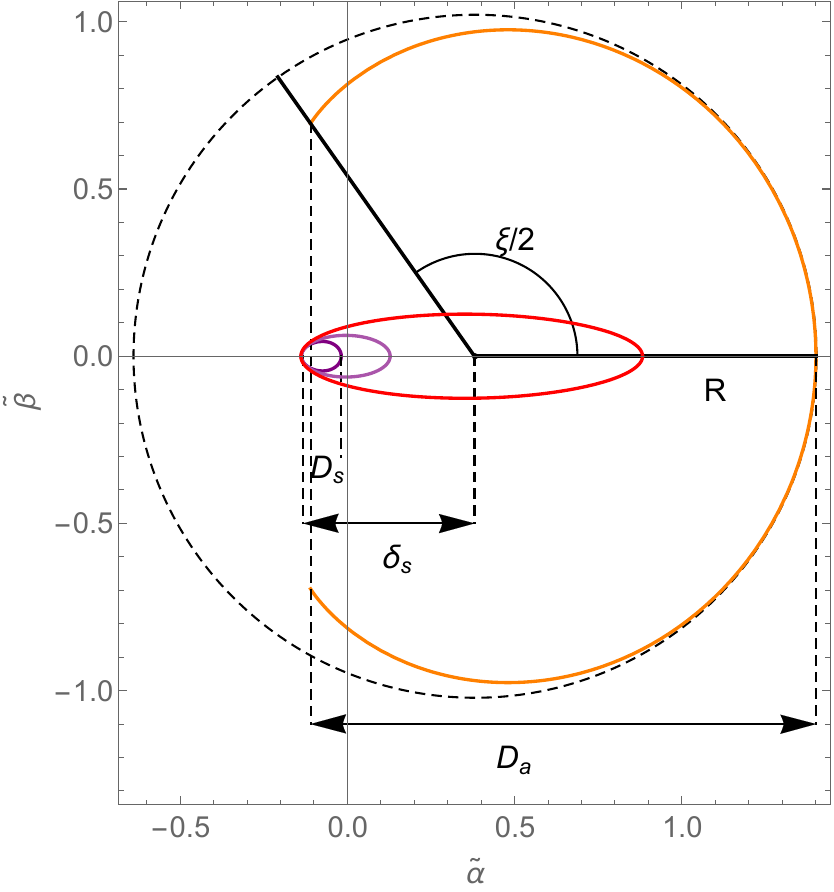}&\includegraphics[width=0.37\textwidth]{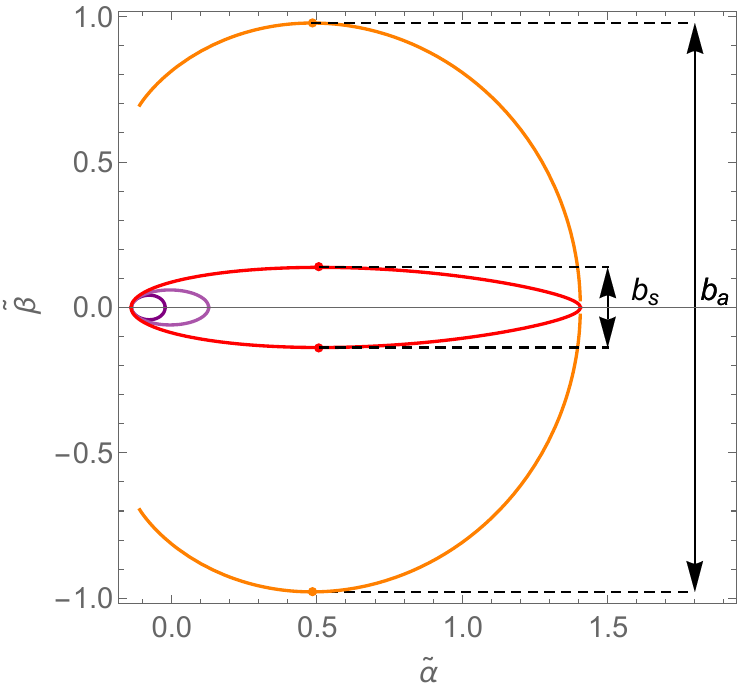}&\includegraphics[width=0.28\textwidth]{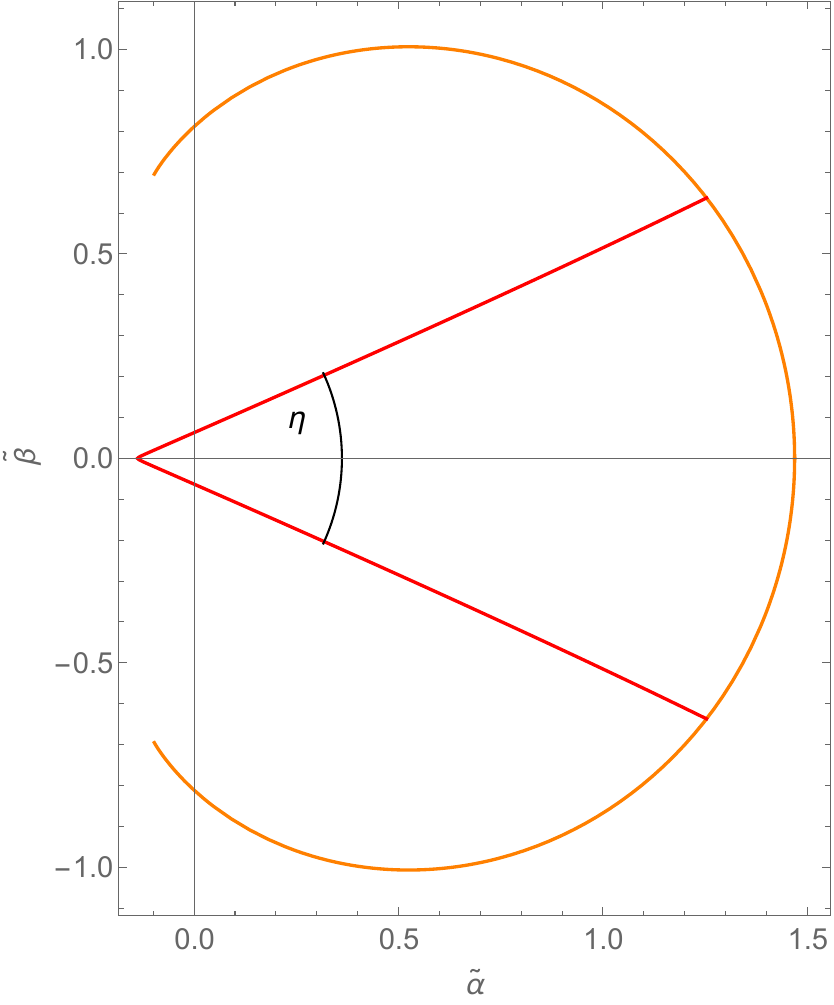}\\
		(a)&(b)&(c)
	\end{tabular}
	\caption{The observables investigated below: (a) the radius $R$ of the osculating circle, approximating the light arc, lining either the shadow of the superspinar with surface at radius $\calr=0.1$ (red curve), or the NS shadow, consisting of the dark spot corresponding to photons escaping to the region of negative radii (darker purple curve) and photons entering the $r<0$ region but repelled back to the $r>0$ region (region between the darker and the light purple curve, that corresponds to photons terminating at the ring singularity); the diameter $D_{s}$ of the dark spot, the diameter $D_{a}$ of the arc, the distance $\delta_{s}$ between the left endpoint of the spot and the centre of the osculating circle, and the central angle $\zeta$ of the arc; (b) the 'vertical' diameter $b_{s}$ of the largest elliptical shadow of the superspinar just reaching the arc with vertical diameter $b_{a}$, observed at some critical latitude $\theta_{max(ell)}$ to be investigated below; (c) the angle $\eta$ of the wedge representing the shadow of the superspinar observed from the equatorial plane.  Note that all the quantities are measured in radians.}\label{captions}
\end{figure*}

\begin{figure*}[h]
	\centering \textbf{ $y=10^{-5}$, $r_{s}=46.4$}\\	
	\begin{tabular}{|ccc|}
		\hline
		Class IVa: $a^2=1.2$& Class IVb: $a^2=2$&\\
		\bet{c}\includegraphics[width=0.4\textwidth]{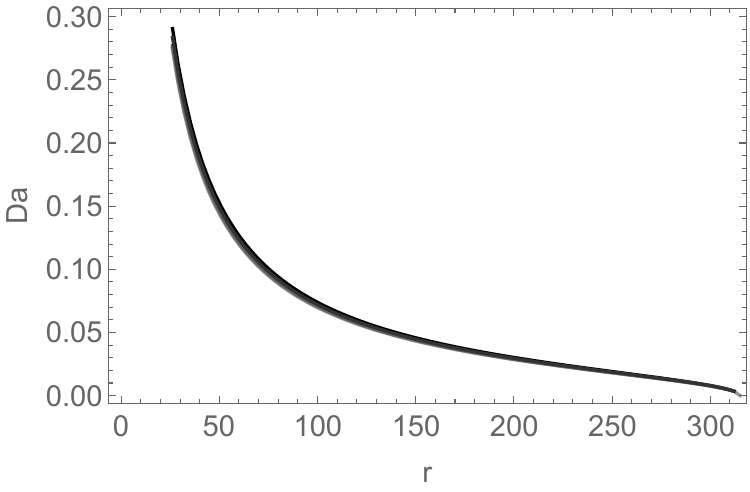} \\ \includegraphics[width=0.4\textwidth]{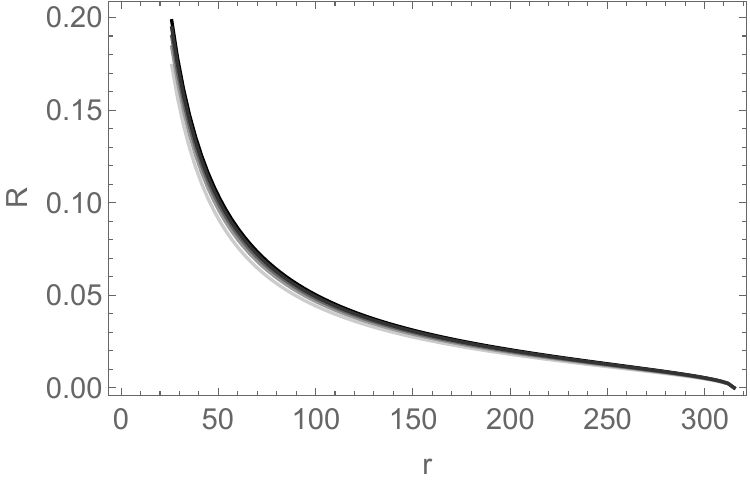}\\ \includegraphics[width=0.4\textwidth]{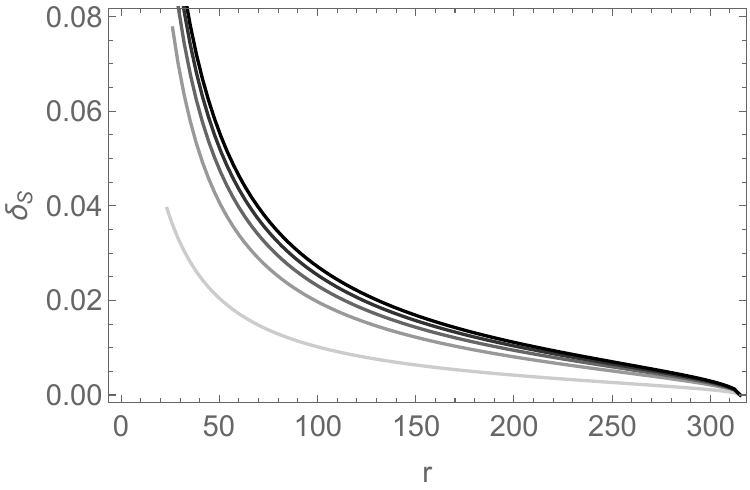}\\ \includegraphics[width=0.4\textwidth]{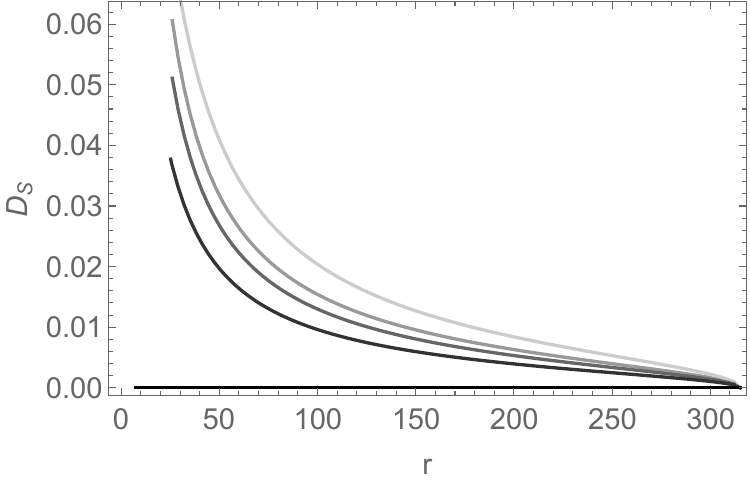}\\
		\ent &
		\bet{c}\includegraphics[width=0.4\textwidth]{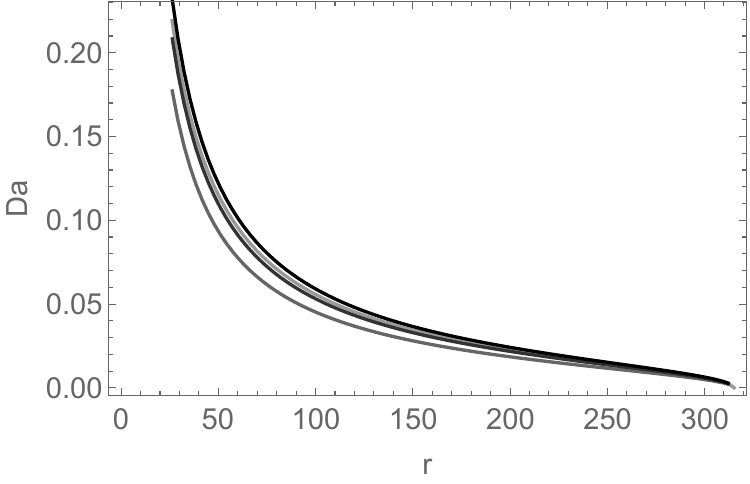} \\ \includegraphics[width=0.4\textwidth]{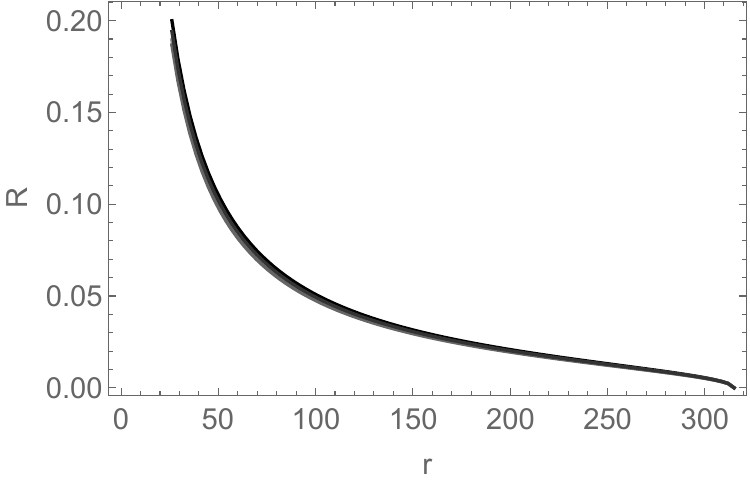}\\ \includegraphics[width=0.4\textwidth]{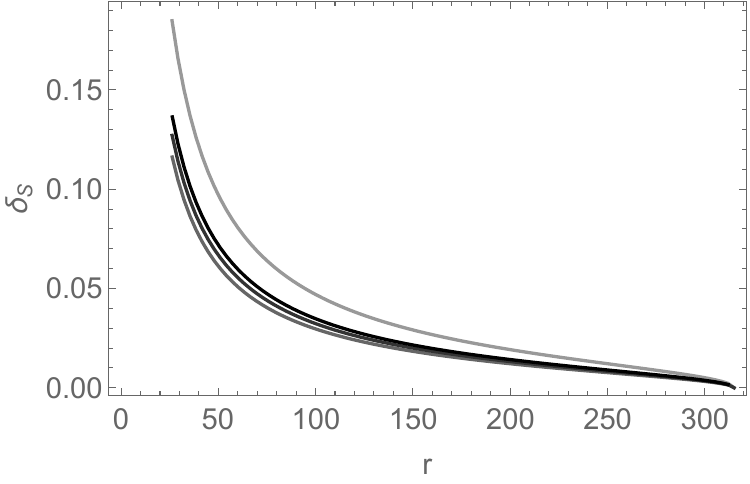}\\ \includegraphics[width=0.4\textwidth]{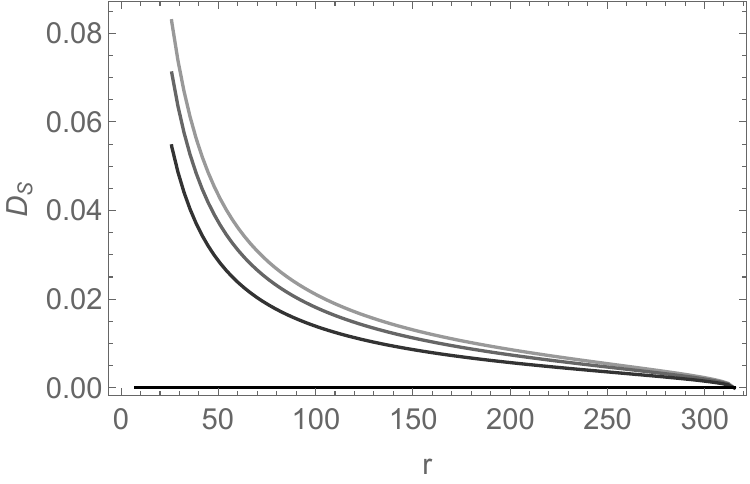}\\
		\ent&
		\raisebox{-3cm}[0pt]{\includegraphics[width=0.15\textwidth]{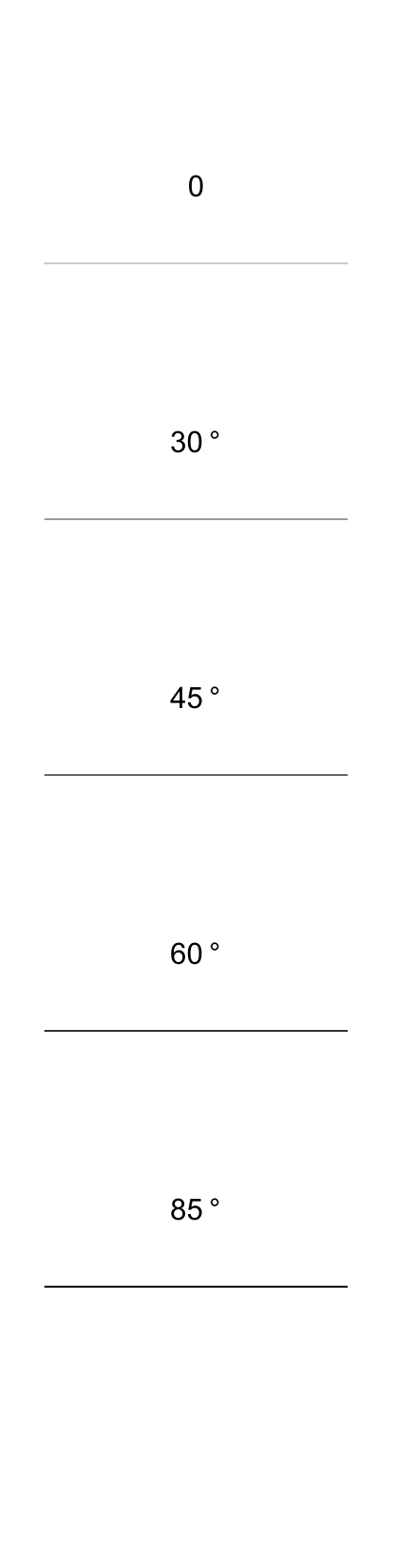}}\\
		\hline
		
	\end{tabular}	
	\caption{Comparison of the dependence of the observables on the radial position of an observer for some selected values of its latitudinal position for the case of the KdS spacetimes with (left) and without (right) polar SPOs.            
	}\label{Fig_observables_ro_yneg5}
\end{figure*}

\begin{figure*}[h]
	
	\centering \textbf{ $y=10^{-2}$, $r_{s}=4.64$}\\	
	\begin{tabular}{|ccc|}
		\hline
		Class IVa: $a^2=1.2$& Class IVb: $a^2=2$&\\
		\bet{c}\includegraphics[width=0.4\textwidth]{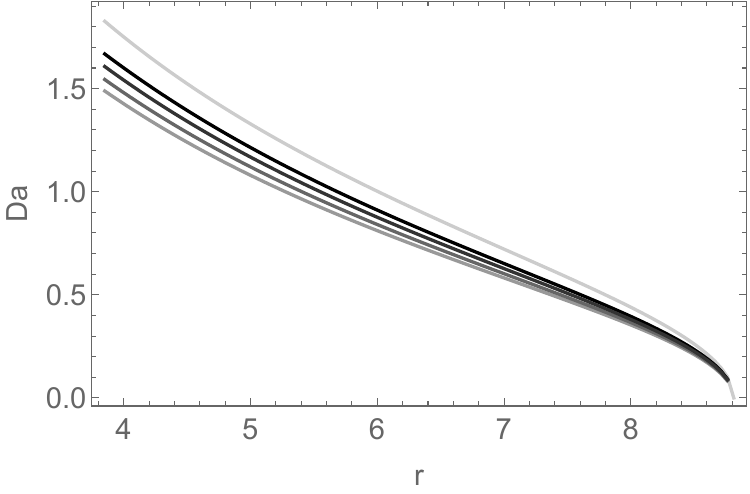} \\ \includegraphics[width=0.4\textwidth]{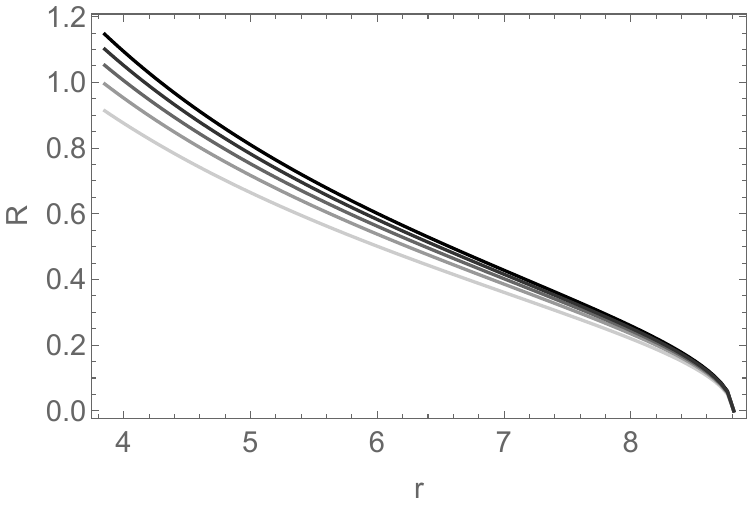}\\ \includegraphics[width=0.4\textwidth]{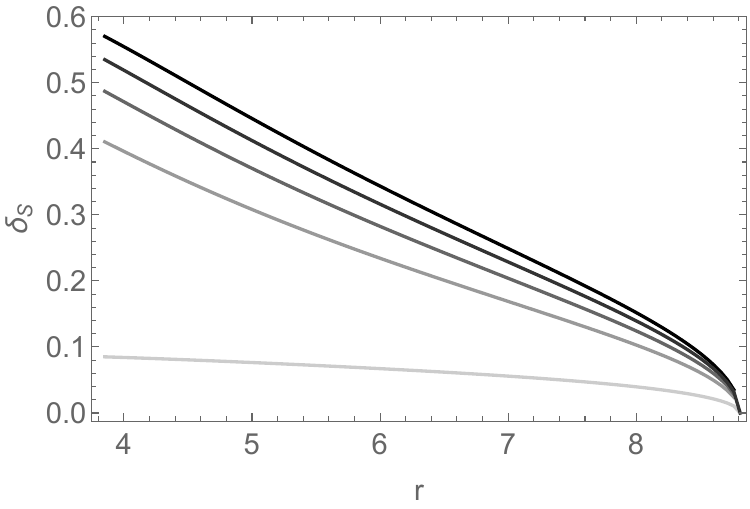}\\ \includegraphics[width=0.4\textwidth]{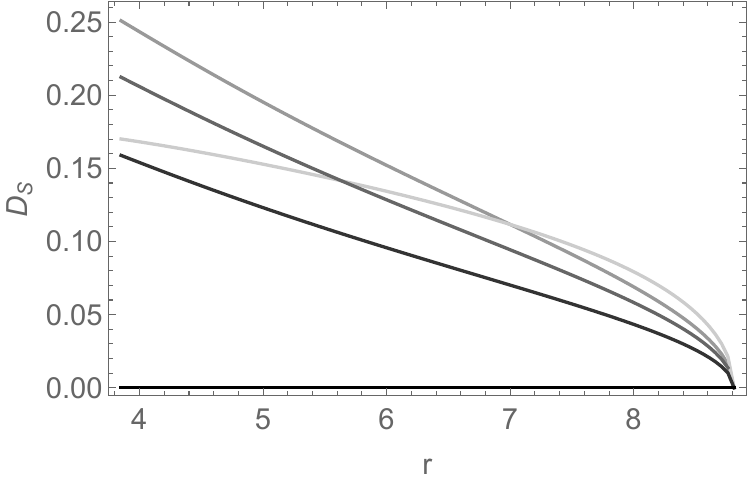}\\
		\ent &
		\bet{c}\includegraphics[width=0.4\textwidth]{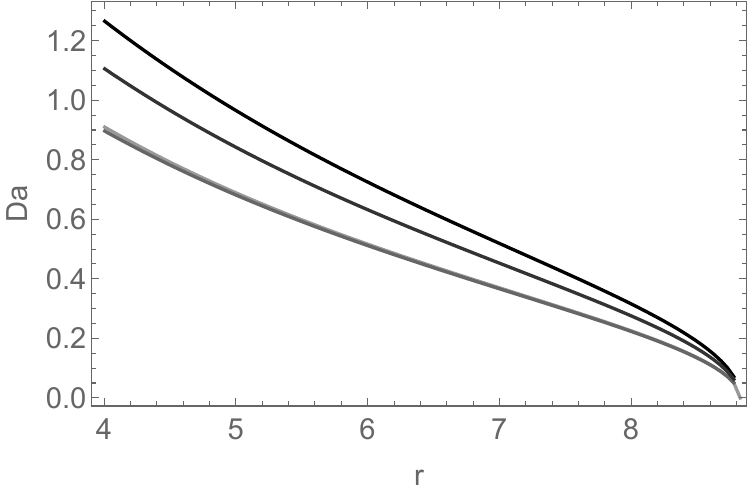} \\ \includegraphics[width=0.4\textwidth]{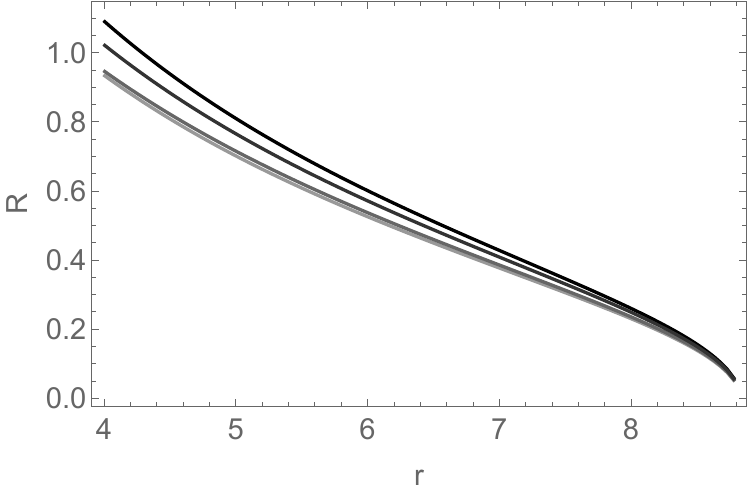}\\ \includegraphics[width=0.4\textwidth]{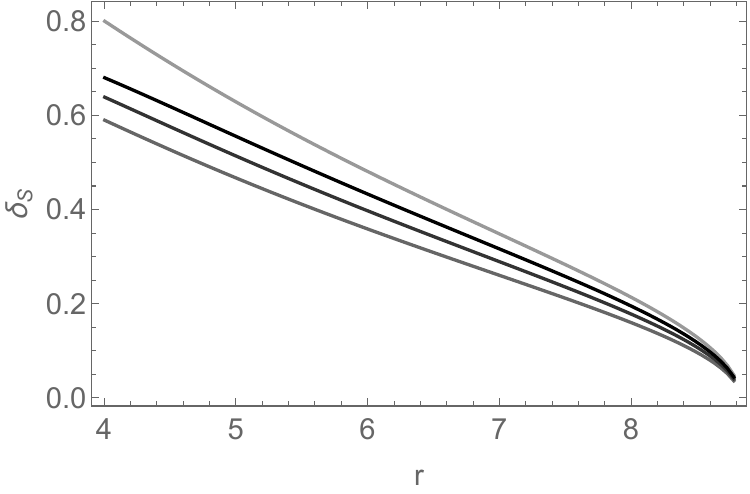}\\ \includegraphics[width=0.4\textwidth]{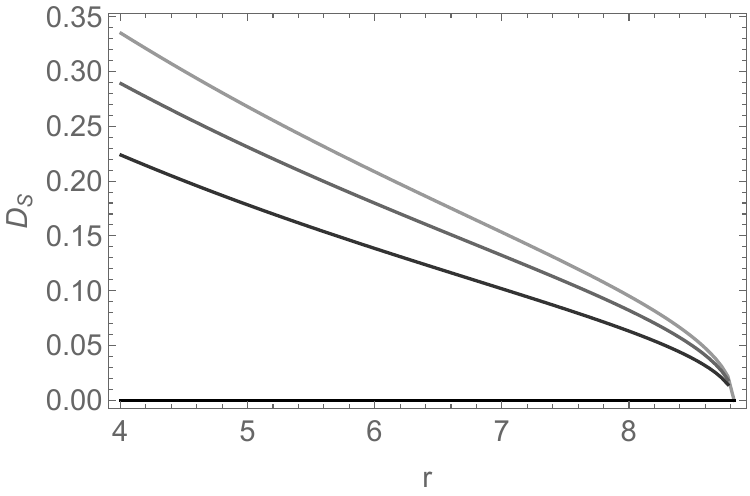}\\
		\ent&
		\raisebox{-3cm}[0pt]{\includegraphics[width=0.15\textwidth]{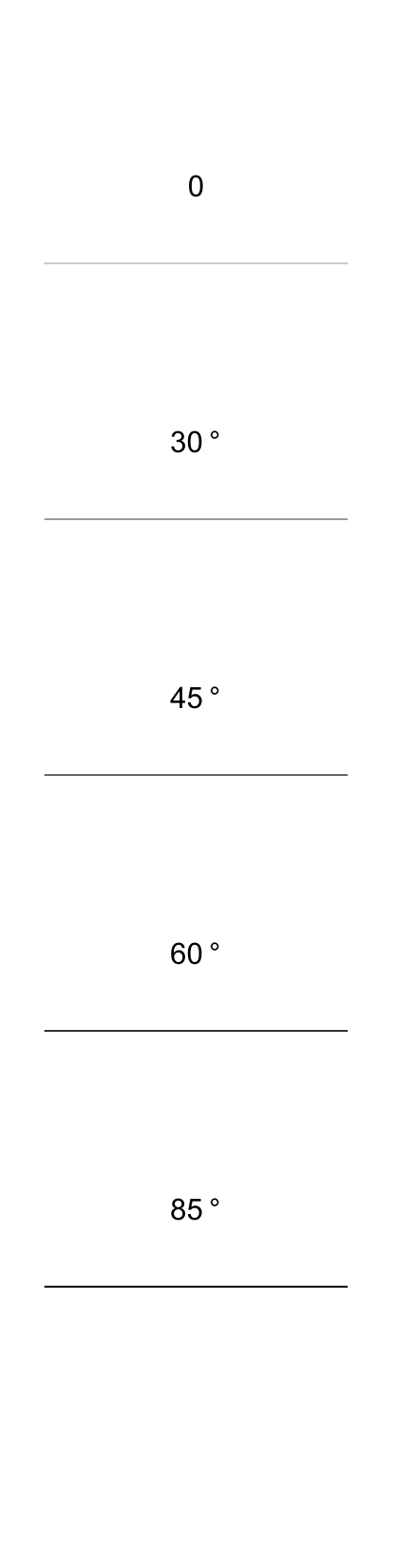}}\\
		\hline
			
	\end{tabular}	
	\caption{Comparison of the dependence of the observables on the radial position of an observer for some selected values of its latitudinal position for the case of the KdS spacetimes with (left) and without (right) polar SPOs.            
	}\label{Fig_observables_ro_yneg2}
\end{figure*}

\begin{figure*}[h]

	\begin{tabular}{cccc}
		\hline
		
		&$y=10^{-5}$, $r_{s}=46.4$&$y=10^{-2}$, $r_{s}=4.64$&\\
		\hline
		\raisebox{3cm}[0pt]{$a^2=1.2$}&\includegraphics[width=0.4\textwidth]{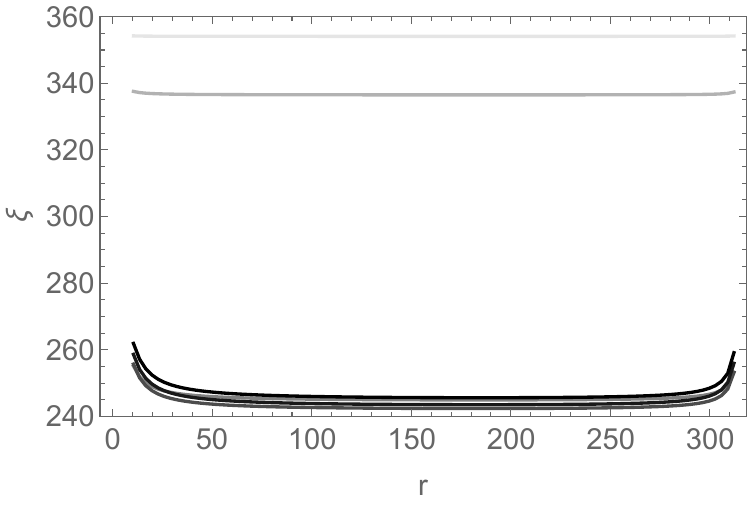}&\includegraphics[width=0.4\textwidth]{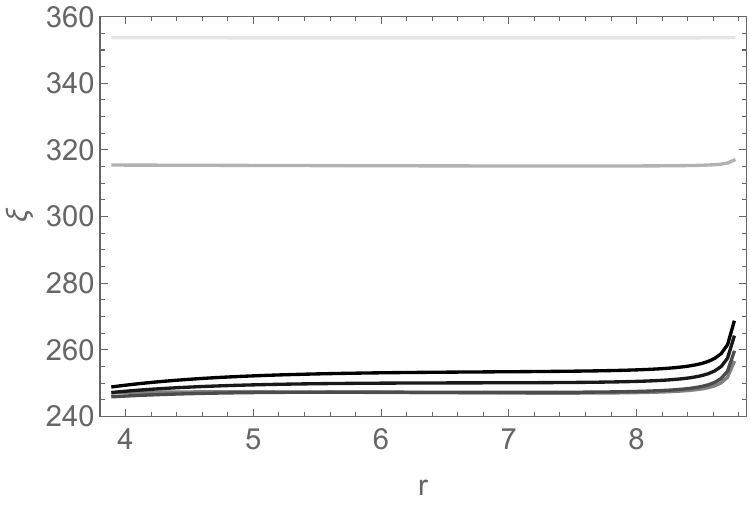}&\includegraphics[width=0.1\textwidth]{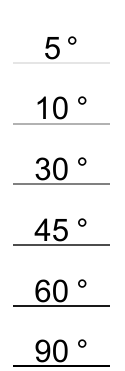}\\
		\hline
		\raisebox{3cm}[0pt]{$a^2=2$}&\includegraphics[width=0.4\textwidth]{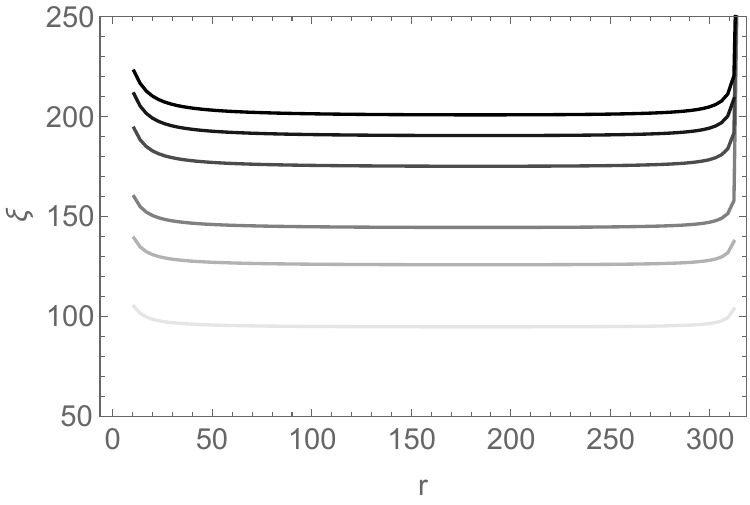}&\includegraphics[width=0.4\textwidth]{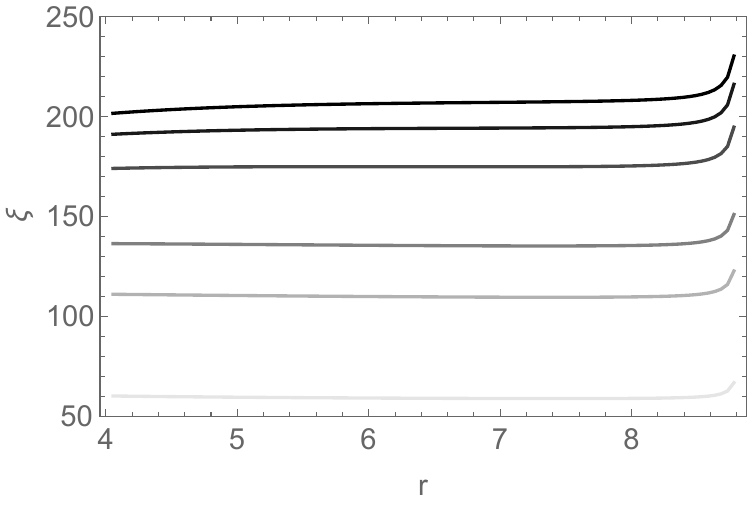}&\includegraphics[width=0.1\textwidth]{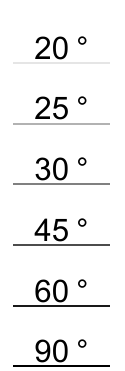}
		
	\end{tabular}	
	\caption{Dependence of the central angle $\xi$ of the arc on the radial position of the observer for some selected values of his latitudinal position. The top row corresponds to the NS spacetimes with polar SPOs, hence for $\theta_{o}<\theta_{max(circ)}(y,a^2)$ it is $\xi(r)=360\dgr$. Here $\theta_{max(circ)}(10^{-5},1.2)=9.74\dgr$, $\theta_{max(circ)}(10^{-2},1.2)=9.04\dgr$. The bottom row corresponds to the NS spacetimes without polar SPOs, hence for $\theta_{o}<\theta_{min(arc)}(y,a^2)$ there is no arc. Here $\theta_{min(arc)}(10^{-5},2)=15.5\dgr$, $\theta_{min(arc)}(10^{-2},2)=18.5\dgr$.         
	}\label{Fig_observables_ro_xi}
\end{figure*}

In Fig. \ref{captions} we present the observables that we shall investigate below. These observables depend on the spacetime parameters and both the latitudinal and radial position of the observer. In Fig.s \ref{Fig_observables_ro_yneg5} - \ref{Fig_observables_ro_xi} we show the radial dependence of some observables for several values of the dimensionless cosmological constant and for two representative spin parameters corresponding to the Class IVa  and VIa KdS superspinar (NS) spacetimes, i.e. both with DRB but the former with the polar SPOs, the latter without the polar SPOs. We shall not deal with the spacetimes with RRB, as they are not astrophysically relevant due to implausibly high values of the dimensionless cosmological constant.
 
\begin{figure*}[h]
	\centering
	\begin{tabular}{cc}
		\includegraphics[width=0.5\textwidth]{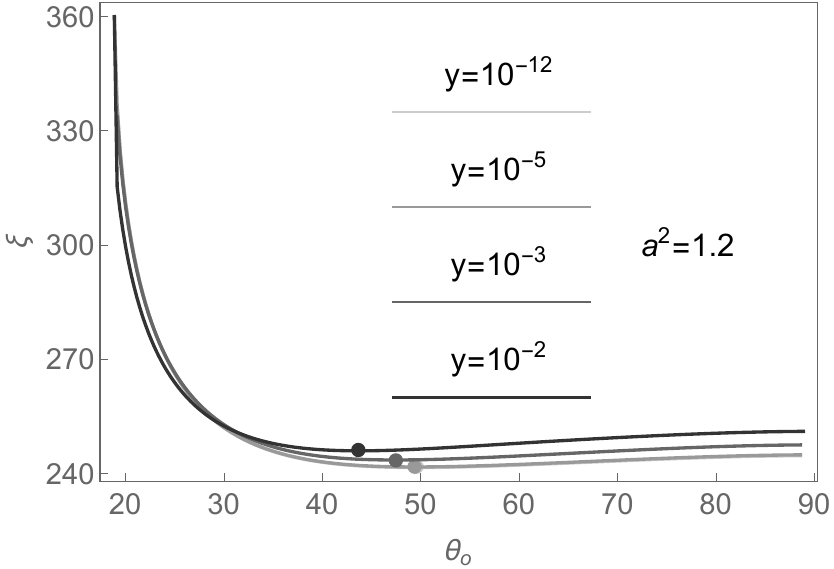}&\includegraphics[width=0.5\textwidth]{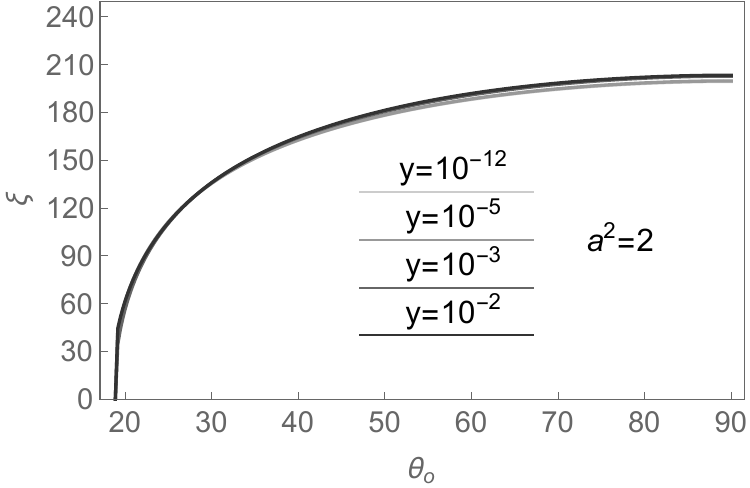}\\
		(a)&(b)		
	\end{tabular}	
	\caption{Comparison of the dependence of the central angle $\xi$ on the latitude of the observer at the appropriate static radius $r_{s}(y)$ for some selected spacetimes with polar SPOs (left) and for spacetimes with the same cosmological parameter $y$ but different spin parameter $a^2$, corresponding to spacetimes with no polar SPOs (right). It can be seen that in the case of spacetimes with polar SPOs there is a certain local minimum $\xi_{min}$ for a certain angle $\theta_{\xi(min)}$ (highlighted by dots), while in the case of spacetimes without polar SPOs the angle $\xi$ increases monotonically up to the value $\xi_{eq}$ corresponding to the latitudinal coordinate in the equatorial plane $\theta=90\dgr$. The magnitude of both angles $\theta$ and $\xi$  is plotted in degrees on both axes.
	}\label{Fig_th_xi}
\end{figure*}

\begin{figure*}[h]
	\centering
	\begin{tabular}{cc}
		\includegraphics[width=0.5\textwidth]{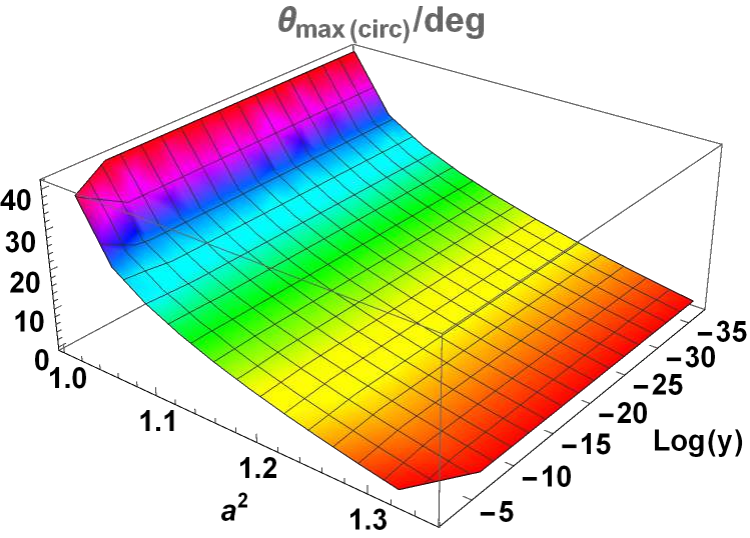}&\includegraphics[width=0.5\textwidth]{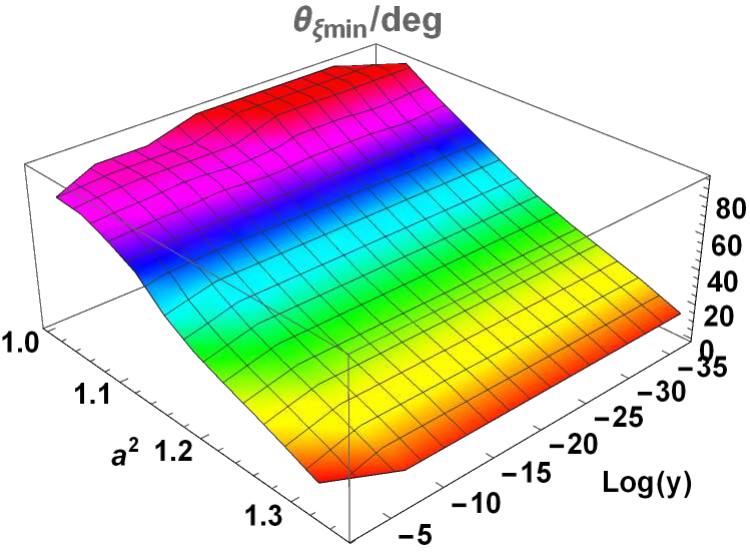}\\
		(a)&(b)\\
		\includegraphics[width=0.5\textwidth]{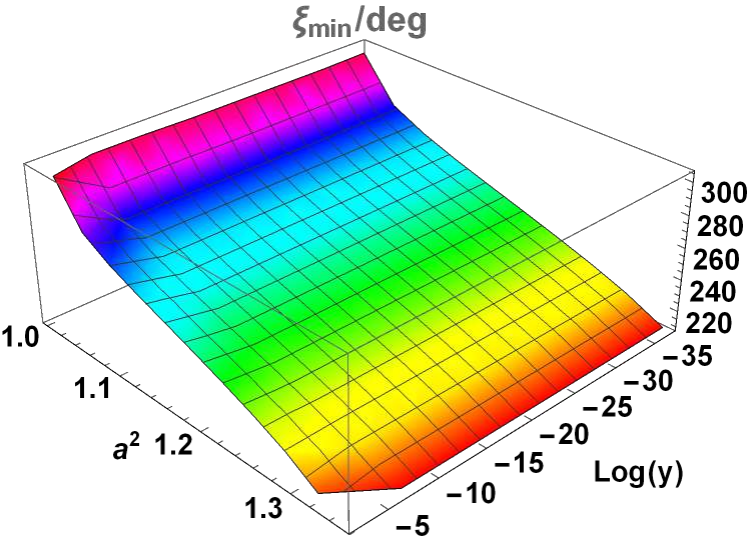}&\includegraphics[width=0.5\textwidth]{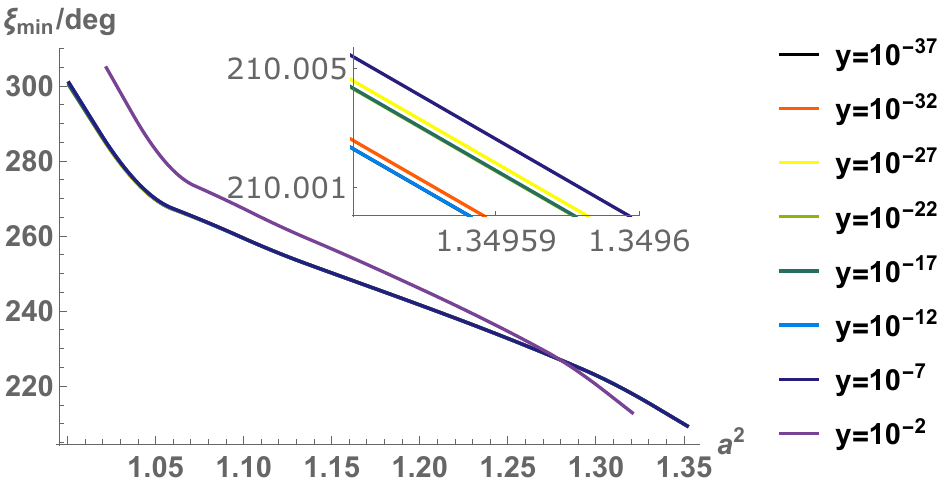}\\
		(c)&(d)		
	\end{tabular}	
	\caption{Dependence of the characteristic angles on the spacetime parameters corresponding to Class IVa: (a) the maximum angle $\theta_{max(circ)}$ under which the observer can see the light circle lining the supespinar/NS shadow, as described e.g. in Fig. \ref{Fig_cones_IVa_4_90deg}b; (b) the observer's latitude $\theta_{\xi(min)}$ at which he sees the local minimum $\xi_{min}$ of the angle $\xi$ of the arc (c.f. Fig. \ref{Fig_th_xi}a); (c) the value $\xi_{min}$ of the minimum angle $\xi$ of the arc; (d) slices of the 3D graph in Fig. (c) through the planes of the constant parameter $y$ for some appropriate values. For $y<<10^{-2}$ the curves almost overlap. 
	}\label{Fig_th_ximin}
\end{figure*}

The fixed value $a^2=1.2\approx a^2_{crit}$ was chosen as a 'medium' spin value of Class IVa KdS spacetimes, allowing the existence of the polar SPOs. In such spacetimes, the shadow consists of the dark spot surrounded by a full 'circle', observed for latitudes $\theta_{o}$ smaller than some critical maximum $\theta_{max(circ)}$ allowing the appearance of the circle. For $\theta_{o}>\theta_{max(circ)}$, the circle splits into an arc (see Fig. \ref{Fig_cones_IVa_4_90deg}). The dependence of the angle $\theta_{max(circ)}$ on the spacetime parameters $a^2, y$ is shown in Fig. \ref{Fig_th_ximin}(a). The central angle $\xi$ of the arc decreases with latitude $\theta_{o}$ and it reaches its minimum value $\xi_{min}$ for some latitude $\theta_{\xi(min)}$, $0<\theta_{\xi(min)}<90\dgr$, from which it grows slightly to its local maximum $\xi_{eq}$ in the equatorial plane $\theta_{o}=90\dgr$ (see Fig. \ref{Fig_th_xi}(a)). The dependence of the angles $\theta_{\xi(min)}$, $\xi_{min}$ on the spacetime parameters $a^2, y$ is shown in Fig. \ref{Fig_th_ximin}(b)-(d).

\begin{figure*}[h]
	\centering
	\begin{tabular}{cc}
		\includegraphics[width=0.5\textwidth]{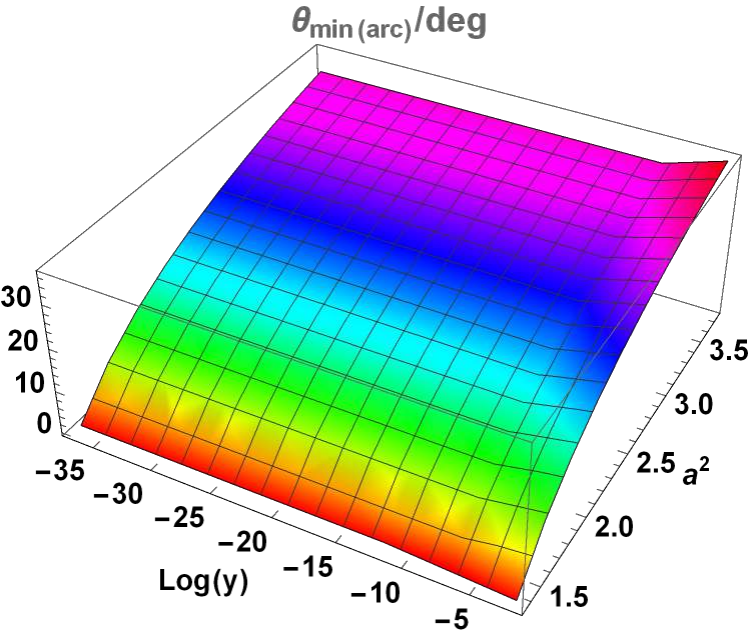}&\includegraphics[width=0.5\textwidth]{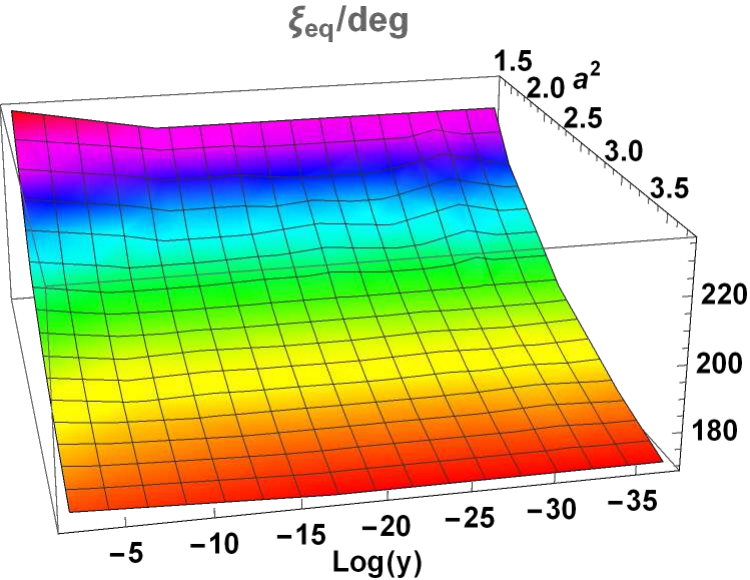}\\
		(a)&(b)
		
	\end{tabular}	
	\caption{Dependence of the characteristic angles on the spacetime parameters corresponding to Class IVb: (a) the minimum observer's latitude $\theta_{min(arc)}$ allowing the observer to see the light arc lining the supespinar/NS shadow, as described e. g. in Fig. \ref{Fig_cones_VIb_rs_22}b; (b) the maximum angle $\xi_{eq}$ of the arc seen in the equatorial plane (c.f. Fig. \ref{Fig_th_xi}b).
	}\label{Fig_th_xi?}
\end{figure*}

The fixed value $a^2=2$ corresponds to a spacetime without the polar SPOs. In such spacetimes there is no arc for latitudes less than some critical minimum $\theta_{min(arc)}$ close to the spin axis, $\theta_{o}<\theta_{min(arc)}$. For $\theta_{o}=\theta_{min(arc)}$ a degenerate arc appears which appears as a spot (see Fig. \ref{Fig_cones_VIb_rs_22}b), i.e. with the central angle $\xi=0$. For $\theta_{min(arc)}<\theta_{o}$ the arc increases and its central angle grows monotonically to its maximum value $\xi_{eq}$ in the equatorial plane $\theta_{o}=90\dgr$ (see Fig. \ref{Fig_th_xi}(b)). The dependence of the angles $\theta_{min(arc)}$, $\xi_{eq}$ on the spacetime parameters $a^2, y$ is shown in Fig. \ref{Fig_th_xi?}.

\begin{figure*}[h]
	\centering
	\begin{tabular}{ccc}
		\includegraphics[width=0.33\textwidth]{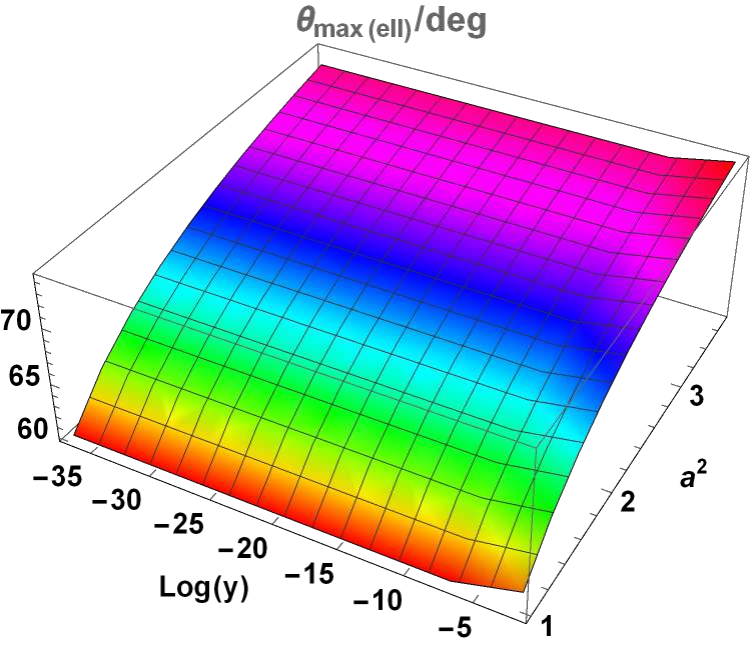}&\includegraphics[width=0.33\textwidth]{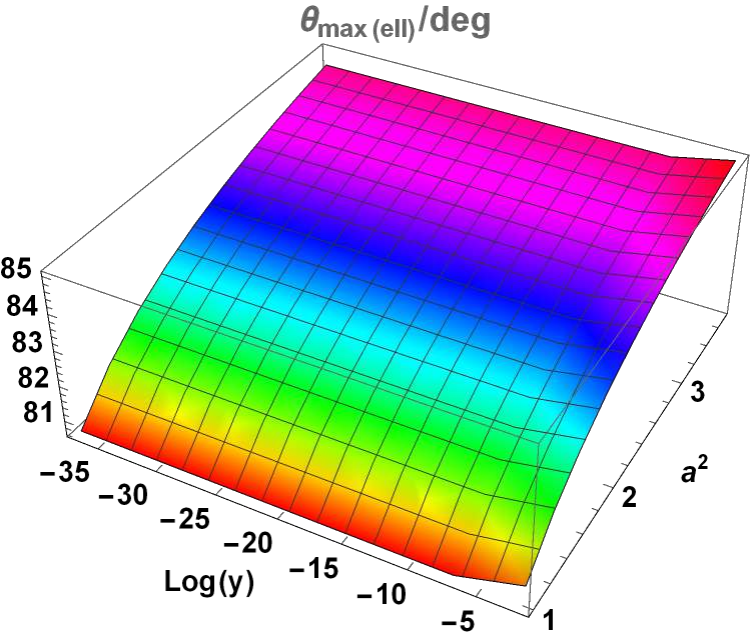}&\includegraphics[width=0.33\textwidth]{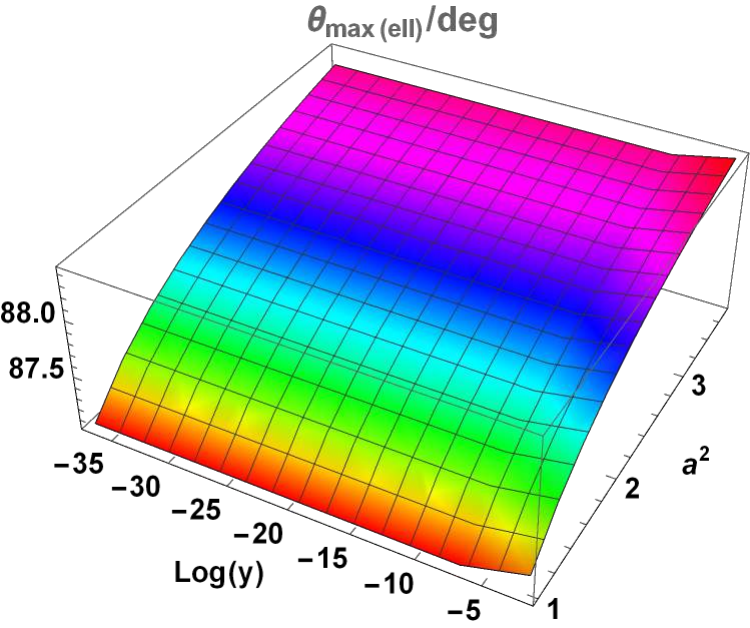}\\
		(a)&(b)&(c)
		
	\end{tabular}	
	\caption{Dependence of the angle $\theta_{max(ell)}$, at which the observer sees the elliptical shadow of the superspinar touching the light arc, on the spacetime parameters corresponding to Class IV. The superspinar boundary surface is set to $\calr=0.1$ (a), $\calr=0.01$ (b) and $\calr=0.001$ (c).
	}\label{Fig_th_max_ell}
\end{figure*}

In Fig. \ref{Fig_th_max_ell} we plot the dependence of the angle $\theta_{max(ell)}$ on the spacetime parameters, which represents the maximum angle at which the observer can still see the elliptical shadow of the superspinar, which appears to be the largest (see Fig. \ref{Fig_cones_IVa_4_90deg}(b)).

The maximum peak angle $\eta$ of the wedge-shaped shadow of the superspinar as seen by the observers in the equatorial plane is plotted as a function of the spacetime parameters in Fig. \ref{Fig_th_eta}.
  
The following Figures \ref{Fig_shadows_y} , \ref{Fig_shadows_t} show the comparison of the superspinar shadows versus the NS shadows for a fixed value of the parameter $a^2/y$ while varying the parameter $y/a^2$, respectively. We used the radial position on the static radius $r_{o}=r_{s}$, as its vicinity represents the best local approximation to the asymptotically flat region of the Kerr spacetime \cite{Stu-Hle:1999:PHYSR4:}, \cite{Stu-Sla:2004:PHYSR4:}, and two representative observer latitudes, i.e., $60\dgr$, representing the position outside the equatorial plane, and the second latitude corresponding to the equatorial plane.

\begin{figure*}[h]

	\begin{tabular}{ccc}
		\includegraphics[width=0.5\textwidth]{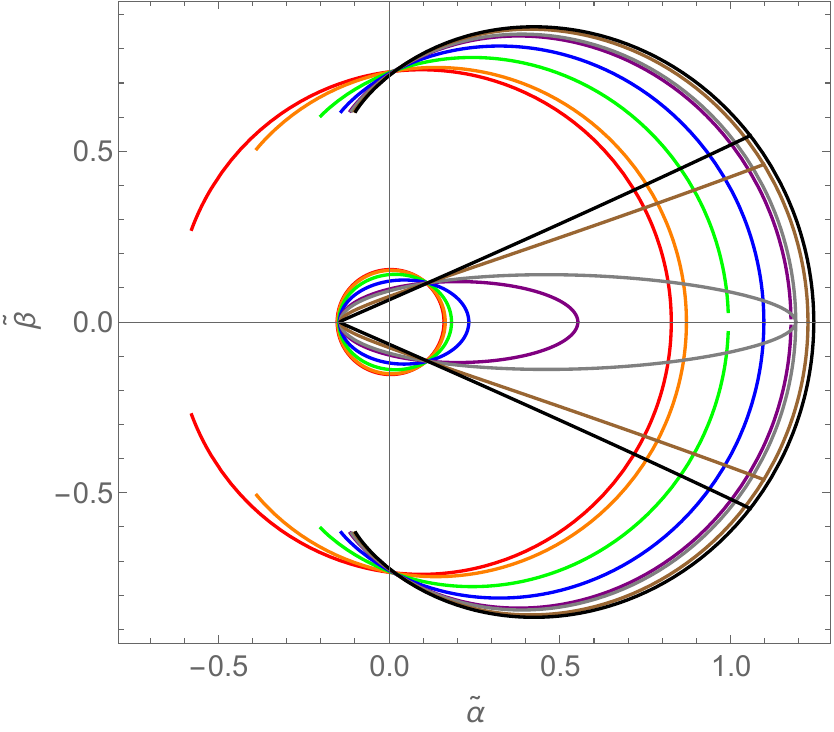}&\raisebox{0.0cm}[0pt]{\includegraphics[height=8cm]{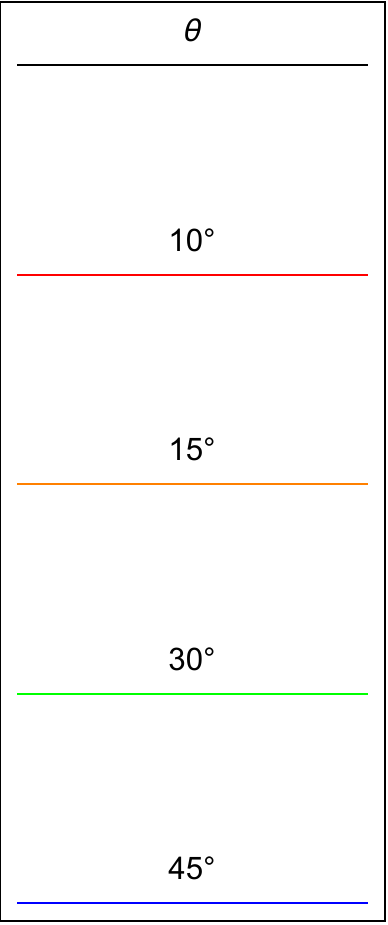}}&\includegraphics[height=8cm]{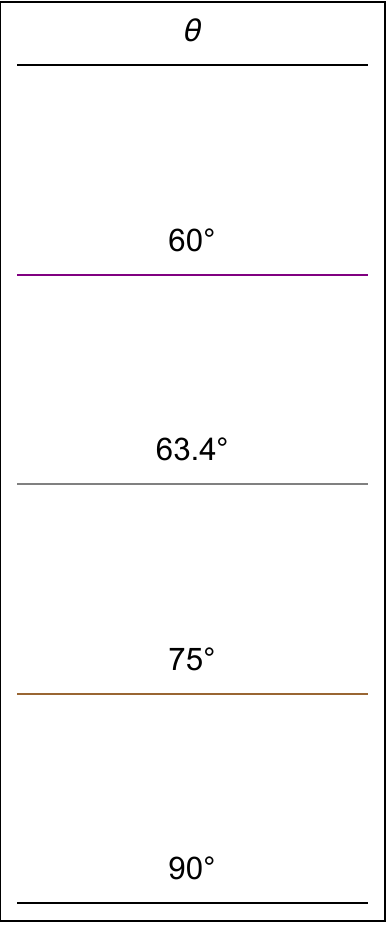}
	
	\end{tabular}	
	\caption{Superspinar  shadows for fixed cosmological parameters $y=0.01$, $a^2=1.2$, corresponding to spacetime IVa with polar SPOs and for different values of the observer's latitude $\theta$, as seen by an observer located at the static radius $r_{s}=0.01^{-1/3}=4.6$. The description of the shadows is the same as in the previous figure. The first value of the observer latitude used is close to the critical value $\theta_{max(circ)}=9.04\dgr$.         
	}\label{Fig_shadows_theta_IVa}
\end{figure*}

\begin{figure*}[h]
	
	\begin{tabular}{ccc}
		\includegraphics[width=0.4\textwidth]{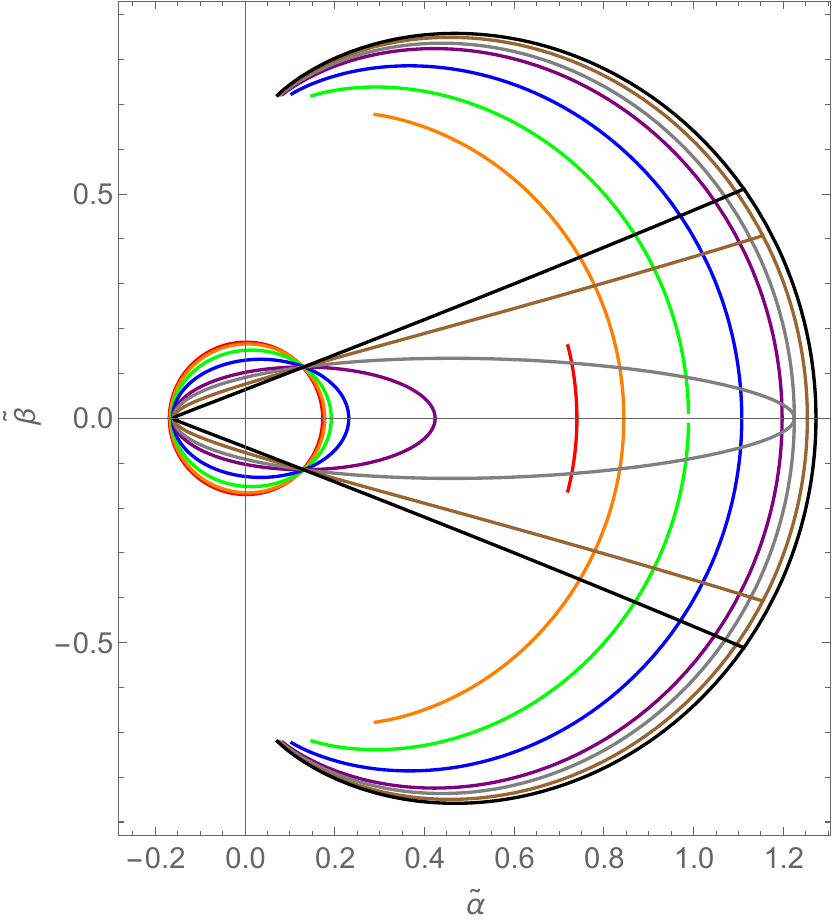}&\raisebox{0.0cm}[0pt]{\includegraphics[height=8cm]{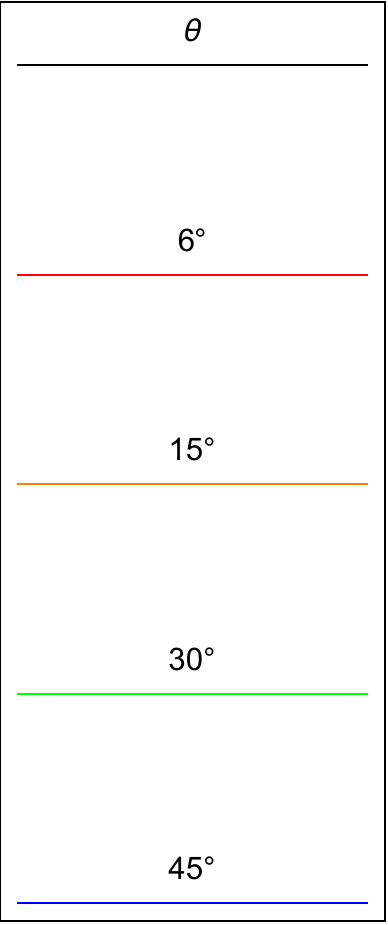}}&\includegraphics[height=8cm]{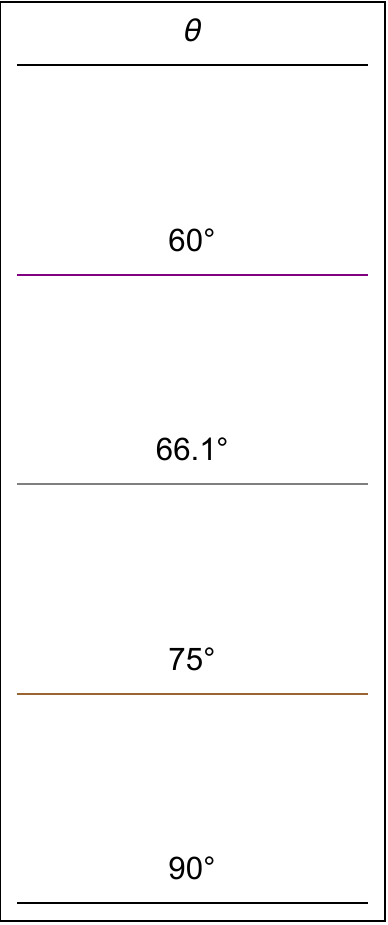}
	\end{tabular}	
	\caption{Superspinar shadows for fixed cosmological parameters $y=0.01$, $a^2=1.5$, corresponding to spacetime IVb without polar SPOs, and for different values of the observer latitude $\theta$ as seen by an observer located at the static radius $r_{s}=0.01^{-1/3}=4.6$. The first value of the observer latitude used is close to the critical value $\theta_{min(arc)}=5.85\dgr$.      
	}\label{Fig_shadows_theta_IVb}
\end{figure*}

In the last two Figures \ref{Fig_shadows_theta_IVa}, \ref{Fig_shadows_theta_IVb} we investigate the influence of the observer latitude on the observed images of the superspinar (NS) for both fixed spacetime parameters $a^2, y$ as seen on the static radius. The spacetime parameters in Fig. \ref{Fig_shadows_theta_IVa} correspond to Class IVa, while in Fig. \ref{Fig_shadows_theta_IVb} they correspond to Class IVb. It can be seen that while the dependence on the spin parameter is significant, the influence of the astrophysically acceptable values of the cosmological parameter is quite negligible.

\twocolumngrid

\begin{figure*}[h]
	\centering
	\begin{tabular}{ccc}
		\includegraphics[width=0.33\textwidth]{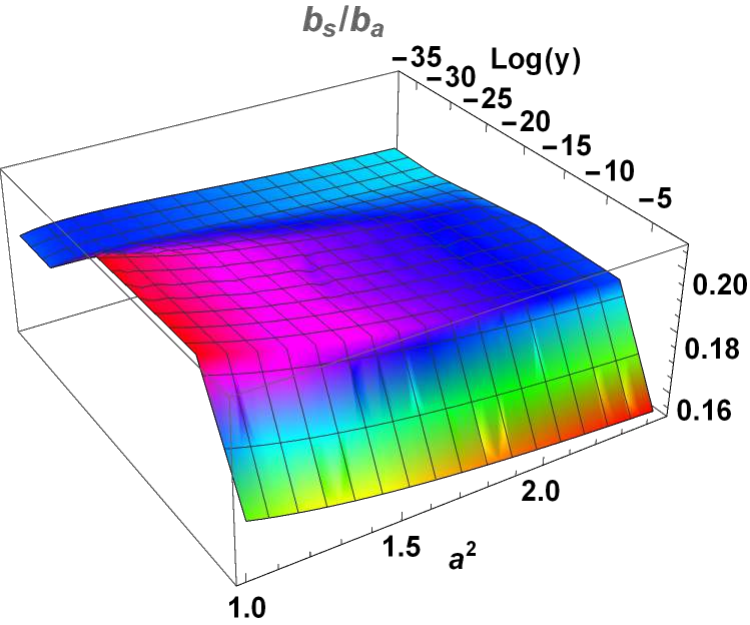}&\includegraphics[width=0.33\textwidth]{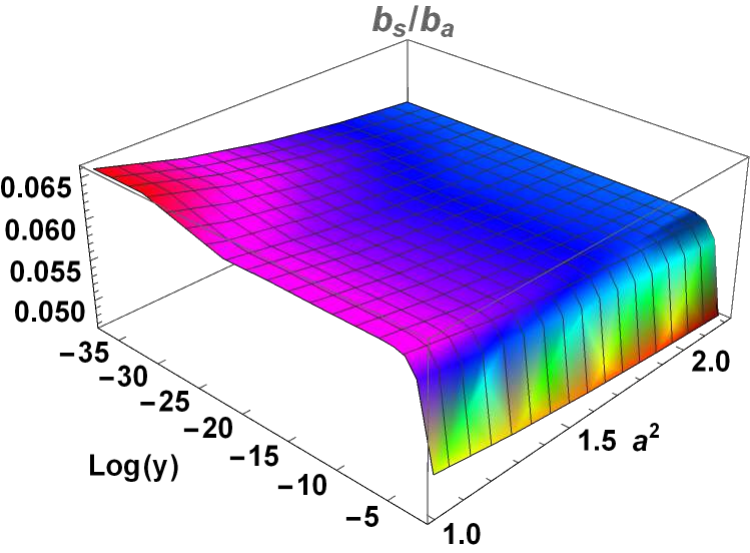}&\includegraphics[width=0.33\textwidth]{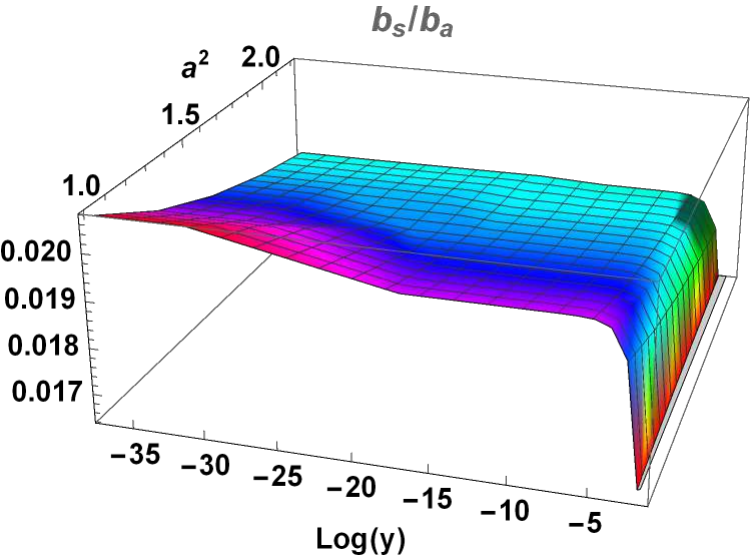}\\
		(a)&(b)&(c)
		
	\end{tabular}	
	\caption{Dependence of the ratio $b_{s}/b_{a}$ as defined in Fig. \ref{captions}(b). The superspinar boundary surface is set to $\calr=0.1$ (a), $\calr=0.01$ (b) and $\calr=0.001$ (c). It can be seen that the smaller the value of the radius $\calr$ of the superspinar surface, the narrower its elliptical shadow relative to the appropriate light arc.
	}\label{Fig_th_bs/ba}
\end{figure*}

\begin{figure*}[h]
	\centering
	\begin{tabular}{ccc}
		\includegraphics[width=0.33\textwidth]{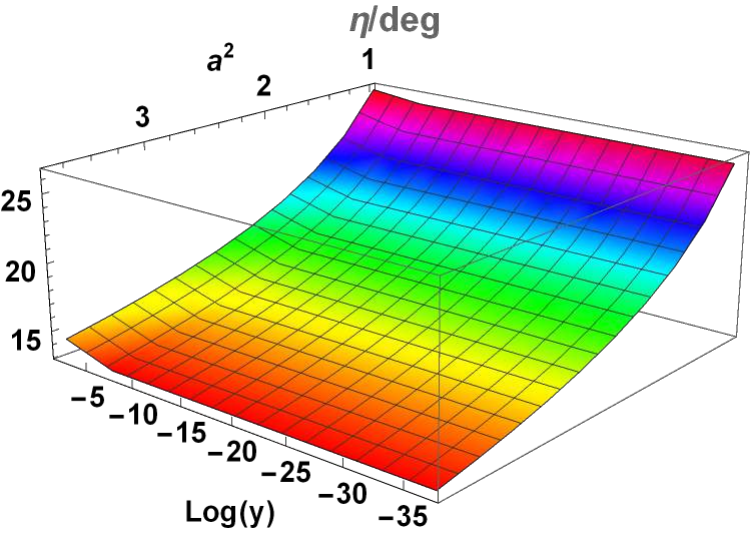}&\includegraphics[width=0.33\textwidth]{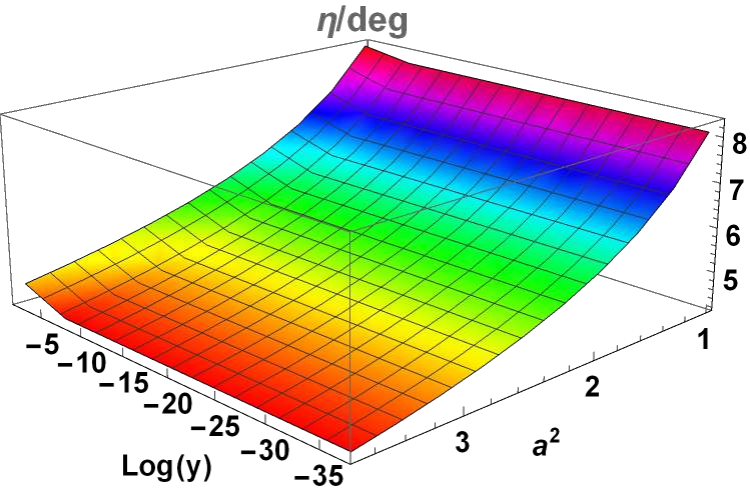}&\includegraphics[width=0.33\textwidth]{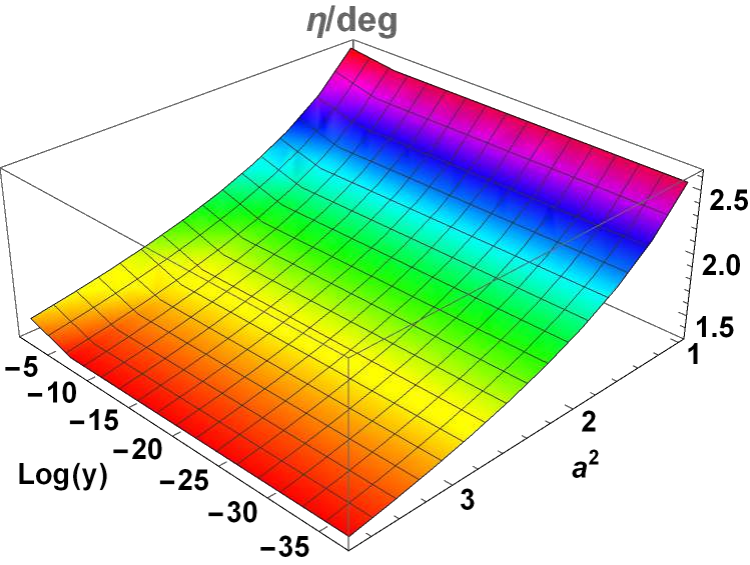}\\
		(a)&(b)&(c)
		
	\end{tabular}	
	\caption{Dependence of the angle $\eta$, which defines the maximum shadow of the superspinar observed in the equatorial plane, on the spacetime parameters corresponding to Class IV. The superspinar boundary surface is set to $\calr=0.1$ (a), $\calr=0.01$ (b) and $\calr=0.001$ (c).
	}\label{Fig_th_eta}
\end{figure*}

\begin{figure*}[h]
		
	\begin{tabular}{ccc}
		
		\includegraphics[width=0.32\textwidth]{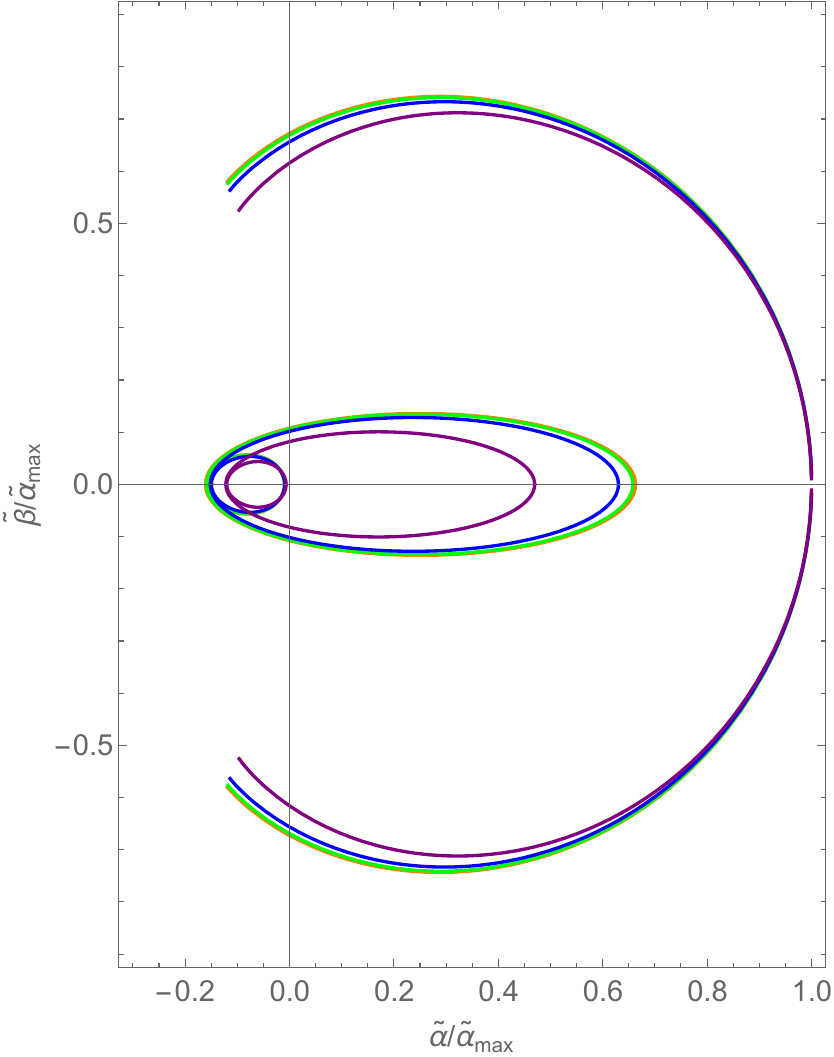}&\includegraphics[width=0.32\textwidth]{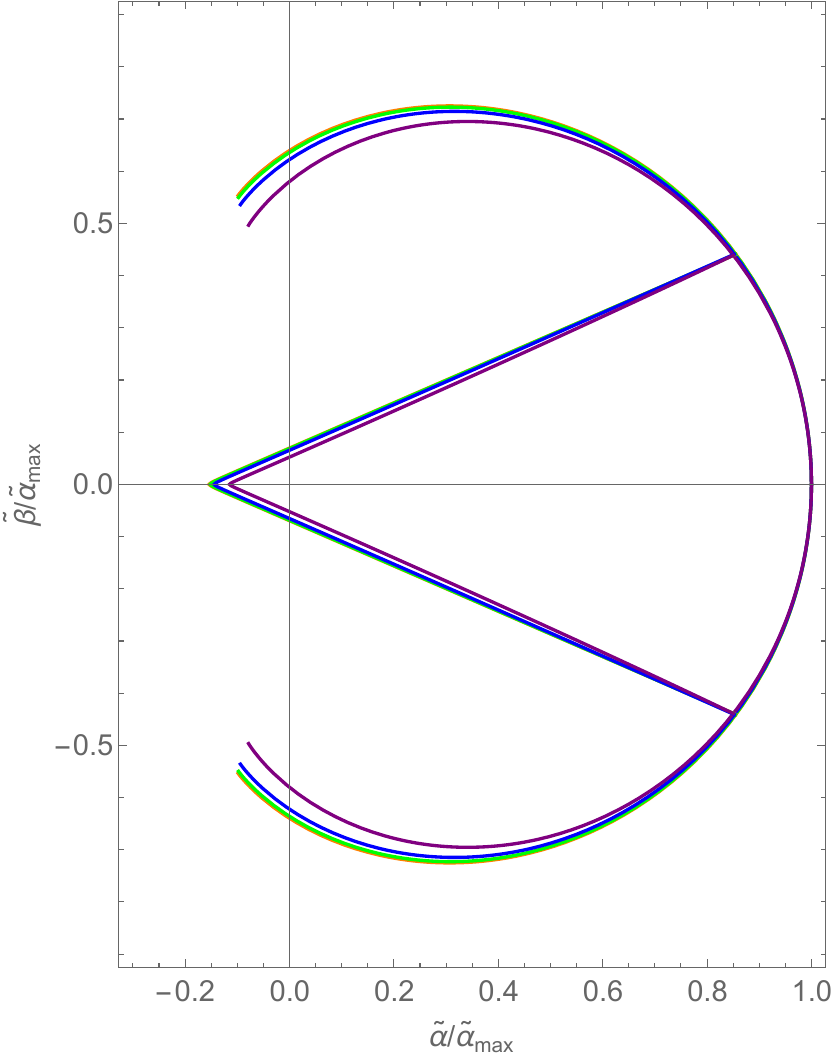}&\raisebox{0.6cm}[0pt]{\includegraphics[width=0.33\textwidth]{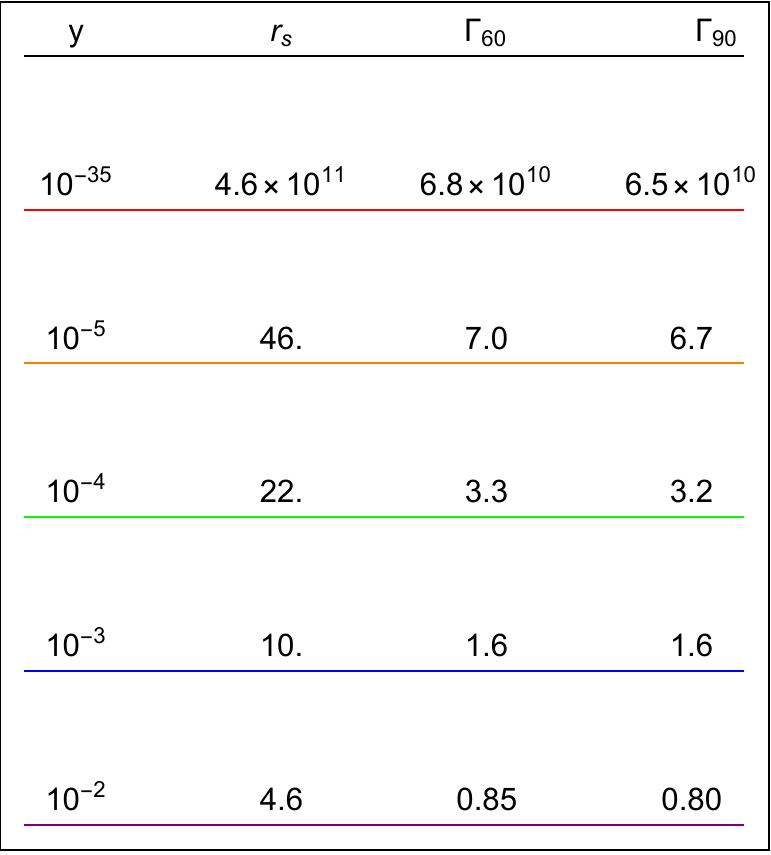}}
			
	\end{tabular}	
	\caption{Superspinar vs. NS shadows for fixed spin parameter $a^2=1.2$ and different values of the cosmological parameter $y$ as seen by an observer located at the static radius $r_{s}=y^{-1/3}$ with latitude $\theta_{o}=60 \dgr$ (left) and at the equatorial plane $\theta_{o}=90 \dgr$ (middle). To clearly show the influence of the cosmological parameter on the shape of the shadow, the images are resized by the multiplicative factor $\Gamma=1/\tilde{\alpha}_{max}$, so that the modified dimensionless coordinate $\tilde{\alpha}/\tilde{\alpha}_{max}$ of their left endpoint is normalized to $1.$ The shadows contain the arcs determined by photons with motion constants $X_{\spo}(r_{\spo}>0), q_{\spo}(r_{\spo}>0)$, which enclose the elliptical (left figure) or wedge-shaped (middle figure) shadows of the superspinar with surface at radius $\calr=0.1$. In the left figure, for comparison with the naked singularity case, we show the images of the dark spots, represented by smaller inner ellipses, through which the light can enter the region of negative radii, enclosed by photons with motion constants $X_{\spo}(r_{\spo}<0), q_{\spo}(r_{\spo}<0)$. In the case $\theta_{o}=90 \dgr$ these spots disappear. Note that the red and orange curves almost overlap, so that for realistic values of the cosmological parameter $y\lesssim 10^{-22}$, where the upper limit corresponds to the most massive objects observed so far, these shadows are indistinguishable from those in the case of pure Kerr spacetime. The right panel gives the corresponding values of $y, r_{s}$ and the factor $\Gamma$ by which each corresponding image should be resized to obtain the real angular coordinate $\tilde{\alpha}, \tilde{\beta}$ in radians.           
	}\label{Fig_shadows_y}
\end{figure*}

\begin{figure*}[h]

	\begin{tabular}{ccc}
		
		\includegraphics[width=0.35\textwidth]{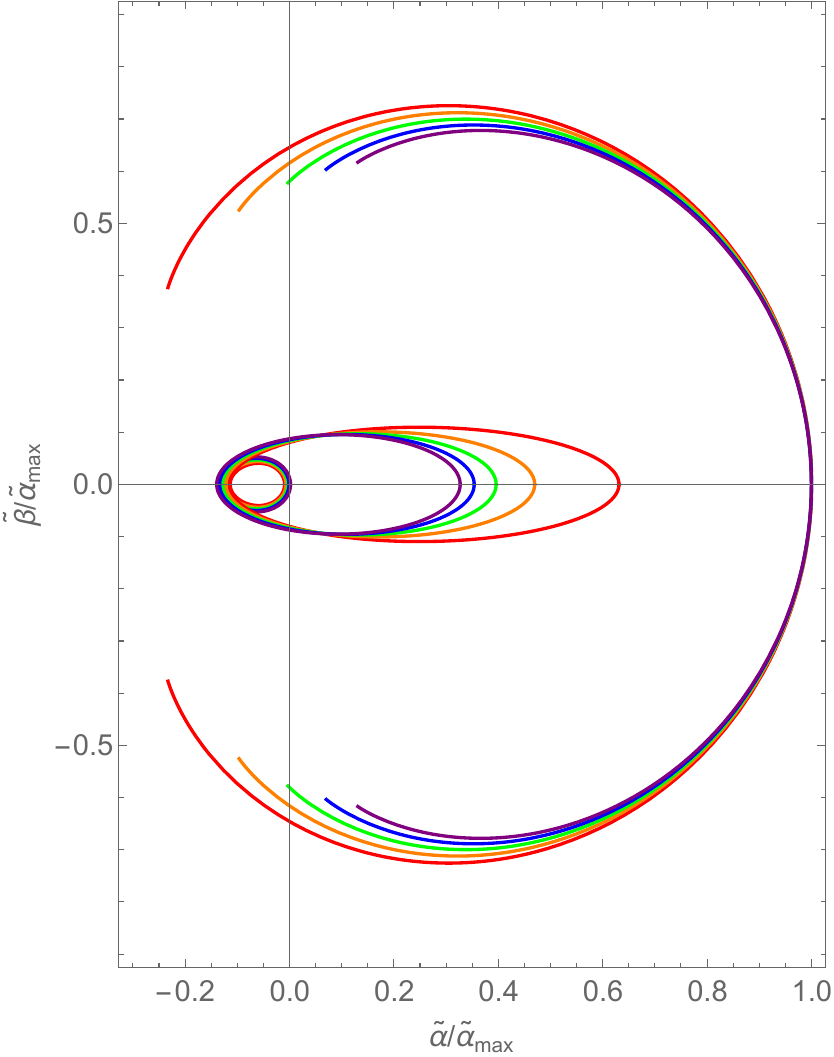}&\includegraphics[width=0.35\textwidth]{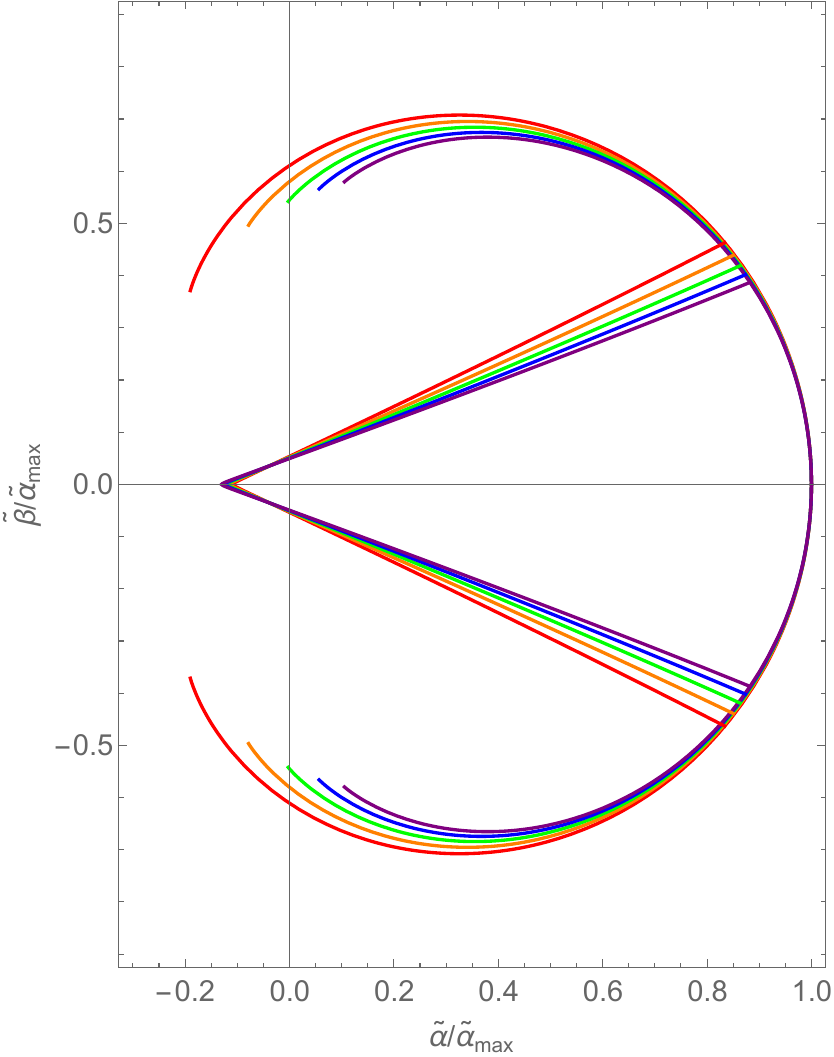}&\raisebox{0.0cm}[0pt]{\includegraphics[width=0.28\textwidth]{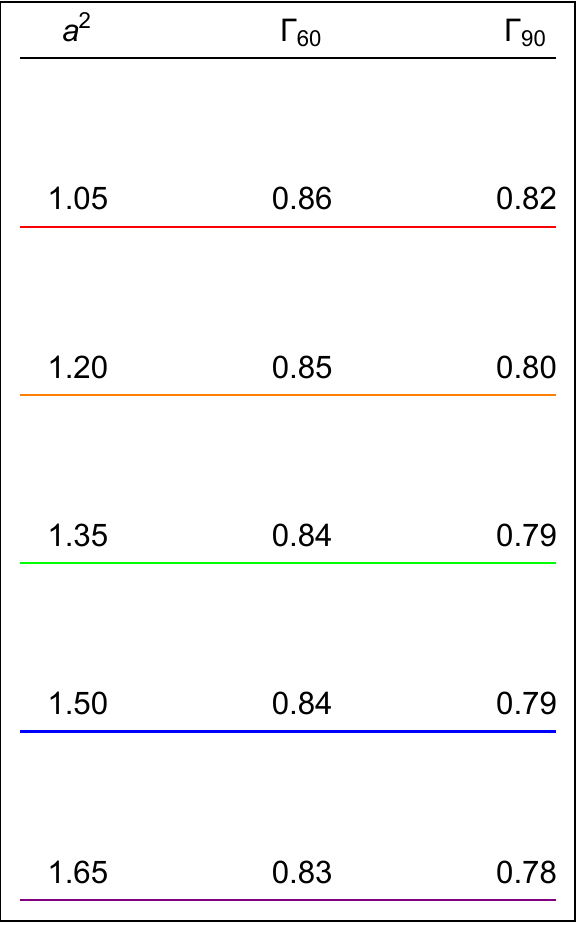}}
		
	\end{tabular}	
	\caption{Superspinar vs. NS shadows for fixed cosmological parameter $y=0.01$ and different values of the spin parameter $a^2$ as seen by an observer located at the static radius $r_{s}=y^{-1/3}=4.6$ with latitude $\theta_{o}=60 \dgr$ (left) and in the equatorial plane $\theta_{o}=90 \dgr$ (middle). The description of the shadows is the same as in the previous figure. In order to clearly display the influence of the spin parameter on the shape of the shadow, the images are resized by multiplicative factor $\Gamma=1/\tilde{\alpha}_{max}$ so that the modified dimensionless coordinate $\tilde{\alpha}/\tilde{\alpha}_{max}$ of their left endpoint is normalized to $1$. The right panel gives the corresponding values of $a^2$ and the factor $\Gamma$ by which each image should be resized to obtain the real angular coordinate $\tilde{\alpha}, \tilde{\beta}$ in radians. The first three values of the parameter $a^2$ correspond to spacetimes with polar SPOs, while the remaining two values correspond to spacetimes without polar SPOs. The middle value is close to the marginal value $a^2_{max(pol)}(y=0.01)\approx 1.355$, where $a^2_{max(pol)}(y)$ denotes the inverse of the function $y_{max(pol)}(a^2)$ defined in (\ref{ypolmax}).         
	}\label{Fig_shadows_t}
\end{figure*}

\clearpage

\section{Concluding remarks}\label{concl}

In the present paper we have performed an analysis allowing us to construct the light escape cones and related shadows in the \KdS\ superspinar (NS) spacetimes. The main results obtained can be summarized as follows. 
\begin{enumerate}
	\item For emitters located under the radii of the equatorial circular photon orbit at $r=r^{-}_{ph}$, there exist a wide variety of shapes of light 'cones' containing the photons emitted to bound orbits, which we call 'trapped cones'. In the 3D depiction they can take the shapes of spherical belts (see Figs. \ref{Fig_cones_IVa_1.25_0_2p4deg}(a), \ref{Fig_cones_IVa_1.8_6.4deg}(a)), spherical bi-angles (see Figs. \ref{Fig_cones_IVa_1.25_5_66deg}(a),  \ref{Fig_cones_IVa_1.8_6.4deg}(b),(c), \ref{Fig_cones_IVa_2.2_3.3deg}(c),\ref{Fig_cones_IVa_2.2_90deg},  \ref{Fig_cones_IVa_2.65_90deg}), or spherical canopy with an elliptical base (Fig. \ref{Fig_cones_V_2.5_60deg}), in dependence on both the radial and latitudinal position coordinates. The crucial role plays the relative radial position with respect to the outer unstable polar SPO at radii $r^{-}_{pol}$, if it exists, and the radius of the equatorial circular photon orbit $r^{-}_{ph}$. The trapped cones are relevant for the study of self-illumination and self-eclipse effects in the presence of radiating matter, e.g. in an accretion disk.
	\item In section \ref{ssec_class_compl_cones} we introduced a subdivision of the complementary cones to the LECs, i.e. the trapped cones, the cones of light escaping to the other infinity, the repelled cones and the engrossed cones, which for $r_{e}<r^{-}_{ph}$ are separated from the LECs by photons emitted towards the unstable SPOs. For emitters at $r_{e}>r^{-}_{ph}$ there are no trapped cones.
	\item In the ergoregion there are photon orbits with negative energy. In the case of spacetimes with DRB of the radial photon motion, there is an inner ergoregion, where the corresponding light cones are depicted in Figs. \ref{Fig_cones_IVa_0.9}, \ref{Fig_cones_IVa_1.25_0_2p4deg}, \ref{Fig_cones_IVa_1.25_5_66deg}, \ref{Fig_cones_IVa_1.25_90deg} in blue, and the outer ergoregion, where the corresponding light cones are shown in Fig. \ref{Fig_cones_IVa_5.75_90deg} in green.   
	\item The 3D representations of the LECs can also be regarded as the star globes showing the celestial sky and the superspinar (NS) shadows viewed from the point where the emitter (observer) resides. The light grey areas, originally representing the trapped cones, should then be understood as the dark sky on the observer's celestial sphere, since no photons from distant radii can come from these directions. Similarly, the red areas on the globe, originally representing the engrossed cones, have the meaning of the superspinar shadow.   
	\item For the astrophysically relevant case of a distant observer at $r_{o}>r^{-}_{ph}$, the superspinar (NS) shadow consists of the dark spot as described above, the silhouette of the ring singularity itself, which is one-dimensional and hence not observable, and an arc or circle on the observer's celestial sky, along which the images of distant stars are distorted, as the photons coming from such directions had to circle many times around the appropriate spherical orbits.
	 
	The appearance of the arc is strongly affected by the spacetime parameters and the observer's latitude $\theta_{o}$. In the spacetimes with the polar SPOs and for observers on and near the spin axis, the shadow constitutes from the spot and a circle, which separates for some critical angle $\theta_{max(circ)}$ into an arc (cf. Figs. \ref{Fig_cones_IVa_4_90deg}(a),(b)). Such separating angle exist even for trapped cones (cf. e.g. Figs.\ref{Fig_cones_IVa_1.25_5_66deg}(a),(b), \ref{Fig_cones_IVa_1.25_90deg}a).
	
	In the spacetimes without the polar SPOs there is neither a circle nor an arc seen from the spin axis or its vicinity, until a minimum angle $\theta_{min(arc)}$ is attained (cf. Figs. \ref{Fig_cones_VIb_rs_22}(b),(c)). For larger latitude $\theta_{o}>\theta_{min(arc)}$ an arc appears. 
 
	\item  We have introduced observable parameters of the shadows (see Fig. \ref{captions}), the measurement of which can in principle be used to identify the observed object and to determine its real parameters. We have studied the influence of the space-time parameters on these observables. We found some influence of the cosmological constant on the shape of the shadows, but the influence of the spin is clearly dominant.  
	 
	For instance, an arc indicates athe presence of the naked singularity, which central angle is highly sensitive on the spin. From Fig. \ref{Fig_th_xi} it follows that for $y<y_{max(pol)}(a^2)$, i.e. in the spacetimes with the polar SPOs, there is some minimum value of the central angle for any latitude of the observation, under which the central angle cannot decrease. At least, the detection of an arc with $\xi \to 360\dgr$ points to a small latitude (high inclination) of the observation and gives a strong upper limit on the spin. On the other hand, under the same conditions, the observation of an arc with $\xi \to 0\dgr$ indicates the spin $y>y_{max(pol)}(a^2)$. Moreover, the latitude of the observation can be inferred from the shape of the superspinar shadow (elliptical or wedge-shaped). Of course, a practical measurement of such an arc from the distortion of the celestial background is rather a task for advanced devices in the future.

\end{enumerate}
  
 \acknowledgements{The authors acknowledge support of the Research Centre for Theoretical Physics and Astrophysics, Institute of Physics, Silesian University in Opava. }         

\bibliography{bibliography}	
\bibliographystyle{abbrv}
\end{document}